\documentclass[preprint,11pt,trackchanges]{aastex61}

\usepackage{grffile}  %grffile allows .'s in filenames
\usepackage{url}                                                                                  
\graphicspath{{AFGL2591/}{G333/}{./G34.4+0.23/}{./GM24/}{./IRAS19410/}{./IRAS20126/}{./NGC6334+GM24/}{./NGC7538/}{./RCW120/}{./W31/}{./W33/}{./W42/}{./W75N/}{./Wd1/}}

\newcommand{\anchorfoot}[2] {\href{#1}{#2}\footnote{\url{#1}}}
\newcommand{\anchorparen}[2]{\href{#1}{#2} (\url{#1})}

\renewcommand{\S}{Section }
\newcommand{\tnm}[1]{\tablenotemark{#1}}
\renewcommand{\dataset}[2][]{{#1}}

\newcommand{\hii}{H{\scriptsize II} }

\newcommand{\UCHIIR}{UCH{\scriptsize II}R }

\newcommand{\Spitzer} {{\em Spitzer~}}
\newcommand{\WISE} {{\em WISE~}}
\newcommand{\Herschel} {{\em Herschel~}}
\newcommand{\XMM}      {{\em XMM~}}
\newcommand{\Chandra} {{\em Chandra~}}
\newcommand{\ACIS}    {{ACIS}}
\newcommand{\CIAO}    {{\em CIAO}}

\newcommand{\DSnine}  {{\em DS9}}
\newcommand{\MARX}    {{\em MARX}}
\newcommand{\AEacro}  {{\em AE}}

\accepted{2 Feb 2018, ApJS}
\shorttitle{MOXC2}
\shortauthors{Townsley et al.} 

\begin{document}

\title{THE MASSIVE STAR-FORMING REGIONS OMNIBUS X-RAY CATALOG, \\ SECOND INSTALLMENT }

%\correspondingauthor{Patrick Broos}
\email{townsley@astro.psu.edu, patrick.broos@icloud.com}

\author{Leisa K. Townsley}
\affil{Department of Astronomy \& Astrophysics, 525 Davey Laboratory, 
Pennsylvania State University, University Park, PA 16802, USA}

\author{Patrick S. Broos}
\affil{Department of Astronomy \& Astrophysics, 525 Davey Laboratory, 
Pennsylvania State University, University Park, PA 16802, USA}

\author{Gordon P. Garmire}
\affil{Huntingdon Institute for X-ray Astronomy, LLC, 10677 Franks Road, Huntingdon, PA 16652, USA}

\author{Gemma E. Anderson}
\affil{International Centre for Radio Astronomy Research, Curtin University, GPO Box U1987, Perth, WA 6845, Australia}

\author{Eric D. Feigelson}
\affil{Department of Astronomy \& Astrophysics, 525 Davey Laboratory, 
Pennsylvania State University, University Park, PA 16802, USA}

\author{Tim Naylor}
\affil{School of Physics, University of Exeter, Stocker Road, Exeter EX4 4QL, UK}

\author{Matthew S. Povich}
\affil{Department of Physics and Astronomy, California State Polytechnic University, 3801 West Temple Ave, Pomona, CA 91768, USA}

\begin{abstract}
We present the second installment of the Massive Star-forming Regions (MSFRs) Omnibus X-ray Catalog (MOXC2), a compilation of X-ray point sources detected in {\em Chandra}/ACIS observations of 16 Galactic MSFRs and surrounding fields.  MOXC2 includes 13 ACIS mosaics, three containing a pair of unrelated MSFRs at different distances, with a total catalog of 18,396 point sources.  The MSFRs sampled range over distances of 1.3~kpc to 6~kpc and populations varying from single massive protostars to the most massive Young Massive Cluster known in the Galaxy.  By carefully detecting and removing X-ray point sources down to the faintest statistically-significant limit, we facilitate the study of the remaining unresolved X-ray emission.  Through comparison with mid-infrared images that trace photon-dominated regions and ionization fronts, we see that the unresolved X-ray emission is due primarily to hot plasmas threading these MSFRs, the result of feedback from the winds and supernovae of massive stars.  The 16 MSFRs studied in MOXC2 more than double the MOXC1 sample, broadening the parameter space of ACIS MSFR explorations and expanding {\em Chandra}'s substantial contribution to contemporary star formation science.
\end{abstract}

% Six keywords are allowed: 
% The list of available keywords is at:
%   http://journals.aas.org/authors/keywords2013.html

% Enclose first mention of each named astronomical object in an \object{} macro; see:
%   http://journals.aas.org/authors/aastex.html
% NOTE -- as of 7 Dec 2017, Greg Schwarz says nobody is using this \object{} information.

% =============================================================================
\section{INTRODUCTION \label{sec:intro}}

To explore basic questions of Galactic star formation and structure, the first step is to identify the stars that make up Massive Star-Forming Regions (MSFRs), where most massive stars form \citep{Motte17}.  This simple goal is surprisingly hard to achieve; MSFRs are typically far away, behind large absorbing columns in the Galactic Plane.  Infrared (IR) surveys are sensitivity-limited by bright extended emission and confusion-limited at longer wavelengths; only IR-excess stars found outside the brightest cluster cores can be reliably catalogued.  This problem is not remedied as massive clusters age; by the time the natal dust and gas in a giant \hii region have been dispersed by massive star feedback, their clusters have expanded in size and their pre-main sequence (pre-MS) stars have dissipated their disks (and their IR excesses), so they are lost in the overwhelming contamination of unrelated field stars.

Ushering in the 21st century, the {\em Chandra X-ray Observatory} brought high spatial resolution to X-ray astronomy, opening a new window into the workings of MSFRs.  \Chandra observations, combined with IR surveys, provide the best stellar census available for MSFRs.  This is because \Chandra has a large field of view that often captures an entire MSFR in a single pointing, it has sub-arcsecond spatial resolution on-axis, the instrumental background is low, and pre-MS stars are strong X-ray emitters independent of their disk status.  Even the faintest X-ray sources that are matched to IR sources usually indicate MSFR membership.  Additionally, \Chandra observations do not suffer the bright, highly-variable backgrounds that plague IR studies.  

Because of its fine spatial resolution and hard X-ray spectral response, \Chandra is good at finding massive stars and intermediate-mass pre-MS stars (IMPS) as well as lower-mass pre-MS stars.  Once all of these X-ray point sources are identified and removed from \Chandra images, it is possible to use the remaining unresolved \Chandra emission to map the hot, shocked interstellar medium (ISM) created by massive star winds and supernova feedback.  This diffuse X-ray emission fills \hii region bubbles and permeates superbubbles; these structures trace the framework of star formation in the Galaxy and in turn define starburst clusters in other galaxies.  \Chandra has established definitively that star formation proceeds in the presence of 1--10 million degree plasmas \citep{Townsley03} and that even the youngest massive stars, still ionizing hyper- and ultra-compact \hii regions (UCH{\scriptsize II}Rs), blast their cold, molecular birth environments with hard X-rays \citep{Anderson11}.  Thus no picture of a MSFR's evolution and environmental impact is complete without the \Chandra perspective.

\Chandra has amassed a large body of MSFR observations, giving new insights into the stellar populations and energetics of individual regions and enabling ambitious multi-target comparison projects.  Examples focusing on individual MSFR complexes include {\em The Astrophysical Journal Supplement Series} special issues on the \Chandra Orion Ultradeep Project (COUP) \citep[Volume 160 Number 2, e.g.,][]{Getman05} observations of the Orion Nebula Cluster and the \Chandra Carina Complex Project (CCCP) \citep[Volume 194 Number 1, e.g.,][]{Townsley11a} observations of the Great Nebula in Carina.  We look forward to the \Chandra Cygnus OB2 Legacy Survey \citep{Wright14}.  Multi-target comparisons include the Massive Young Star-Forming Complex Study in Infrared and X-ray \citep[MYStIX, e.g.,][]{Feigelson13} and the Star Formation in Nearby Clouds project \citep[SFiNCs, e.g.,][]{Getman17}.  

The immediate precursor to the current work was the first installment of the MSFRs Omnibus X-ray Catalog \citep[MOXC1,][]{Townsley14} which presented \Chandra point source lists and images of diffuse X-ray emission for 12 MSFRs (seven for MYStIX and five more distant regions).  This paper (hereafter MOXC2) closely follows the structure of MOXC1 and does not repeat the methodologies and analysis details explained there, thus we refer readers to MOXC1 for more background on the MOXC2 effort.  All of the \Chandra data used for our MSFR studies comes from the Advanced CCD Imaging Spectrometer (ACIS) instrument \citep{Garmire03}.

With MOXC2, we expand our analysis of {\em Chandra}/ACIS data to include off-axis CCDs and archival data on fields adjacent to our target MSFRs, which were observed for other reasons.  These ancillary pointings often go well beyond most cluster radii and provide the Galactic context for our MSFRs, informing our efforts to understand the ecology of large-scale star formation and feedback.  In particular, these wide-field ACIS mosaics create new opportunities to understand distributed populations of young stars, multi-epoch star formation, and the extent of hot plasmas surrounding massive molecular filaments and threading giant molecular clouds (GMCs).  The influence of evolving MSFRs extends over 10--100~pc scales and includes supernova remnants (SNRs), pulsars and pulsar wind nebulae (PWNe), gamma-ray sources, X-ray binaries, and diffuse plasmas from massive star feedback (both winds and supernovae).  We are using these new wide-field X-ray analysis capabilities to improve our understanding of the energetics, populations, and environments of GMCs, their networks of massive filaments, and their multiple MSFRs, and how they drive Galactic evolution.

The MSFR targets featured in this paper are shown in Table~\ref{targets.tbl}.  As in MOXC1, we include a rough limiting luminosity $L_{tc}$ and the corresponding limiting mass $M_{50\%}$ where the brighter half of the X-ray population is sampled, based on COUP results \citep{Preibisch2005}.  This $L_{tc}$ comes from a \anchorfoot{http://asc.harvard.edu/toolkit/pimms.jsp}{{\em PIMMS}} calculation for a 5-count detection on-axis, for a source with an {\em apec} thermal plasma with $kT$=2.7~keV and abundance 0.4*Z$_{\odot}$.  These values are typical for a pre-MS star \citep{Preibisch2005}.

We have assembled and analyzed 13 wide-field ACIS mosaics for MOXC2, amassing a catalog of 18,396 X-ray point sources and adaptively-smoothed images of the remaining unresolved X-ray emission in each mosaic.  In three of these regions, MOXC2 exploits chance superpositions of unrelated MSFRs that just happen to be superposed on the sky; these are IRAS~19410+2336 and NGC~6823, W42 and RSGC1, and W33 and Cl~1813-178. 

The number of X-ray point sources found for each target (Table~\ref{targets.tbl} Column 10) varies substantially between targets.  This is due to a combination of field of view and sensitivity; short observations and/or single pointings naturally result in fewer source detections than deeper observations and/or wide-field mosaics.  Additionally, the MOXC2 sample of MSFRs includes a wide variety of contexts in which we find massive stars, from sparse clumps containing just a handful of young stars to Young Massive Clusters (YMCs) containing tens of thousands of members.

\noindent
\setlength{\tabcolsep}{0.05in}
\renewcommand{\arraystretch}{0.9}  % to change row spacing in table

\begin{deluxetable}{lrcDDDrccrl}
\tablecaption{MOXC2 Targets \label{targets.tbl}}
\tabletypesize{\tiny}

\tablehead{
\colhead{MOXC2 \rule{0mm}{3ex}} & 
\colhead{Galactic} &  
\colhead{Celestial J2000} & 
\twocolhead{Distance} & 
\twocolhead{Scale} & 
\twocolhead{$\langle A_V \rangle$} & 
\colhead{Nominal} & 
\colhead{$\log L_{tc}$} & 
\colhead{$M_{50\%}$}   & 
\colhead{X-ray Srcs} &
\colhead{Distance Reference} \\
\colhead{Target}  & 
\colhead{($l,b$)}  &                                                                         
\colhead{($RA,Dec$)} & 
\twocolhead{(kpc)}    & 
\twocolhead{(arcmin/pc)} & 
\twocolhead{(mag)} & 
\colhead{Exp (ks)} &
\colhead{(erg/s)} & 
\colhead{($M_{\sun}$)} & 
\colhead{(\#)} &
\colhead{} 
}
\decimalcolnumbers
\startdata
%         Target              Galactic coords           Celestial coords        Distance     Scale    <A_V>      Exp    lim L_tc    M_50%    X-ray srcs     Dist ref
 \object{NGC 6334         }&   350.99  $+$0.62    &    17 19 47.1  $-$36 08 35    & 1.3      &  2.64  &    4     &  40   & 29.64   &   0.3   &   4674    &   \citet{Wu14,Chibueze14} \\ %Wu14 get 1.35kpc; Chibueze14 get 1.26kpc.
 \object{W75N             }&    81.89  $+$0.79    &    20 38 36.5  $+$42 38 46    & 1.30     &  2.64  &   50     &  30   & 30.53   &   1.2   &    356    &   \citet{Rygl12}          \\ %avg A_V from Handbook
 \object{RCW 120          }&   348.22  $+$0.46    &    17 12 20.8  $-$38 29 31    & 1.34     &  2.57  &   40     &  79   & 30.08   &   0.6   &   1692    &   \citet{Zavagno07}       \\ %avg A_V from Zavago07, but highly variable.
 \object{IRAS 20126$+$4104}&    78.13  $+$3.62    &    20 14 29.1  $+$41 13 33    & 1.64     &  2.10  &   70     &  39   & 30.79   &   1.8   &    493    &   \citet{Moscadelli11}    \\ %avg A_V from Anderson11; Nagayama15 get D=1.33kpc from VERA maser parallax.
 \object{W31N             }&    10.32  $-$0.15    &    18 08 59.1  $-$20 05 08    & 1.75     &  1.96  &   18     &  39   & 30.36   &   1.0   &    974    &   \citet{Deharveng15}     \\ %avg A_V from Deharveng15
 \object{IRAS 19410$+$2336}&    59.78  $+$0.06    &    19 43 11.4  $+$23 44 06    & 2.16     &  1.59  &   20     &   9   & 31.14   &  bright &    411    &   \citet{Xu09}            \\ %avg A_V from Beuther02 X$-$ray fits.  Net exp on S3 (where cluster is found) is 8.7ks after flare filtering.
 \object{W42              }&    25.36  $-$0.19    &    18 38 14.6  $-$06 49 19    & 2.2      &  1.56  &   11     &  54   & 30.27   &   0.8   &    840    &   \citet{Blum00}          \\ %Dewangan15 papers use D=3.8kpc.
 \object{NGC 6823         }&    59.40  $-$0.15    &    19 43 11.0  $+$23 17 45    & 2.3      &  1.49  &    3     &   8   & 30.63   &   1.3   &see IRAS 19410&   \citet{Massey95}     \\ %In IRAS 19410 ACIS-S mosaic; A_V from Riaz12
 \object{W33              }&    12.81  $-$0.20    &    18 14 13.5  $-$17 55 42    & 2.4      &  1.43  &   20     &  38   & 30.68   &   1.5   &   1960    &   \citet{Immer13}         \\ 
 \object{NGC 7538         }&   111.54  $+$0.81    &    23 13 45.5  $+$61 28 17    & 2.65     &  1.32  &   11     &  29   & 30.66   &   1.4   &    487    &   \citet{Moscadelli09}    \\ % 
 \object{G333             }&   333.03  $-$0.44    &    16 20 39.5  $-$50 39 59    & 2.6      &  1.32  &   15     &  60   & 30.44   &   1.1   &   3935    &   \citet{Figueredo05}     \\ %Garcia14, Nguyen15 say 3.6 kpc; many others use this distance, which goes back to Lockman79.
 \object{AFGL 2591        }&    78.89  $+$0.71    &    20 29 24.9  $+$40 11 21    & 3.33     &  1.03  &  100     &  30   & 31.66   &  bright &    288    &   \citet{Rygl12}          \\
 \object{G34.4$+$0.23     }&    34.40  $+$0.23    &    18 53 18.5  $+$01 24 48    & 3.7      &  0.93  &  200     &  63   & 31.79   &  bright &    565    &   \citet{Rathborne05}     \\ %Distance from Kurayama11 (1.56 kpc) is not used in recent papers -- see Foster14.
 \object{Cl 1813$-$178    }&    12.74  $-$0.01    &    18 13 24.3  $-$17 53 31    & 3.8      &  0.90  &    9     &  59   & 30.65   &   1.4   &   see W33 &   \citet{Messineo11}      \\ %In W33 ACIS mosaic.
 \object{Wd1              }&   339.55  $-$0.40    &    16 47 04.0  $-$45 51 05    & 4.0      &  0.86  &   11     &  55   & 30.69   &   1.5   &   1721    &   \citet{Gennaro11}        \\%A_V from Damineli17.
 \object{RSGC1            }&    25.27  $-$0.16    &    18 37 58.0  $-$06 53 00    & 6        &  0.57  &   23     &  74   & 31.24   &  bright &   see W42 &   \citet{Froebrich13}     \\ %In W42 ACIS mosaic.
\enddata

\tablecomments{
Col.\ (5):  Image scale assuming the distance given in Col.\ (4).
\\Col.\ (6):  Approximate average absorption to the target, estimated from a variety of literature sources.  Most MSFRs have highly variable and spatially complex obscuration, so this value 
should be used only as a rough indicator.
\\Col.\ (7):  A typical exposure time for the main MSFRs.  Most mosaics have a wide range of exposures; detailed exposure maps are shown in Section~\ref{sec:targets}.
\\Col.\ (8):  A rough limiting luminosity where the brighter half of the X-ray population is sampled.  Subscripts mean ``{\bf t}otal'' band 0.5--8~keV, {\bf c}orrected for extinction.
\\Col.\ (9): The corresponding limiting mass, where the brighter half of the X-ray population is captured.  For shallow observations, this limit is higher than pre-MS masses, so only ``bright'' sources (some massive stars and IMPS) are expected.
\\Col.\ (10):  Total number of X-ray sources found across the entire mosaic.
}
\end{deluxetable} 

%\end{document}

% This is now 16 targets, with 3 ACIS mosaics including 2 unrelated MSFRs each (IRAS 19410 + NGC 6823; W42 + RSGC1; W33 + Cl 1813-178).
% Total of 18,410 sources.

% Scale (arcmin/pc) = 360 degrees * 60'/deg / 2*pi*D(pc) = 3437.75'/D(pc)
                                  
% Code to convert celestial to Galactic coordinates.
% 
% line = strarr(21)
% read, line
% 
% forprint, strmid(line,0,12), strmid(line,12)
% 
% ra  = tenv(strmid(line,0,12))
% dec = tenv(strmid(line,12))
% ind = where((ra EQ 0) and (dec EQ 0))
% ra [ind] = !VALUES.F_NAN
% dec[ind] = !VALUES.F_NAN
% 
% glactc, ra, dec, 2000, gl, gb, 1
% 
% forprint, gl, gb, line, ra, dec, F='(%"(%6.2f,%6.2f) & %22s \% %11.6f %11.6f ")', /NoComm, TEXTOUT='temp.txt'
% 
%                                                    
% save, FILE='target_table_2013.sav', /VERB
% 

%\clearpage
%=============================================================================
\section{{\em CHANDRA} OBSERVATIONS AND DATA ANALYSIS \label{sec:data}}

%-----------------------------------------------------------------------------
\subsection{Observations}
The {\em Chandra}/ACIS observations used for MOXC2 are summarized in Table~\ref{tbl:obslog} and are identified by a unique Observation Identification (ObsID) number.  All of these datasets are available in the \Chandra archive.  Observations are ordered by date for each target; target names as used in this paper are given in Column (1) in bold, followed by the original name assigned by the study's principal investigator (PI), noted in Column (11).  Those observations where Gordon Garmire is listed as PI came from the ACIS Instrument Team Guaranteed Time Observation (GTO) program; those listing Stephen Murray as PI came from the HRC Instrument Team GTO program.  We note this fact because the \Chandra GTO programs provided many of the original seed observations for the MSFRs studied here and for many other Galactic Plane observations that we use in our wide-field mosaics.

Virtually all of the \ACIS\ data we have previously published were taken with the Observatory in the \ACIS\ imaging array (ACIS-I) configuration \citep{Garmire03}, with the optical axis near the center of a 2$\times$2 array of 1024$\times$1024-pixel CCDs covering roughly $17\arcmin \times 17\arcmin$ on the sky (with 0.492$\arcsec$ pixels).  
Although such observations commonly include data from two CCDs lying far off-axis in the \ACIS\ spectroscopy array (ACIS-S), we previously chose to discard data from those CCDs due to the poor angular resolution of the \Chandra mirrors at large off-axis angles\footnote {See Figure~4.12 in the \anchorparen{http://asc.harvard.edu/proposer/POG/}{\Chandra Proposers' Observatory Guide}.}.

In order to widen the field coverage around MSFRs and to include more archival data, MOXC2 includes the off-axis CCDs in ACIS-I observations and imaging observations taken in the ACIS-S configuration \citep{Garmire03}, as shown in the MOXC2 exposure map mosaics in Section~\ref{sec:targets}.  Four of our MSFRs were observed with the ACIS-S imaging configuration.  One side effect of this more liberal data selection policy is that most observations presented here include both front-side illuminated (FI) and back-side illuminated (BI) CCD detectors \citep{Garmire03}.
Because FI and BI detectors exhibit very different background spectra, their analysis must be done separately, thus we split each \Chandra ObsID into separate FI and BI ``virtual'' observations within our data analysis workflow.  

%There are NO grating observations in MOXC2, confirmed with
%foreach file ( */data/extract/obs*/validation.evt )
%dmlist $file |grep GRATING
%end
%\clearpage
%  Mon Apr 17 08:28:43 2017
%  acis_extract, version 5109  2017-03-12; ae_flatten_collation, version 5096  2017-02-24; hmsfr_tables, version 5096  2017-02-24

\renewcommand{\arraystretch}{0.9}  % to change row spacing in table
\setlength{\tabcolsep}{0.03in}

%\begin{longrotatetable}
\startlongtable
\begin{deluxetable}{crrccrccrcllh}
\centering 
\tabletypesize{\tiny} %\tablewidth{6in}
%\tablecolumns{8}

\tablecaption{ Log of {\em Chandra} Observations 
 \label{tbl:obslog}}

\tablehead{
\colhead{Target} & 
\colhead{ObsID} & 
\colhead{Start Time} & 
\colhead{Exposure\tablenotemark{a}} & 
\multicolumn{3}{c}{Aimpoint\tablenotemark{b}} & 
\colhead{Roll} & 
\colhead{Mode\tablenotemark{c}} &  
\colhead{Detectors\tablenotemark{d}} & 
\colhead{PI}  &
\colhead{TGAIN\tablenotemark{e},\tablenotemark{f}} &
%\colhead{CONTAM\tablenotemark{f}}  
\\
\cline{5-7}
\colhead{} & 
\colhead{} & 
\colhead{(UT)} & 
\colhead{(s)} & 
\colhead{CCD} & 
\colhead{$\alpha_{\rm J2000}$} & 
\colhead{$\delta_{\rm J2000}$} & 
\colhead{(\arcdeg)} & 
\colhead{} & 
\colhead{} &
\colhead{} &
\colhead{} 
}
\colnumbers
\startdata
  \multicolumn{1}{l}{\bf NGC 6334} \\ % 
        NGC 6334 REGION 1 & \dataset[  2574]{  ADS/Sa.CXO\#obs/02574} & 2002-08-31T12:49 &   39648 & I3 & 17:20:54.00 & -35:47:03.9 &  269 &            TE-F &  012367 & Yuichiro Ezoe  & \path{acisD2002-08-01t_gainN0006}  \\ %                                                                                    
        NGC 6334 REGION 2 & \dataset[  2573]{  ADS/Sa.CXO\#obs/02573} & 2002-09-02T00:12 &   25473 & I3 & 17:20:01.00 & -35:56:07.0 &  268 &            TE-F &  012367 & Yuichiro Ezoe  & \path{acisD2002-08-01t_gainN0006}  \\ % 
               G351.2+0.1 & \dataset[  3844]{  ADS/Sa.CXO\#obs/03844} & 2003-10-04T19:21 &   14773 & S3 & 17:22:24.70 & -36:10:59.9 &  261 &           TE-VF &  235678 & David Helfand  & \path{acisD2003-08-01t_gainN0006}  \\ % 
               G351.2+0.1 & \dataset[  4591]{  ADS/Sa.CXO\#obs/04591} & 2004-07-31T00:30 &   38012 & S3 & 17:22:28.00 & -36:10:59.9 &  278 &           TE-VF &  235678 & Stephen Murray & \path{acisD2004-05-01t_gainN0006}  \\ % 
          IGR J17204-3554 & \dataset[  8975]{  ADS/Sa.CXO\#obs/08975} & 2009-01-26T23:19 &    1015 & I3 & 17:20:25.00 & -35:53:31.2 &   98 &            TE-F &   01237 & Mariano Mendez & \path{acisD2008-11-01t_gainN0006}  \\ % 
                 NGC 6334 & \dataset[ 13436]{  ADS/Sa.CXO\#obs/13436} & 2011-07-13T09:37 &   62810 & I3 & 17:19:47.10 & -36:08:35.0 &  288 &           TE-VF &  012367 & Scott Wolk     & \path{acisD2011-05-01t_gainN0006}  \\ % 
                 NGC 6334 & \dataset[ 12382]{  ADS/Sa.CXO\#obs/12382} & 2011-07-17T19:19 &   32045 & I3 & 17:19:47.10 & -36:08:35.0 &  288 &           TE-VF &  012367 & Scott Wolk     & \path{acisD2011-05-01t_gainN0006}  \\ %
                    GM 24 & \dataset[ 18876]{  ADS/Sa.CXO\#obs/18876} & 2016-06-27T12:52 &   16744 & I3 & 17:17:06.90 & -36:21:27.2 &  302 &           TE-VF &  012367 & Leisa Townsley & \path{acisD2016-02-01t_gain_biN0002}  \\ % 
                    GM 24 & \dataset[ 18082]{  ADS/Sa.CXO\#obs/18082} & 2016-06-30T16:45 &   22658 & I3 & 17:17:06.90 & -36:21:27.2 &  302 &           TE-VF &  012367 & Leisa Townsley & \path{acisD2016-02-01t_gain_biN0002}  \\ % 
           G350.776+0.831 & \dataset[ 18081]{  ADS/Sa.CXO\#obs/18081} & 2016-07-23T18:50 &   37566 & I3 & 17:18:19.30 & -36:12:09.6 &  281 &           TE-VF &   01237 & Leisa Townsley & \path{acisD2016-02-01t_gain_biN0002}  \\ % 
\\
  \multicolumn{1}{l}{\bf W75N} \\ % 
                     W75N & \dataset[  8893]{  ADS/Sa.CXO\#obs/08893} & 2008-02-02T01:15 &   29740 & I3 & 20:38:36.49 & +42:38:46.2 &    5 &           TE-VF &  012367 & Gordon Garmire & \path{acisD2008-02-01t_gainN0006}  \\ % 
\\
  \multicolumn{1}{l}{\bf RCW 120} \\ % 
                  CTB 37A & \dataset[  6721]{  ADS/Sa.CXO\#obs/06721} & 2006-10-07T05:13 &   19905 & I3 & 17:14:35.80 & -38:31:24.5 &  259 &           TE-VF &    0123 &Hiroshi Nakajima& \path{acisD2006-08-01t_gainN0006}  \\ % 
          HESS J1713--381 & \dataset[  6692]{  ADS/Sa.CXO\#obs/06692} & 2007-02-02T11:55 &   24836 & I3 & 17:14:04.09 & -38:11:05.3 &   96 &           TE-VF &  012367 & James Hinton   & \path{acisD2007-02-01t_gainN0006}  \\ % 
                   RCW120 & \dataset[ 13621]{  ADS/Sa.CXO\#obs/13621} & 2012-06-30T12:28 &   49117 & I3 & 17:12:20.80 & -38:29:30.5 &  303 &           TE-VF &    0123 &Konstantin Getman& \path{acisD2012-05-01t_gainN0006}  \\ % 
                   RCW120 & \dataset[ 13276]{  ADS/Sa.CXO\#obs/13276} & 2013-02-11T07:39 &   29688 & I3 & 17:12:20.80 & -38:29:30.5 &   93 &           TE-VF &    0123 & Gordon Garmire & \path{acisD2013-02-01t_gainN0006}  \\ % 
\\
  \multicolumn{1}{l}{\bf IRAS 20126+4104} \\ % 
          IRAS 20126+4104 & \dataset[  3758]{  ADS/Sa.CXO\#obs/03758} & 2003-03-13T02:43 &   38761 & I3 & 20:14:29.05 & +41:13:32.4 &   58 &           TE-VF &  012367 & Peter Hofner   & \path{acisD2003-02-01t_gainN0006}  \\ % 
\\
  \multicolumn{1}{l}{\bf W31N} \\ % 
              SGR 1806-20 & \dataset[  1827]{  ADS/Sa.CXO\#obs/01827} & 2000-07-24T18:22 &    4393 & S3 & 18:08:40.32 & -20:24:41.1 &  265 &            TE-F &  235678 & Gautam Vasisht & \path{acisD2000-05-01t_gainN0006}  \\ % 
              SGR 1806-20 & \dataset[  6224]{  ADS/Sa.CXO\#obs/06224} & 2005-02-09T07:01 &   18620 & I3 & 18:08:39.29 & -20:24:38.9 &   88 &            TE-F &  012367 & Derek Fox      & \path{acisD2005-02-01t_gainN0006}  \\ % 
          AX J180816-2021 & \dataset[  8151]{  ADS/Sa.CXO\#obs/08151} & 2007-10-26T07:15 &    2107 & S3 & 18:08:16.79 & -20:21:43.1 &  272 &            TE-F &  235678 & Stephen Murray & \path{acisD2007-08-01t_gainN0006}  \\ % 
          AX J180857-2004 & \dataset[ 10518]{  ADS/Sa.CXO\#obs/10518} & 2009-02-10T08:00 &    6467 & S3 & 18:08:57.40 & -20:04:33.6 &   88 &            TE-F &  235678 & Bryan Gaensler & \path{acisD2009-02-01t_gainN0006}  \\ % 
               G9.95-0.81 & \dataset[ 10713]{  ADS/Sa.CXO\#obs/10713} & 2009-02-10T10:21 &    9914 & I3 & 18:10:41.09 & -20:43:33.4 &   88 &           TE-VF &  012367 & Gordon Garmire & \path{acisD2009-02-01t_gainN0006}  \\ % 
                W31 North & \dataset[ 18452]{  ADS/Sa.CXO\#obs/18452} & 2016-08-14T05:31 &   39243 & I3 & 18:08:59.09 & -20:05:08.1 &  268 &           TE-VF &  012367 & Gordon Garmire & \path{acisD2016-02-01t_gain_biN0002}  \\ % 
\\
  \multicolumn{1}{l}{\bf IRAS 19410+2336;} \\ % 
      {\bf NGC 6823} \\ % 
               19410+2336 & \dataset[  1868]{  ADS/Sa.CXO\#obs/01868} & 2001-10-15T22:25 &    8679 & S3 & 19:43:11.40 & +23:44:05.9 &  276 &           TE-VF &  235678 & Henrik Beuther & \path{acisD2001-08-01t_gainN0006}  \\ % 
          AX J194152+2251 & \dataset[  8164]{  ADS/Sa.CXO\#obs/08164} & 2007-07-16T04:55 &    2714 & S3 & 19:41:52.99 & +22:51:43.1 &  179 &            TE-F &  235678 & Stephen Murray & \path{acisD2007-05-01t_gainN0006}  \\ % 
          AX J194332+2323 & \dataset[ 10517]{  ADS/Sa.CXO\#obs/10517} & 2009-02-15T10:18 &    5879 & S3 & 19:43:32.59 & +23:23:52.7 &   45 &            TE-F &  235678 & Bryan Gaensler & \path{acisD2009-02-01t_gainN0006}  \\ % 
          AX J194310+2318 & \dataset[ 10502]{  ADS/Sa.CXO\#obs/10502} & 2009-02-26T09:52 &    2054 & S3 & 19:43:10.30 & +23:18:50.4 &   58 &            TE-F &  235678 & Bryan Gaensler & \path{acisD2009-02-01t_gainN0006}  \\ % 
\\
  \multicolumn{1}{l}{\bf W42; RSGC1} \\ % 
           HESS J1837-069 & \dataset[  6719]{  ADS/Sa.CXO\#obs/06719} & 2006-08-19T16:30 &   19897 & I3 & 18:37:43.00 & -06:54:20.9 &  259 &           TE-VF &  012367 & Gerd Puehlhofer& \path{acisD2006-08-01t_gainN0006}  \\ % 
                      W42 & \dataset[ 16673]{  ADS/Sa.CXO\#obs/16673} & 2016-02-13T16:26 &   53865 & I3 & 18:38:14.59 & -06:49:18.9 &   78 &           TE-VF &    0123 & Leisa Townsley & \path{acisD2016-02-01t_gain_biN0002}  \\ % 
\\
  \multicolumn{1}{l}{\bf W33; Cl 1813-178} \\ % 
                G12.7-0.0 & \dataset[  6685]{  ADS/Sa.CXO\#obs/06685} & 2006-09-15T00:56 &   29568 & I3 & 18:13:28.80 & -17:52:00.8 &  270 &           TE-VF &  012367 & David Helfand  & \path{acisD2006-08-01t_gainN0006}  \\ % 
                      W33 & \dataset[ 16674]{  ADS/Sa.CXO\#obs/16674} & 2015-07-27T21:36 &   38188 & I3 & 18:14:13.49 & -17:55:41.9 &  262 &           TE-VF &   01237 & Leisa Townsley & \path{acisD2015-05-01t_gainN0006}  \\ % 
              Cl 1813-178 & \dataset[ 17695]{  ADS/Sa.CXO\#obs/17695} & 2016-05-29T22:12 &   12886 & I3 & 18:13:24.29 & -17:53:30.8 &  103 &           TE-VF &   01237 & Gordon Garmire & \path{acisD2016-02-01t_gain_biN0002}  \\ % 
              Cl 1813-178 & \dataset[ 17440]{  ADS/Sa.CXO\#obs/17440} & 2016-06-05T21:05 &   16725 & I3 & 18:13:24.29 & -17:53:30.8 &  103 &           TE-VF &  012367 & Gordon Garmire & \path{acisD2016-02-01t_gain_biN0002}  \\ % 
\\
  \multicolumn{1}{l}{\bf NGC 7538} \\ % 
                  NGC7538 & \dataset[  5373]{  ADS/Sa.CXO\#obs/05373} & 2005-03-25T22:55 &   28785 & I3 & 23:13:45.49 & +61:28:16.6 &   19 &           TE-VF &  012367 & Gordon Garmire & \path{acisD2005-02-01t_gainN0006}  \\ % 
\\
  \multicolumn{1}{l}{\bf G333} \\ % 
          IGR J16195-4945 & \dataset[  5471]{  ADS/Sa.CXO\#obs/05471} & 2005-04-29T17:25 &    4750 & I3 & 16:19:30.00 & -49:45:00.0 &   42 &            TE-F &  012367 & John Tomsick   & \path{acisD2005-02-01t_gainN0006}  \\ % 
          AX J162246-4946 & \dataset[  8161]{  ADS/Sa.CXO\#obs/08161} & 2007-06-13T15:01 &    2900 & S3 & 16:22:46.99 & -49:46:55.1 &  329 &            TE-F &  235678 & Stephen Murray & \path{acisD2007-05-01t_gainN0006}  \\ % 
          AX J162208-5005 & \dataset[  8141]{  ADS/Sa.CXO\#obs/08141} & 2007-06-21T02:36 &    1529 & S3 & 16:22:08.39 & -50:05:41.9 &  316 &            TE-F &  235678 & Stephen Murray & \path{acisD2007-05-01t_gainN0006}  \\ % 
          AX J162011-5002 & \dataset[  9602]{  ADS/Sa.CXO\#obs/09602} & 2008-05-28T19:52 &    1531 & S3 & 16:20:11.50 & -50:02:09.6 &  358 &            TE-F &  235678 & Stephen Murray & \path{acisD2008-05-01t_gainN0006}  \\ % 
          AX J162046-4942 & \dataset[ 10507]{  ADS/Sa.CXO\#obs/10507} & 2009-01-26T20:06 &    3333 & S3 & 16:20:46.60 & -49:42:46.8 &   97 &            TE-F &  235678 & Bryan Gaensler & \path{acisD2008-11-01t_gainN0006}  \\ % 
               G333.6-0.2 & \dataset[  9911]{  ADS/Sa.CXO\#obs/09911} & 2009-06-14T12:19 &   60096 & I3 & 16:22:09.19 & -50:06:03.4 &  327 &           TE-VF &  012367 & Leisa Townsley & \path{acisD2009-05-01t_gainN0006}  \\ % 
             PSR J1622-49 & \dataset[ 10929]{  ADS/Sa.CXO\#obs/10929} & 2009-07-10T07:21 &   19879 & I3 & 16:22:52.90 & -49:49:35.1 &  292 &            TE-F &  012367 & Nanda Rea      & \path{acisD2009-05-01t_gainN0006}  \\ % 
               G333.3-0.4 & \dataset[ 15617]{  ADS/Sa.CXO\#obs/15617} & 2013-02-26T01:12 &   26131 & I3 & 16:21:32.59 & -50:24:20.3 &   79 &           TE-VF &  012367 & Leisa Townsley & \path{acisD2013-02-01t_gainN0006}  \\ % 
               G333.3-0.4 & \dataset[ 14532]{  ADS/Sa.CXO\#obs/14532} & 2013-02-28T16:50 &   28204 & I3 & 16:21:32.59 & -50:24:20.3 &   79 &           TE-VF &  012367 & Leisa Townsley & \path{acisD2013-02-01t_gainN0006}  \\ % 
               G333.1-0.4 & \dataset[ 14531]{  ADS/Sa.CXO\#obs/14531} & 2013-06-25T01:41 &   60271 & I3 & 16:20:39.49 & -50:39:58.6 &  310 &           TE-VF &   01237 & Leisa Townsley & \path{acisD2013-05-01t_gainN0006}  \\ % 
                  RCW 106 & \dataset[ 15393]{  ADS/Sa.CXO\#obs/15393} & 2013-07-03T00:22 &   59149 & I3 & 16:20:02.80 & -50:58:18.6 &  294 &           TE-VF &   01237 & Gordon Garmire & \path{acisD2013-05-01t_gainN0006}  \\ % 
\\
  \multicolumn{1}{l}{\bf AFGL 2591} \\ % 
                AFGL 2591 & \dataset[  6442]{  ADS/Sa.CXO\#obs/06442} & 2006-02-08T01:26 &   29791 & S3 & 20:29:24.89 & +40:11:21.0 &   16 &           TE-VF &  235678 & Arnold Benz    & \path{acisD2006-02-01t_gainN0006}  \\ % 
\\
  \multicolumn{1}{l}{\bf G34.4+0.23} \\ % 
          IRDC G34.4+0.23 & \dataset[ 14541]{  ADS/Sa.CXO\#obs/14541} & 2013-06-17T12:03 &   28125 & I3 & 18:53:18.49 & +01:24:47.9 &  146 &           TE-VF &   01236 & Jonathan Tan   & \path{acisD2013-05-01t_gainN0006}  \\ % 
          IRDC G34.4+0.23 & \dataset[ 15664]{  ADS/Sa.CXO\#obs/15664} & 2013-08-12T17:44 &   34526 & I3 & 18:53:18.49 & +01:24:47.9 &  239 &           TE-VF &  012367 & Jonathan Tan   & \path{acisD2013-08-01t_gainN0006}  \\ % 
\\
  \multicolumn{1}{l}{\bf Wd1} \\ % 
              Westerlund1 & \dataset[  6283]{  ADS/Sa.CXO\#obs/06283} & 2005-05-22T20:38 &   18808 & S3 & 16:47:05.40 & -45:50:36.7 &   25 &            TE-F &  234678 & Stephen Skinner& \path{acisD2005-05-01t_gainN0006}  \\ % 
              Westerlund1 & \dataset[  5411]{  ADS/Sa.CXO\#obs/05411} & 2005-06-18T16:09 &   36596 & S3 & 16:47:05.40 & -45:50:36.7 &  325 &            TE-F &  234678 & Stephen Skinner& \path{acisD2005-05-01t_gainN0006}  \\ % 
        0FGL J1648.1-4606 & \dataset[ 11836]{  ADS/Sa.CXO\#obs/11836} & 2010-01-24T03:00 &   10035 & I3 & 16:48:34.59 & -46:07:10.9 &  102 &           TE-VF &  012367 & Gordon Garmire & \path{acisD2009-11-01t_gainN0006}  \\ % 
\enddata

\tablenotetext{a}{Exposure times are the net usable times after various filtering steps are applied in the data reduction process. 
For the following ObsIDs, we discarded exposure time as noted to remove periods of high instrumental background: 
% NGC6334
2573 (35\%),
4591 (7\%),
% GM24% W75N
% IRAS20126
% RCW120
% W31
6224 (1\%),
% IRAS19410
1868 (56\%),
% W42
% W33
% NGC7538
5373 (2\%),
% G333
% AFGL2591
% G34.4+0.23
% Wd1
5411 (4\%).
The time variability of the ACIS background is discussed in \S6.16.3 of the \anchorparen{http://asc.harvard.edu/proposer/POG/}{\Chandra Proposers' Observatory Guide} and in the ACIS Background Memos at \url{http://asc.harvard.edu/cal/Acis/Cal_prods/bkgrnd/current/}.
}

\tablenotetext{b}{The aimpoints (given in celestial coordinates) are obtained from the satellite aspect solution before astrometric correction is applied.  Units of right ascension ($\alpha$) are hours, minutes, and seconds; units of declination ($\delta$) are degrees, arcminutes, and arcseconds.}

\tablenotetext{c}{ACIS observing modes are described in \S6.12 of the \anchorparen{http://asc.harvard.edu/proposer/POG/}{\Chandra\ Proposers' Observatory Guide}.}

\tablenotetext{d}{The layout of the ten CCD detectors (numbered 0 thru 9 here) in the ACIS focal plane is shown in \S6.1 of the \anchorparen{http://asc.harvard.edu/proposer/POG/}{\Chandra\ Proposers' Observatory Guide}.}

\tablenotetext{e}{The ACIS Time-Dependent Gain file used for calibration of event energies.}

\tablenotetext{f}{Version N0010 of the Optical Blocking Filter model was used for calibration of Ancillary Response Files and exposure maps.}
\end{deluxetable}
%\end{longrotatetable} 
        
% MOXC2 uses 9 ACIS GTO observations!

% 2017 Dec 8 PSB  All exposure times updated to reflect flare filtering.  Confirmed numbers in Note A.  

% The version of the OBF contamination model used is shown as a footnote in observing log.  

%\clearpage
%-----------------------------------------------------------------------------
\subsection{Data Analysis}
Our data reduction, diffuse emission analysis, and point source detection, extraction, and masking techniques employ several innovations beyond standard \Chandra procedures, as discussed at length by \citet{Broos10}. These techniques were standardized for MSFRs by the CCCP and improved during the MYStIX and MOXC1 studies.  Many of the CCCP data analysis steps were implemented by the \anchor{http://personal.psu.edu/psb6/TARA/ae_users_guide.html}{{\em ACIS Extract}} (AE) software package\footnote{ The {\em ACIS Extract} software package and User's Guide are available at \url{http://personal.psu.edu/psb6/TARA/ae_users_guide.html}. } \citep{AE12,AE16} and are described in detail in \citet{Broos10}.  Nearly identical procedures were applied to the MOXC2 MSFRs, so we do not provide an exhaustive review of those procedures here.  A few of the basic steps and improvements to methodologies previously published are described below.

The \Chandra data analysis system, \anchorfoot{http://cxc.harvard.edu/ciao/}{\CIAO} \citep{Fruscione06}, the \anchorfoot{http://ds9.si.edu}{\it SAOImage DS9} visualization tool \citep{Joye03}, and the \anchorfoot{www.harrisgeospatial.com/ProductsandTechnology/Software/IDL.aspx}{{\it Interactive Data Language}} (IDL) are used throughout our data analysis workflow, from data preparation through science analysis.  Models of the local point spread function (PSF) of each source, built by the \anchorfoot{http://space.mit.edu/ASC/marx/index.html}{\MARX}
%\footnote
%{Event positions are calculated (in the \CIAO\ tool {\em acis\_process\_events}) by the EDSER algorithm (\url{http://cxc.harvard.edu/ciao/why/acissubpix.html}).  }
observatory simulator \citep{Davis12}, play a central role in source finding, constructing extraction apertures, accounting for source crowding when extracting local backgrounds, calculating energy-dependent aperture corrections for calibration data products, modeling detector pile-up effects (Section~\ref{sec:pileup}), and masking point sources for diffuse analysis (Section~\ref{sec:diffuse}).  As recommended by the Chandra X-ray Center\footnote{\url{http://cxc.harvard.edu/ciao/why/aspectblur.html}}, 
we tuned two \MARX\ parameters to produce PSFs that match our own data:  AspectBlur=0.07\arcsec\ and {\em pix\_adj}=NONE for ACIS-I observations; AspectBlur=0.07\arcsec\ and {\em pix\_adj}=EDSER for ACIS-S observations.  Note that this tuning applies only to \MARX\ version 5.3.0.

\clearpage
%-------------------
\subsubsection{Point Source Detection Strategy \label{sec:detection}}
As in previous studies, our point source detection workflow first identifies a liberal set of candidate sources, derived mostly from maximum likelihood image reconstruction, % using local models of the {\em Chandra}/ACIS PSF, 
then iteratively prunes candidates found to be insignificant after extraction and careful local background estimation.  Extraction apertures are normally sized to contain 90\% of the PSF (at 1.5~keV), but are reduced when necessary to minimize overlap among crowded sources.  (Aperture correction is discussed in \S\ref{sec:catalog}.)  Iteration is necessary because backgrounds and extraction apertures depend upon neighboring sources, so as insignificant source candidates are pruned from the catalog, remaining source candidates must be re-extracted and re-evaluated for validity.

The goals of {\em completeness} and {\em validity} of a source list are always in conflict.  Our point source detection procedure is designed to be aggressive, emphasizing sensitivity and accepting a reasonable occurrence of possibly-spurious detections to achieve that sensitivity.  In the CCCP study, \citet[][Figure~9]{Broos11} showed that when deep near-IR catalogs are available, the fraction of X-ray detections without apparent near-IR counterparts rises only slowly as detection significance falls; this is indirect evidence that our procedures do not lead to a large number of false sources. \citet[][\S6.2]{Broos11} discuss the impracticality of quantifying the false detection rate in our  X-ray catalog, and point out that such an estimate would be nearly useless because most science analyses select subsets of the X-ray catalog (e.g., sources with IR photometry available, or sources classified as young stars).

Detection of source candidates has changed in various ways since MOXC1.  Since different ACIS pointings now commonly overlap considerably in our wide-field mosaics, pointing-based tiling has been eliminated.  Now the whole field is tiled, tile sections are extracted from every ObsID, and only tile extractions with similar off-axis angles are merged, to avoid combining data with very different angular resolutions.  These merged tile images, along with tile images from each constituent ObsID, are independently searched for sources.

We now use a more complex method to estimate the background in a reconstructed image, separately modeling particle and X-ray backgrounds.  This accounts for the energy-dependent and time-dependent distribution of the X-ray background across the detector, caused by the spatially non-uniform and time-dependent contamination on the \anchorfoot{http://cxc.harvard.edu/ciao/why/acisqecontamN0010.html}{\ACIS\ Optical Blocking Filter}.  

Our procedure for searching reconstructed tile images for source candidates now involves several steps, in order to improve our ability to detect faint sources.  After source candidates are identified in each reconstructed tile image, those candidates are removed, the resulting image is smoothed, and the smoothed image is searched again for fainter sources.  At the end of the source-finding process, the resultant candidate source lists are combined (and duplicates removed); the final candidate source list for the entire target is then assessed for validity.

%-------------------
\subsubsection{Astrometric Alignment \label{sec:alignment}}
Since tremendous resources have been invested to achieve {\em Chandra}'s superb angular resolution, we make considerable efforts to preserve that angular resolution by carefully aligning overlapping ACIS observations and assigning absolute astrometry by using the Two Micron All Sky Survey \citep[2MASS,][]{Skrutskie06} catalog as the astrometric reference (this near-IR catalog works best for our obscured targets).  This alignment work is iterative.  The first round of alignment is performed by finding bright X-ray sources in each ObsID, then matching those independent single-ObsID catalogs of bright X-ray sources to each other and to the 2MASS catalog.  (This matching scheme, considering relative astrometry shifts between ACIS ObsIDs as well as absolute astrometry shifts to align to 2MASS, is very helpful for aligning ACIS observations with a wide range of integration times; short ObsIDs are best aligned to longer ObsIDs, then the longer ObsIDs provide the absolute alignment to 2MASS.)  A weighted analysis of the resulting offsets between pairs of matched sources produces recommended shifts for each ObsID; no rotational correction is attempted.  Those shifts are applied, then the process is repeated until no further shifts are recommended.  The final cumulative shifts are applied to the aspect calibration files for each ObsID and the event lists are reprojected onto the sky.  Those single-ObsID catalogs of bright sources, which now have obsolete coordinates, are not considered further.  

A fresh list of candidate sources (\S\ref{sec:detection}) is obtained from these aligned data, and the iterative process of extracting and validating those candidates begins.  During this process we  assess the alignment of ObsIDs at least twice more.  As above, shift recommendations for each ObsID are obtained by combining error-weighted estimates of all pairs of inter-ObsID offsets and error-weighted estimates of single-ObsID offsets with respect to 2MASS; no rotational correction is attempted.   We believe this second round of alignment work is worthwhile for two reasons.  First, estimates of inter-ObsID shifts made at this point should be more accurate, because a single target-wide catalog is extracted from every ObsID---no matching is required.  Second, the (single target-wide) catalog of source candidates available at this point, built after our first alignment of the data (described above), is expected to be a better representation of the sky (e.g., crowded sources are more likely to be resolved).
%Only extractions that meet several relevant criteria (e.g., brighter sources not too far off-axis) participate in the astrometry calculations. 

%-------------------
\subsubsection{Validation of Source Candidates \label{sec:validation}}
For sources with multiple observations, \citet[][Section~6.2]{Broos10} point out that a subset of the observations may produce higher quality measurements of source properties than the full set of observations would.  This is often true when the observations have very different off-axis angles, and thus very different angular resolutions and degrees of crowding.  As in previous studies, MOXC2 builds spectra and photometry from a set of observations that favors the photometric signal-to-noise ratio\footnote{The ``MERGE\_FOR\_PHOTOMETRY'' option for combining ObsIDs in \AEacro\ tries to balance the goals of maximum signal-to-noise ratio and zero photometry bias.} and estimates the source position from a set of observations that favors small position uncertainty.
          
In previous studies, source {\em validation} followed the same strategy; the validity of a source (in each energy band) was calculated from the set of observations in which the source was most valid.  \citet[][Section~6.2]{Broos10} point out that this selection process increases sensitivity to variable sources at the cost of an increased false detection rate arising from the additional number of data sets that are searched.

Recently, while reducing a \Chandra target with a large number of overlapping ObsIDs, we found this source validation strategy to be inadequate, since the number of possible ways to combine $N$ ObsIDs grows rapidly with $N$.  Thus, in MOXC2 we have adopted a different strategy to balance false detection rate and sensitivity to variable sources.  Source validity is now evaluated on a pre-defined, small set of ObsID combinations.  Those combinations do not depend on characteristics of the data; they consist simply of (1) each ObsID by itself and (2) ObsIDs with similar off-axis angles.  Empirically-determined off-axis angle ranges are 0\arcmin--3\arcmin, 2\arcmin--6\arcmin, 5\arcmin--9\arcmin, 8\arcmin--14\arcmin, and $>$13\arcmin.  These ranges overlap to avoid missing sources at the boundaries.

As described in MOXC1, our procedures account for the \Chandra \anchorfoot{http://cxc.harvard.edu/ciao/caveats/psf_artifact.html}{PSF hook}, a structure that extends $\sim$0.8$\arcsec$ from the main PSF peak and contains ${\sim}$5\% of the flux.  
%PSB cannot find a published citation for the hook.  
This artifact is significant for us, since our detection efforts emphasize faint sources and hooks around bright sources can be reconstructed as fainter neighboring candidate sources.  We mark the hook location around bright sources, then visually examine those locations for source candidates that are consistent with the expected hook brightness.  Such spurious ``hook sources'' are removed from the catalog.

%-------------------
\subsubsection{Diffuse Emission \label{sec:diffuse}} 
As always, our strategy for studying diffuse X-ray emission is to mask (remove) events that are likely to be associated with point sources, to subtract particle background, and to normalize by the exposure map \citep[][Section~9]{Townsley03,Broos10}.  The resulting diffuse image represents observed surface brightness, which has units of photon~cm$^{-2}$~s$^{-1}$~arcsec$^{-2}$.  Masked data products (observed event list, particle background event list, and exposure maps) are built independently for each ObsID, and then combined into target-level images.  A surface brightness image is computed from those observed, background, and exposure map images using an adaptive smoothing algorithm \citep[][Section~9.1]{Townsley03,Broos10}, in which the smoothing kernel is sized to achieve a target signal-to-noise ratio in the surface brightness.
% We subtract the Stowed Data from the observed data; Stowed Data represents only particle background.

The adaptively-smoothed X-ray flux images presented below (Section~\ref{sec:targets}) cover the energy band 0.5--7.0~keV\footnote{For diffuse images, the ``full band'' omits the energy range 7--8~keV to avoid a bright emission line in the instrumental background. } and are smoothed to a signal-to-noise ratio of 15.  The exposure map used in these images represents the instrument response at 1~keV. 

\clearpage
%=============================================================================
\section{MOXC2 DATA PRODUCTS \label{sec:products}}

%-----------------------------------------------------------------------------
\subsection{The MOXC2 Chandra Point Source Catalog \label{sec:catalog}}

The primary data product of MOXC2 is a catalog of properties for 18,396 point sources found in our 13 ACIS mosaics of MSFRs and their surrounding Galactic Plane environments.  This catalog provides the same source information as that given in MOXC1 and is similar to catalogs from CCCP \citep{Broos11}, MYStIX \citep{Kuhn13}, and SFiNCs \citep{Getman17}.  

% The FITS file to submit is /bulk/hiawatha1/targets/MOXC2/targets/xray_properties.fits
Table~\ref{xray_properties.tbl} defines the columns of the MOXC2 point source catalog.  This catalog is available in FITS format from the electronic edition of this article and may be available in many other formats from VizieR \citep{Ochsenbein00}.  All photometric quantities in this table are apparent (not corrected for absorption).  The suffixes ``\_t'', ``\_s'', and ``\_h'' on names of photometric quantities designate the {\em total} (0.5--8~keV), {\em soft} (0.5--2~keV), and {\em hard} (2--8~keV) energy bands.
Energy-dependent correction for finite extraction apertures is applied to the ancillary reference file (ARF) calibration products \citep[see][\S5.3]{Broos10}; the SrcCounts and NetCounts quantities characterize the extraction and are not aperture-corrected.
The only calibrated quantities in the catalog are apparent {\em photon} flux in units of photon~cm$^{-2}$~s$^{-1}$ \citep[see][\S7.4]{Broos10}, and an estimate for apparent {\em energy} flux in units of erg~cm$^{-2}$~s$^{-1}$ \citep{Getman10}.  Table notes provide additional information regarding the definition of source properties.

We caution potential users of these results that we have not attempted to clean the point source catalog for sources that might be spurious reconstruction peaks, for example over-reconstructions of diffuse emission.  This can be a problem for bright PWNe, SNRs, and dust-scattering halos around X-ray binaries.  As we describe each target mosaic in Section~\ref{sec:targets} below, we will point out areas of particular concern for such spurious point sources.  

%\clearpage
%\renewcommand{\arraystretch}{0.7}  % to change row spacing in table

%\todo{I believe that the chi-square variability index is adequately described by table note ``c''.  }

%\begin{longrotatetable}
\startlongtable
\begin{deluxetable*}{lll}
\tablecaption{MOXC X-ray Sources and Properties \label{xray_properties.tbl}
}
\tablewidth{7in}
\tabletypesize{\scriptsize}
\tablecolumns{3}
%\tablenum{Table 2}

\tablehead{   
  \colhead{Column Label} & 
  \colhead{Units} & 
  \colhead{Description} 
}
\colnumbers
\startdata
\cutinhead{Name and position, derived from the ObsIDs that minimize the position uncertainty \citep[][\S6.2 and 7.1]{Broos10}}
%Seq                        & \nodata              & sequence number \\
 RegionName                 & \nodata              & name of the MSFR \\ 
 Name                       & \nodata              & \parbox[t]{4.0in}{X-ray source name in IAU format; prefix is CXOU~J} \\
%a X-ray Observatory} Great Nebula in Carina)} \\
 Label\tnm{a}               & \nodata              & X-ray source name used within the project \\
 RAdeg                      & deg                  & right ascension (J2000) \\
 DEdeg                      & deg                  & declination (J2000) \\
 PosErr                     & arcsec               & 1-$\sigma$ error circle around (RAdeg,DEdeg) \\
 PosType                    & \nodata              & algorithm used to estimate position \citep[][\S7.1]{Broos10}  \\
\\[0.10em]
\cutinhead{Validity metrics, derived from a pre-defined set of ObsID combinations (Section~\ref{sec:validation})}
 ProbNoSrc\_MostValid       & \nodata              & smallest of ProbNoSrc\_t, ProbNoSrc\_s, ProbNoSrc\_h, ProbNoSrc\_v \\
%Merge\_MostValid           & \nodata              & merge that produced ProbNoSrc\_MostValid \\
%Band\_MostValid            & \nodata              & energy band that produced ProbNoSrc\_MostValid \\
%SrcCounts\_MostValid       & count                & observed counts used in calculaton of ProbNoSrc\_MostValid \\
 ProbNoSrc\_t               & \nodata              & \parbox[t]{4.0in}{smallest {\em p}-value\tnm{b} under the no-source null hypothesis \citep[][\S4.3]{Broos10} among validation merges}\\
 ProbNoSrc\_s               & \nodata              & \parbox[t]{4.0in}{smallest {\em p}-value under the no-source null hypothesis among validation merges}\\
 ProbNoSrc\_h               & \nodata              & \parbox[t]{4.0in}{smallest {\em p}-value under the no-source null hypothesis among validation merges}\\
%ProbNoSrc\_v               & \nodata              & \parbox[t]{4.0in}{smallest {\em p}-value under the no-source null hypothesis among validation merges}\\
%NumValidObsIDs             & \nodata              & number of individual ObsIDs in which the source was valid \\
%ValidObsIDList             & \nodata              & list of the individual ObsIDs in which the source was valid\\
%IsOccasional               & boolean              & \parbox[t]{4.0in}{flag indicating that source validation failed in all multi-ObsID merges; source validation comes from a single ObsID}\\
\\[0.10em]
\cutinhead{Variability indices, derived from all ObsIDs}
 ProbKS\_single\tnm{c}      & \nodata              & \parbox[t]{4.0in}{smallest {\em p}-value under the null hypothesis (no variability within each single ObsID) for the Kolmogorov--Smirnov test on the timestamps of each ObsID's event list } \\
 ProbKS\_merge\tnm{c}       & \nodata              & \parbox[t]{4.0in}{{\em p}-value under the null hypothesis (no variability) for the Kolmogorov--Smirnov test on the timestamps of the multi-ObsID event list } \\
 ProbChisq\_PhotonFlux      & \nodata              & \parbox[t]{4.0in}{{\em p}-value under the null hypothesis (no variability) for the $\chi^2$ test on the single-ObsID measurements of PhotonFlux\_t } \\
\\[0.10em]
\cutinhead{Observation details and photometric quantities, derived from the set of ObsIDs that optimizes photometry \citep[][\S6.2 and 7]{Broos10}}
 ExposureTimeNominal        & s                    & total exposure time in merged ObsIDs \\
 ExposureFraction\tnm{d}    & \nodata              & fraction of ExposureTimeNominal that source was observed \\
 RateIn3x3Cell\tnm{e}       &  count/frame        & 0.5:8 keV, in 3$\times$3 CCD pixel cell \\
 NumObsIDs                  & \nodata              & total number of ObsIDs extracted \\
 NumMerged                  & \nodata              & \parbox[t]{4.0in}{number of ObsIDs merged to estimate photometry properties} \\
 MergeBias                  & \nodata              & fraction of exposure discarded in merge \\
\\[0.10em]
 Theta\_Lo                  & arcmin               & smallest off-axis angle for merged ObsIDs \\
 Theta                      & arcmin               &  average off-axis angle for merged ObsIDs \\
 Theta\_Hi                  & arcmin               &  largest off-axis angle for merged ObsIDs \\
\\[0.10em]
 PsfFraction                & \nodata              & average PSF fraction (at 1.5 keV) for merged ObsIDs \\
 SrcArea                    & (0.492 arcsec)$^2$   & average aperture area for merged ObsIDs \\
 AfterglowFraction\tnm{f}   & \nodata              & suspected afterglow fraction\\
% Correction\_t        & \nodata              & pile-up correction to SrcCounts\_t, NetCounts\_t \\
\\[0.10em]
 SrcCounts\_t               & count                & observed counts in merged apertures \\
 SrcCounts\_s               & count                & observed counts in merged apertures \\
 SrcCounts\_h               & count                & observed counts in merged apertures \\
\\[0.10em]
 BkgScaling                 & \nodata              & scaling of the background extraction \citep[][\S5.4]{Broos10} \\
\\[0.10em]
 BkgCounts\_t               & count                & observed counts in merged background regions \\
 BkgCounts\_s               & count                & observed counts in merged background regions \\
 BkgCounts\_h               & count                & observed counts in merged background regions \\
\\[0.10em]
 NetCounts\_t               & count                & net counts in merged apertures \\
 NetCounts\_s               & count                & net counts in merged apertures \\
 NetCounts\_h               & count                & net counts in merged apertures \\
\\[0.10em]
 NetCounts\_Lo\_t\tnm{g}    & count                & 1-$\sigma$ lower bound on NetCounts\_t \\
 NetCounts\_Hi\_t           & count                & 1-$\sigma$ upper bound on NetCounts\_t \\
\\[0.10em]
 NetCounts\_Lo\_s           & count                & 1-$\sigma$ lower bound on NetCounts\_s \\
 NetCounts\_Hi\_s           & count                & 1-$\sigma$ upper bound on NetCounts\_s \\
\\[0.10em]
 NetCounts\_Lo\_h           & count                & 1-$\sigma$ lower bound on NetCounts\_h \\
 NetCounts\_Hi\_h           & count                & 1-$\sigma$ upper bound on NetCounts\_h \\  
\\[0.10em]
 MeanEffectiveArea\_t\tnm{h}& cm$^2$~count~photon$^{-1}$ & mean ARF value \\
 MeanEffectiveArea\_s       & cm$^2$~count~photon$^{-1}$ & mean ARF value \\
 MeanEffectiveArea\_h       & cm$^2$~count~photon$^{-1}$ & mean ARF value \\
\\[0.10em]
 MedianEnergy\_t\tnm{i}     & keV                  & median energy, observed spectrum \\
 MedianEnergy\_s            & keV                  & median energy, observed spectrum \\
 MedianEnergy\_h            & keV                  & median energy, observed spectrum \\
\\[0.10em]
 PhotonFlux\_t\tnm{j}       & photon~cm$^{-2}$~s$^{-1}$     & apparent photon flux \\
 PhotonFlux\_s              & photon~cm$^{-2}$~s$^{-1}$     & apparent photon flux \\
 PhotonFlux\_h              & photon~cm$^{-2}$~s$^{-1}$     & apparent photon flux \\
\\[0.10em]
 EnergyFlux\_t              & erg~cm$^{-2}$~s$^{-1}$ & max(EnergyFlux\_s,0) + max(EnergyFlux\_h,0) \\
 EnergyFlux\_s\tnm{k}       & erg~cm$^{-2}$~s$^{-1}$ & apparent energy flux \\
 EnergyFlux\_h\tnm{k}       & erg~cm$^{-2}$~s$^{-1}$ & apparent energy flux \\
\enddata
                                                                                 and
%\tablecomments{Rows are sorted by R.A.}
                                                                                                                                              
\tablecomments{
These X-ray columns are produced by the
{\em ACIS Extract} (AE) software package \citep{Broos10,AE12}.  Similar column labels were previously published by the CCCP \citep{Broos11} and by MOXC1 \citep{Townsley14}.
The AE software and User's Guide are available at \anchor{http://personal.psu.edu/psb6/TARA/ae_users_guide.html}{\url{http://personal.psu.edu/psb6/TARA/ae_users_guide.html}}.
%\anchorparen{http://personal.psu.edu/psb6/TARA/ae_users_guide.html}{{\it ACIS Extract} software}.}
}
\tablecomments{
The suffixes ``\_t'', ``\_s'', and ``\_h'' on names of photometric quantities designate the {\em total} (0.5--8~keV), {\em soft} (0.5--2~keV), and {\em hard} (2--8~keV) energy bands. 
}
\tablecomments{
Source name and position quantities (Name,RAdeg, DEdeg, PosErr,PosType) are computed using a subset of each source's extractions chosen to minimize the position uncertainty \citep[][\S6.2 and 7.1]{Broos10}.   
Source significance quantities (ProbNoSrc\_MostValid, ProbNoSrc\_t, ProbNoSrc\_s, ProbNoSrc\_h) are computed using a pre-defined set of ObsID combinations, which do not depend on the data observed (Section~\ref{sec:validation}).
Variability indices (ProbKS\_single, ProbKS\_merge, ProbChisq\_PhotonFlux ) are computed using all ObsIDs.
All other quantities are computed using a subset of ObsIDs chosen, independently for each source, to balance the conflicting goals of minimizing photometric uncertainty and of avoiding photometric bias \citep[][\S6.2 and 7]{Broos10}. 
}

\tablenotetext{a}{Source ``labels'' identify each source during data analysis, as the source position (and thus the Name) is subject to change.}

\tablenotetext{b}{In statistical hypothesis testing, the {\em p}-value is the probability of obtaining a test statistic at least as extreme as the one that was actually observed, when the null hypothesis is true.
The {\em p}-value of the observed extraction under the no-source hypothesis is calculated by the method described by \citet[][Appendix~A2]{Weisskopf07}, which is derived under the assumption that X-ray extractions follow Poisson distributions.}

\tablenotetext{c}{See \citet[][\S7.6]{Broos10} for a description of the ProbKS\_single and ProbKS\_merge variability indexes, and caveats regarding possible spurious indications of variability using the ProbKS\_merge index.   The ProbChisq\_PhotonFlux variability index is the {\em p}-value under the null hypothesis (no variability) for the standard $\chi^2$ test on the single-ObsID measurements of PhotonFlux\_t. \\  }

\tablenotetext{d}{Due to dithering over inactive portions of the focal plane, a \Chandra source often is not observed during some fraction of the nominal exposure time.  (See \url{http://cxc.harvard.edu/ciao/why/dither.html}.)  The reported quantity is FRACEXPO, produced by the \CIAO\ tool {\em mkarf}.}

\tablenotetext{e}{ACIS suffers from a non-linearity at high count rates known as {\em photon pile-up}, described in Section~\ref{sec:pileup} below. 
RateIn3x3Cell is an estimate of the observed count rate falling on an event detection cell of size 3$\times$3 \ACIS\ pixels, centered on the source position.
When RateIn3x3Cell $> 0.05$ (count/frame), the reported source properties may be biased by pile-up effects.
See Table~\ref{pile-up_risk.tbl} for a list of MOXC2 sources with significant pile-up.
All source properties in this table are {\em not} corrected for pile-up effects.
}

\tablenotetext{f}{Some background events arising from an instrumental effect known as ``afterglow'' (\url{http://cxc.harvard.edu/ciao/why/afterglow.html}) may contaminate source extractions, despite careful procedures to identify and remove them during data preparation \citep[][\S3]{Broos10}.
After extraction, we attempt to identify afterglow events using the AE tool {\em ae\_afterglow\_report}, and report the fraction of extracted events attributed to afterglow; see the \anchorparen{http://personal.psu.edu/psb6/TARA/ae_users_guide.html}{{\it ACIS Extract} manual}.} 

\tablenotetext{g}{Confidence intervals (68\%) for NetCounts quantities are estimated by the \CIAO\ tool {\em aprates} (\url{http://asc.harvard.edu/ciao/ahelp/aprates.html}).}

\tablenotetext{h}{The ancillary response file (ARF) in \ACIS\ data analysis represents both the effective area of the observatory and the fraction of the observation for which data were actually collected for the source (ExposureFraction).}

\tablenotetext{i}{MedianEnergy is the median energy of extracted events, corrected for background \citep[][\S7.3]{Broos10}. } % \AEacro\ quantity ENERG\_PCT50\_OBSERVED

\tablenotetext{j}{PhotonFlux = (NetCounts / MeanEffectiveArea / ExposureTimeNominal) \citep[][\S7.4]{Broos10}. }

\tablenotetext{k}{EnergyFlux = $1.602 \times 10^{-9} {\rm (erg/keV)} \times$ PhotonFlux $\times$ MedianEnergy \citep[][\S2.2]{Getman10}. }

\end{deluxetable*}
%\end{longrotatetable}

%-----------------------------------------------------------------------------
%\subsection{Characteristics of the X-ray Catalog}

%The cumulative distribution plots shown in MOXC1 are not very practical, because we need a region of constant exposure time for each target.

\clearpage
%-----------------------------------------------------------------------------
\subsection{Piled Sources \label{sec:pileup}}

As described in detail in MOXC1, ACIS detections of bright X-ray sources can suffer from a non-linearity known as \anchorfoot{http://cxc.harvard.edu/ciao/why/pileup_intro.html}{{\em photon pile-up}}, where multiple X-ray photons are mistakenly detected as a single event because they fell close to each other on the CCD and arrived during the same readout frame.  Pile-up causes photometry to be underestimated and the spectrum to be hardened.  We check for pile-up in every observation of a source by estimating the observed count rate in a 3$\times$3 pixel detection cell centered on the source position.  (The highest rate found in all observations of the source is reported in the column RateIn3x3Cell in Table~\ref{xray_properties.tbl}.)

For each source extraction in which RateIn3x3Cell exceeded a threshold of 0.05~count/frame, we modeled pile-up using a Monte Carlo approach that reconstructs a pile-up-free spectrum from a piled ACIS observation \citep{Broos11}.  Table~\ref{pile-up_risk.tbl} lists those extractions; column (8) characterizes the inferred level of pile-up\footnote{We choose not to use the terms ``pile-up fraction'' or ``pile-up percentage'' because the \ACIS\ community has several conflicting definitions for those terms; see \S1.2 in \anchorparen{http://cxc.harvard.edu/ciao/download/doc/pileup_abc.pdf}{The \Chandra ABC Guide to Pileup}.}.
For those sources, several entries in Table~\ref{xray_properties.tbl} are expected to be biased by pile-up effects (in an energy-dependent way).  Users are warned that photometric quantities for piled-up sources should be used with caution; higher pile-up corrections and narrower bandpasses should evoke the most caution.  Alternatively, we provide pileup-corrected spectra for all source/ObsID entries in Table~\ref{pile-up_risk.tbl}, as described below.  Fitting those spectra will result in more meaningful source properties than possible from pileup-distorted photometry.

% Find pile-up reports in extraction logs, e.g.
%  foreach  dir ( NGC6334 GM24 W75N IRAS20126 RCW120 W31 IRAS19410 W42 W33 NGC7538 G333 AFGL2591 G34.4+0.23 Wd1 )
%    printf "\n%s\n" $dir 
%    ls -1               $dir/data/extract/point_sources.noindex/ae_*.*I.log
%    egrep -h 'ct/frame' $dir/data/extract/point_sources.noindex/ae_*.*I.log
%  end
%
%  foreach  dir ( NGC6334 GM24 W75N IRAS20126 RCW120 W31 IRAS19410 W42 W33 NGC7538 G333 AFGL2591 G34.4+0.23 Wd1 )
%    printf "\n%s\n" $dir 
%    grep 'Event rate correction' $dir/data/extract/point_sources.noindex/*/EPOCH*/recon_spectrum.log
%  end

%\todo{Columns 2, 7, and 8 were checked/updated on 2017 Oct 31, after catalog patching.}

\renewcommand{\arraystretch}{1.0}  % to change row spacing in table

\begin{deluxetable}{llllcccc}
\tablecaption{Sources Exhibiting Photon Pile-up \label{pile-up_risk.tbl}}
\tablewidth{0pt}
\tabletypesize{\scriptsize}

\tablehead{
\colhead{MSFR} & \colhead{Name} & \colhead{Label} & \colhead{Identifier} & \colhead{ObsID} & \colhead{$\theta$} & \colhead{PsfFraction}& \colhead{Correction} \\ % & \colhead{cell CR} 
               & \colhead{(CXOU J)} &&&& \colhead{(\arcmin)}                                                 % & \colhead{(count frame$^{-1}$)}
}
\colnumbers
\startdata
NGC 6334      & 171701.53-362100.6 & GM24$\_$c717 & VVV J171701.53-362100.67 & 18876 & 1.9 & 0.64 & 1.027 \\ %     206 (ct)   0.066 (ct/frame)  VISTA source VVV J171701.53-362100.67 in Saito12.  No 2MASS match in VizieR.
              & 171946.16-360552.2 & c1934 & HD 319703A & 12382 & 3.1 & 0.89 & 1.015 \\ %     266 (ct)   0.055 (ct/frame)  this is HD 319703A, an O7V((f))z star according to Sota14 (GOSSS); 2MASS J17194616-3605522; GLIMPSE G351.0294+00.6517.  Maiz Apellaniz 2016 says it's an O7V((f)) + O9.5V, according to Skiff catalog on VizieR.
              & 171946.16-360552.2 & c1934 & HD 319703A & 13436 & 3.1 & 0.89 & 1.033 \\ %     709 (ct)   0.073 (ct/frame)
              & 172001.73-355816.2 & c3105 & 2MASS J17200173-3558162 &  2573 & 1.9 & 0.90 & 1.024 \\ %     388 (ct)   0.055 (ct/frame)  Russeil12 Table A.2 (catalog of R mags) recno=15980 ; Willis13 call this a YSO
              & 172031.75-355111.4 & c5437 & 2MASS J17203178-3551111 &  8975 & 2.5 & 0.90 & 1.121 \\ %      46 (ct)   0.140 (ct/frame)   ; 2MASS ref is Cutri03.  This may be the X-ray counterpart to IGR J17204-3554.
%
%W75N
%IRAS20126
\hline
RCW 120       & 171405.73-381031.4 & c2911 & pulsar CXOU J171405.7-381031 &  6692 & 0.4 & 0.89 & 1.226 \\ %    1007 (ct)   0.272 (ct/frame)
\hline
W31N          & 180839.35-202439.9 & c633  & SGR 1806-20 &  1827 & 0.6 & 0.89 & 2.543 \\ %     594 (ct)   0.485 (ct/frame)
              & 180839.35-202439.9 & c633  & SGR 1806-20 &  6224 & 9.8 & 0.50 & 1.056 \\ %    3712 (ct)   0.079 (ct/frame)
\hline
IRAS 19410+2336 & 194154.60+225112.3 & c19 & ChI J194152+2251$\_$2\tablenotemark{a} &  8164 & 0.7 & 0.46 & \nodata \\ %      30 (ct)   0.080 (ct/frame)  Name is from Anderson14 ChIcAGO Chandra catalog.  Also UKIDSS src from Lucas12, UGPS J194154.61+225112.3.
\hline
W42; RSGC1    & 183803.17-065533.7 & c989  & PSR J1838-0655 & 16673 & 7.2 & 0.90 & 1.031 \\ %    6732 (ct)   0.069 (ct/frame)
              & 183803.17-065533.7 & c989  & PSR J1838-0655 &  6719 & 5.3 & 0.90 & 1.126 \\ %    1866 (ct)   0.152 (ct/frame)
\hline
W33; Cl 1813-178 & 181335.16-174957.4 & c2292 & PSR J1813-1749 &  6685 & 2.8 & 0.90 & 1.066 \\ %     703 (ct)   0.078 (ct/frame)
%
%NGC7538
\hline
G333          & 161932.20-494430.6 & c439  & HMXB IGR J16195-4945 &  5471 & 0.4 & 0.91 & \nodata \\ %      66 (ct)   0.129 (ct/frame)
              & 162006.69-500158.8 & c1594 & ChI J162011-5002$\_$1\tablenotemark{a} &  9602 & 1.0 & 0.90 & 1.174 \\ %      42 (ct)   0.103 (ct/frame)  Name is from Anderson14 ChIcAGO Chandra catalog.
              & 162048.81-494215.0 & c3690 & ChI J162046-4942$\_$1\tablenotemark{a} & 10507 & 0.6 & 0.89 & 1.137 \\ %     112 (ct)   0.120 (ct/frame)  Name is from Anderson14 ChIcAGO Chandra catalog.
              & 162244.91-495052.8 & c8972 & magnetar PSR J1622-4950 &  8161 & 4.0 & 0.89 & 1.146 \\ %     174 (ct)   0.105 (ct/frame)
%
%AFGL2591
%G34.4+0.23
\hline
Wd1           & 164710.18-455216.8 & c1081 & magnetar CXOU J164710.2-455216 &  5411 & 1.3 & 0.89 & 1.062 \\ %     700 (ct)   0.069 (ct/frame)
              & 164710.18-455216.8 & c1081 & magnetar CXOU J164710.2-455216 &  6283 & 1.8 & 0.89 & 1.055 \\ %     379 (ct)   0.077 (ct/frame)
\enddata
\tablenotetext{a}{``ChI'' sources are from the ChIcAGO project \citep[Chasing the Identification of ASCA Galactic Objects,][]{Anderson14}.}
\tablecomments{
Col.\ (1): Name of the MSFR.
\\Col.\ (2): X-ray source name in IAU format ({\em Name} in Table~\ref{xray_properties.tbl}).
\\Col.\ (3): X-ray source name used within the project ({\em Label} in Table~\ref{xray_properties.tbl}).
\\Col.\ (4): Source counterpart from VizieR or SIMBAD.
\\Col.\ (5): \Chandra Observation Identification.
\\Col.\ (6): Off-axis angle ({\em Theta} in Table~\ref{xray_properties.tbl}).
\\Col.\ (7): Fraction of the PSF (at 1.497 keV) enclosed within the extraction region ({\em PsfFraction} in Table~\ref{xray_properties.tbl}). A reduced PSF fraction (significantly below 90\%)  indicates that the source is in a crowded region. 
\\Col.\ (8): Estimated ratio of pile-up-free to observed (piled) count rates in the 0.5--8~keV energy band.  This correction can be applied to the quantities SrcCounts\_t, NetCounts\_t, and PhotonFlux\_t in Table~\ref{xray_properties.tbl}.   No value is reported when pile-up correction could not be performed.  Note that the core of any source is more piled than the wings, so this correction factor depends on the aperture size (Column 7).
}
\end{deluxetable}

%Regarding src c5437 in NGC 6334.  In MOXC1, we list it as being piled in Obs2574 as well as Obs8975, but here it's only listed as piled in Obs8975.  See Pat's e-mail from 8 Dec 2017 in MOXCs e-mail folder -- RateIn3x3Cell is noisy and falls below pile-up threshold for Obs2574 in MOXC2.

%\clearpage
%-----------------------------------------------------------------------------
\subsection{Archive of Reduced Data Products \label{sec:repository}}

The \anchorfoot{https://zenodo.org}{Zenodo data repository} archives many \anchorfoot{https://doi.org/10.5281/zenodo.1067749}{MOXC2 reduced data products} \citep{Townsley17}, including astrometrically aligned event lists and exposure maps, \DSnine\ region files representing point source extraction apertures and PSF hooks, source photometry in 16 energy bands, reconstructed spectra for sources that suffer from photon pile-up (Section~\ref{sec:pileup}), lightcurve plots for the brighter sources, event lists and exposure maps with point sources masked, and smoothed images of diffuse emission (Section~\ref{sec:targets}).  For the piled sources in Table~\ref{pile-up_risk.tbl}, we provide reconstructed spectra for each ObsID in which the source suffered from pile-up.  The ``README'' file in the Zenodo archive describes the files there.

%-----------------------------------------------------------------------------
\subsection{MOXC2 Sources in Published \Chandra Catalogs \label{sec:other_analyses}}

Many of the brighter X-ray point sources in MOXC2 also appear in the \anchorfoot{http://cxc.cfa.harvard.edu/csc/}{\Chandra Source Catalog} \citep{Evans10}.  Some may also have data available from the {\em XMM-Newton} mission's EPIC camera; catalogs of \XMM sources are available from the \anchorfoot{http://xmmssc.irap.omp.eu}{{\em XMM-Newton} Survey Science Centre}.  In crowded regions such as MSFRs, care should be taken in matching MOXC2 sources to \XMM sources due to the mismatch in spatial resolution between the \Chandra and \XMM telescopes.

Several short archival \Chandra observations in and around MSFRs were obtained by the snapshot survey Chasing the Identification of ASCA Galactic Objects \citep[ChIcAGO,][]{Anderson14}.  We use these datasets in the MOXC2 mosaics and the brighter X-ray sources in these fields are contained in the ChIcAGO catalog.  Several examples are mentioned in the descriptions of individual MSFRs below.  

We have one target (RCW~120) in common with SFiNCs \citep{Getman17} and two targets (NGC~6334 and G333) for which a subset of MOXC2 ObsIDs were included in MOXC1; comparisons of these catalogs are mentioned briefly in the specific target descriptions of Section~\ref{sec:targets}.  A few other individual MOXC2 MSFRs have published \Chandra point source catalogs.  MOXC2 typically recovers all or most of these sources and goes on to find additional faint sources; again details are given in Section~\ref{sec:targets}.

\clearpage
%=============================================================================
\section{X-RAY CHARACTERIZATIONS OF MOXC2 TARGETS \label{sec:targets}}

Following the format we established in MOXC1, we now provide short vignettes of each MSFR included in MOXC2, to illustrate the distribution of X-ray point sources in the \Chandra mosaics and to give a qualitative sense of the diffuse X-ray emission present in and around these fields.  We do not attempt comprehensive reviews of the literature, which is often vast and multi-generational for these famous targets; rather we provide only the most cursory set of references, focusing on recent papers and X-ray studies.  These references (and the papers they cite) provide a much better introduction to the MOXC2 MSFRs than space allows here.  We also do not attempt extensive quantitative analysis either of the MSFR X-ray source populations or diffuse X-ray emission; such work is beyond the scope of MOXC2 and must await future efforts by ourselves and the wider star formation community.

For each target we show two basic figures (in celestial J2000 coordinates):  the distribution of brighter ($\geq$5 net counts) X-ray sources on the exposure map of our ACIS mosaic (with ObsID numbers given in blue) and an adaptively-smoothed image of unresolved X-ray emission presented in the context of mid-IR data.  In the first of these, X-ray sources are represented as colored dots, with the color indicating the median energy of the extracted X-rays.  The legend shows the number of sources plotted in each median energy range.  The hardest sources were plotted last, so some softer sources may be covered by the symbols for harder sources.  Fainter sources are not included because their spatial distribution is strongly dependent on \Chandra sensitivity (their numbers necessarily fall off sharply with off-axis angle).  The number of faint sources can be calculated as the total number of detected sources (noted on the figure below the target name) minus the numbers given in the legend.  Since we have made great efforts to detect the faintest X-ray sources possible, most targets have well over half of their ACIS detections absent from these plots.  They do, however, serve to indicate important clumps and clusters of X-ray sources that trace the structure of the MSFRs they sample.  Often those groupings of X-ray sources are shown in more detail in further images that show soft and hard X-ray events in the context of unresolved X-ray emission or {\em Spitzer}/IRAC 8~$\mu$m structures.

The second basic figure shown for each target is a three-color image intended to place unresolved X-ray emission in the context of the cold ISM traced by mid-IR images from \Spitzer or {\em WISE}.  In many instances, we include zoomed panels of this three-color image to highlight particularly interesting regions, with more informative image scaling tailored specifically for the diffuse X-ray emission in those regions.  For diffuse emission, we consider ``full-band'' to be 0.5--7~keV, not the 0.5--8~keV used for point source photometry, because the \anchorfoot{http://cxc.cfa.harvard.edu/contrib/maxim/stowed/}{ACIS particle background} rises sharply above 7~keV.

Since MSFRs are full of stars and all of our \Chandra observations are too shallow to trace the complete initial mass function of a MSFR, we expect some of the unresolved X-ray emission (especially in cluster centers) to come from pre-MS stars.  It is clear from these images, however, that we are teasing out faint, truly diffuse X-ray emission in these complexes, as demonstrated by the dramatic anticoincidence of the unresolved X-ray emission with extended IR structures.  Thus, for convenience, we will refer to unresolved X-ray emission as ``diffuse,'' keeping in mind that this is simply shorthand for ``a mix of unresolved point source emission and truly diffuse X-ray structures.''

As mentioned above, we additionally show the ACIS binned event data in two false-color images for selected regions of each target.  These are often cluster centers or other interesting clumps of X-ray sources.  For all such images, ACIS soft (0.5--2~keV) events are shown in red, hard (2--7~keV) events in green; the binsize is one ``sky pixel'' (0.5$\arcsec$; since ACIS X-ray events have real-valued positions, an image can be made with whatever binsize is helpful to the eye).  In the superposition of these two event images, pixels that contain both soft and hard events appear yellow.
Point source extraction regions are outlined with blue polygons.  Those polygons may come from different ObsIDs, so they do not all have the same default size and orientation.
% See Pat's e-mail from 8 Dec 2017 in ``MOXCs'' mail folder -- which polygon is shown is complicated -- it encodes the deepest observation except for crowded sources.
Superposed on the event images is a blue image; this is either the ACIS full-band diffuse emission or {\em Spitzer}/IRAC 8~$\mu$m emission.

Throughout this section, we occasionally note X-ray spectral fit parameters for massive stars or other bright X-ray sources of particular significance in MSFRs.  Massive stars generate X-rays via a variety of emission mechanisms \citep[e.g.,][]{Gudel09}.  Individual massive stars emit soft X-rays ($<$1~keV) from shocks caused by velocity differences in their line-driven winds \citep[see CCCP X-ray spectral fit examples in][]{Gagne11,Naze11}.  Magnetic massive stars can generate harder X-rays (up to a few keV) when their winds are channeled along magnetic field lines and collide \citep{Babel97}.  In massive binaries, hard X-rays (several keV) can be generated when the powerful winds from the two massive stars collide; these ``colliding-wind binaries'' (CWBs) can be bright X-ray sources and have been studied extensively with {\em XMM-Newton} and \Chandra \citep[e.g.,][]{Rauw16}.

These X-ray spectral fits were performed with {\em XSPEC} \citep{Arnaud96} usually employing the absorption model {\em TBabs} with solar abundances and the thermal plasma emission model {\em apec}; the model form used is {\em TBabs*apec} unless otherwise noted.  Other bright X-ray sources are often collapsed objects with synchrotron X-ray emission; their X-ray spectra are fitted with absorbed power law models.  The X-ray luminosities ($L_X$) are intrinsic (absorption-corrected) and calculated for full-band (now 0.5--8~keV, the typical band used for ACIS point source fits in the literature).  We intentionally do not give errors on fit parameters to emphasize that these are rough, preliminary characterizations of source spectra.  

Since soft X-rays are readily absorbed by intervening material and such absorption may be strong (and spatially complex) in MSFRs, our estimates of X-ray luminosities are often lower limits.  Especially in massive stars, soft X-ray plasma components are likely to be present but may be completely absorbed by heavy obscuration, thus they remain uncharacterized by our ACIS spectra.  The same holds true for diffuse X-ray emission; its spatial complexity is probably due to a combination of absorption by intervening material and displacement of hot plasmas by colder ISM structures.

% I'm saving spectral fits for point sources as moxc2.xcm in each source's /photometry directory.

As described above, our analysis machinery uses image reconstruction to find candidate X-ray point sources.  Regions of bright diffuse emission (e.g., PWNe or SNRs) can be erroneously over-reconstructed in this process, leading to spurious point sources in our catalogs.  We have chosen to leave such suspicious point sources in place, as long as they satisfy all point source validity criteria, because PWNe or SNRs can certainly have real X-ray point sources superposed on them.  We caution users of our catalog to be aware of this decision and to employ more sophisticated means (such as searching for multiwavelength counterparts) to assess the validity of our sources that lie in close proximity to pulsars or in other regions of highly-structured, bright diffuse X-ray emission.

\clearpage
%-----------------------------------------------------------------------------
\subsection{NGC~6334 \label{sec:n6334}}
% NGC 6334 (including 2 GM24 ptgs) -- 4674 point sources
% SF Handbook reviews:
% NGC 6334 -- {Persi08} 
% GM 24 -- {Tapia08} 
% More than a cluster of clusters -- rather an ensemble of major MSFRs, like G333.
% At 1.3 kpc, 4*pi*D^2 = 2.0225e44.

Using the example of the massive filamentary infrared dark cloud (IRDC) known as ``Nessie,'' \citet{Jackson10} suggested that multiple star formation sites can form at regular intervals along a cylindrical IRDC via the ``sausage instability.''  This instability was originally described by \citet{Chandra53} and is essentially the analog of Jeans collapse in a self-gravitating sphere, applied now to a self-gravitating fluid cylinder.  The dominant path for the intense star formation that results in MSFRs is now recognized to be such massive molecular filaments \citep{Goodman14}, structures up to 100~pc long formed by supersonic flows that compress the gas of the ISM into a web of dense, cylindrical structures criss-crossing GMCs \citep{Andre14}.  These pressurized ``ridges'' are crucibles for creating MSFRs; material flows down the filaments, feeding hot cores and clumps that eventually collapse to become massive stars and star clusters \citep{Tackenberg14}.

Our closest example is the G352 GMC, a 90-pc-long massive filament made famous by \Herschel \citep{Russeil13}.  G352 hosts the two multi-MSFR complexes NGC~6357 and NGC~6334 \citep{Persi08}; such complexes are known as ``clusters of clusters'' \citep[e.g.,][]{Bastian07,Elmegreen08}.  Each of these sports a string of MSFRs with multiple bubbles and degree-sized ``bowls'' filled with hot plasma (MOXC1); they are situated on opposite sides of G352's main filament.

NGC~6334 is the best nearby example we have of the ongoing transfiguration of such a massive molecular filament into stars; it is thought to contain $\sim$175 O--B3 stars \citep{Russeil12} distributed among several MSFRs.  It hosts a large number of \hii regions of various sizes and evolutionary states, spread across $>2^{\circ}$ of the Galactic Plane \citep{Russeil16}.
%The earliest spectral type is HD319699, O5V((f)) in Russeil12 Table 2; it is WAY northwest of the main ridge, far off our ACIS mosaic.
Many of its most massive clusters line up along a dense molecular ridge with a complicated filament morphology and many subfilaments.  In some of these subfilaments, densities are high enough that gravity should dominate the energetics and they should be collapsing to form stars \citep{Russeil13}.  This is substantiated by a recent IR study \citep{Willis13} that identifies 2283 Class I and II young stellar objects in NGC~6334; these sources tend to cluster along the subfilaments and extend many parsecs beyond the main ridge.  As shown in MOXC1 and earlier studies, the three ACIS pointings along the main ridge capture several collapsing subfilaments and add a large number of Class III (diskless) pre-MS stars to the census of young stars populating those regions \citep{Broos13, Feigelson09, Ezoe06}. 

Recent maser parallax measurements by the VLBA BeSSeL Survey \citep{Wu14} and VERA \citep{Chibueze14} both yield a distance of $\sim$1.3~kpc for NGC~6334, which we adopt for MOXC2.  Readers should note, however, that some recent papers \citep[e.g.,][]{Russeil16,Tige17} use a distance of $\sim$1.75~kpc.

NGC~6334 serves as a transition target for us, illustrating the changes in our ACIS data analysis code and procedures between the MOXC1 analysis (circa 2013) and the current analysis for MOXC2.  The same three ObsIDs analyzed for MOXC1 are re-analyzed here, augmented by several surrounding archival ACIS-I and ACIS-S pointings and our recent \Chandra Cycle~17 General Observer (GO) data on the western side of NGC~6334's main filament, including the MSFR GM~24 (GM~1-24 in SIMBAD) \citep{Tapia08,Tapia09}.  

\begin{figure}[htb]
\centering
\includegraphics[width=0.95\textwidth]{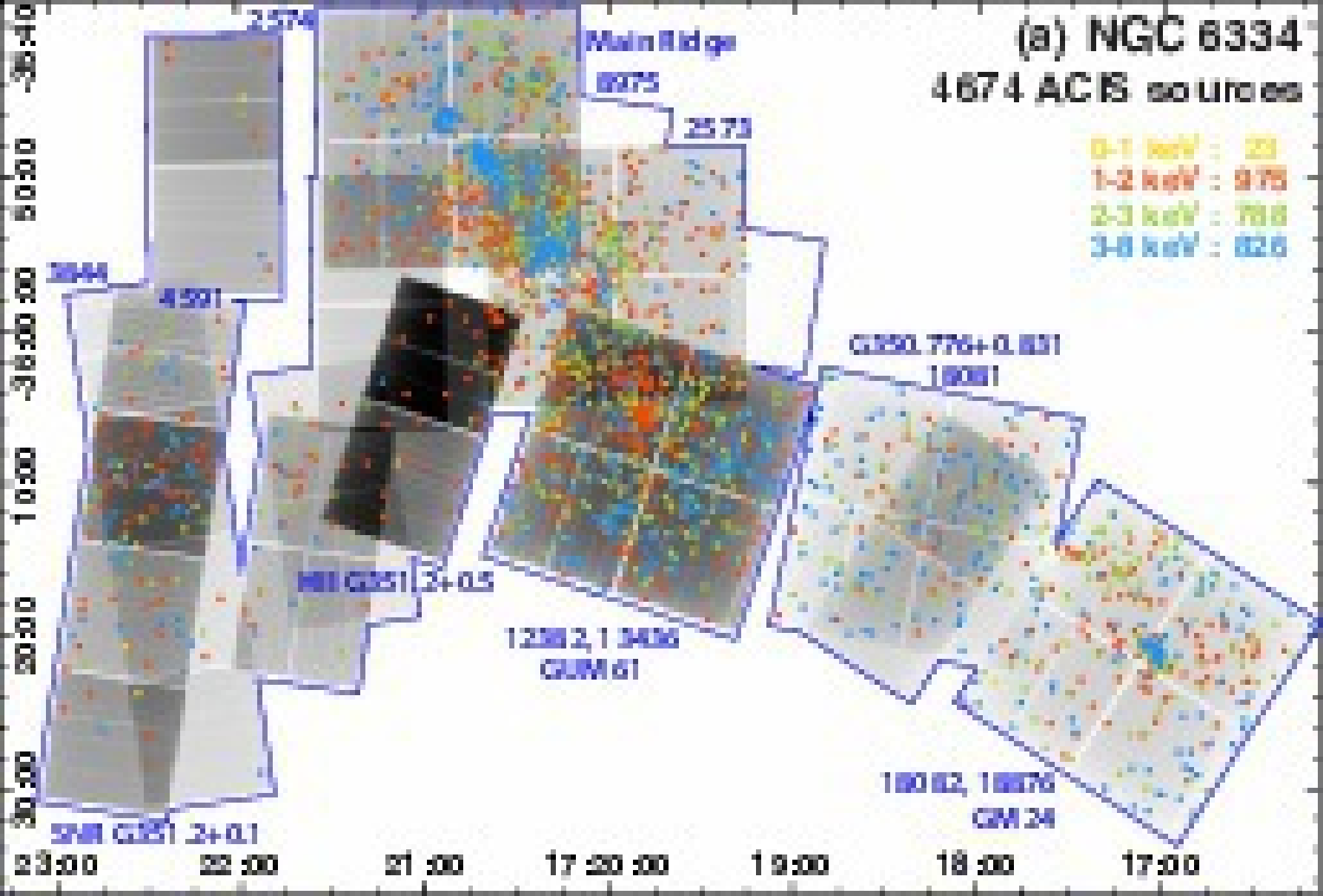}
\includegraphics[width=0.95\textwidth]{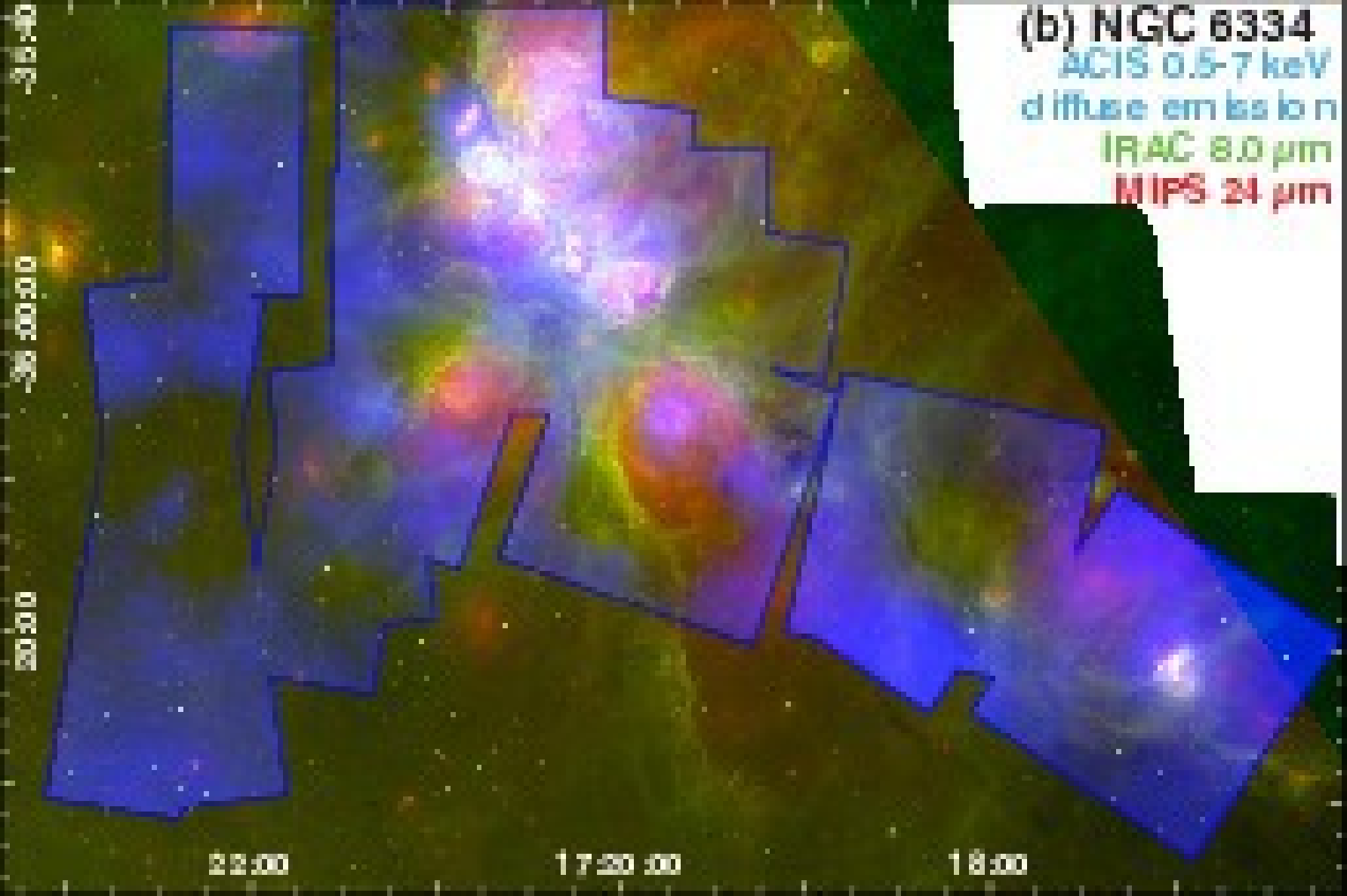}
\caption{NGC~6334.
(a) ACIS exposure map with brighter ($\geq$5 net counts) ACIS point sources overlaid; colors denote median energy for each source.  ObsID numbers and regions named in the text are shown in blue.
(b) ACIS diffuse emission in the \Spitzer context.  
\label{ngc6334.fig}}
\end{figure}

Our final $1.4^{\circ}$-wide ACIS mosaic is shown in Figure~\ref{ngc6334.fig}.  \Chandra typically characterizes diskless young stellar populations, both older pre-MS stars distributed across MSFRs and younger sources still embedded in their natal clouds.  That generalization appears to hold true for NGC~6334; widely-distributed X-ray sources are seen across the entire ACIS mosaic, extending far from the main ridge of MSFRs typically studied in NGC~6334.  

%\clearpage

\begin{figure}[htb]
\centering
\includegraphics[width=0.50\textwidth]{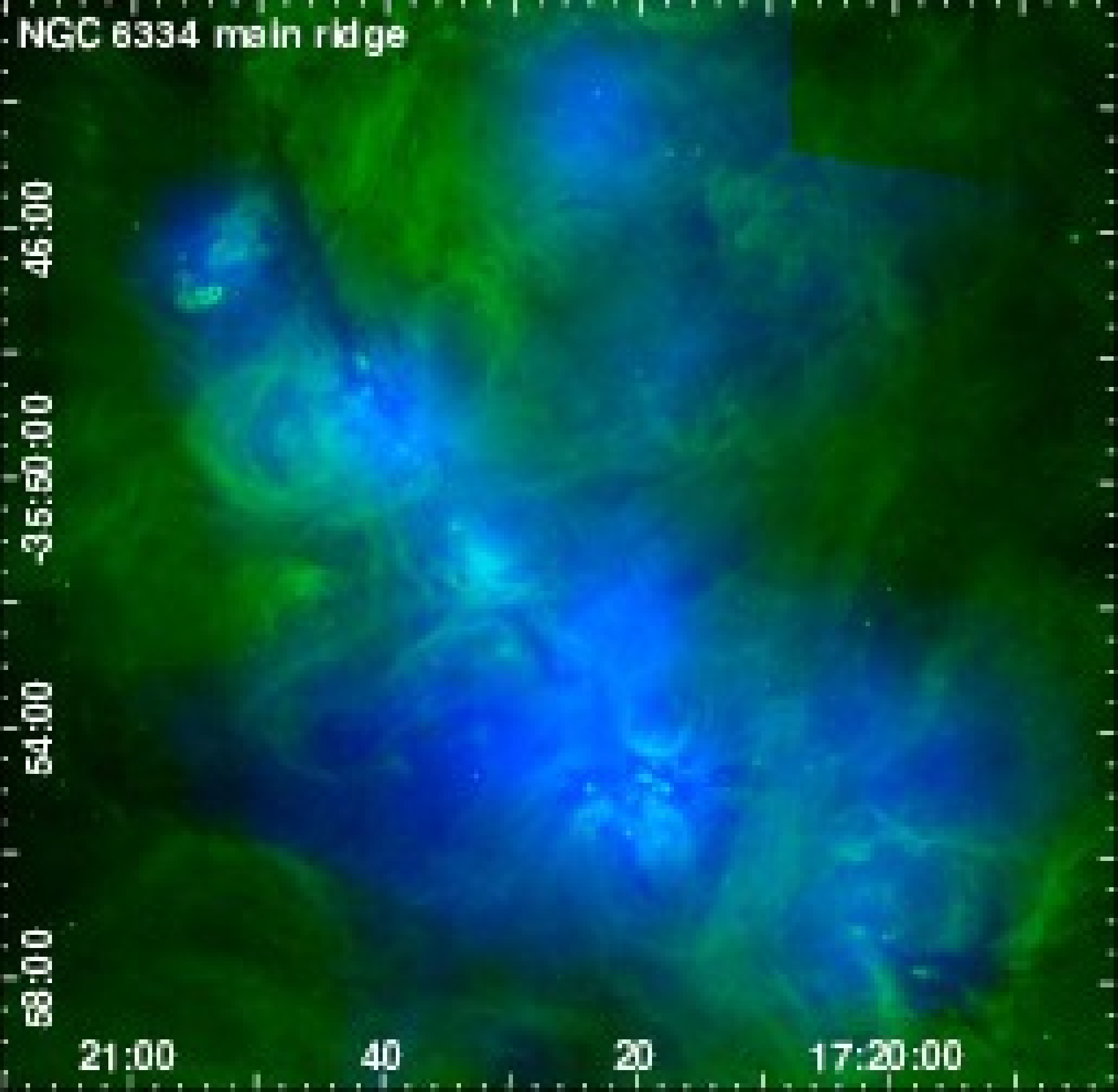}
\caption{NGC~6334's main ridge of MSFRs.
Zoomed version of Figure~\ref{ngc6334.fig}(b) for the central clusters on the main star-forming ridge, now with the MIPS data omitted because they are largely saturated.  This figure is included to emphasize the extensive diffuse X-ray emission associated with this famous star-forming ridge.
%(b) Comparing the MOXC1 analysis of one of the MSFRs in NGC~6334's main ridge to the MOXC2 analysis.  The greyscale image shows full-band ACIS events; the blue polygons are MOXC2 point source extraction regions (centered on MOXC2 source positions).  Sources found by both algorithms (green) are shown with MOXC1 positions.  Magenta sources were found only by MOXC1; cyan sources were found only by MOXC2.
\label{ngc6334ridge.fig}}
\end{figure}

The original ACIS pointings on the main ridge have been described at length \citep{Feigelson09, Ezoe06} and were presented in MOXC1, so we do not discuss them further here, except to note that they are pervaded by diffuse X-ray emission (Figure~\ref{ngc6334ridge.fig}).  %These original pointings also give us the opportunity to compare our MOXC1 source list to that from MOXC2 (Figure~\ref{ngc6334ridge.fig}(b)).  The sample region shown is a crowded part of the main ridge, observed at a range of off-axis angles.  Both algorithms recovered 79 sources in this sample field; an additional 13 sources were unique to MOXC1 and 17 unique to MOXC2.  We believe that the MOXC2 source list is more robust, but many factors contribute to the validity of faint sources, so both epochs of analysis should be considered by investigators interested in the most complete catalog of NGC~6334 members.
Comparing the MOXC1 and MOXC2 catalogs in this region (where the two epochs of analysis used the same data), we find that both recover the same brighter sources ($\geq$12 net counts) at all off-axis angles.  For faint sources, the catalogs differ in several ways.  Sometimes one analysis split a group of counts into a close pair of sources while the other found a single source.  Sometimes faint sources were deemed valid in one analysis but not the other; this is especially true at large ($>$7$\arcmin$) off-axis angles.  These results are expected, given the strong interdependencies between faint candidate sources, the iterative nature of source validation, and the many decisions made, both algorithmic and by the analyst during visual reviews.  Close examination of source positions shows that the inter-ObsID alignment is improved in MOXC2.  Given this and the algorithmic improvements made over the years between MOXC1 and MOXC2, we recommend the use of the MOXC2 source list.

%Pat says:  1254 of the 1627 MOXC2 sources in the ROI formally match to MOXC1.  That leaves 373 MOXC2 sources not formally matched and 75 MOXC1 sources not formally matched.  As we saw in NGC6334+GM24/data/counterparts/counterparts_log.txt, only ~10% of these unmatched sources (shown in the table) have more than 10 net counts. 

%\clearpage
\subsubsection{GUM~61}
\begin{figure}[htb]
\centering
\includegraphics[width=0.49\textwidth]{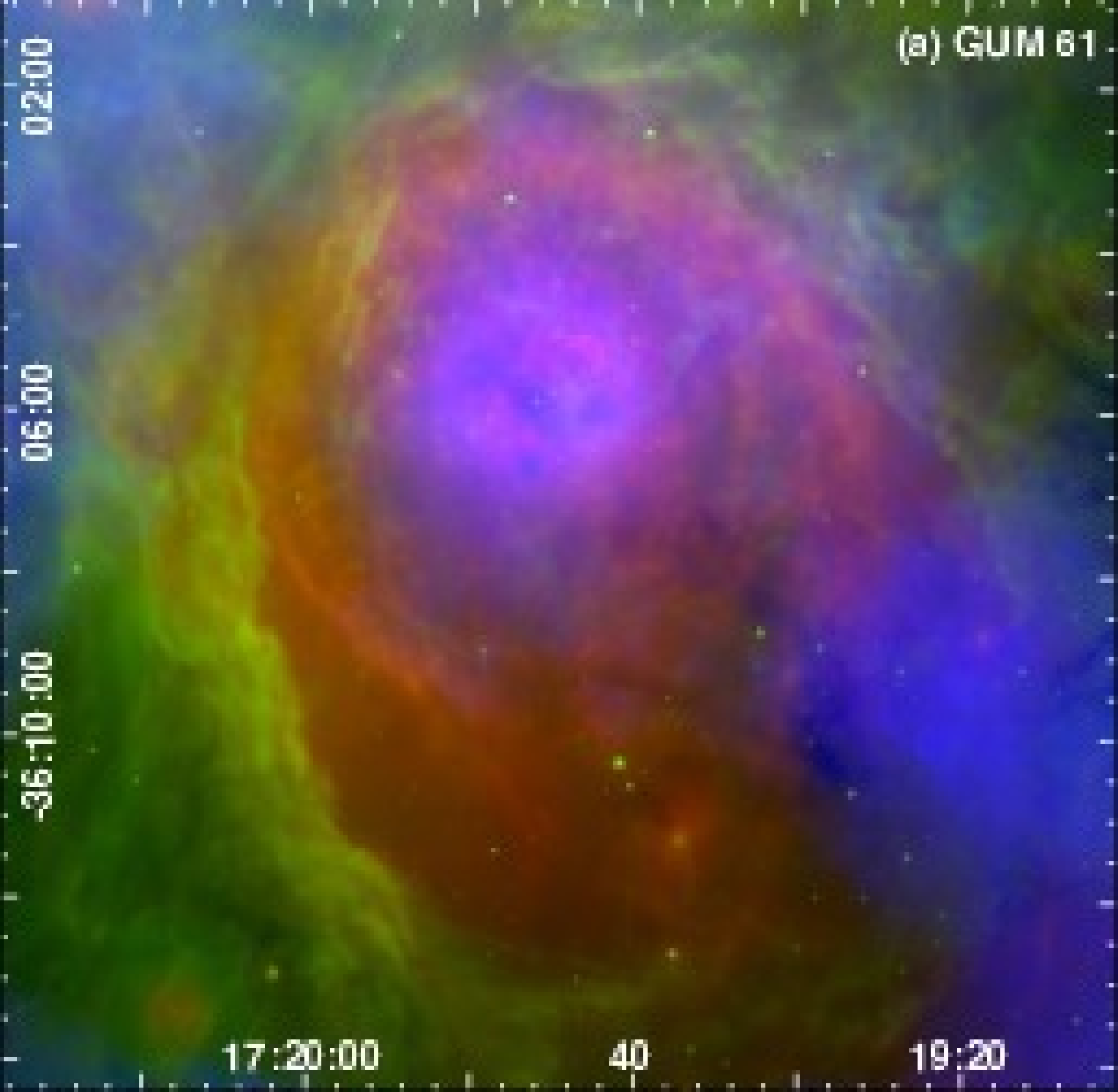} % Gum 61
\includegraphics[width=0.50\textwidth]{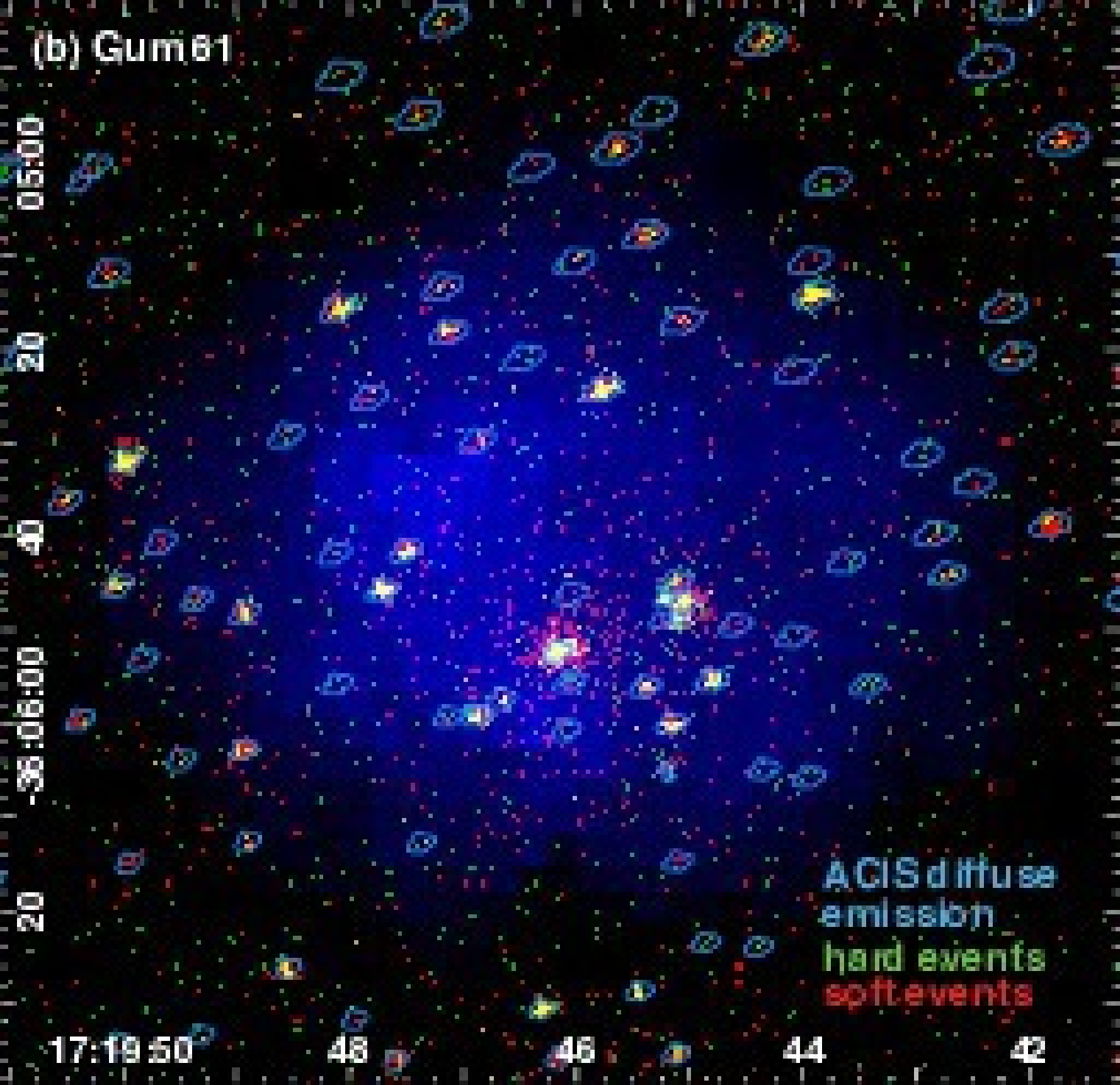}
\caption{GUM~61.
(a) Zoomed version of Figure~\ref{ngc6334.fig}(b) for the large \hii region GUM~61.  % Name from Russeil16
(b) ACIS event data and diffuse emission for the center of GUM~61. 
\label{GUM61.fig}}
\end{figure}

Continuing down the G352 molecular filament southwest of the main ridge in NGC~6334, we find deep archival ACIS data on the large \hii region GUM~61 \citep{Russeil16} (Figure~\ref{GUM61.fig}).  This hosts a populous diffuse cluster, as shown by the large number of X-ray sources seen across the entire ACIS-I field in this pointing (Figure~\ref{ngc6334.fig}(a)).

At the center of the cluster (Figure~\ref{GUM61.fig}(b)), the piled ACIS source c1934 (CXOU~J171946.16-360552.2) matches HDE~319703A, a spectroscopic binary of type O7V((f)) + O9.5V \citep{Maiz16}.  A spectral fit to the pileup-corrected spectrum of c1934 (from ObsID~13436) requires two thermal plasma components, {\em TBabs(apec1 + apec2)}, for an acceptable fit.  This yields $N_H = 1.1 \times 10^{22}$~cm$^{-2}$, $kT1 = 0.7$~keV, $kT2 = 1.9$~keV, and $L_X = 1.6 \times 10^{32}$~erg~s$^{-1}$.  The hard thermal plasma component dominates the spectrum.  In ObsID~12382, obtained four days after ObsID~13436, c1934 shows slightly different spectral parameters ($N_H = 1.4 \times 10^{22}$~cm$^{-2}$, $kT1 = 0.6$~keV, $kT2 = 2.5$~keV) and is brighter, with $L_X = 2.0 \times 10^{32}$~erg~s$^{-1}$.  The softer plasma component dominates the spectrum below 2~keV.  A hard, variable X-ray spectrum and changing X-ray luminosity often indicate wind interactions in binary systems; this may be yet another CWB discovered by {\em Chandra}.

Nearby is HD~319703B, a visual binary of type O6V((f))z \citep{Sana14}, matching ACIS source c1800 (CXOU~J171945.05-360547.0).  It has 432 net counts and a median energy of 1.3~keV.    % Lightcurve not variable.
The spectral fit yields $N_H = 1.4 \times 10^{22}$~cm$^{-2}$, $kT = 0.5$~keV, and $L_X = 9 \times 10^{31}$~erg~s$^{-1}$.  Source c1800 has four close neighbors in the ACIS data. 
%HD~319703C, an early-B star that sits just north of HD~319703B \citep{Sana14}, may be ACIS source ??? (the neighbor to the north), but I don't have a position for HD~319703C!
We also detect HD~319703D, spectral type O9.5:Vn \citep{Sana14}, as ACIS source c2162 (CXOU~J171948.94-360602.8), with 98 net counts and a median energy of 1.4~keV.  For the spectral fit, we find $N_H = 1.9 \times 10^{22}$~cm$^{-2}$, $kT = 0.6$~keV, and $L_X = 3 \times 10^{31}$~erg~s$^{-1}$.  % Lightcurve not variable.

There is prominent diffuse X-ray emission near these massive stars and at the southwest (open) edge of the IR bubble (Figure~\ref{GUM61.fig}(a)), where \citet{Russeil16} note outflow in the kinematics of this region.  This is probably a good example of diffuse X-ray emission tracing hot gas from massive star feedback in an \hii region, flowing out of that region to enrich and heat the surrounding ISM.  It might be compared to M17, a spectacular example of this phenomenon \citep[][MOXC1]{Townsley03}.
%Is the soft diffuse emission in Wolk's pointing just due to spill-over from the piled source?  It's clearly not just unresolved pre-MS stars, or there would be a hard component to it.  Compare spectrum of piled src to spectrum of diffuse ball-of-fuzz.
%Cite paper suggesting that GUM 61 or its outflow has seen a SN?

%\clearpage
\subsubsection{Southwest Pointings}
\begin{figure}[htb]
\centering
\includegraphics[width=0.49\textwidth]{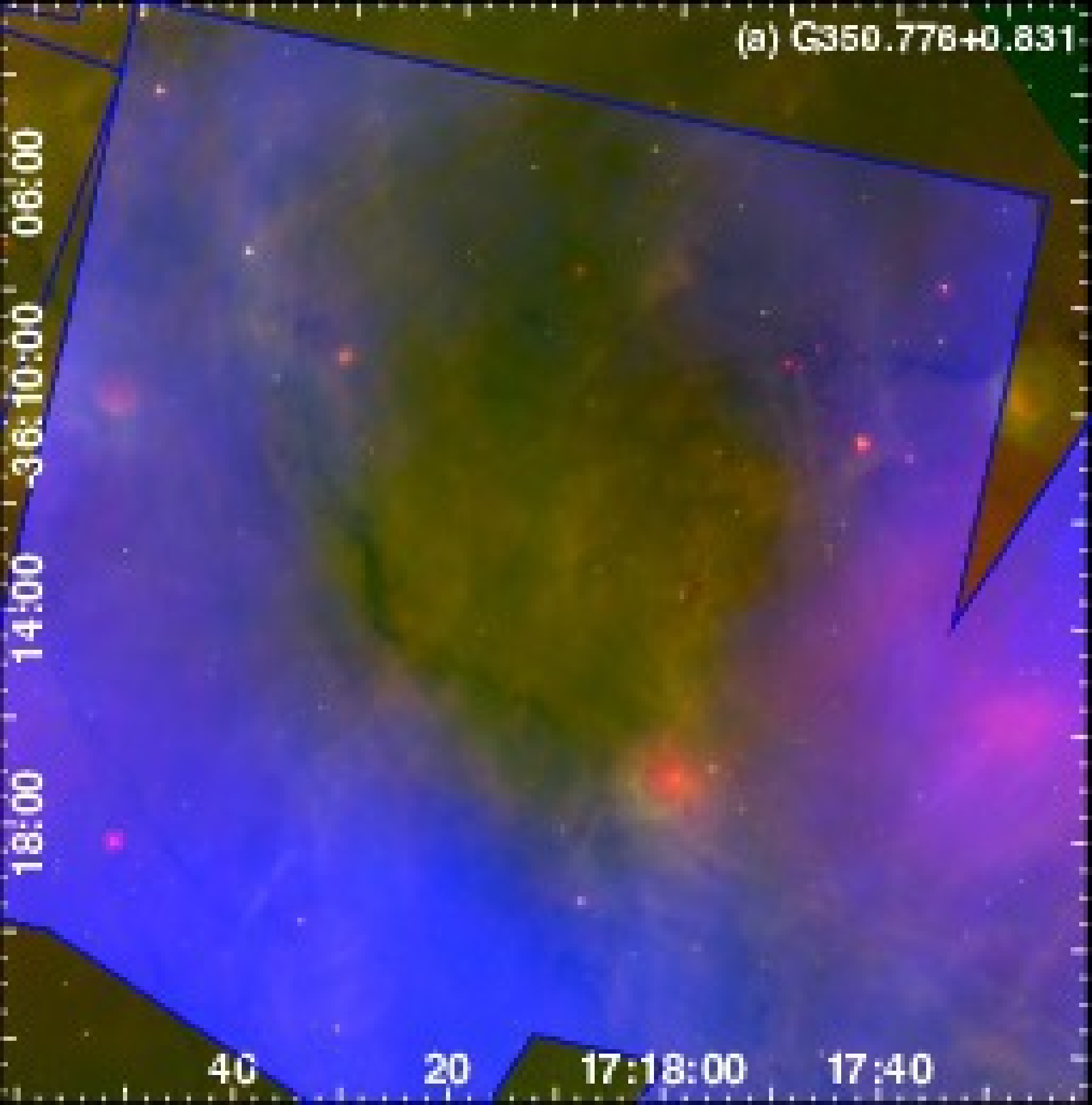}
\includegraphics[width=0.49\textwidth]{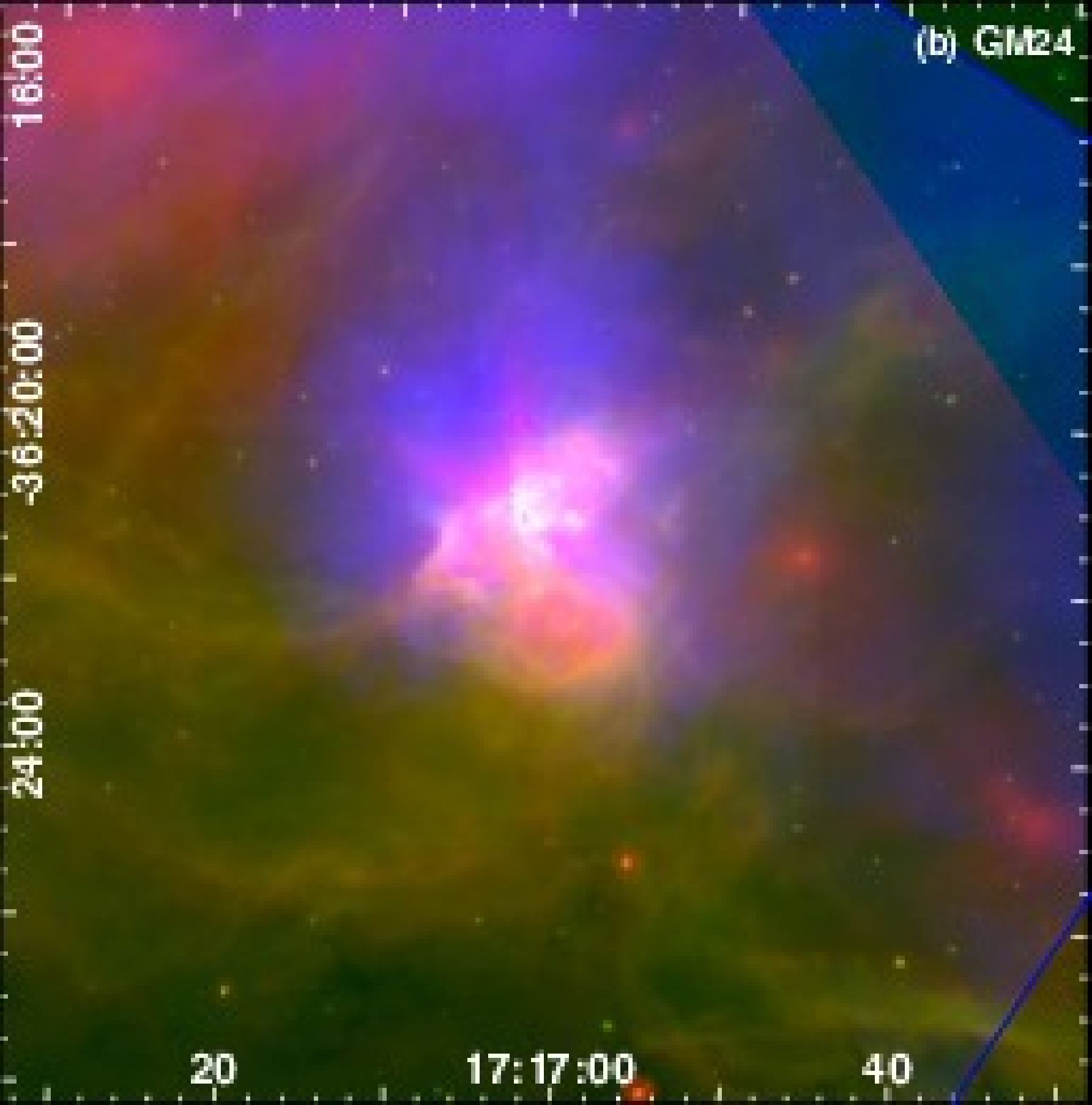}
\includegraphics[width=0.50\textwidth]{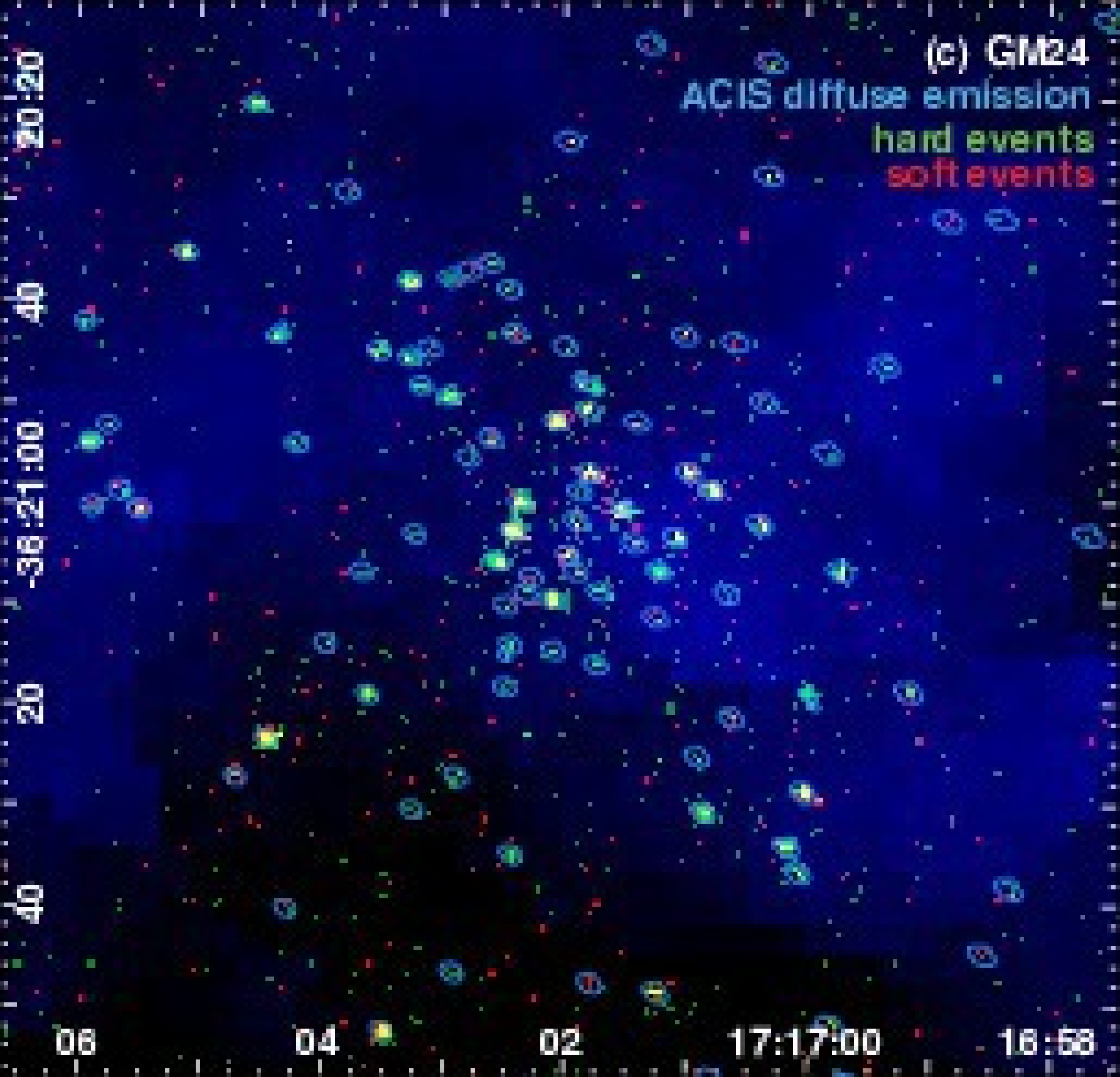}
\caption{NGC~6334 southwest, along the giant filament.
(a) Zoomed version of Figure~\ref{ngc6334.fig}(b) for the G350.776+0.831 region.    
(b) Zoomed version of Figure~\ref{ngc6334.fig}(b) for the isolated, very young MSFR GM~24.
(c) ACIS event data and diffuse emission for the embedded cluster in GM~24. 
\label{GM24.fig}}
\end{figure}

Extending the ACIS mosaic southwest along NGC~6334's natal molecular filament past GUM~61, our ACIS pointing named G350.776+0.831 (ObsID~18081) captures a complex network of subfilaments (Figure~\ref{GM24.fig}(a)) and known clumps of pre-MS stars \citep{Russeil13,Willis13}.  It is centered on the base of a large ``bowl'' of bright 8~$\mu$m emission seen opening to the northwest in Figure~\ref{ngc6334.fig}(b); the main molecular filament forming NGC~6334's backbone makes up the southern edge of this 30$\arcmin$-diameter structure.  This pointing shows substantial shadowing or displacement of hot plasma in a large oval region at the base of the bowl, but other parts of the bowl contain diffuse X-ray emission, as does the large elongated cavity on the southern side of the main filament, opening from Gum~61 to the southwest.

The ACIS mosaic continues down the main filament to the isolated, embedded IR cluster GM~24 (Figures~\ref{GM24.fig}(b) and (c)).  This very young, midsized cluster ($\sim$250~M$_{\odot}$) \citep{Tapia09} is ionized by an O8 star \citep{Bik06}; it sits at the southern edge of the \Spitzer bowl, separated from the cluster-of-clusters complex by over 40$\arcmin$.  GM~24 could have formed as a result of the bowl's dynamical influence on NGC~6334's main molecular filament.  GM~24 appears to support a strong bipolar outflow that is carving cavities in the surrounding GMC \citep{Tapia09}.  ACIS finds a strong concentration of X-ray point sources coincident with GM~24, plus a large distributed population.  Hot plasma clearly fills the western bowl above the main filament; this pointing also captures the edge of the southwestern elongated cavity and again finds diffuse X-ray emission there.

The piled ACIS source GM24$\_$c717 (CXOU~J171701.53-362100.6), near the center of GM~24, has a reduced extraction aperture due to a close, faint neighbor (c715) to the northwest.   
% This is IRS~28 according to SIMBAD.  Bik06 gives a position for the IRAS 17136-3617 O8 star as 17:17:01.5 -36:20:57.7.  This is closer to my piled source than to the IRS3 complex, which Tapia09 claim to be "the nucleus of the complex."  
The position for the O8 star from \citet{Bik06} is 3\arcsec\ north of GM24$\_$c717, so it is not clear that this X-ray source is the counterpart of the O8 star.  There is no X-ray source close to the O8 star's position.

We fit GM24$\_$c717's pileup-corrected spectrum (from ObsID~18876) with the same simple spectral model used for other massive stars.  Unlike the GUM~61 sources described above, GM24$\_$c717 has a very hard spectrum, with $N_H = 4 \times 10^{22}$~cm$^{-2}$, $kT = 9$~keV, and $L_X = 2 \times 10^{32}$~erg~s$^{-1}$.  Any additional soft thermal plasma component could easily be absorbed away by this high column, so the luminosity we report is a lower limit.  In the second GM~24 dataset taken three days later (ObsID~18082), the X-ray luminosity of GM24$\_$c717 is lower by a factor of 20 but the spectrum remains very hard.  Such extremely hard X-ray spectra and variability have certainly been seen in other young stars \citep[e.g., W51~IRS2E in MOXC1 and two early-B stars in IRAS~20126+4104,][]{Anderson11}, but this is still an exceptional source due to its high luminosity and extreme variability; it deserves further attention.

%\clearpage
\subsubsection{\hii 351.2+0.5}
\begin{figure}[htb]
\centering
\includegraphics[width=0.48\textwidth]{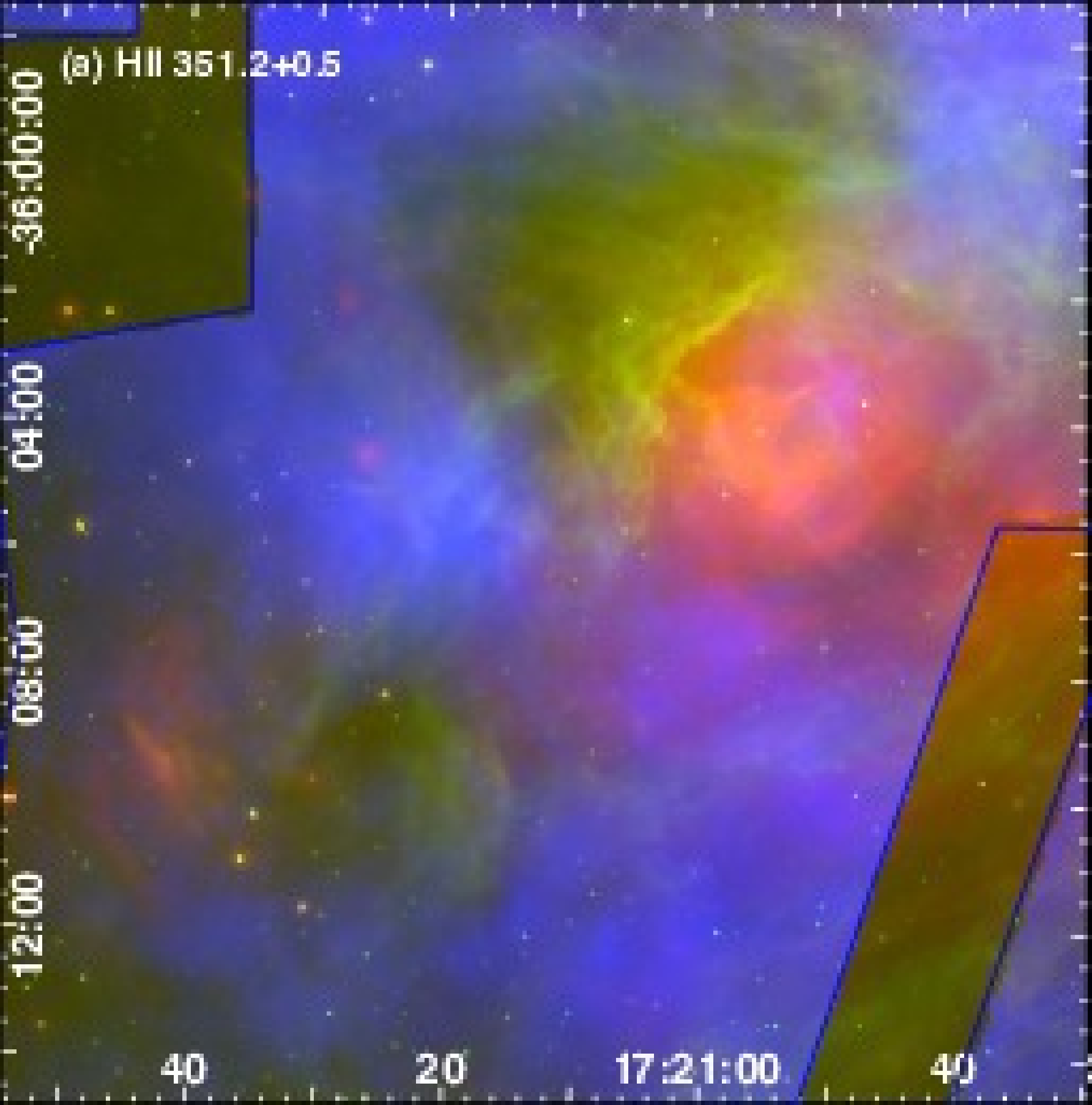}
\includegraphics[width=0.50\textwidth]{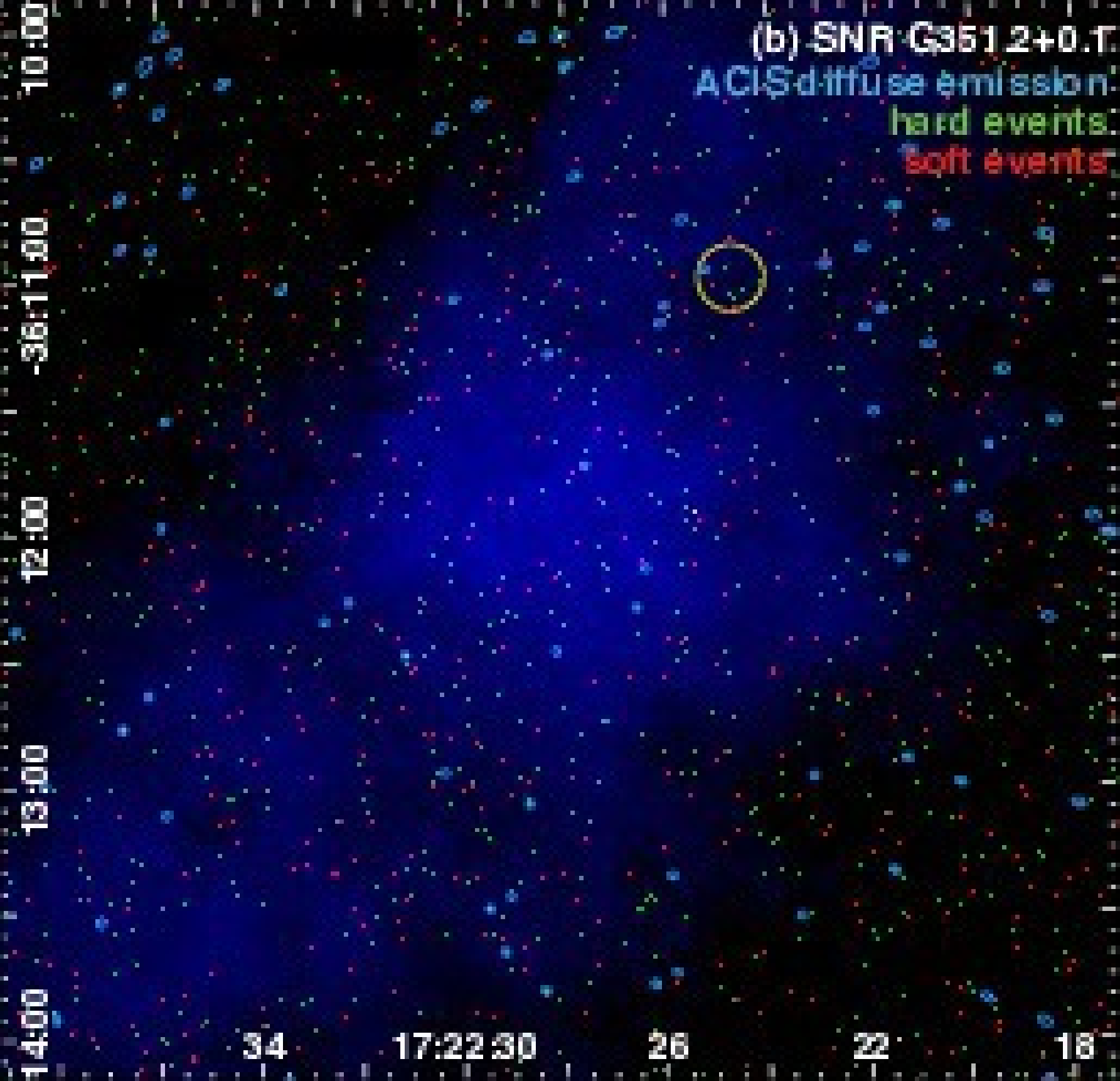}
\caption{NGC~6334 southeast.
(a) Zoomed version of Figure~\ref{ngc6334.fig}(b) for \hii 351.2+0.5.  
(b) ACIS event data and diffuse emission for the SNR~G351.2+0.1 region.  A yellow circle indicates the central radio object \citep{Becker88}. 
\label{ngc6334SE.fig}}
\end{figure}

East of GUM~61 lies another prominent \hii region south of NGC~6334's main ridge, called \hii 351.2+0.5 by \citet{Russeil16} (Figure~\ref{ngc6334SE.fig}(a)).  Comparatively few X-ray point sources are found in this part of our ACIS mosaic, even though it is captured by three ACIS pointings, because it is only observed far off-axis in each of these pointings.  This region does serve to illustrate the usefulness of including off-axis CCDs in our mosaics; diffuse X-ray emission is seen throughout this area.  It appears at the center of \hii 351.2+0.5 and at the bubble rim, perhaps indicating hot plasma flowing out its southeast side.

The brightest X-ray source in this part of the ACIS mosaic is c6709 (CXOU~J172052.63-360420.6); it is the X-ray counterpart to the massive binary HD~156738 \citep{Sana14}, spectral type O6III((f)) \citep{Sota14}, located near the center of \hii 351.2+0.5.  Source c6709 has 948 net counts and a median energy of 1.2~keV.  % Lightcurve not variable.
As for ACIS source c1934 above, c6709 requires a two-temperature fit, {\em TBabs(apec1 + apec2)}, although the plasmas are much softer here.  Fit parameters are $N_H = 1.4 \times 10^{22}$~cm$^{-2}$, $kT1 = 0.2$~keV, $kT2 = 0.5$~keV, and $L_X = 5.5 \times 10^{32}$~erg~s$^{-1}$.  These soft plasmas are consistent with instability-driven wind shocks in the individual stars.

\subsubsection{SNR~G351.2+0.1}
Two archival ACIS-S observations of SNR~G351.2+0.1 are included in our mosaic, totaling $\sim$53~ks exposure.  This composite SNR sits to the southeast of NGC~6334 and was observed by \Chandra to search for its neutron star and pulsar wind nebula; no clear detection was made.
% No papers on the Chandra data.
Radio observations show a $4\arcmin \times 6\arcmin$ shell, flattened along the northern side, with a 16$\arcsec$-diameter central object \citep{Becker88} (shown in (Figure~\ref{ngc6334SE.fig}(b)).  \citet{Dubner93} note that the flattened northern edge of the remnant may indicate that it is encountering dense molecular material that is slowing its expansion in this direction.  Perhaps this is an indication that the SNR is associated with the G352 GMC, although the SNR distance is not well-determined.

There is no bright ACIS source that is clearly the counterpart of the central radio object.  The wider field, however, shows a large number of X-ray sources concentrated in the vicinity of the SNR (Figure~\ref{ngc6334.fig}(a)).  Perhaps these constitute another young cluster; this should be investigated at longer wavelengths.  Excising those point sources leaves some faint, soft diffuse X-ray emission apparent in Figure~\ref{ngc6334SE.fig}(b); this may be the feeble X-ray signature of the SNR.  A large region surrounding this faint diffuse emission is conspicuously lacking in diffuse X-rays (Figure~\ref{ngc6334.fig}(b)); cold material in the G352 GMC may be shadowing any hot plasma generated by the SNR or NGC~6334 there.

\clearpage
%-----------------------------------------------------------------------------
\subsection{W75N \label{sec:w75n}}
% W75N -- 356 point sources
% Mentioned on p.58 of SF Handbook North (in Cygnus)
% At 1.3 kpc, 4*pi*D^2 = 2.0225e44.

W75N is part of the great concentration of star formation in the Cygnus~X~North molecular cloud complex \citep{Reipurth08}.  It contains several UCH{\scriptsize II}Rs ionized by early-B stars \citep{Shepherd04} and exhibits complicated dynamics from multiple molecular outflows \citep{Shepherd03,Davis07}.  \citet{Persi06} find at least 25 cluster members based on their IR excess; we find that several of these have X-ray counterparts.  \citet{Kumar07}, also using near- and mid-IR data, find a distributed population of young stars across the region, tracing the molecular filaments.  %A recent near-IR study \citep{Maia16} suggests an excess of intermediate-mass stars compared to a standard initial mass function.

\begin{figure}[htb]
\centering
\includegraphics[width=0.465\textwidth]{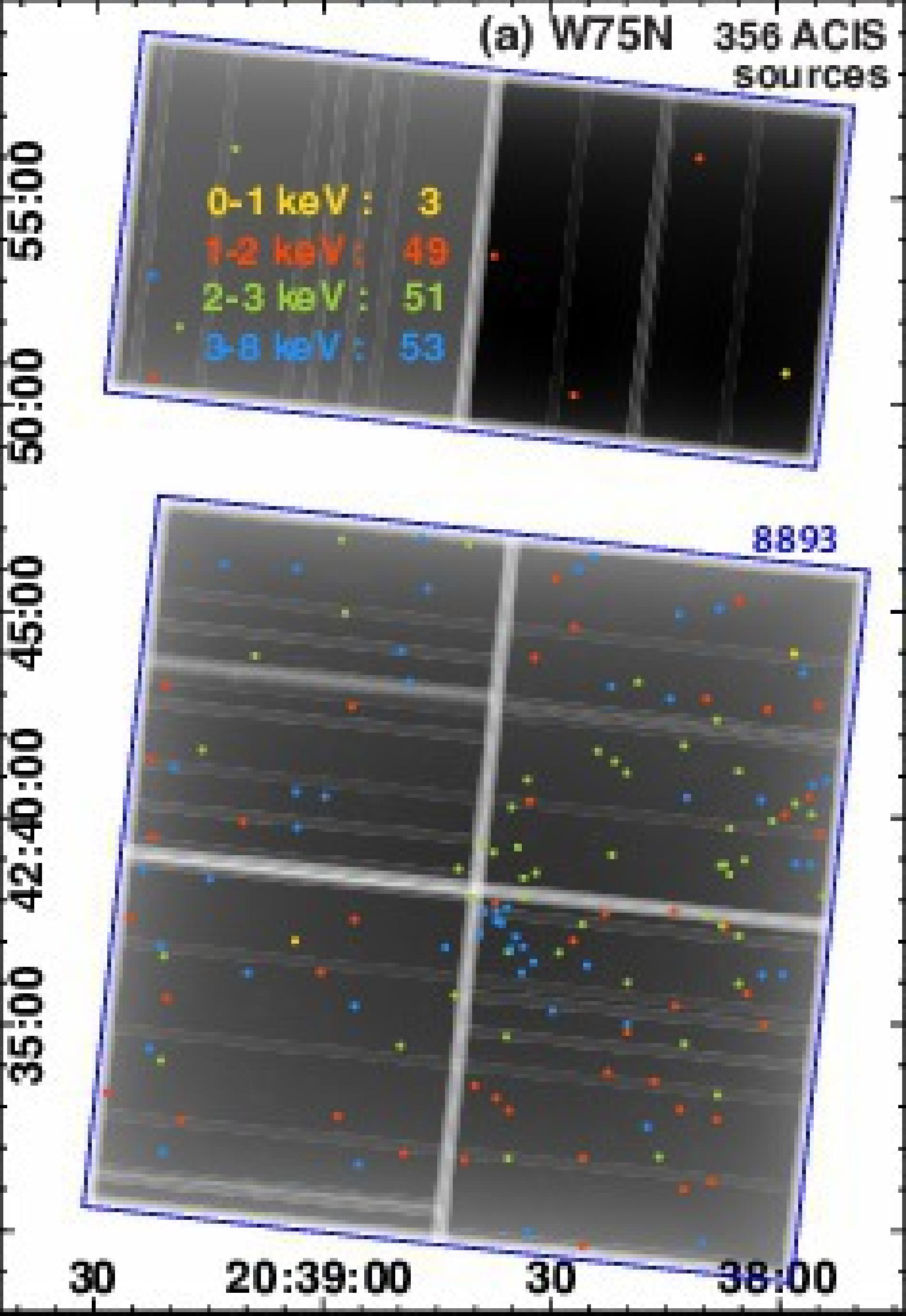}
\includegraphics[width=0.48\textwidth]{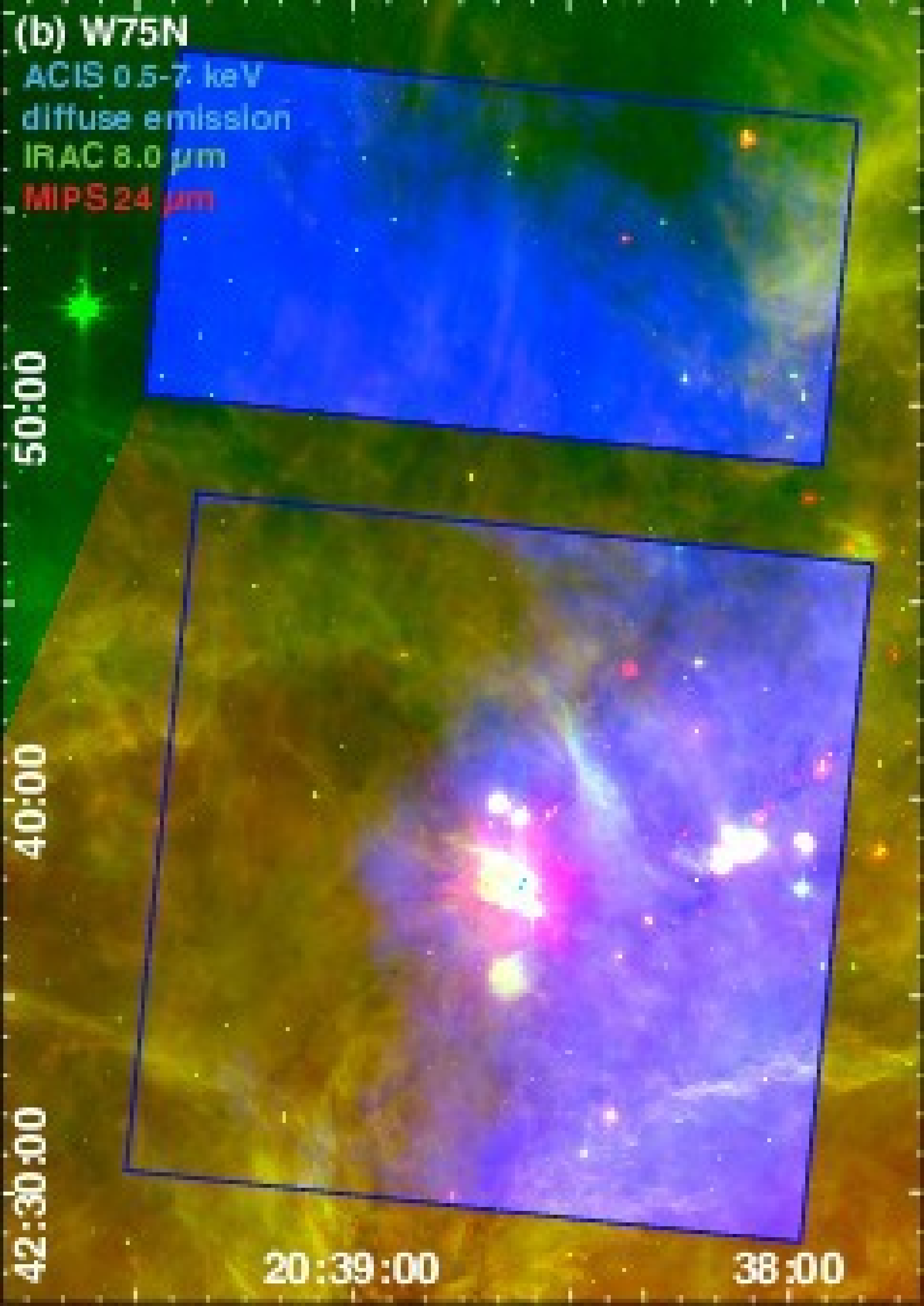}
\includegraphics[width=0.48\textwidth]{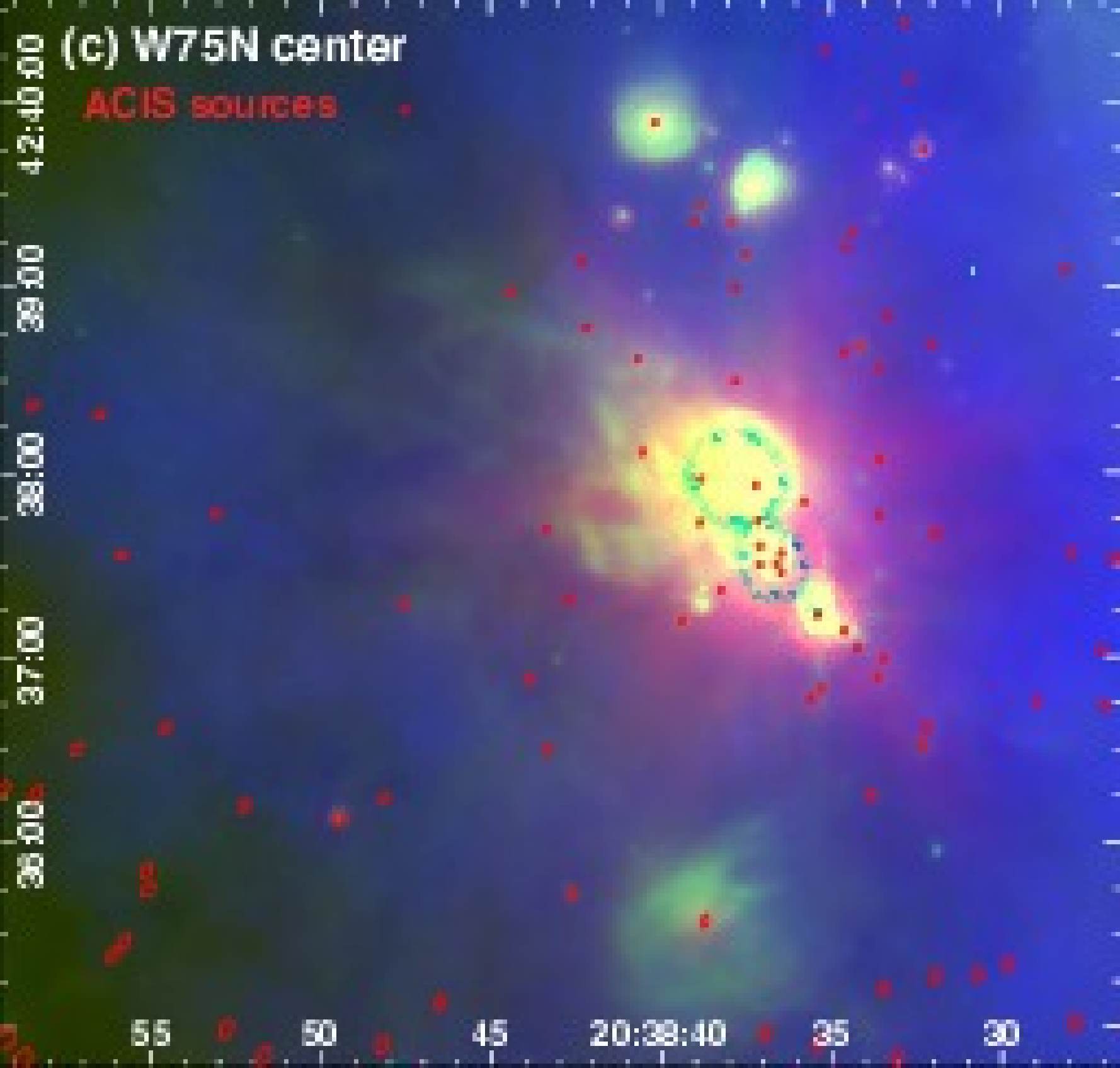}
\includegraphics[width=0.48\textwidth]{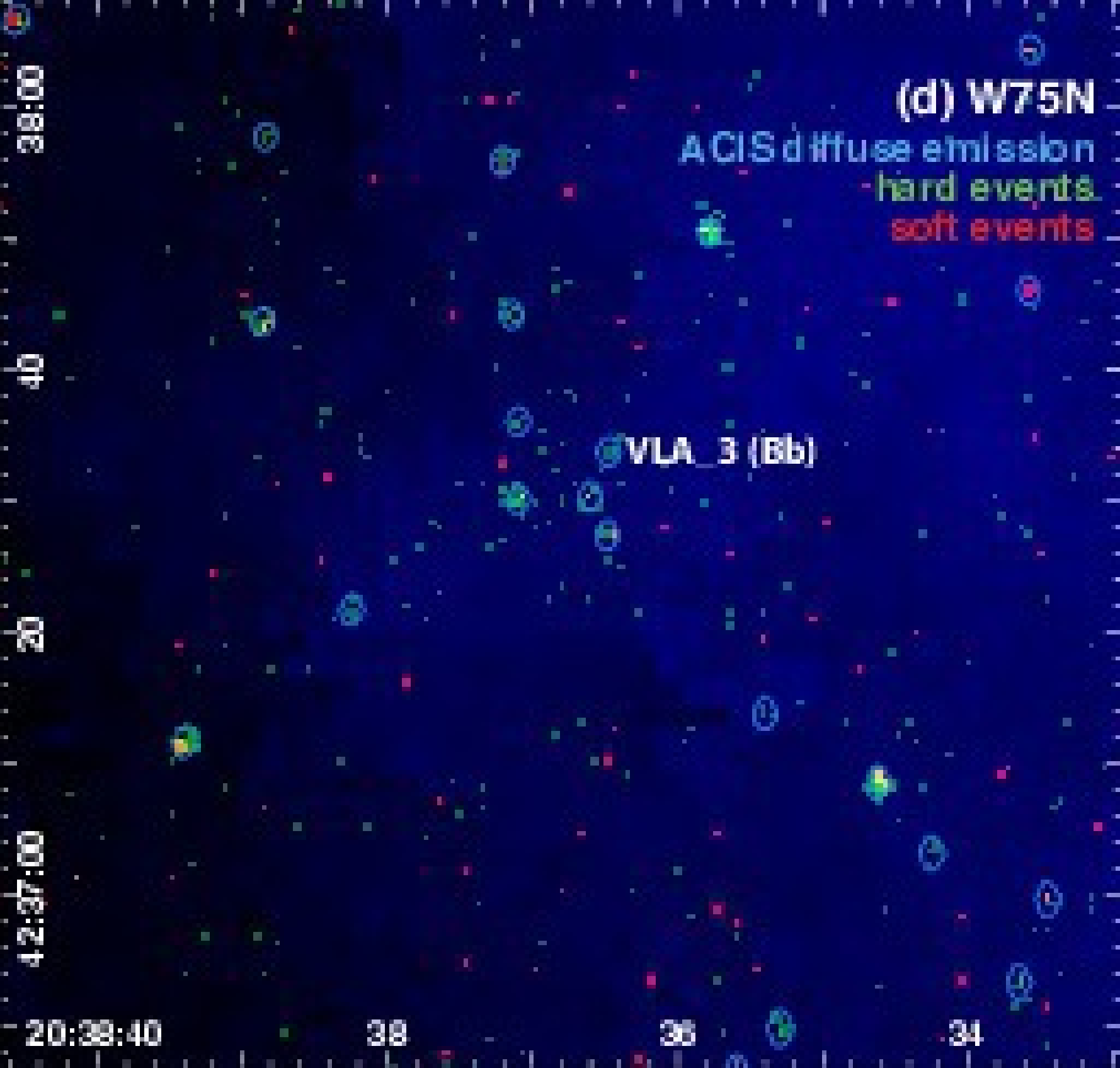}
\caption{W75N.
(a) ACIS exposure map with brighter ($\geq$5 net counts) ACIS point sources overlaid; colors denote median energy for each source.  The ObsID number is shown in blue.
(b) ACIS diffuse emission in the \Spitzer context.  
(c) Zoom of (b) showing the center of W75N and surrounding diffuse X-ray emission.  Green rings are artifacts due to MIPS saturation.  Red polygons show extraction regions for ACIS sources.
(d) ACIS event data and diffuse emission at the center of W75N.  The hard X-ray counterpart to the radio \UCHIIR VLA\_3 (Bb) is marked.
\label{w75n.fig}}
\end{figure}

The single ACIS observation of W75N (Figure~\ref{w75n.fig}) shows a concentration of hard sources near the aimpoint.  ACIS source c202 (CXOU~J203836.47+423733.7) appears to be the X-ray counterpart to VLA\_3~(Bb), an \UCHIIR with Lyman continuum flux consistent with ionization by a B1 star \citep{Shepherd04}.  Source c202 has only five counts, but its median energy is 6.7~keV and all five counts have energies $>$6~keV.  The spectral fit %using cstat
gives $N_H \sim 30 \times 10^{22}$~cm$^{-2}$, $kT \sim 7$~keV, and $L_X \sim 8 \times 10^{30}$~erg~s$^{-1}$ but the fit parameters are not well-constrained.  This is just a rough characterization given the limited counts, but there is no doubt that this is an extreme X-ray source, exceptionally hard and highly obscured; such emission is not expected from a single B1 star. 

Diffuse X-ray emission is seen on the west side of W75N despite the high obscuration and short ($<$30~ks) observation.  It is brightest in the southwest corner of the ACIS-I image, outside the large dark arc in the \Spitzer images that runs vertically at RA $\sim$ 20:38:15.  In Figure~\ref{w75n.fig}(b) we have scaled the ACIS image to show the fainter diffuse emission surrounding the central cluster and filling apparent cavities in the \Spitzer emission.  Bright diffuse emission is seen far off-axis on the ACIS-S CCDs; this may be associated with the wider Cygnus~X star-forming complex.  The central region also shows extensive diffuse X-ray emission (Figure~\ref{w75n.fig}(c)) extending all the way down to the cluster center (Figure~\ref{w75n.fig}(d)).

%Should be able to match to a few catalogs, at least to see if we detect massive members.
%In /literature, Pat made Shepherd_2004_ApJ_601_952.table2.reg and Persi06.table3.reg.
%Kumar07's Table 2 gives IRAC mags.
%Maia16's near-IR catalog is on VizieR.  From their Sec. 5.7, W75N includes the cluster BDB59.  Cluster BDB60 lies a couple of arcmins to the north, but doesn't seem to be associated with W75N.  NOTE that they find DR21 (Sec. 5.8, BDB58) to be at 3kpc, not at the distance (1.5kpc) used in MYStIX!  They ref Davis07 and Pestalozzi05 for this large distance too.  Also see Gottschalk12 -- 3 separate layers of SF in Cyg X North, at <1kpc, ~1.5kpc, and >2.5kpc.

%Maia16 near-IR sources:
%http://vizier.u-strasbg.fr/viz-bin/VizieR-3?-source=J/MNRAS/458/3027/table&-out.max=50&-out.form=HTML%20Table&-out.add=_r&-out.add=_RAJ,_DEJ&-sort=_r&-oc.form=sexa
%downloaded to /literature/Maia16.fits.  See also Maia16_membership_notes.pdf -- code 5 is what we want for W75N.
%Position uncertainties:  they say "rms residuals ~0.05" internally and ~0.12" in relation to 2MASS."  Data appear to be registered to 2MASS.

\clearpage
%-----------------------------------------------------------------------------
\subsection{RCW~120 \label{sec:rcw120}}
% RCW 120 -- 1692 point sources in 3-ptg field.  999 in just RCW 120 ACIS-I.
% In SFiNCs -- this is the only target that MOXC2 has in common with SFiNCs.
% Handbook South article is Deharveng08.
% At 1.34 kpc, 4*pi*D^2 = 2.1489e44.

The innocuous visual \hii region RCW~120 was catapulted to fame by the \Spitzer GLIMPSE survey \citep{Churchwell06,Zavagno07}, which revealed it to be a canonical example of a single, dusty bubble hosting massive star formation \citep{Deharveng08}.  It has since received much attention at long wavelengths, especially from {\em Herschel}, which detects several very young stars forming in dense, cold material around the edge of the bubble \citep{Zavagno10,Figueira17}.  The bubble appears to be disrupted at its northern edge and is leaking photons at several points around its rim \citep{Zavagno07,Deharveng09,Anderson15}.  Such ``leaky'' \hii regions may demonstrate that ionizing radiation from massive stars extends well beyond classical \hii region boundaries, influencing star formation over a much wider area \citep{Zavagno07}.  

\begin{figure}[htb]
\centering
\includegraphics[width=0.487\textwidth]{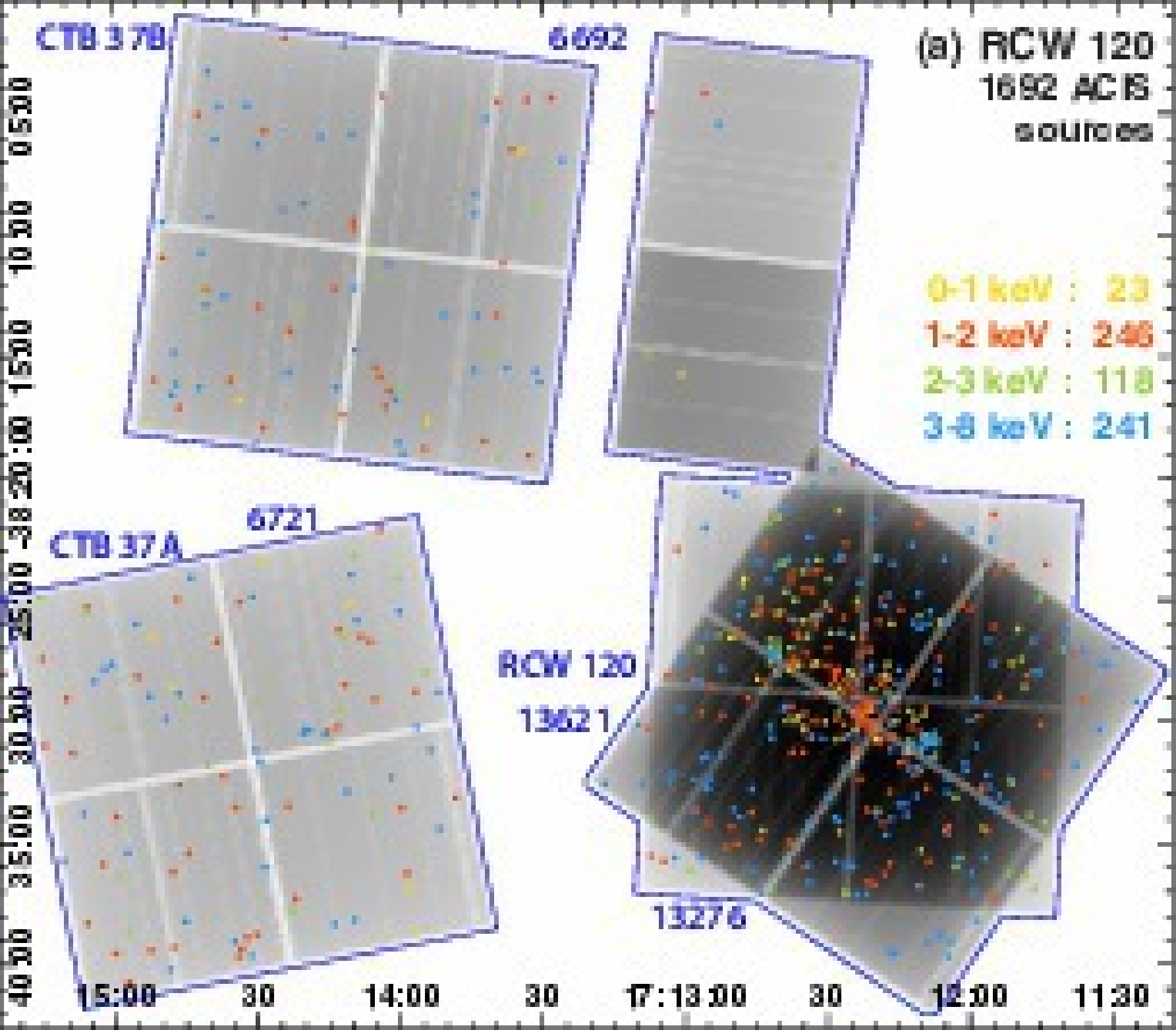}
\includegraphics[width=0.49\textwidth]{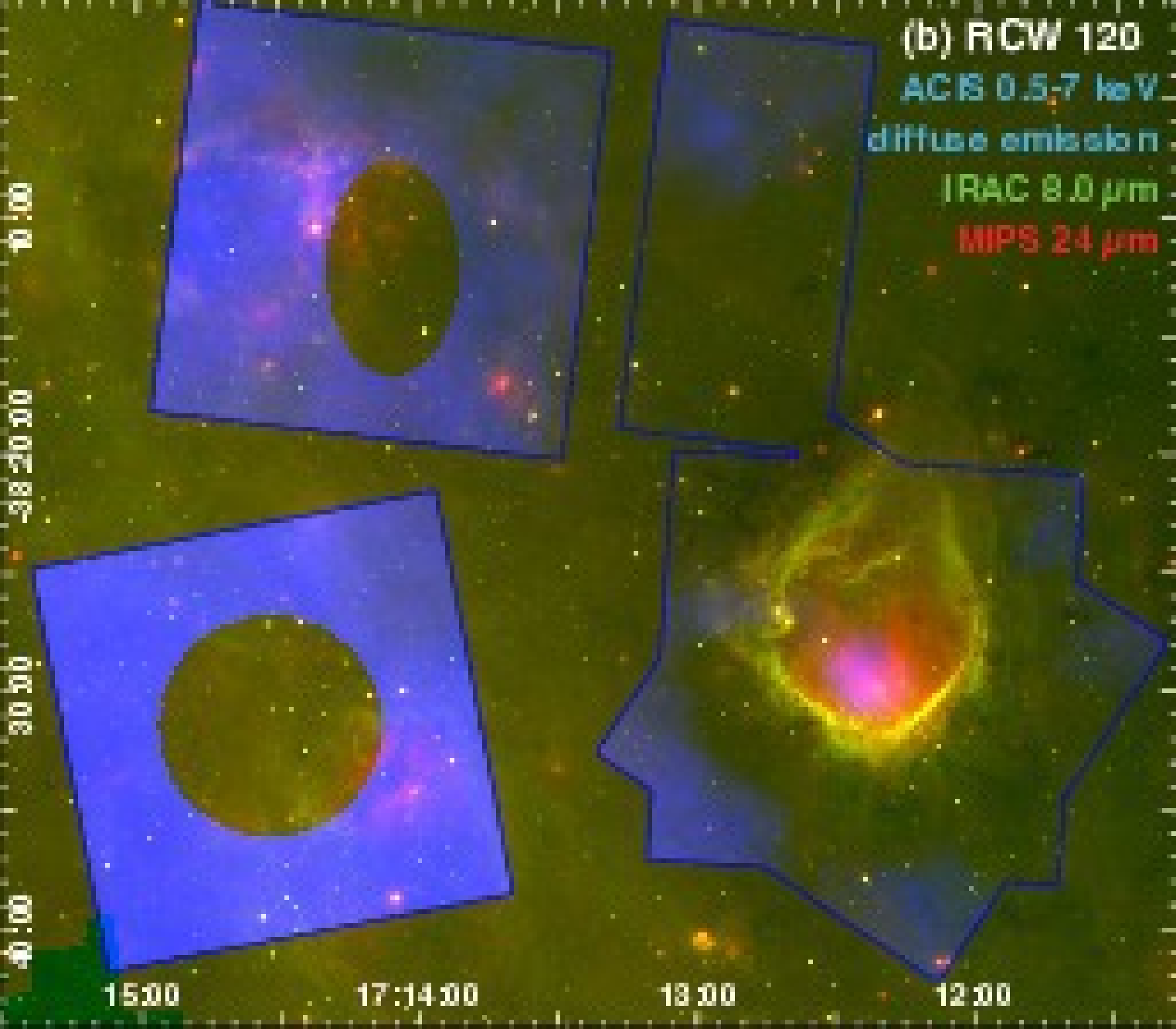}
\includegraphics[width=0.4\textwidth]{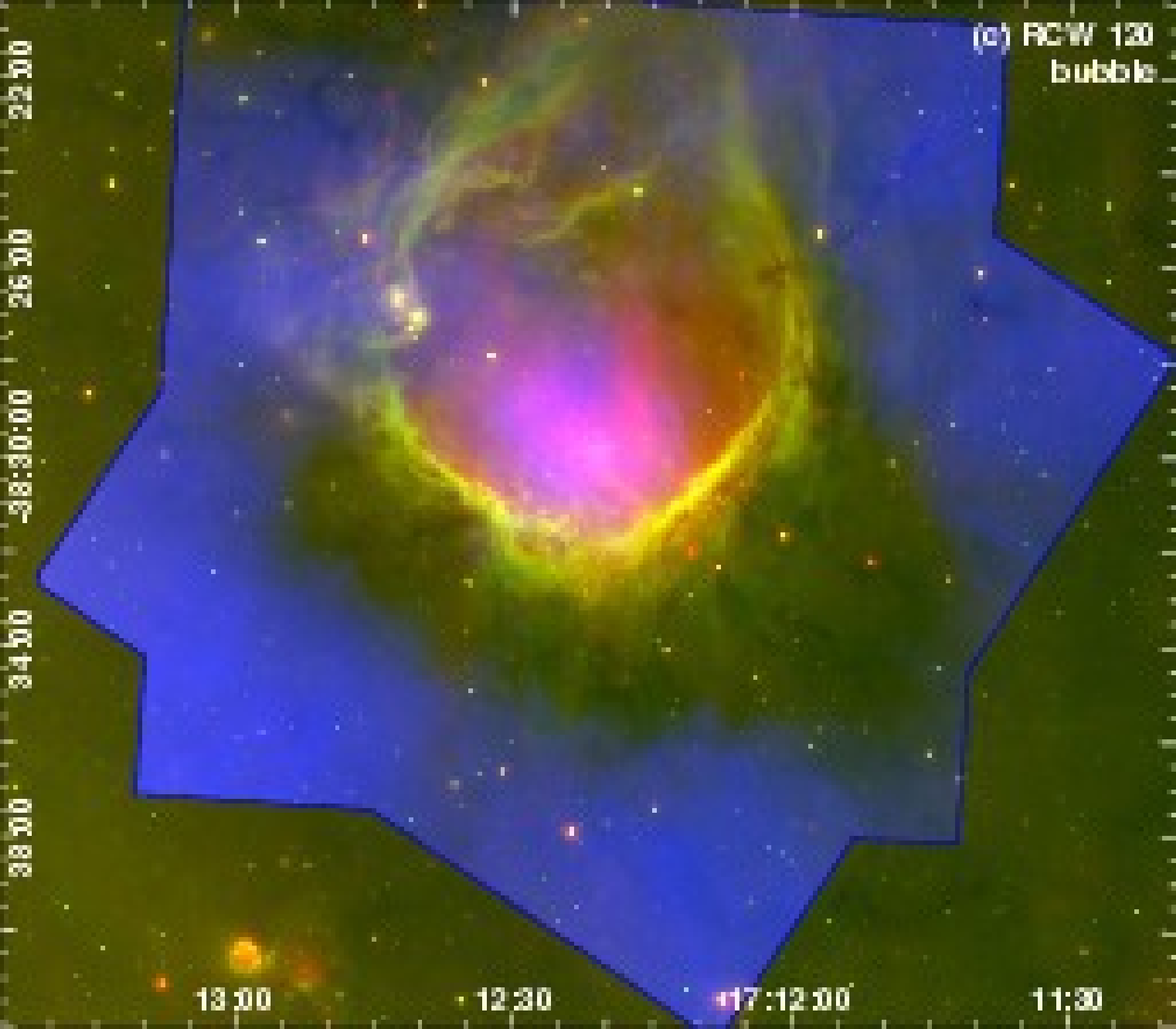}
\includegraphics[width=0.4\textwidth]{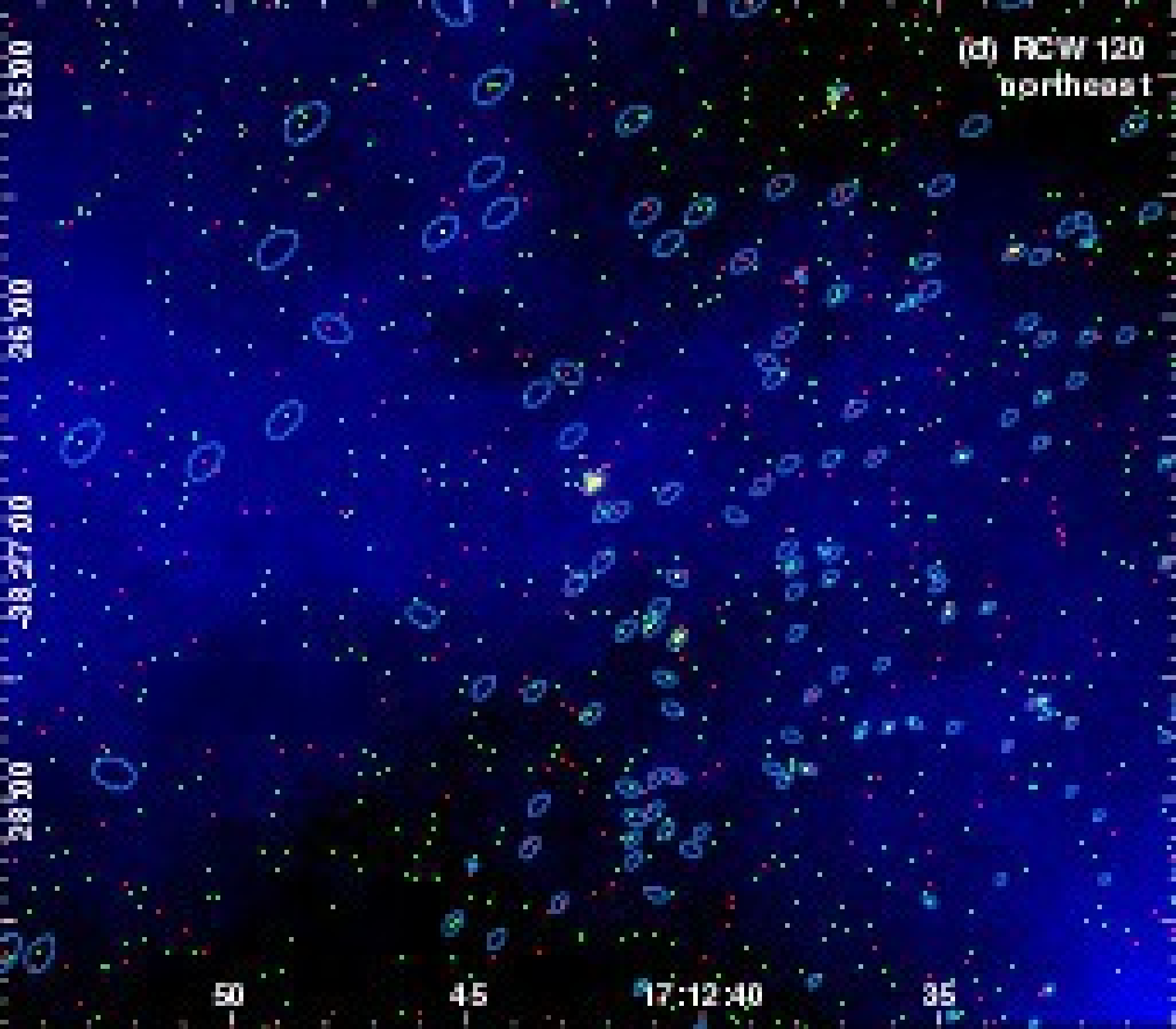}
\includegraphics[width=0.4\textwidth]{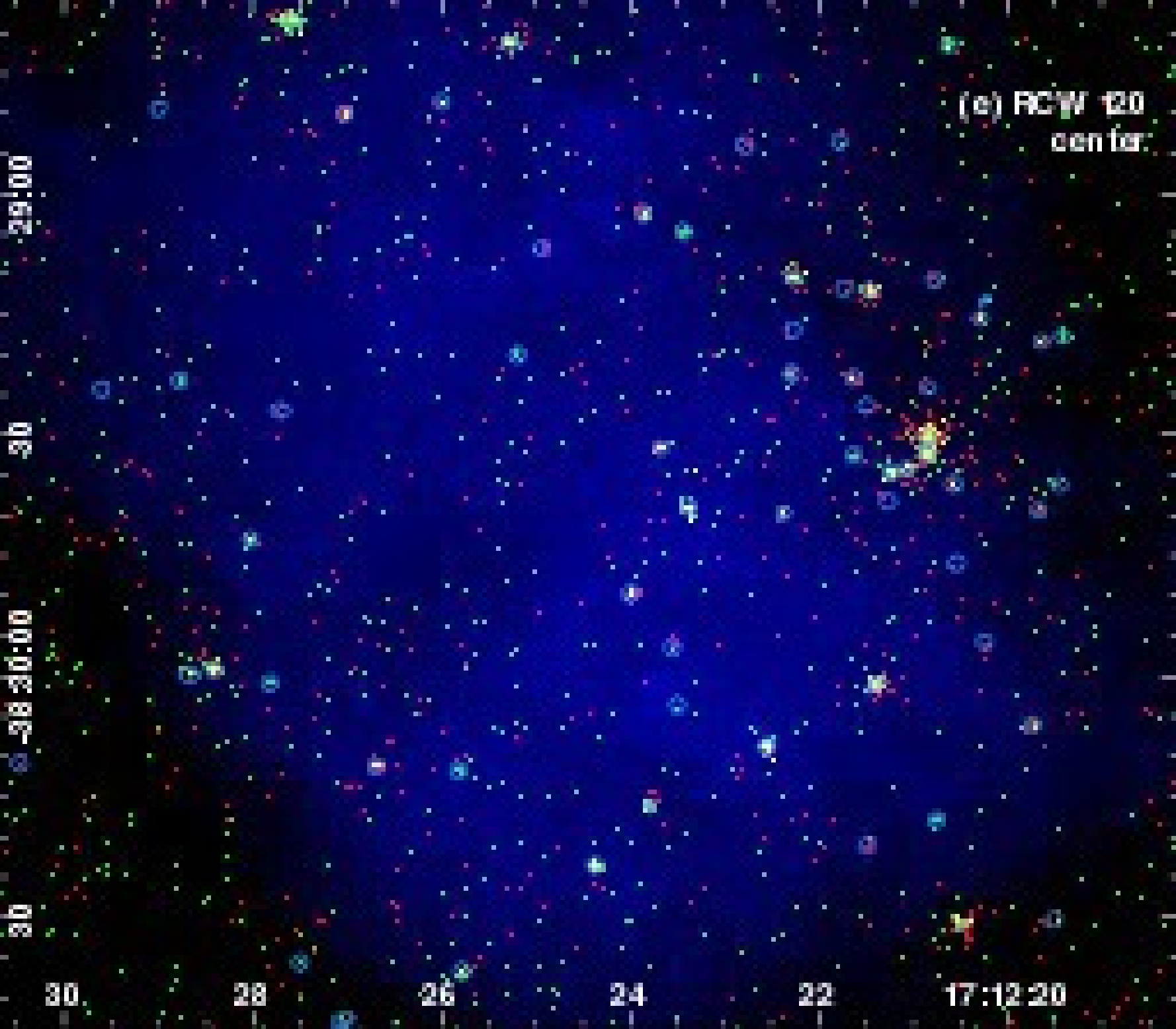}
\includegraphics[width=0.4\textwidth]{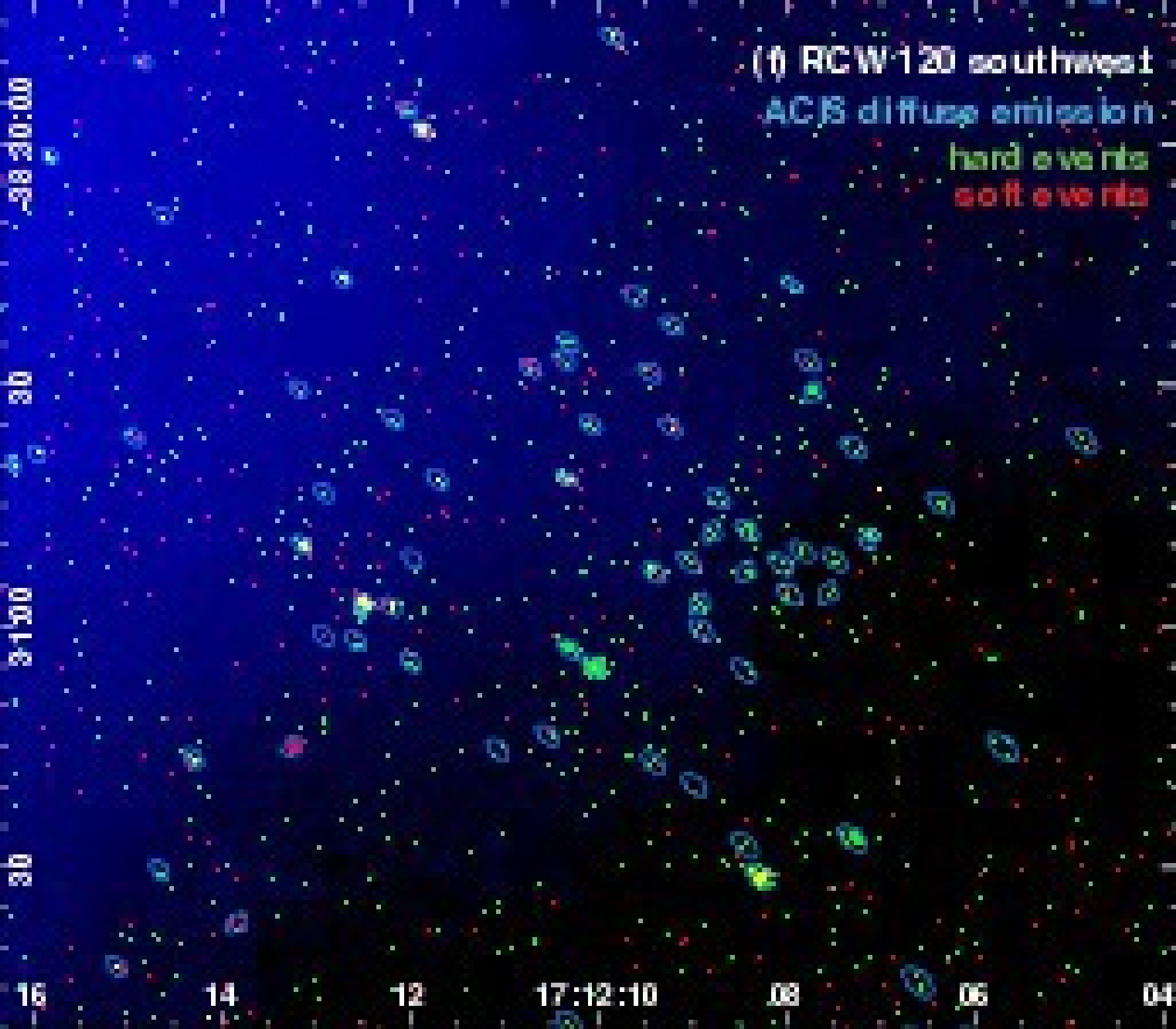}
\caption{RCW~120.
(a) ACIS exposure map with brighter ($\geq$5 net counts) ACIS point sources overlaid; colors denote median energy for each source.  ObsID numbers and regions named in the text are shown in blue.
(b)  ACIS diffuse emission in the \Spitzer context.    
(c) Zoomed version of (b) for the RCW~120 bubble.
(d)--(f) ACIS event data and diffuse emission detailing clumps of X-ray point sources in RCW~120.
(d) On the northeast rim of the bubble, where diffuse X-ray emission appears to breach the bubble wall.
(e) In the south-central part of the bubble, associated with the brightest diffuse X-ray emission in the bubble.  The ionizing O8 star is the bright X-ray source on the western side of this group.
(f) On the southwest rim of the bubble.
\label{rcw120.fig}}
\end{figure}

The \Chandra observations of RCW~120 were presented in the SFiNCs (Star Formation in Nearby Clouds) survey \citep{Getman17}.  SFiNCs studies X-ray point sources in \Chandra observations of nearby, typically lower-mass star-forming regions, so this is the only target that MOXC2 and SFiNCs have in common.  We included RCW~120 in MOXC2 (Figure~\ref{rcw120.fig}) primarily to facilitate study of the diffuse X-ray emission in this iconic \Spitzer bubble, but this common target does give us the opportunity to compare our point source catalog with that of SFiNCs.  This is a useful check, since the two projects use similar (but not identical) analysis methods.  The SFiNCs analysis is a few years older than the MOXC2 analysis.

In MOXC2, we find 999 X-ray sources in the main ACIS-I array observations of RCW~120 itself; SFiNCs found 678 X-ray sources in the same data.  Matching the two catalogs, we find 604 formal matches, 395 MOCX2 sources not matched, and 74 SFiNCs sources not matched.  One goal of MOXC2 is to detect fainter sources, thus we expect to have many unmatched MOXC2 sources.  In several cases, either MOXC2 or SFiNCs found a pair of sources while the other catalog found a single source.  Many of the unmatched sources (in both catalogs) are far off-axis and faint.  For such sources, the source validity metric\footnote{The source validity metric is  a {\em p}-value for the no-source (``null'') hypothesis, i.e., that all extracted counts are background.} can be sensitive to slight source position differences, algorithmic changes to the background calculation, and differences in the catalogs (neighboring sources affect the background calculation).  Because source candidates are pruned by a simple threshold on this noisy validity metric, two reductions of the same data will not produce identical catalogs near the detection limit.

For our ACIS mosaic of the RCW~120 region, we included neighboring observations of SNRs to get a wider context for RCW~120's diffuse X-ray emission:  ObsID~6721 on CTB~37A \citep{Pannuti14} and ObsID~6692 on CTB~37B \citep{Aharonian08} (Figure~\ref{rcw120.fig}(a)).  Large masks were applied to these SNRs to keep them from dominating our smoothed diffuse X-ray image of the mosaic (Figure~\ref{rcw120.fig}(b)).  For completeness, we include the point sources we found in these fields in our X-ray source list for RCW~120.

Extensive diffuse X-ray emission is seen in RCW~120 (Figure~\ref{rcw120.fig}(c)).  It appears to trace hot plasma through a breach in the northeast side of the bubble and out into the surrounding ISM.  The whole bubble apparently shadows background diffuse X-ray emission.  
%Maybe say something about the presence/morphology of diffuse X-ray emission in light of Ochsendorf14 -- would it be worthwhile to ask Jesus to spin off a paper using our diffuse X-ray data products?

Clumps of X-ray point sources are shown in Figures~\ref{rcw120.fig}(d)--(f).  Panels (d) and (f) show groups of sources on the bubble rim; panel (e) shows the central cluster.  The main ionizing source for RCW~120 is LSS~3959 (CD-28~11636 in SIMBAD), an O8V star \citep{Zavagno07}.  Its X-ray counterpart is ACIS source c1121 (CXOU~J171220.84-382930.3, prominent in Figure~\ref{rcw120.fig}(e)), with 682 net counts and a median energy of 1.4~keV.  The spectral fit requires two thermal plasma components, {\em TBabs(apec1 + apec2)}; the softer plasma dominates the spectrum.  Parameters are $N_H = 1.4 \times 10^{22}$~cm$^{-2}$, $kT1 = 0.5$~keV, $kT2 = 2$~keV, and $L_X = 1.4 \times 10^{32}$~erg~s$^{-1}$.  The harder plasma component is not expected from a single late-O star and is perhaps suggesting magnetic activity or binarity.
%Zavagno07:  they locate the O8 star at 17 12 20.6 -38 29 26.  SIMBAD gives coords as 17 12 20.854 -38 29 30.50.  Those are pretty different.  Our c1121 is at 17:12:20.837 -38:29:30.51, matching the SIMBAD position well.  We have nothing close to the former position.  There is a second X-ray src, c1125, just south of c1121 a couple of arcsec.  They can be seen in the "center" image below.  

%\clearpage
%-----------------------------------------------------------------------------
\subsection{IRAS~20126+4104 \label{sec:iras20126}}
% IRAS 20126+4104 -- 493 point sources
% Mentioned on p.78 of SF Handbook North (in Cygnus)
% Anderson11's I20S match is our c200.  Their I20Var match is our c208.
% Spitzer data only cover a small area at the center of the cluster and the 24um data are badly saturated.  ACIS diffuse emission makes more sense with WISE images.
% At 1.64 kpc, 4*pi*D^2 = 3.219e44.

Like W75N described above, the IRAS~20126+4104 MSFR is another component of the extensive and diverse star formation complex in Cygnus \citep{Reipurth08}.  Its distance is well-determined by maser parallax \citep{Moscadelli11}.  IRAS~20126+4104 itself is a famous massive young stellar object (MYSO) with a prominent molecular outflow that has long been the subject of radio study \citep[e.g.,][]{Cesaroni05}.  \citet{Cesaroni14} have imaged this MYSO at 1.4~mm, finding a distorted disk and associated jet rotating around a proto-B star.  \citet{DeBuizer17} presents a recent SOFIA detection of the MYSO; he finds that SOFIA and \Herschel data show elongation in the direction of the outflow.  The \Chandra data were first presented by \citet{Anderson11}, with details on the X-ray counterparts to radio sources I20S and I20Var \citep{Hofner07}.  These authors note the detection of 150 point sources in the \Chandra data and suggest that IRAS~20126+4104 is surrounded by a young stellar cluster.  Those sources are catalogued in a recent paper by the same group \citep{Montes15}, who determine that $\sim$80 of the 150 X-ray sources are likely members of this MSFR.  They find that the surface density of X-ray sources peaks on the MYSO, thus IRAS~20126+4104 is a young cluster forming an early-B star at its center.

\begin{figure}[htb]
\centering
\includegraphics[width=0.495\textwidth]{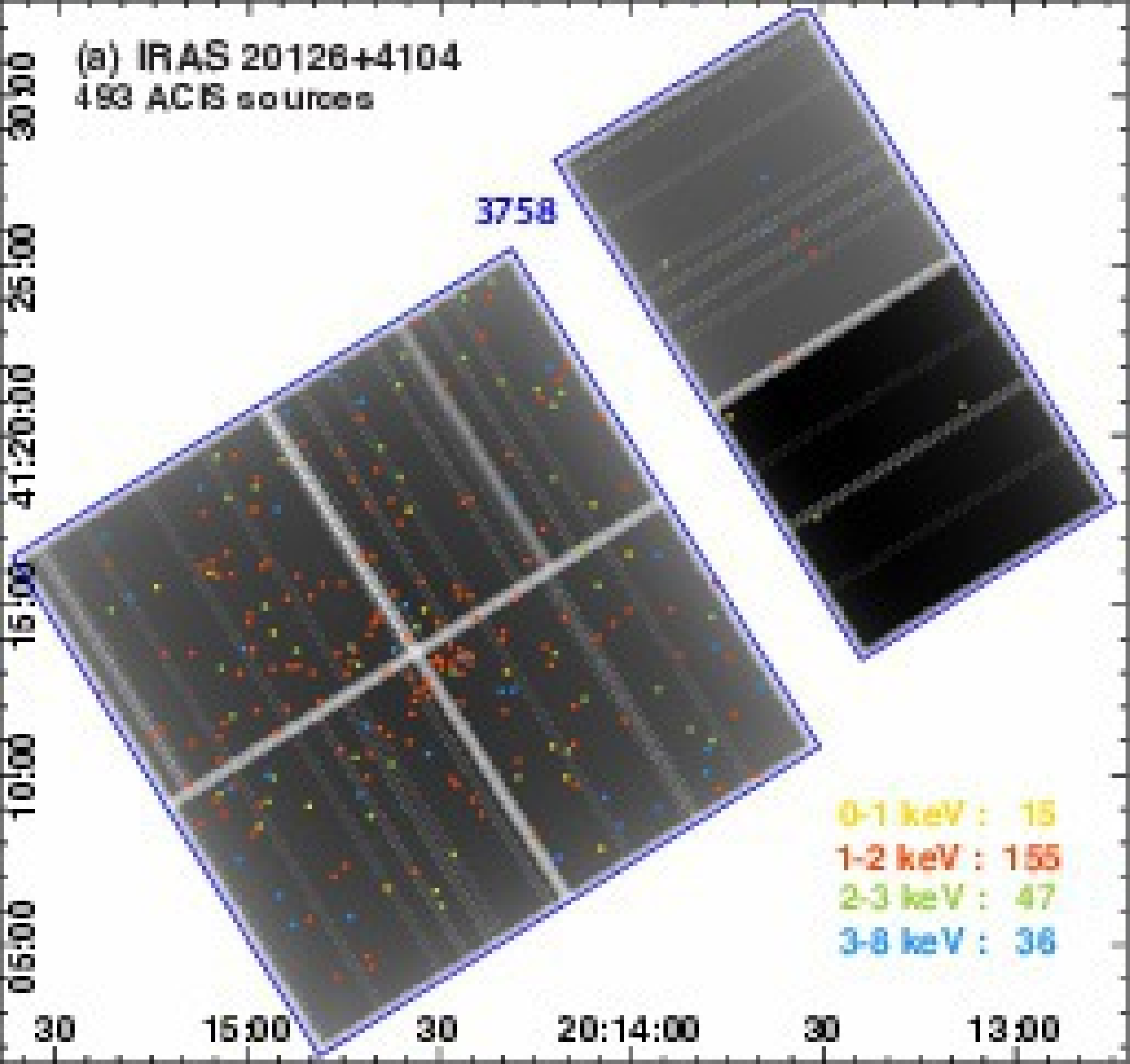}
\includegraphics[width=0.48\textwidth]{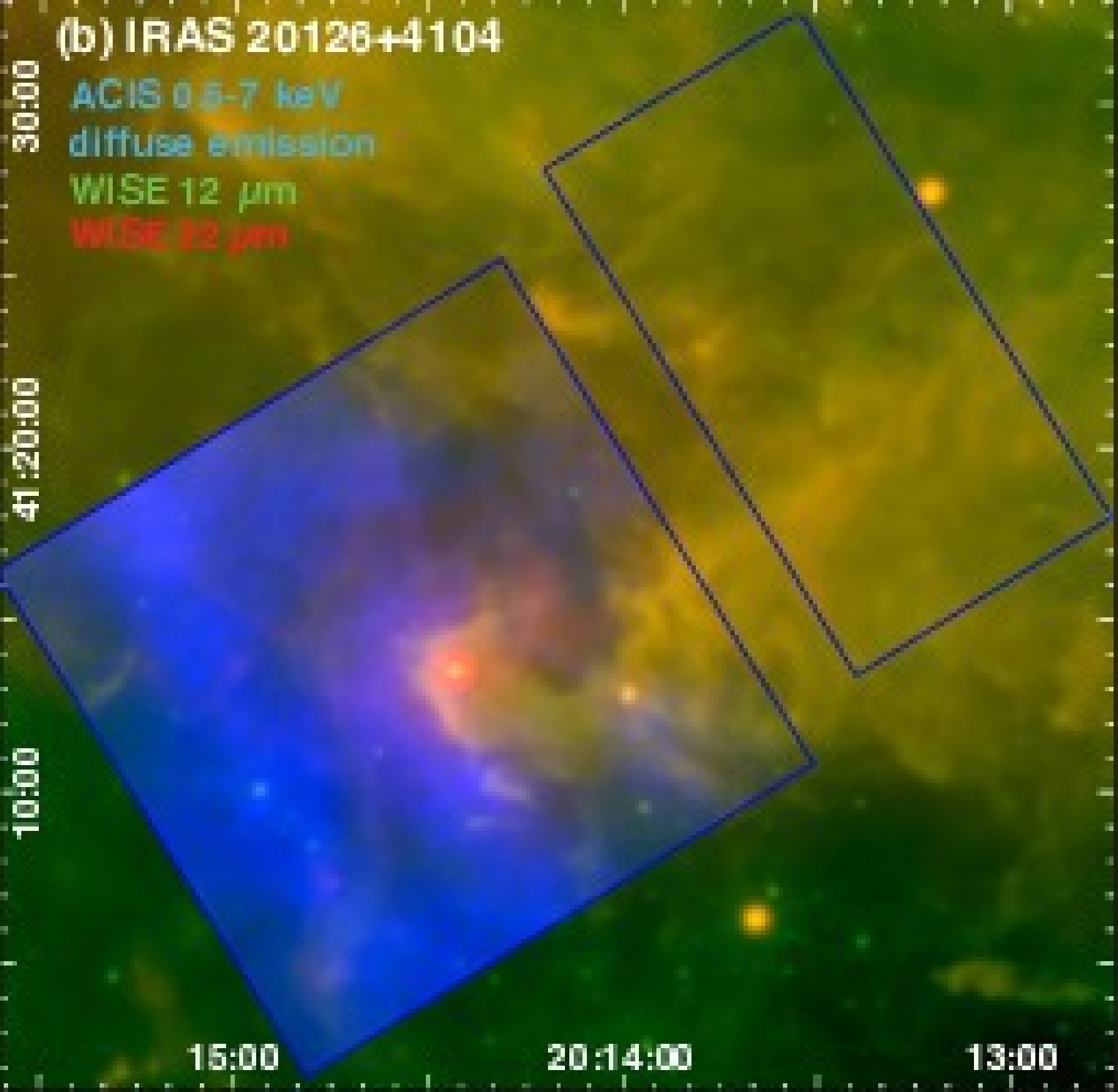}
\includegraphics[width=0.49\textwidth]{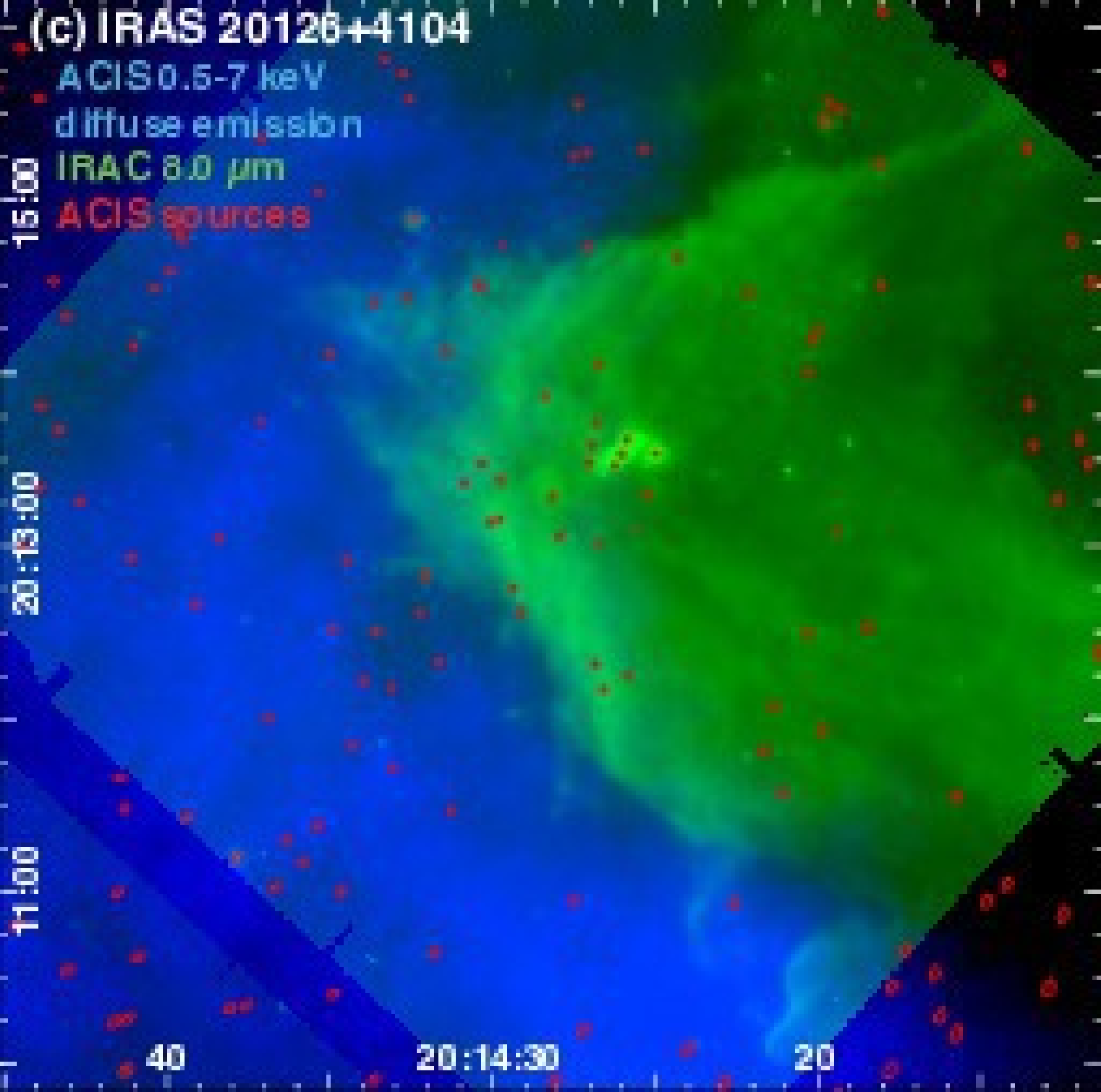}
\includegraphics[width=0.49\textwidth]{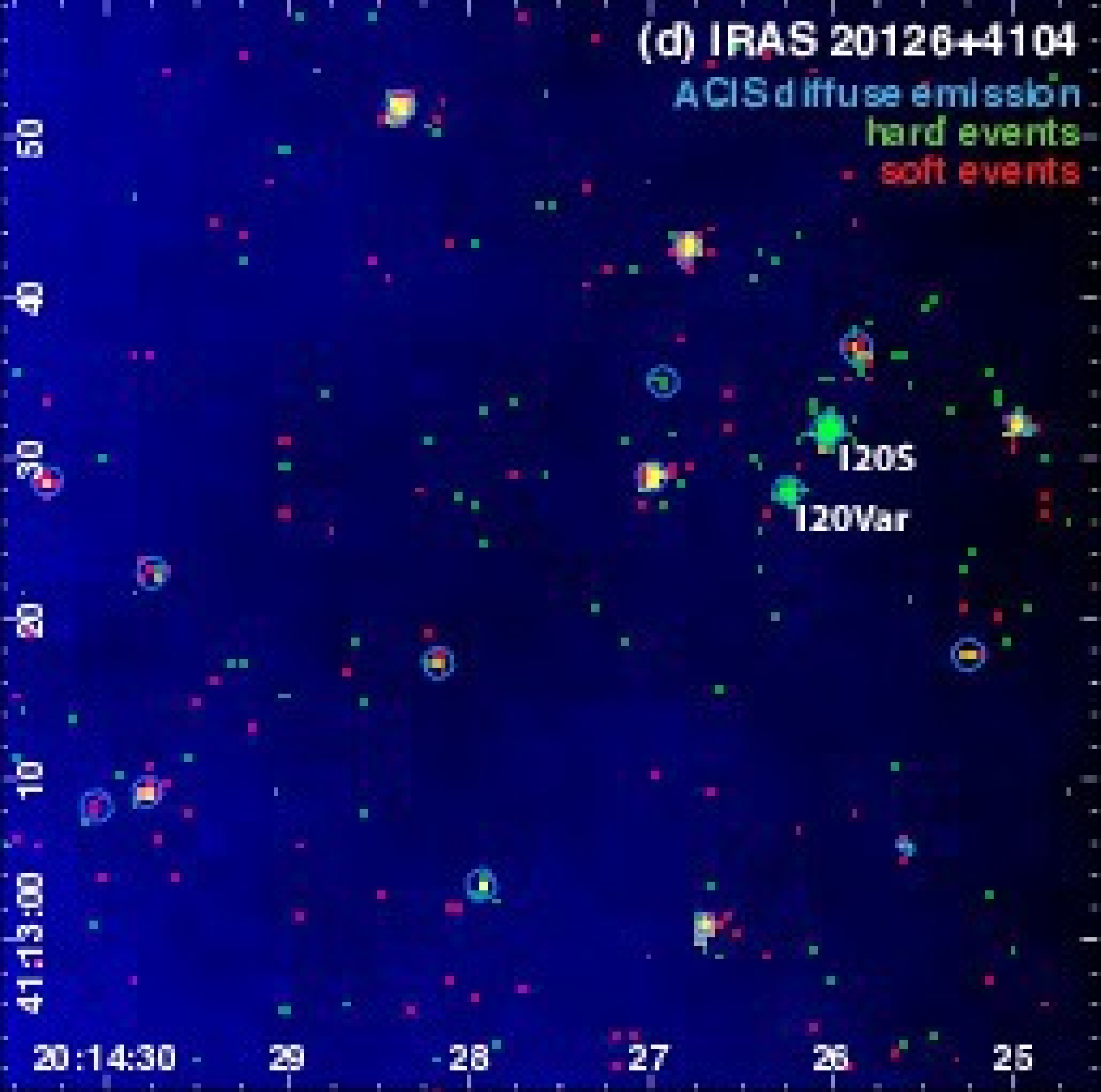}
\caption{IRAS~20126+4104.
(a) ACIS exposure map with brighter ($\geq$5 net counts) ACIS point sources overlaid; colors denote median energy for each source.  The ObsID number is shown in blue.
(b) ACIS diffuse emission in the \WISE context.  \Spitzer data lack full field coverage for this target.
(c) ACIS diffuse emission around the main pillar of IRAS~20126+4104, shown with {\em Spitzer}/IRAC data (badly saturated MIPS data are omitted).  Red polygons show extraction regions for ACIS sources.
(d) ACIS event data and diffuse emission at the center of IRAS~20126+4104.  Zoom to see close pairs of X-ray sources.  Hard X-ray counterparts to radio sources I20S and I20Var are marked.
\label{IRAS20126.fig}}
\end{figure}

Our analysis of the same \Chandra data found more faint point sources, as designed (Figure~\ref{IRAS20126.fig}); we recovered all 150 X-ray sources reported by \citet{Montes15}.  Many X-ray sources near the center of the cluster resolve into close pairs in our analysis (Figure~\ref{IRAS20126.fig}(d)).  \citet{Anderson11} described the exceptional X-ray spectra of the ACIS counterparts to I20S and I20Var; we find similar spectral fit results.  They concluded that I20S was an early-B MYSO and I20Var a likely IMPS.  We concur with these findings and emphasize their implications:  in some cases, stars that will become X-ray-quiet on the main sequence are extraordinarily hard, bright X-ray emitters in their youth, perhaps due to magnetic activity similar to that found in lower-mass pre-MS stars \citep{Povich11,Gregory16}.

% Spectral fit results are in point_sources.noindex/moxc2_spfit_notes.txt.

We find prominent diffuse X-ray emission in this ACIS field (Figure~\ref{IRAS20126.fig}(b)--(d)), with a bright ridge running the length of the ACIS-I array to the east of the cluster (Figure~\ref{IRAS20126.fig}(b)).  The image center shows an arc of X-ray emission just to the east of the large pillar that hosts the embedded cluster (Figure~\ref{IRAS20126.fig}(c)); between this and the eastern ridge is an absence of diffuse X-rays coincident with faint \WISE 22~$\mu$m emission.  This may be material shadowing the diffuse X-rays.  Faint diffuse X-ray emission pervades the pillar as far west as the central cluster (Figure~\ref{IRAS20126.fig}(d)), then drops off sharply.  The western side of the ACIS-I field shows a complex mix of fainter diffuse X-ray emission anticoincident with bright \WISE structures; this fades to only minimal diffuse X-ray emission on the off-axis ACIS-S CCDs.  Again shadowing or displacement of hot gas is likely here.

\clearpage
%-----------------------------------------------------------------------------
\subsection{W31N \label{sec:w31n}}
% W31N -- 974 point sources
% At 1.75 kpc, 4*pi*D^2 = 3.665e44.

\begin{figure}[htb]
\centering
\includegraphics[width=0.48\textwidth]{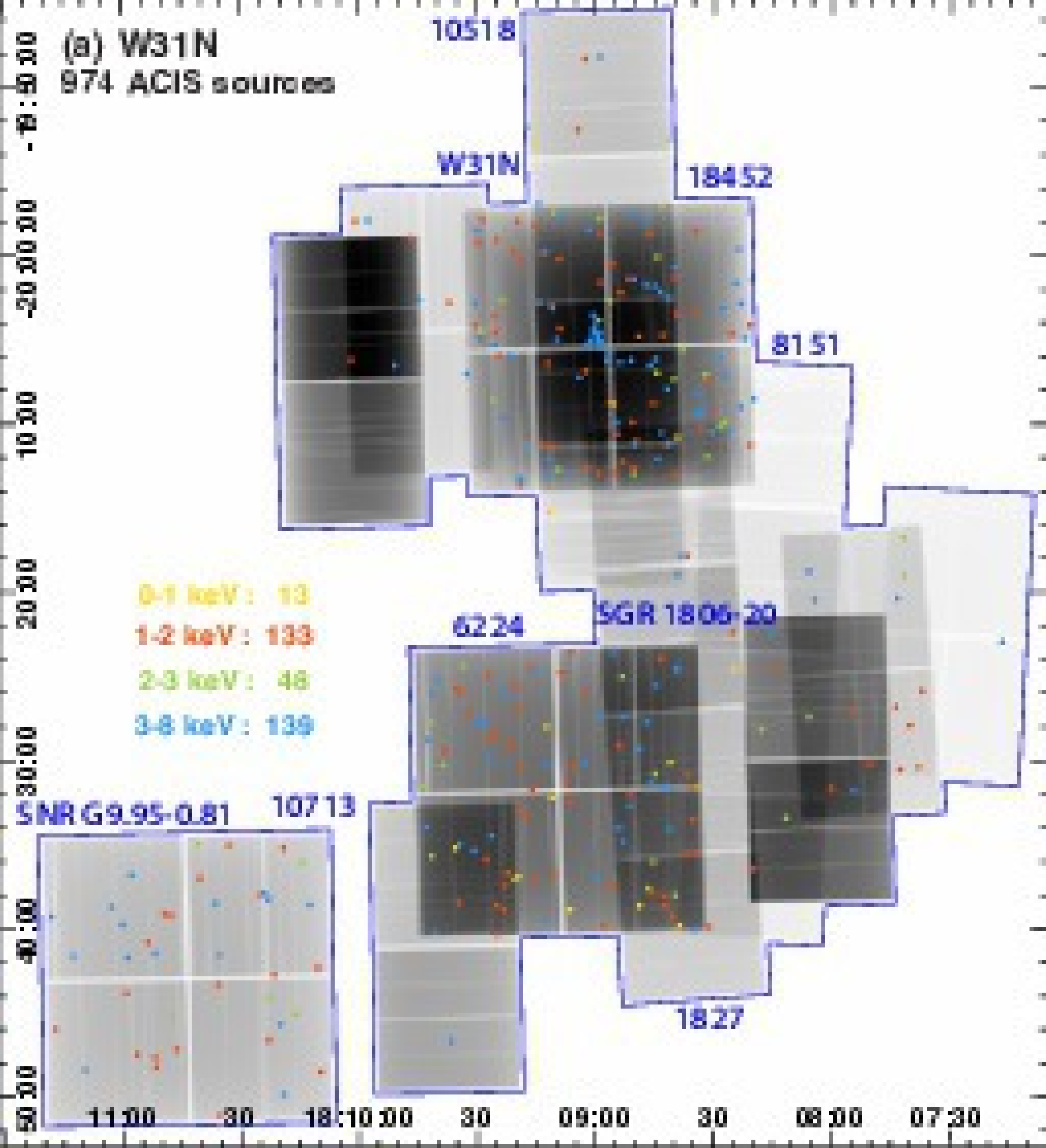}
\includegraphics[width=0.49\textwidth]{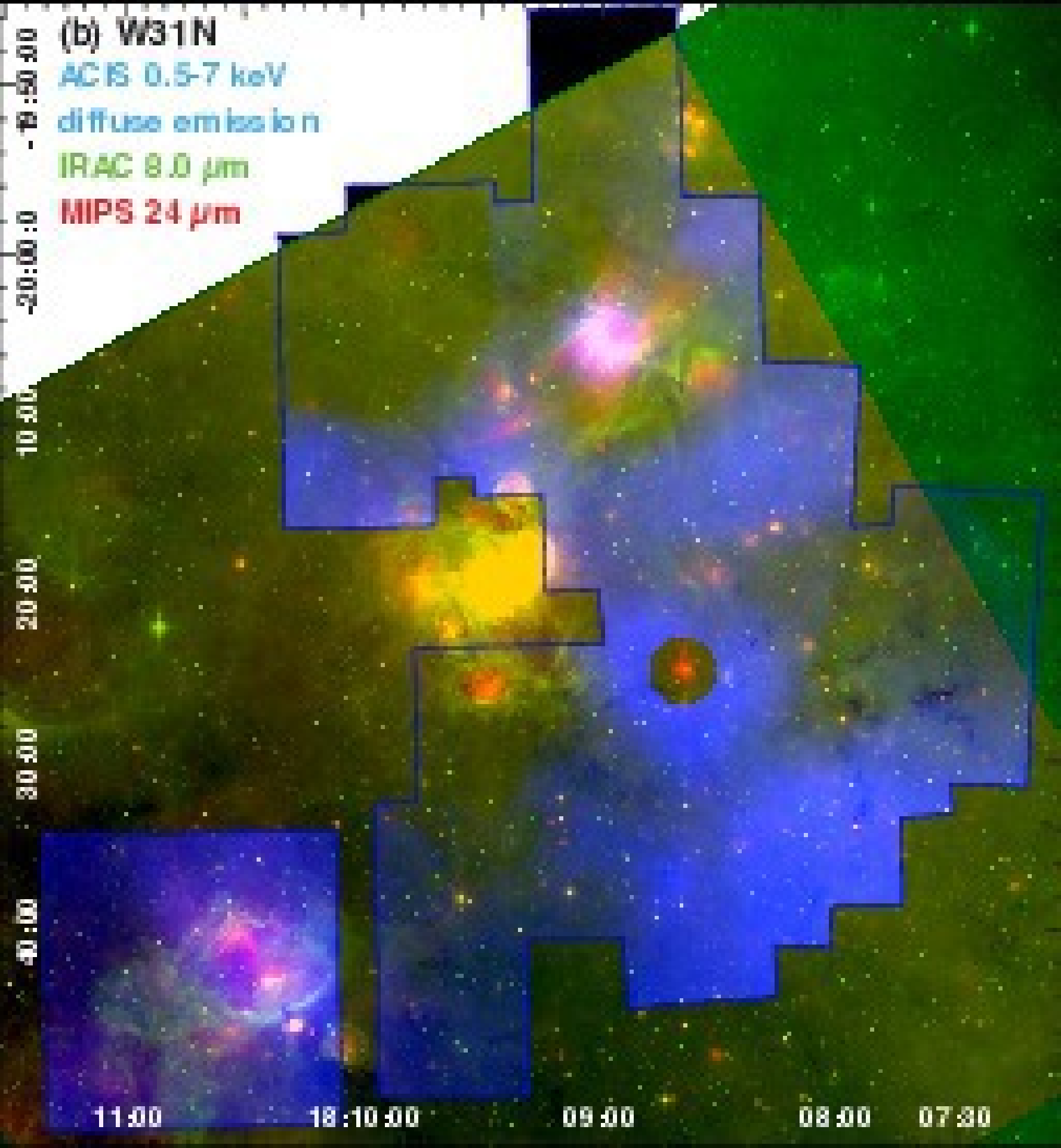}
\includegraphics[width=0.32\textwidth]{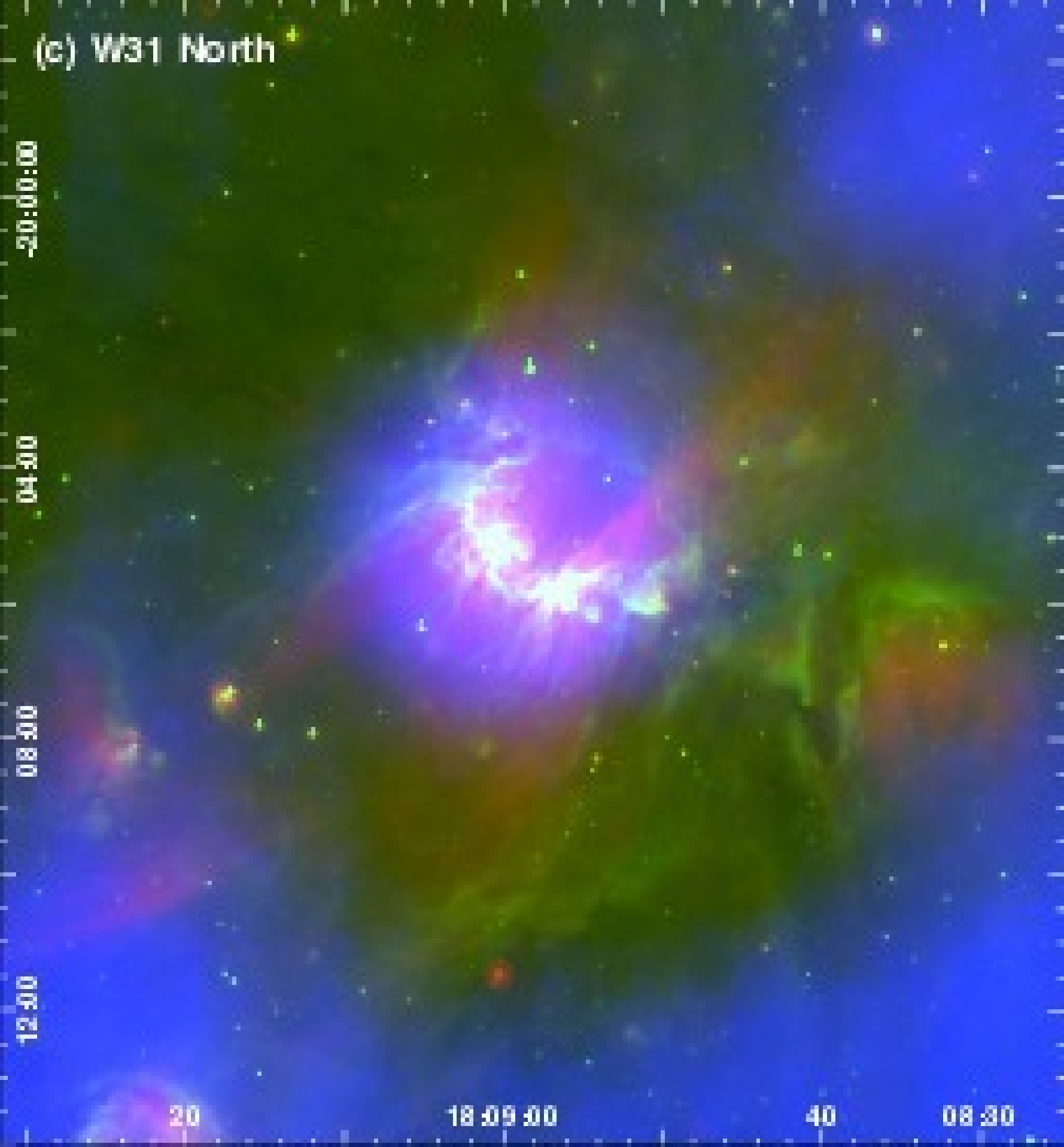}
\includegraphics[width=0.32\textwidth]{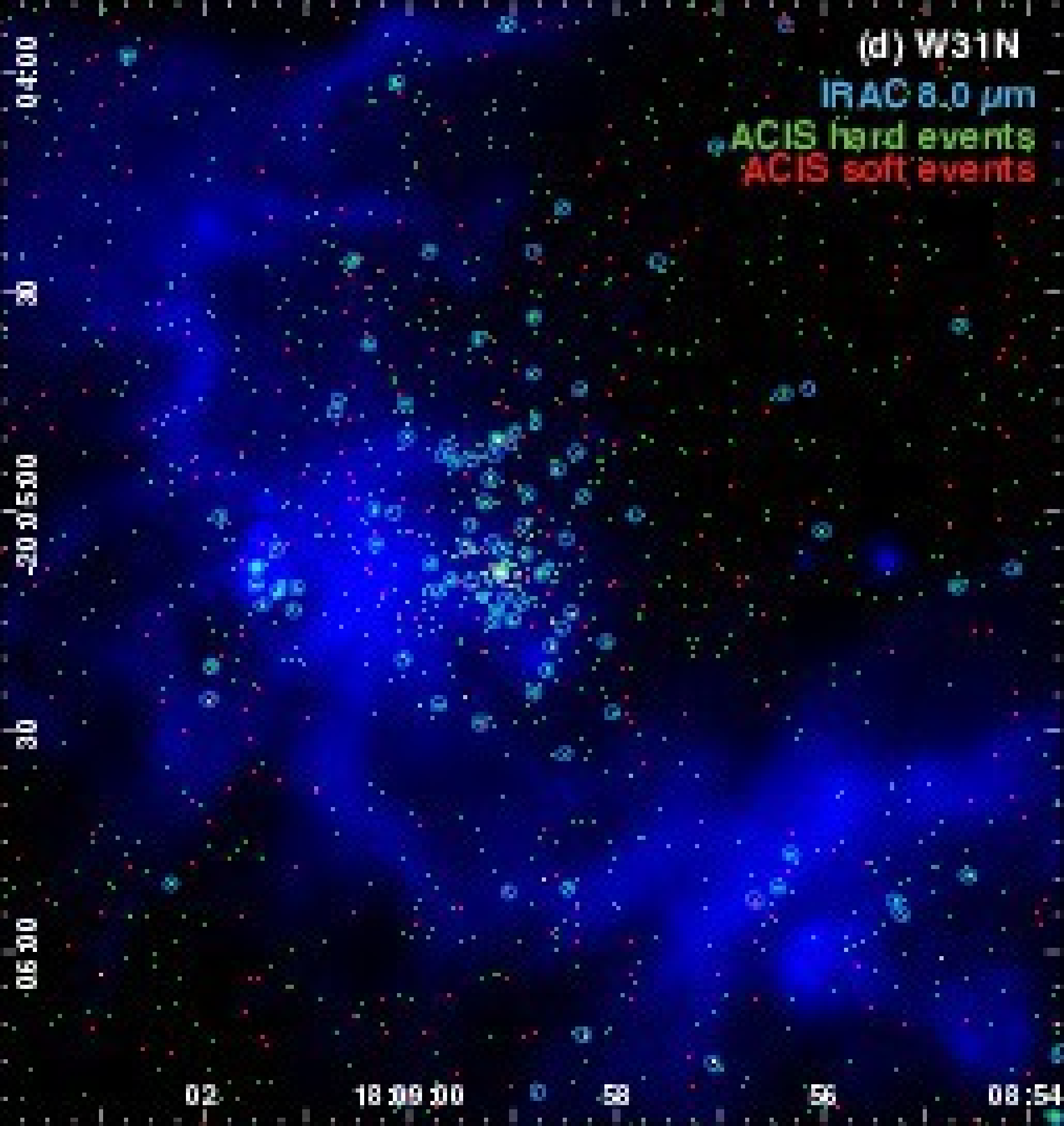}
\includegraphics[width=0.32\textwidth]{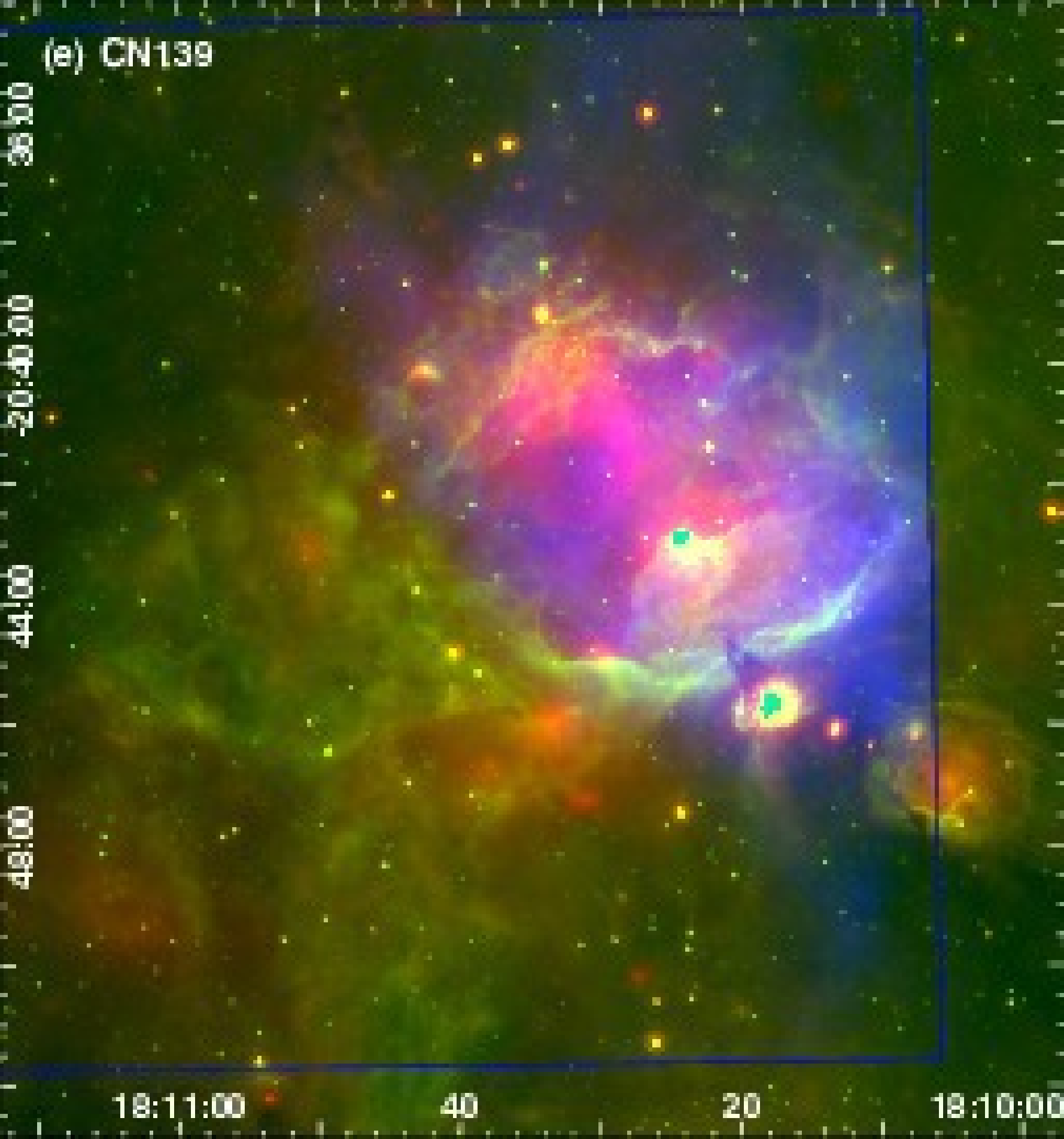}
\caption{W31.
(a) ACIS exposure map with brighter ($\geq$5 net counts) ACIS point sources overlaid; colors denote median energy for each source.  ObsID numbers and regions named in the text are shown in blue.
(b) A wide-field ACIS mosaic of the W31 region, featuring ACIS diffuse emission in the \Spitzer context.  A region around the bright X-ray source SGR~1806-20 has been masked so that fainter diffuse X-ray emission across the field can be displayed.
(c) Zoomed version of (b) for the nearby MSFR W31N.
(d) ACIS event data on W31N.  For context, {\em Spitzer}/IRAC 8~$\mu$m emission is shown in blue.
(e) Zoomed version of (b) for the GLIMPSE bubble CN139 at the southeast corner of our W31 ACIS mosaic.  
\label{W31.fig}}
\end{figure}

The vicinity of the famous giant \hii region W31 has several {\em Chandra}/ACIS observations of a variety of targets at a wide range of distances.  As for other MOXC2 targets, we have combined them into a wide-field mosaic (Figures~\ref{W31.fig}(a) and (b)) to sample the X-ray point source populations and diffuse emission present in large Galactic Plane fields.

\citet{Deharveng15} showed that the northern component of the W31 complex (W31N, which they call G010.32-0.15) is actually a foreground MSFR unassociated with the other major MSFRs along this sightline, at a distance of just 1.75~kpc.  W31N (G10.3-0.1) is a bipolar \hii region ionized by an O5--O6 star \citep{Bik05} and containing an IR cluster; it is triggering a menagerie of second-generation massive stars and their UCH{\scriptsize II}Rs in the dense filament at its waist \citep{Dewangan15b,Deharveng15}.  

At this modest distance, it makes a rich \Chandra target; we find $>$50 X-ray point sources in the central cluster and 490 sources across the ACIS-I field centered on W31N.  As demonstrated below, with these data we can study both the first-generation ionizing cluster and the onset of X-ray emission in massive stars just formed.  Figure~\ref{W31.fig}(c) shows that W31N also contains diffuse X-ray emission and strongly shadows (or displaces) the diffuse emission seen in the wider W31 field.

The original 6.5~ks ACIS-S observation (ObsID~10518) of W31N detected the ionizing O5 star \citep[source ChI~J180857-2004\_2 in][]{Anderson14}; these authors suggest that it could be a CWB.  In our data (Figure~\ref{W31.fig}(d)), this is source c1345 (CXOU~J180859.12-200508.4), with 553 net counts and a median energy of 3.3~keV.  A spectral fit gives $N_H = 4.9 \times 10^{22}$~cm$^{-2}$, $kT = 3.8$~keV, and $L_X = 2.4 \times 10^{32}$~erg~s$^{-1}$.  Once again, this high column could be hiding an additional soft plasma component, so this luminosity is a lower limit.  We concur with \citet{Anderson14}; this hard spectrum suggests a CWB.
%SIMBAD calls the ionizing star [BKH2005] 18060nr1733.  Bik05 doesn't give coordinates.  Name is derived from an IRAS source name.  SIMBAD's position is 18 08 58.6 -20 05 24 but I don't know where they get it.
%The ionizing star is Deharveng15's #1.  They give coords as (l,b) = 10.31798 -0.15185, which is 18 08 59.116 -20 05 08.55 as converted by SIMBAD.  This is consistent with our source c1345 (CXOU~J180859.12-200508.4).
%Anderson14 Table 1 gives the position of their source J180857-2004_2 as 18 08 59.11 -20 05 08.2.  This is c1345.

Our source c1447 (CXOU~J180901.48-200507.5) is the MYSO (YSO \#4) in clump C2 of \citet{Deharveng15}.  It has 43 net counts, a median energy of 4.6~keV, and is variable; fit parameters are $N_H = 21 \times 10^{22}$~cm$^{-2}$, $kT = 5.5$~keV, and $L_X = 5 \times 10^{31}$~erg~s$^{-1}$.  This extremely high column explains the high median energy.  The hard thermal plasma and variability demonstrate that dramatic X-ray emission turns on early in the formation process of at least some massive stars.

Near the center of our ACIS mosaic of the W31 complex (Figure~\ref{W31.fig}(b)), the conspicuous MSFR not yet observed by \Chandra is G10.2-0.3, a giant \hii region at a distance of 3.4~kpc \citep{Blum01}.  It contains four O stars and four MYSOs in a very young (0.6~Myr) cluster \citep{Furness10}, several UCH{\scriptsize II}Rs \citep{Ghosh89}, and \XMM source detections \citep{Nebot_Gomez-Moran15}; it would make an excellent \Chandra target.  Southwest of G10.2-0.3, our mosaic includes the piled-up soft gamma repeater SGR~1806-20; we have masked its bright X-ray halo of dust-scattered light \citep{Kaplan02} in our image of diffuse X-ray emission.

At the southeast corner of our ACIS mosaic, we have included a 9.9-ks ACIS-I observation of the radio-bright SNR~G9.95-0.81 \citep{Brogan06} (ObsID~10713).  This field also includes the large GLIMPSE bubble CN139, with a near kinematic distance of 4.3~kpc \citep{Churchwell07,Watson09}.  Quite surprisingly (given the short observation), Figure~\ref{W31.fig}(e) clearly shows diffuse X-ray emission coincident with this bubble, somewhat offset to the northwest from the center of the radio SNR.  Spectral analysis to determine the absorbing column to this X-ray emission should help to determine whether it is really associated with the bubble.

Through further serendipity, the globular cluster 2MASS~GC02 is captured far off-axis on the S3 CCD in ObsID~10713, visible (by zooming) at the bottom of the {\em Spitzer}/IRAC 8~$\mu$m image in Figure~\ref{W31.fig}(b).  ACIS source c2200 (CXOU~J180936.55-204645.1) is centered on 2MASS~GC02.  It has 38 net counts and a median energy of 4.0~keV.  It is most likely a composite of emission from multiple members of the globular cluster, but its properties might help to establish the feasibility for a longer ACIS observation of this target.

%Small IRDCs in west-central part of ACIS mosaic:
%Source c22 may be a match to the bright Spitzer source SSTGLMC G009.8474-00.0322, a YSO according to SIMBAD.  This IRDC is included in Ragan09 (ApJ 698, 324-349 (2009)).
%Source c106 may be a match to the Spitzer source 2MASS J18075514-2027112 -- Young Stellar Object Candidate according to SIMBAD.

\clearpage
%-----------------------------------------------------------------------------
\subsection{IRAS~19410+2336 and NGC~6823 \label{sec:iras19410}}
% IRAS 19410+2336 -- 411 point sources
% At 2.16 kpc, 4*pi*D^2 = 5.584e44.
% Beuther02a used all 20 ks of ACIS data and say that there are no background flares.

%NGC 6823 is mentioned in Handbook North article by Prato, Rice, and Dame
%At 2.3 kpc, 4*pi*D^2 = 6.331e44.

%  NOTE -- observing_log.pdf is years older than observing_log.tex -- why is that????  This file looks WRONG -- time used for the main observation should be much shorter than the ~20ks reported here, due to removal of bkgd flares.  Pat agrees -- use first page of xray_properties.pdf for a more recent observing log, but it also is WRONG.  It's reporting the exposure time of I3 apparently, not the aimpoint device.

\begin{figure}[htb]
\centering
\includegraphics[width=0.487\textwidth]{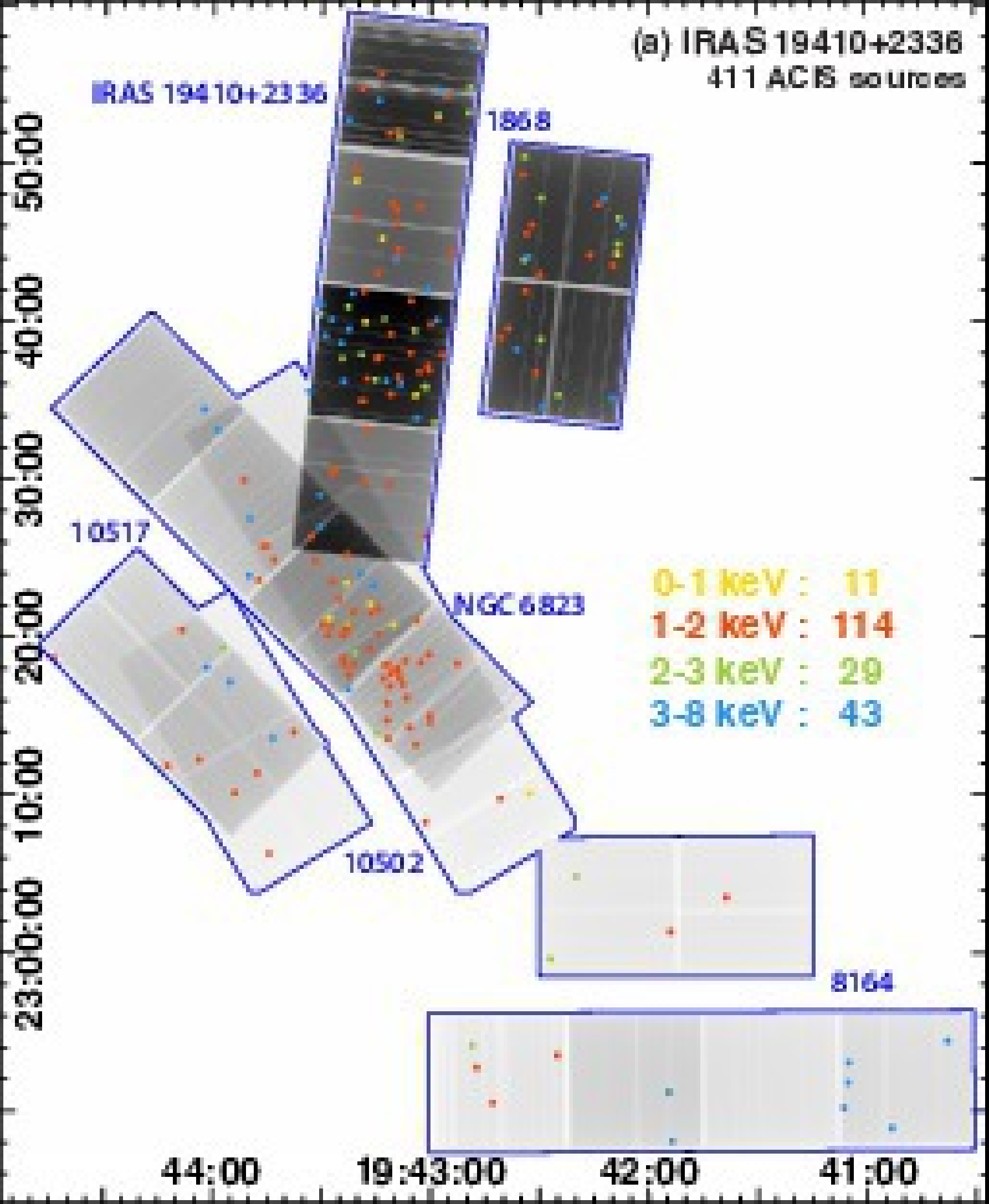}
\includegraphics[width=0.49\textwidth]{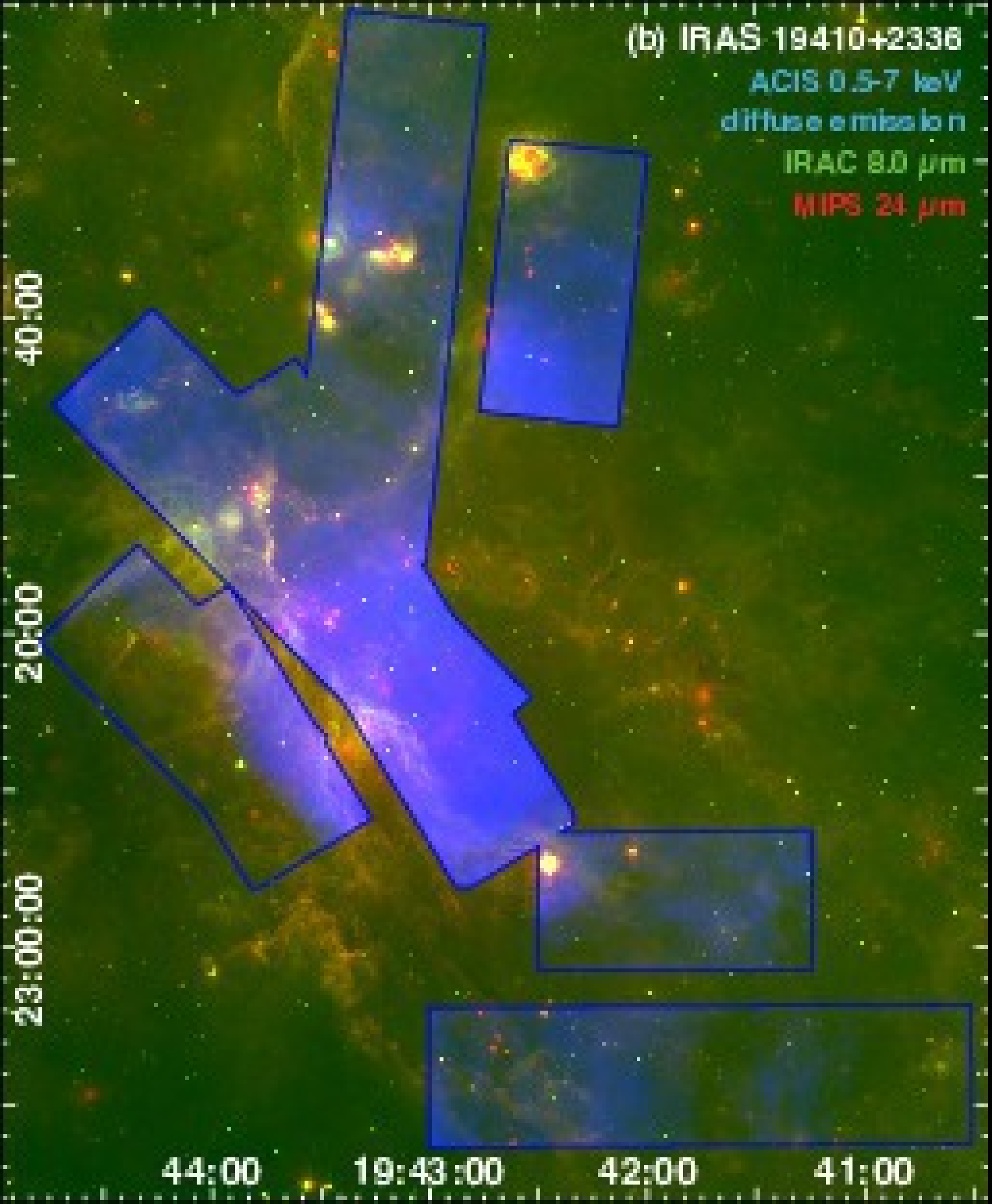}
\includegraphics[width=0.24\textwidth]{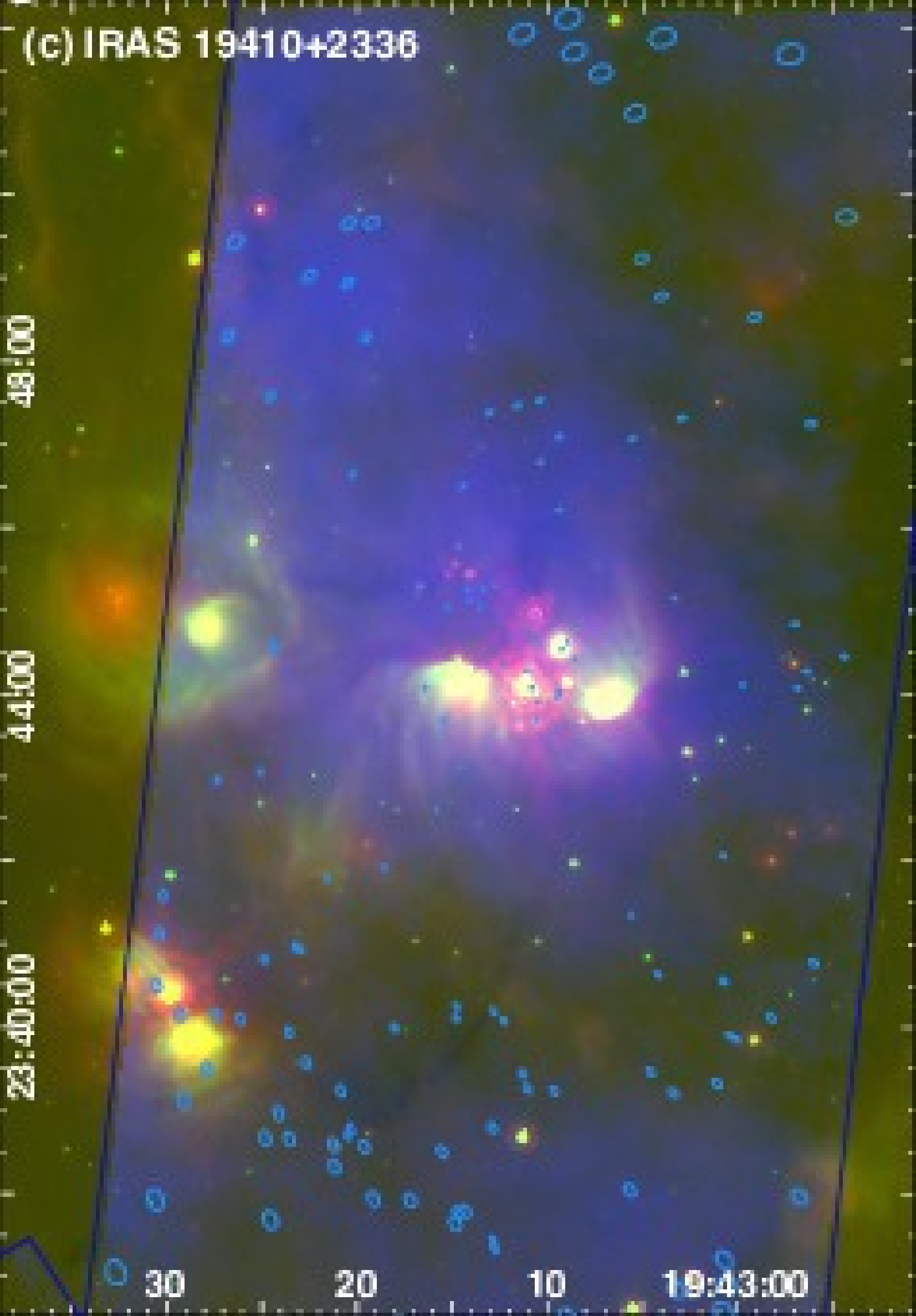}
\includegraphics[width=0.375\textwidth]{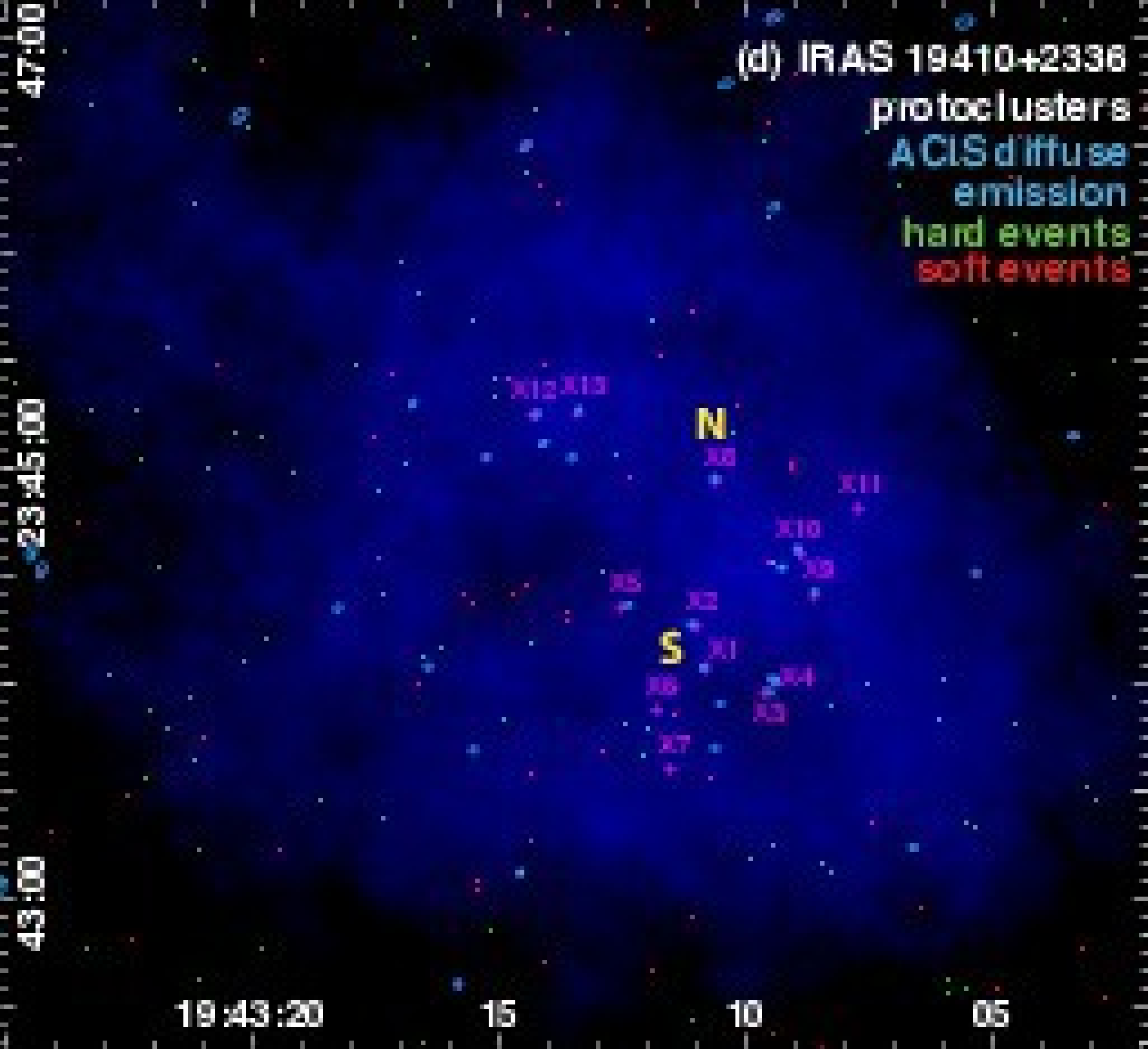}
\includegraphics[width=0.37\textwidth]{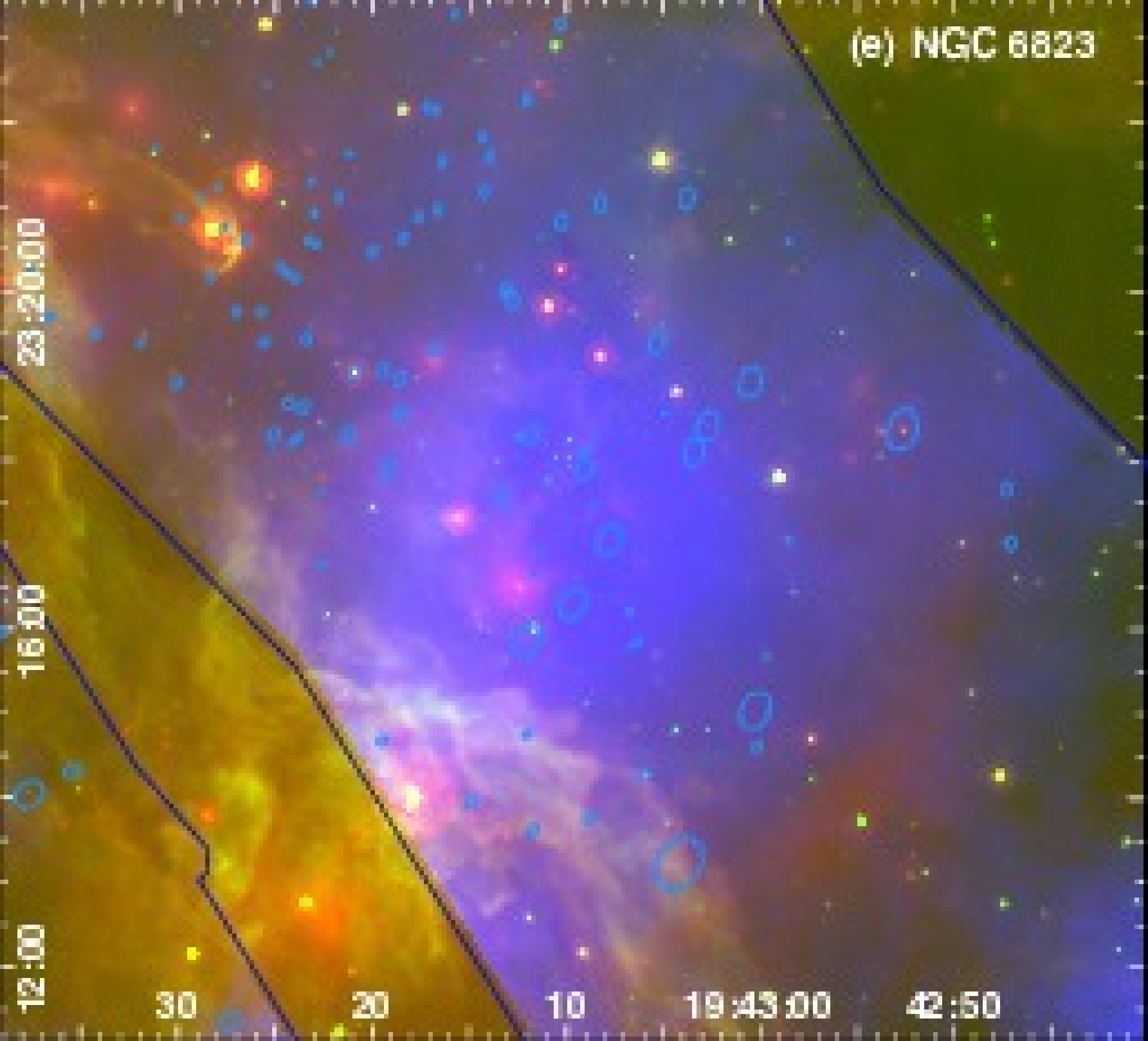}
\caption{IRAS~19410+2336.
(a) ACIS exposure map with brighter ($\geq$5 net counts) ACIS point sources overlaid; colors denote median energy for each source.  ObsID numbers and regions named in the text are shown in blue.
(b) ACIS diffuse emission in the \Spitzer context.  
(c) Zoomed version of (b) for IRAS~19410+2336, with MOXC2 point source extraction regions overlaid.
(d) ACIS event data and diffuse emission at the center of IRAS~19410+2336.  MOXC2 point source extraction regions are blue; X-ray sources from \citet{Beuther02a} are noted in magenta.  Approximate centers of the northern and southern star-forming clumps \citep{Rodon12} are marked with N and S respectively.
(e) Zoomed version of (b) for NGC~6823, with MOXC2 point source extraction regions overlaid.
\label{IRAS19410.fig}}
\end{figure}

Starting with a 20-ks observation of the MSFR IRAS~19410+2336, we built a 4-pointing ACIS-S mosaic of overlapping short observations from archival \Chandra data in this region (Figures~\ref{IRAS19410.fig}(a) and (b)).  This is the shallowest mosaic in MOXC2 (just 2--20~ks).  This mosaic also features a second MSFR, NGC~6823, presumably not associated with IRAS~19410+2336 although at a similar distance.  This is the first of three examples of ACIS mosaics in MOXC2 that feature two unrelated MSFRs that just happen to be projected near each other on the sky.

IRAS~19410+2336 is a young, compact, embedded MSFR made up of two main star-forming clumps \citep{Beuther02b,Rodon12}, both exhibiting masers \citep{Beuther02c} and a large number of outflows \citep{Beuther03}.  The \Chandra observation of IRAS~19410+2336 (ObsID~1868, 20~ks) was analyzed by \citet{Beuther02a}, who tabulated 13 X-ray sources in the vicinity of the two clumps (Figure~\ref{IRAS19410.fig}(d)).  They find these sources to have hard X-ray spectra and to be heavily obscured, with fairly high intrinsic luminosities.  They conclude that the X-ray sources are mainly IMPS; as noted above, subsequent work has confirmed that such sources can be strong X-ray emitters \citep{Povich11,Gregory16}.

\citet{Beuther02a} included the full 20~ks exposure in their analysis; our standard processing of this dataset leaves just 8.7~ks of usable data on the aimpoint CCD S3 and on S1, the other backside-illuminated device, due to space weather resulting in high backgrounds for these chips.  The ACIS frontside-illuminated devices in operation for this observation (I2, I3, S2, and S4) were unaffected by these background flares, so they retain the full integration time in our analysis.  Unfortunately, the two main embedded clumps in IRAS~19410+2336 are imaged on CCD S3, so we report source detections in only 8.7~ks of data for these clumps.  As we will see below, this is comparable to the ACIS integration time we have on the second MSFR imaged in this multi-pointing ACIS-S mosaic, NGC~6823.

As shown in Figures~\ref{IRAS19410.fig}(c) and (d), we find X-ray sources in the same regions as \citet{Beuther02a}, although not always the same sources.  The southern clump is well-populated with X-ray sources but the northern one is not; rather a second group of X-ray sources sits to the east of the northern clump.  Diffuse X-ray emission appears between the two groups of X-ray sources.

% From Rodon12, northern clump is centered at RA(J2000) = 19h43m10.7s; Dec(J2000) = 23 44 58.4; southern clump at RA(J2000) = 19h43m11.2s; Dec(J2000) = 23 44 03.2.

About 26$\arcmin$ south of IRAS~19410+2336 we find the other MSFR in our mosaic, NGC~6823 \citep{Prato08}, captured on two short ACIS-S observations (Figure~\ref{IRAS19410.fig}(e)).  It features a tight grouping of massive stars at its center \citep{Riaz12,Shi99}, with pre-MS stars distributed out to the cluster radius of 16.2$\arcmin$ \citep{Kharchenko05}.  Its earliest star is HDE~344784~A, with a spectral type of O6.5~V((f))z \citep{Arias16}.  \citet{Anderson14} detect its X-ray counterpart as their source ChI~J194310+2318\_5.  In our analysis, this is \Chandra source c228 (CXOU~J194310.96+231745.5), with 72 net counts and a median energy of 1.1~keV.  Our simple thermal plasma fit yields $N_H = 0.9 \times 10^{22}$~cm$^{-2}$, $kT = 0.6$~keV, and $L_X = 1.9 \times 10^{32}$~erg~s$^{-1}$.  

\citet{Anderson14} list several other X-ray sources in this region; we also find many other sources.  Despite having just 7.9~ks total integration on this target and with the cluster center placed 8$\arcmin$ off-axis on the 6-ks ObsID, ACIS still clearly detects this MSFR.  Several of our X-ray sources are counterparts to young stellar objects tabulated in \citet{Riaz12}.

%Anderson14 tabulates J194310+2318_(1–10) and J194332+2323_(1–8) in this region.  
%Pat made /Volumes/hiawatha1/targets/IRAS19410/data/counterparts/Rias12/riaz12_cat.reg -- he says no matching was done.

In the fourth pointing of this mosaic (ObsID 8164, 2.7~ks) we have two sources at the aimpoint, the piled-up source c19 (CXOU~J194154.60+225112.3) and its close neighbor c20 (CXOU~J194154.60+225112.8).  Due to crowding, extraction apertures are reduced to 45\% for c19 (25 net counts, median energy 3.3~keV) and just 36\% for c20 (10 net counts, median energy 2.4~keV).  With so few counts, we cannot correct c19 for pile-up.  Rough spectral fits to these sources show them both to be very hard (kT $\sim$ 10~keV) but with very different absorbing columns and absorption-corrected full-band fluxes:  $N_H = 4 \times 10^{22}$~cm$^{-2}$, $F_X \sim 8 \times 10^{-13}$~erg~s$^{-1}$~cm$^{-2}$ for c19 and $N_H = 1 \times 10^{22}$~cm$^{-2}$, $F_X \sim 2 \times 10^{-13}$~erg~s$^{-1}$~cm$^{-2}$ for c20.
%From Vizier -- there's a UKIDSS counterpart to c19 with J=18.2, no K-band measurement.  Not much else.
\citet{Anderson14} find a single X-ray source (ChI~J194152+2251\_2) at this location (a blend of our c19 and c20), using this same dataset; they fit it with a power law with $N_H = 0.9 \times 10^{22}$~cm$^{-2}$, slope $\Gamma \sim 0.4$, and absorption-corrected full-band flux $F_X \sim 7 \times 10^{-13}$~erg~s$^{-1}$~cm$^{-2}$.  They noted that the near-IR image of this source appears extended, consistent with a blend of sources.

%Xu12 (in /targets/NGC6823/) says the progenitor of SNR G59.5+0.1, located at l,b = 59.58, +0.12 or 19 42 32.24  +23 35 08.9, could have triggered SF in this region.

Despite the very short integration times, diffuse X-ray emission is seen throughout this mosaic.  It is brightest around NGC~6823 (Figure~\ref{IRAS19410.fig}(e)), likely indicating a mix of unresolved pre-MS stars in the cluster and perhaps hot plasma from the winds of the O6 star.  We see highly structured diffuse emission suffusing the deeply embedded IRAS~19410+2336 (Figure~\ref{IRAS19410.fig}(d)) and pervading the surrounding field (Figure~\ref{IRAS19410.fig}(c)).  The wider field contains SNR~G59.5+0.1 \citep{Xu12}, which may also be contributing to diffuse X-ray emission in our mosaic.

%\clearpage
%-----------------------------------------------------------------------------
\subsection{W42 and RSGC1 \label{sec:w42}}
%W42 + RSGC1 -- 840 point sources
%At 2.2 kpc, 4*pi*D^2 = 5.792e44.
%At 3.8 kpc (alternate distance for W42), 4*pi*D^2 = 1.728e45.
%At 6 kpc (RSGC1), 4*pi*D^2 = 4.31e45.

In a recent \Chandra GO+GTO program, we targeted the newly-formed MSFRs W42 and W33, both offering rich samples of very young ($<$0.2~Myr) massive stars ionizing UCH{\scriptsize II}Rs, to study the emergence of X-ray emission in these objects and their influence on the early evolution of MSFRs.  At the same time we explored two of the Galaxy's extremely rare YMCs, with $>$10$^{4}$~M$_{\odot}$ \citep{Portegies10}, Red Supergiant Cluster 1 (RSGC1), age 10~Myr \citep{Froebrich13}, and Cl~1813-178, age 4.5~Myr \citep{Messineo11}.  YMCs persist for $>$10~Myr because of their extraordinarily large masses; they allow us to explore changes in X-ray emission from massive stars as they approach the ends of their lives, along with the neutron stars and SNRs that they leave behind.

\begin{figure}[htb]
\centering
\includegraphics[width=0.50\textwidth]{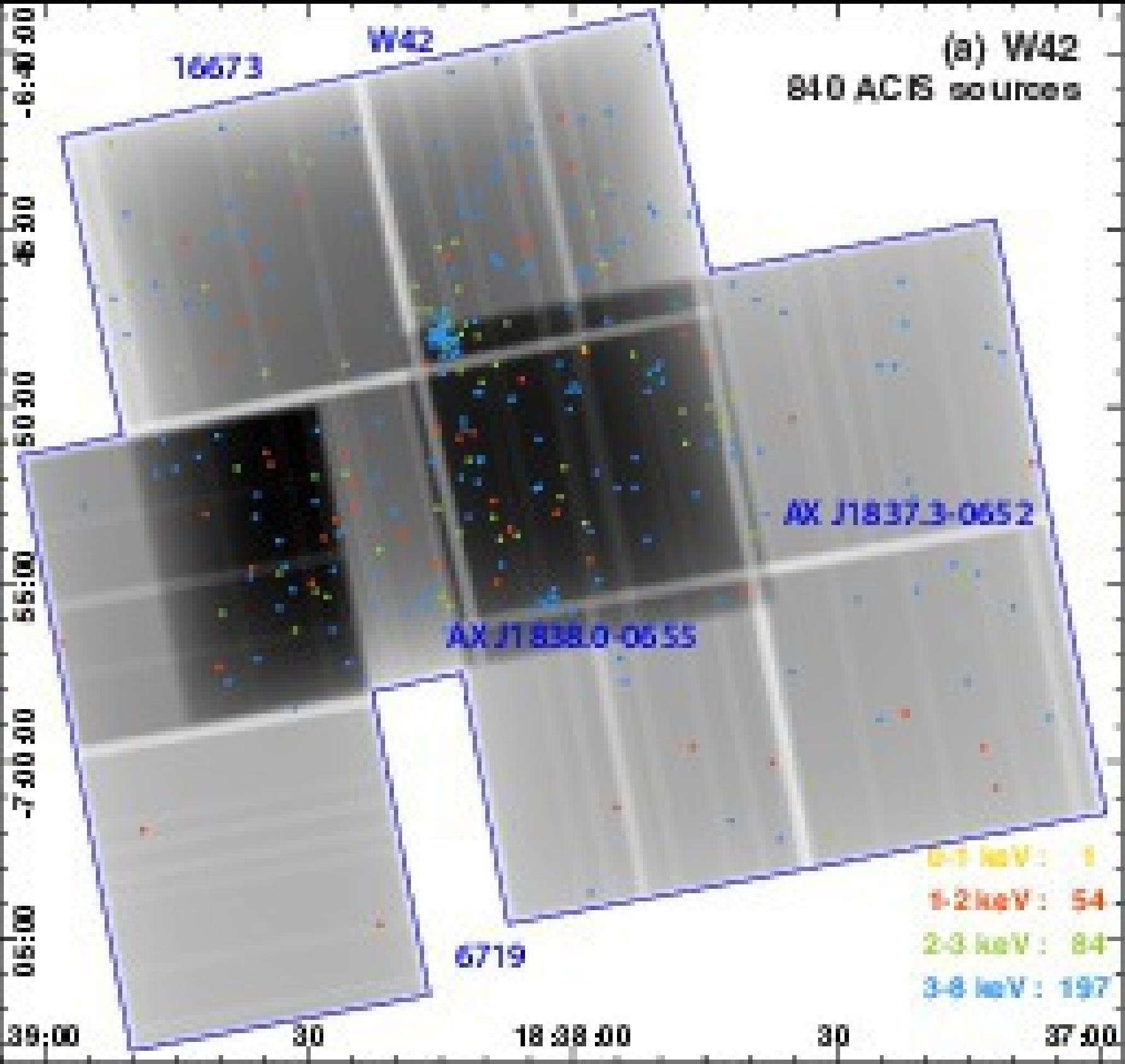}
\includegraphics[width=0.49\textwidth]{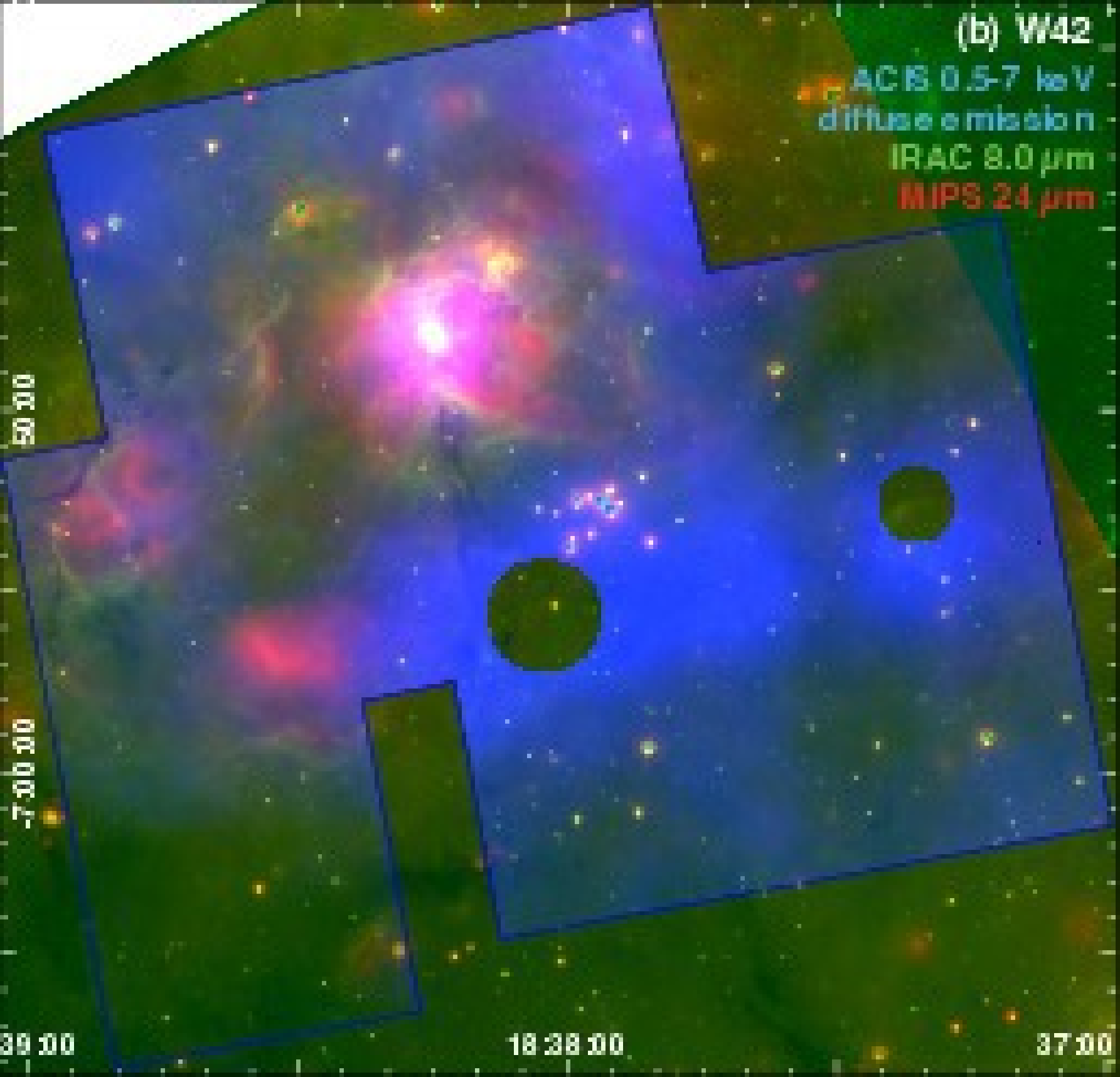}
\includegraphics[width=0.49\textwidth]{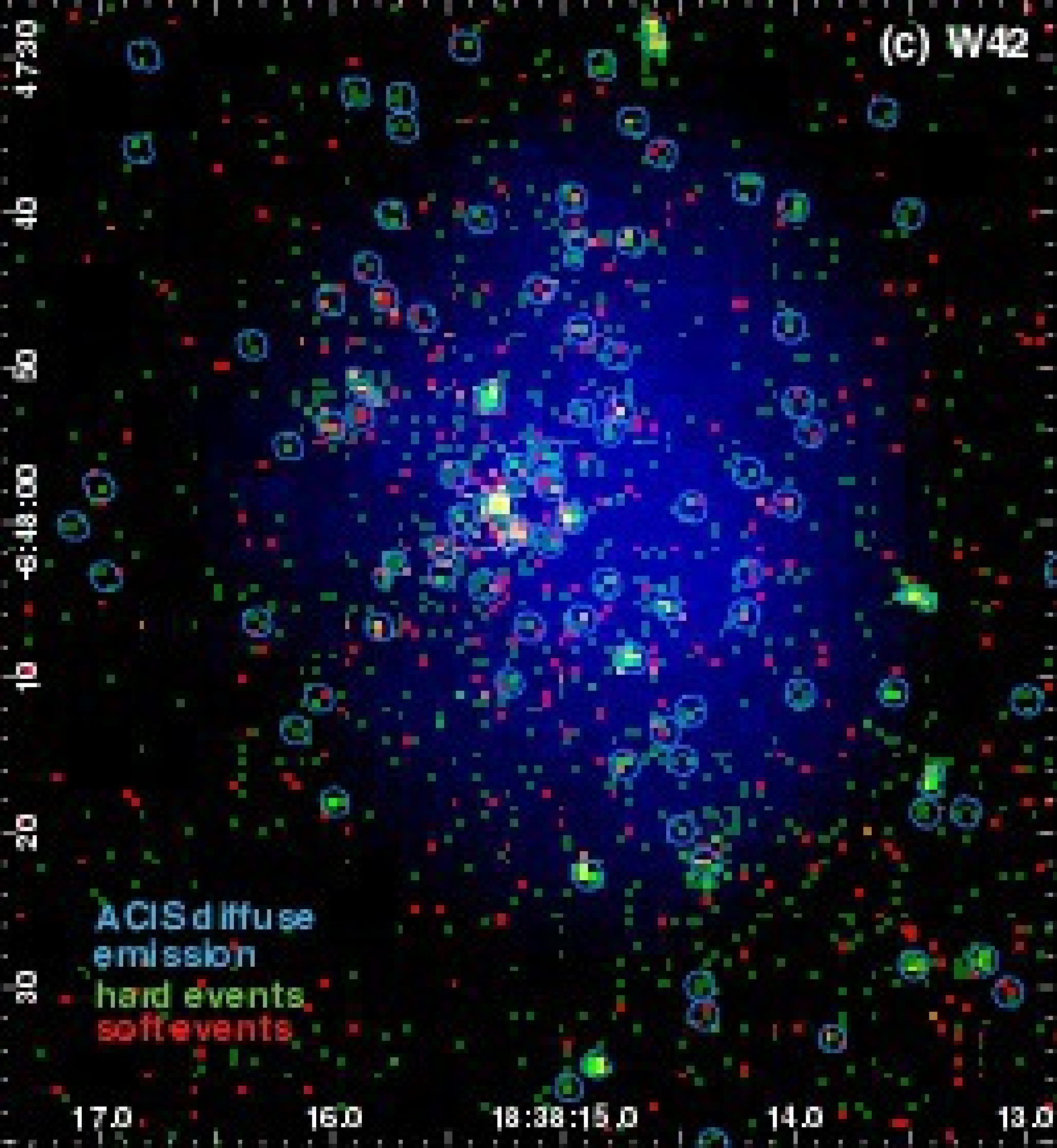}
\caption{W42 and RSGC1.
(a) ACIS exposure map with brighter ($\geq$5 net counts) ACIS point sources overlaid; colors denote median energy for each source.  ObsID numbers and regions named in the text are shown in blue.
(b) ACIS diffuse emission in the \Spitzer context.  The two pulsars in the field, AX~J1838.0$-$0655 and AX~J1837.3$-$0652 \citep{Gotthelf08}, along with the brightest parts of their associated diffuse X-ray emission, have been masked so that fainter diffuse X-ray emission across the field can be displayed.
(c) ACIS event data and diffuse emission at the center of W42. 
\label{W42.fig}}
\end{figure}

Here our ACIS-I mosaic (Figure~\ref{W42.fig}) captures both the nearby MSFR W42 \citep{Blum00} and the unrelated background YMC RSGC1 \citep{Figer06}, along with two pulsars and their PWNe \citep{Gotthelf08}.  We recover all X-ray sources tabulated by \citet{Gotthelf08}.  We find a dense cluster of X-ray sources at the center of W42 (Figure~\ref{W42.fig}(c)).

W42 hosts at least 4 UCH{\scriptsize II}Rs \citep{Urquhart13} and contains over 500 candidate young stellar objects based on IR-excesses \citep{Dewangan15a}.  The brightest near-IR source, W42~No.~1, is an O5--O6 star \citep{Blum00}.  W42's edge-on orientation illustrates the connection between molecular filaments and MSFRs:  part of a massive filament has collapsed to form the MSFR that is now blowing a bipolar bubble normal to the filament's long axis \citep{Deharveng10,Dewangan15a}.  We find few X-ray sources along this filament south of the W42 cluster. 

The bright yellow X-ray source slightly northeast of field center in Figure~\ref{W42.fig}(c) is c1575 (CXOU~J183815.27-064758.8), the counterpart to W42~No.~1 \citep{Blum00}.  It has 148 net counts and a median energy of 2.4~keV; the spectral fit gives $N_H = 2.3 \times 10^{22}$~cm$^{-2}$, $kT = 3.3$~keV, and $L_X = 5.5 \times 10^{31}$~erg~s$^{-1}$.  Again this thermal plasma temperature is too hard for a single mid-O star, although the luminosity is modest compared to many CWBs.  %Perhaps a strong magnetic field should also be considered for this source.  
Alternatively, the W42 complex may be more distant than the 2.2~kpc \citep{Blum00} we have assumed; \citet{Dewangan15a} use a distance of 3.8~kpc.  That distance increases the X-ray luminosity of source c1575 to $L_X = 1.6 \times 10^{32}$~erg~s$^{-1}$.
%Gotthelf08's source #17 had contributions from many X-ray sources in the cluster core.

RSGC1 \citep{Figer06} is a 10-Myr-old obscured cluster \citep{Froebrich13,Davies08}, the fifth most massive Galactic YMC known \citep{Richards12}.  It has $>$200 massive stars, far more than the other known RSGCs \citep{Froebrich13}, that form a compact cluster.  The X-ray-bright pulsar AX~J1838.0$-$0655 is thought to be a member of RSGC1 \citep{Gotthelf08}.  Despite almost 74~ks of ACIS-I integration, we still do not detect any of the red supergiants in RSGC1 \citep[Table~1 in][]{Figer06}.  There are many X-ray sources in this region of the mosaic, however, and several are quite hard.  These highly-absorbed sources may well be members of RSGC1.

\citet{Gotthelf08} analyzed ObsID~6719 and noted that their soft X-ray source \#12 is the high-proper-motion spectroscopic binary HD~171999.  Our source c747 (CXOU~J183758.77-064822.2) is their source \#12.  Nearby in our data is the similarly soft X-ray source c738 (CXOU~J183758.67-064826.0).  ObsID~16673 was obtained 9.5 years after ObsID~6719.  Using SIMBAD proper motions for HD~171999, 
% -129.65 mas/yr in RA, -398.31 mas/yr in Dec
we calculate that this source should have moved about -1.2$\arcsec$ in RA, -3.8$\arcsec$ in Dec, in the 9.5 years between ACIS observations.  Separations between c747 and c738 are about -1.5$\arcsec$ in RA, -3.8$\arcsec$ in Dec.  Thus we conclude that our sources c747 and c738 are in fact the same X-ray source, the counterpart to HD~171999.

Diffuse X-ray emission around the pulsars AX~J1838.0$-$0655 and AX~J1837.3$-$0652 \citep{Gotthelf08} extends far beyond the brightest regions that we masked and dominates our ACIS mosaic of this field (Figure~\ref{W42.fig}b).  The bright PWN around pulsar AX~J1838.0$-$0655 (our piled source c989) may have been over-reconstructed into point sources by our machinery, thus the six sources surrounding c989 may be spurious.  

Fainter X-ray emission pervades much of the wider field, filling bubbles and voids in the \Spitzer images.  The brightest unresolved emission at the center of the W42 cluster (Figure~\ref{W42.fig}c) undoubtedly includes a substantial contribution from faint cluster members; spectral analysis will be required to establish if any of this emission is truly diffuse.

%\clearpage
%-----------------------------------------------------------------------------
\subsection{W33 and Cl~1813-178 \label{sec:w33}}
% W33 + Cl 1813-178 -- 1960 point sources
% W33 is a cluster of clusters, like NGC 7538.
% At 2.4 kpc (W33), 4*pi*D^2 = 6.893e44.
% At 3.8 kpc (Cl 1813-178), 4*pi*D^2 = 1.728e45.
% See data/counterparts/notes.txt for details on counterparts.

%Messineo15 -- massive stars in W33.  They appear not to have looked at W33 Main.
%Messineo11 Table 5 gives positions of clusters around W33.  It doesn't include W33 Main for some strange reason -- I used Immer14's Table 1 position for that cluster.  Messineo11's Table 6 gives candidate SNRs -- they are big, so they encompass many X-ray sources.  In the future, look especially for soft things that could be neutron stars.  See w33.reg.

\begin{figure}[htb]
\centering
\includegraphics[width=0.98\textwidth]{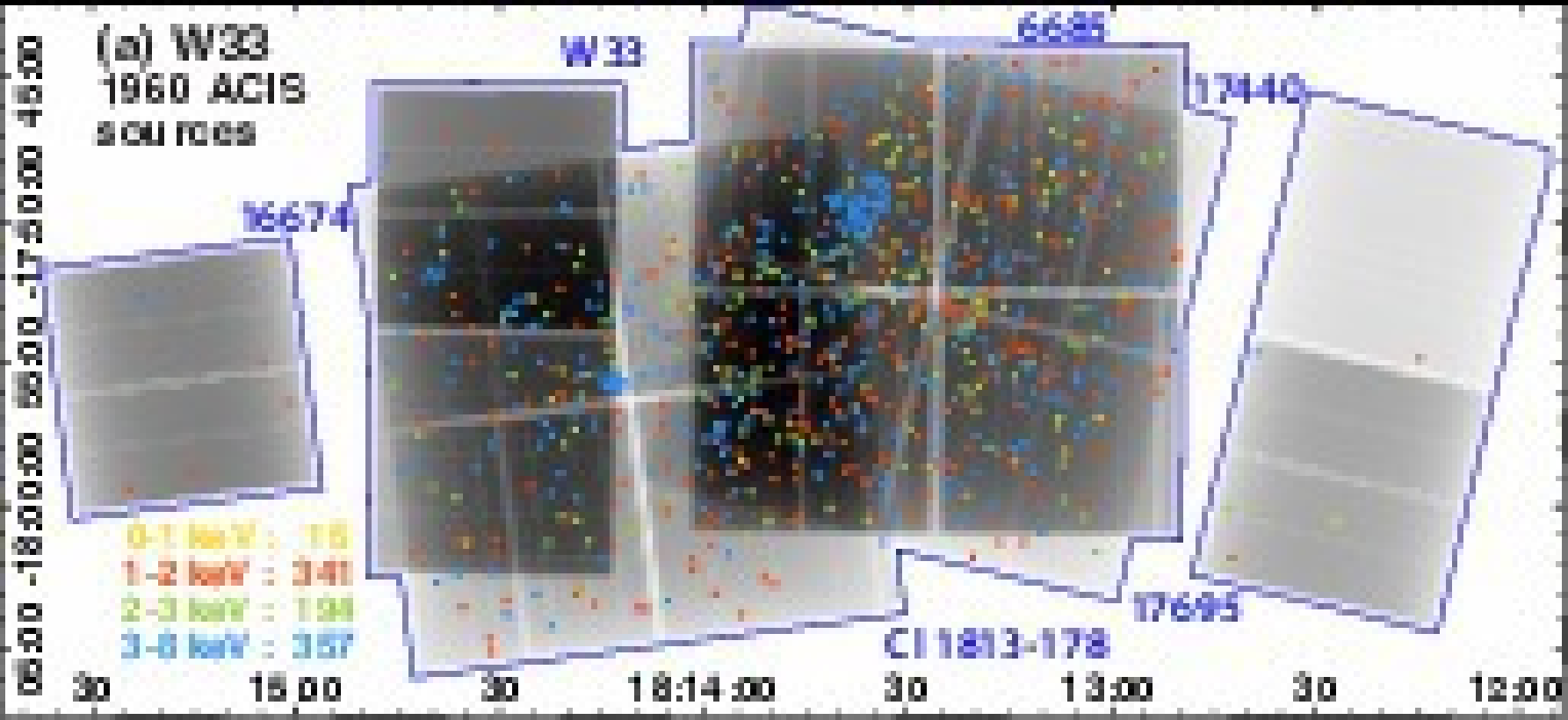}
\includegraphics[width=0.98\textwidth]{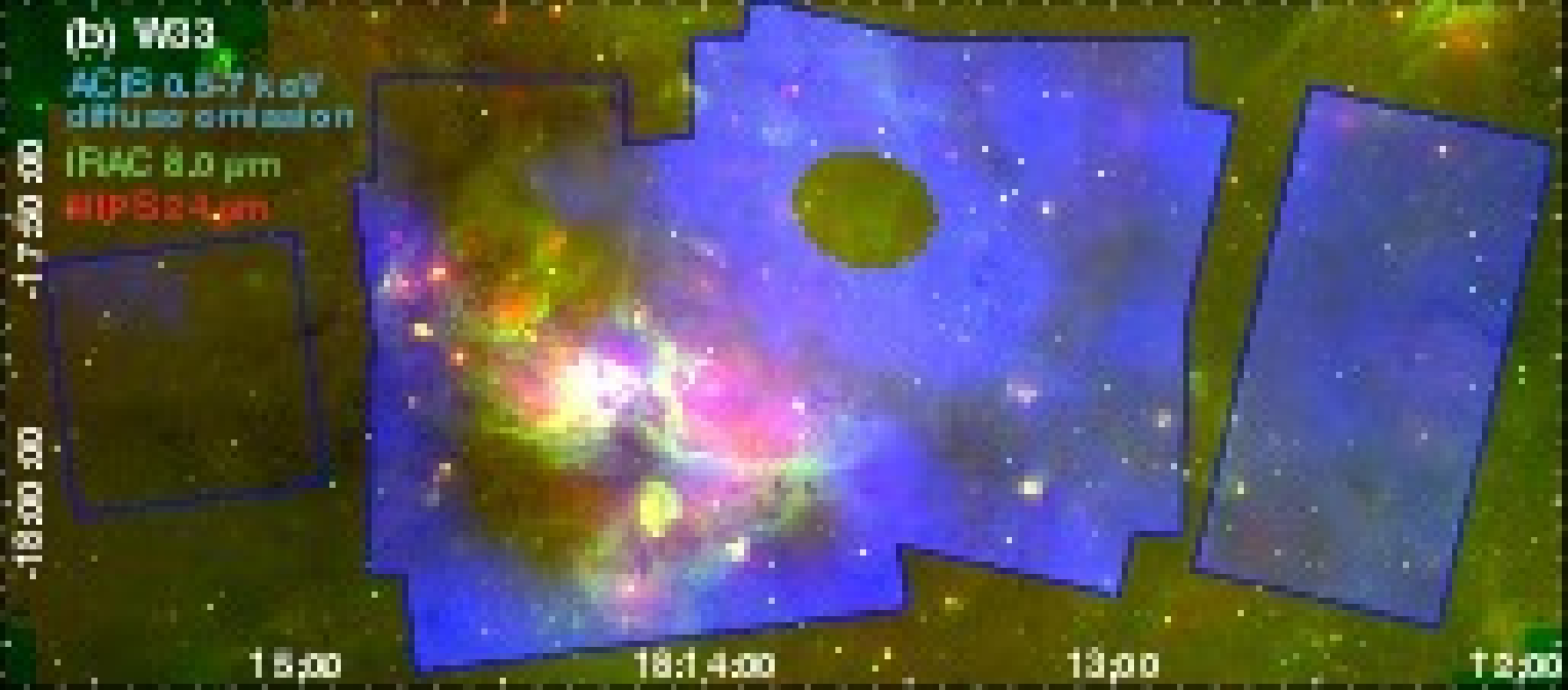}
\caption{W33.
(a) ACIS exposure map with brighter ($\geq$5 net counts) ACIS point sources overlaid; colors denote median energy for each source.  ObsID numbers and regions named in the text are shown in blue.
(b) ACIS diffuse emission in the \Spitzer context.  The bright SNR G12.82$-$0.02 has been masked.
\label{W33.fig}}
\end{figure}

The other targets in our recent GO program to study MSFRs at a range of ages were the very young W33 complex and the older, more distant YMC Cl~1813-178.  As we found in the ACIS mosaic of W42 and RSGC1, here again we have a superposition of unrelated MSFRs separated by just a few arcminutes on the sky (Figure~\ref{W33.fig}).  In this case, Cl~1813-178 is so big and so well-populated with X-ray sources that it will be a challenge to assign membership in the part of the field where it overlaps with the foreground W33 complex. 

The W33 MSFR consists of a chain of very young stellar clumps and sparse clusters.  It appears that these clusters formed separately and nearly simultaneously \citep{Beck98}; this may indicate that they formed at sausage-instability density enhancements along a massive molecular filament \citep{Jackson10,Chandra53}.  The distance to W33 has recently been firmly established at 2.4~kpc via maser trigonometric parallax \citep{Immer13}.

\begin{figure}[htb]
\centering
\includegraphics[width=0.49\textwidth]{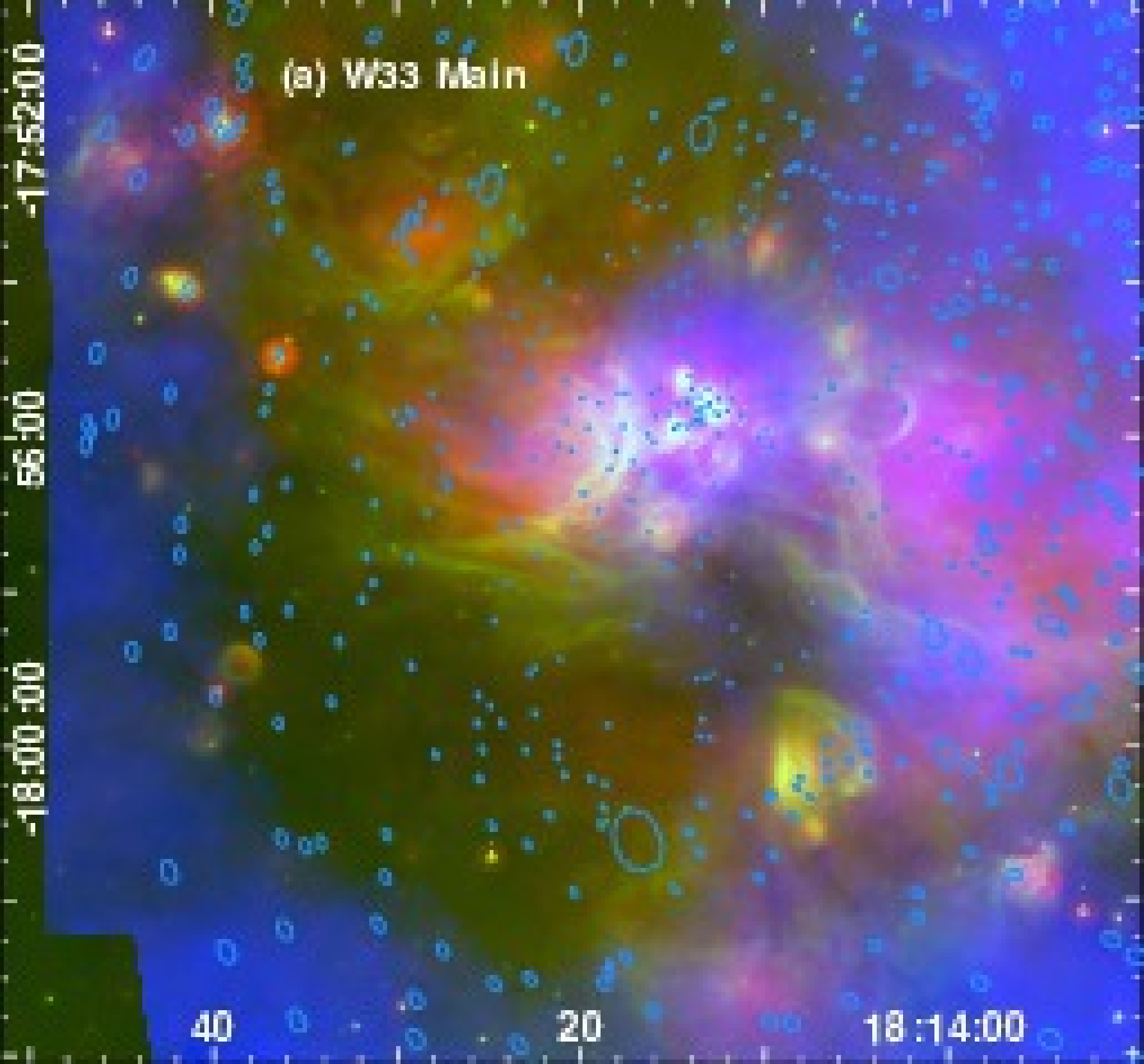}
\includegraphics[width=0.50\textwidth]{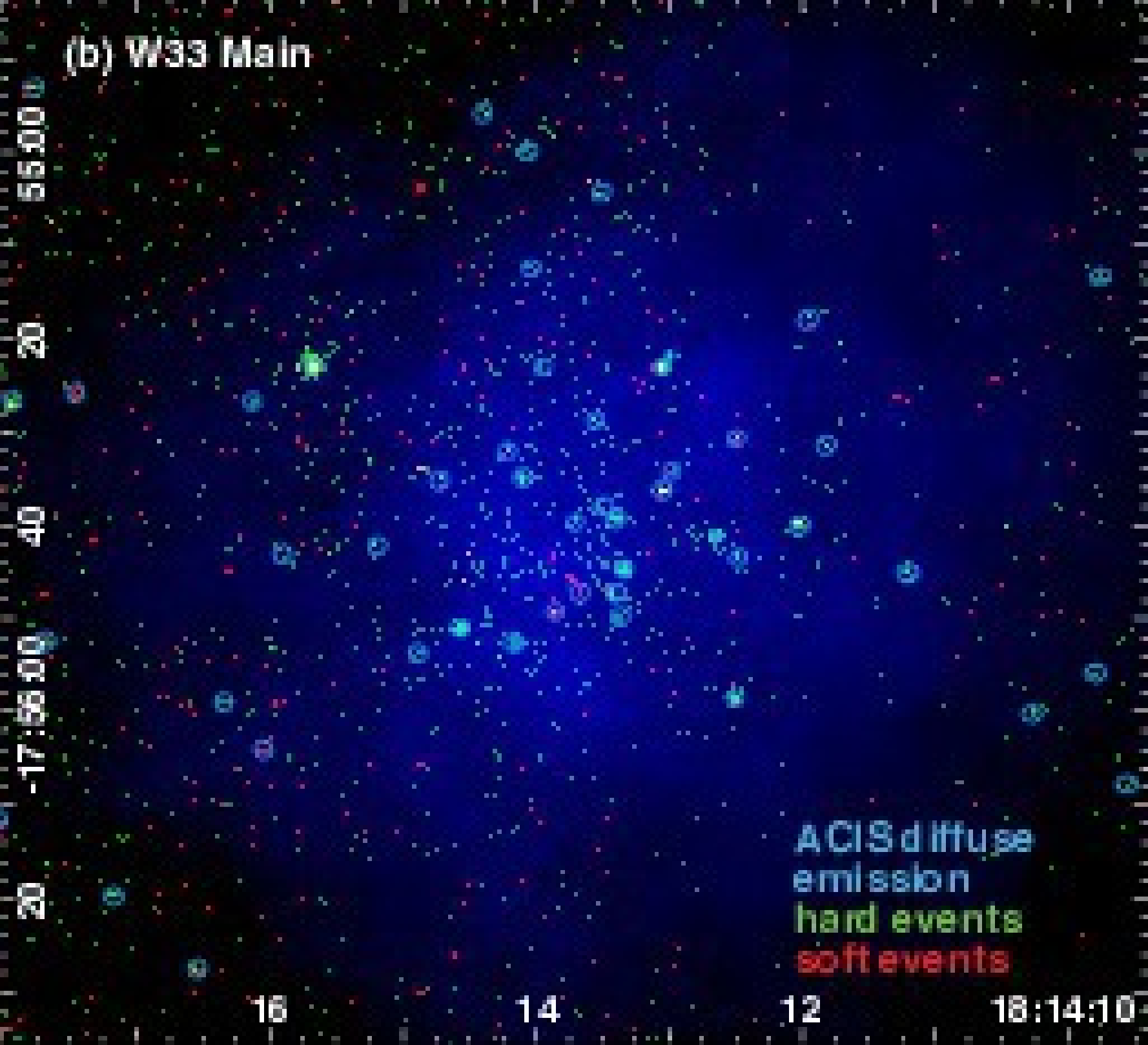}
\caption{W33 Main.
(a) Zoomed version of Figure~\ref{W33.fig}(b) for W33 Main, with point source extraction regions.
(b) ACIS event data and diffuse emission on W33 Main.  
\label{W33Main.fig}}
\end{figure}

The ACIS-I aimpoint for our study of W33 was centered on W33~Main (Figure~\ref{W33Main.fig}), which contains at least 5 mid-O stars \citep{Beck98} ionizing at least 3 distinct UCH{\scriptsize II}Rs, separated by dense material.  We find a clear concentration of X-ray sources consistent with the cluster position as given by \citet{Immer14} (Figure~\ref{W33Main.fig}(b)), but no counterparts to the infrared sources IRS1, IRS2, or IRS3 from \citet{Dyck77} (using SIMBAD positions).  

W33A, a deeply-embedded MSFR that features a well-studied MYSO \citep{Davies10}, is captured at the northeast corner of the ACIS-I field.  W33A may be forming due to the intersection of two massive molecular filaments \citep{Galvan10}.  We find five faint X-ray sources in a clump at the location of W33A; all have median energies above 3~keV, indicating that they are highly obscured.  If these sources are associated with star formation in W33A, they indicate either that X-ray emission turns on very early in the star formation process or that W33A has experienced multiple epochs of star formation, with today's MYSO and dusty cores representing current and future star formation while the X-ray sources represent an older population.
%From DeWit10 Fig2, W33A is at 18 14 39.51 -17 52 00.2.  Chandra sources are c4333, c4325, c4318, c4336, and c4341

The ACIS data cover four more clusters associated with \hii regions across the W33 complex:  radio region W33B and IR clusters cl1, cl2, and Mercer~1 \citep{Messineo11}.  While all of these regions contain X-ray sources, none of them shows a clear central concentration indicative of an X-ray clump, as we saw in W33~Main and W33A.

The monolithic YMC Cl~1813-178 was discovered serendipitously \citep{Messineo08} just $\sim$13$\arcmin$ west of W33~Main, but \citet{Messineo11} estimate that it is much more distant (3.8--4.8~kpc), hence unrelated to W33.  Cl~1813-178 is one of fewer than 20 YMCs known in the Milky Way \citep{Clark13}, with a mass of $10^{4}$~M$_{\odot}$ and many known massive stars \citep{Messineo11}.  

\begin{figure}[htb]
\centering
\includegraphics[width=0.49\textwidth]{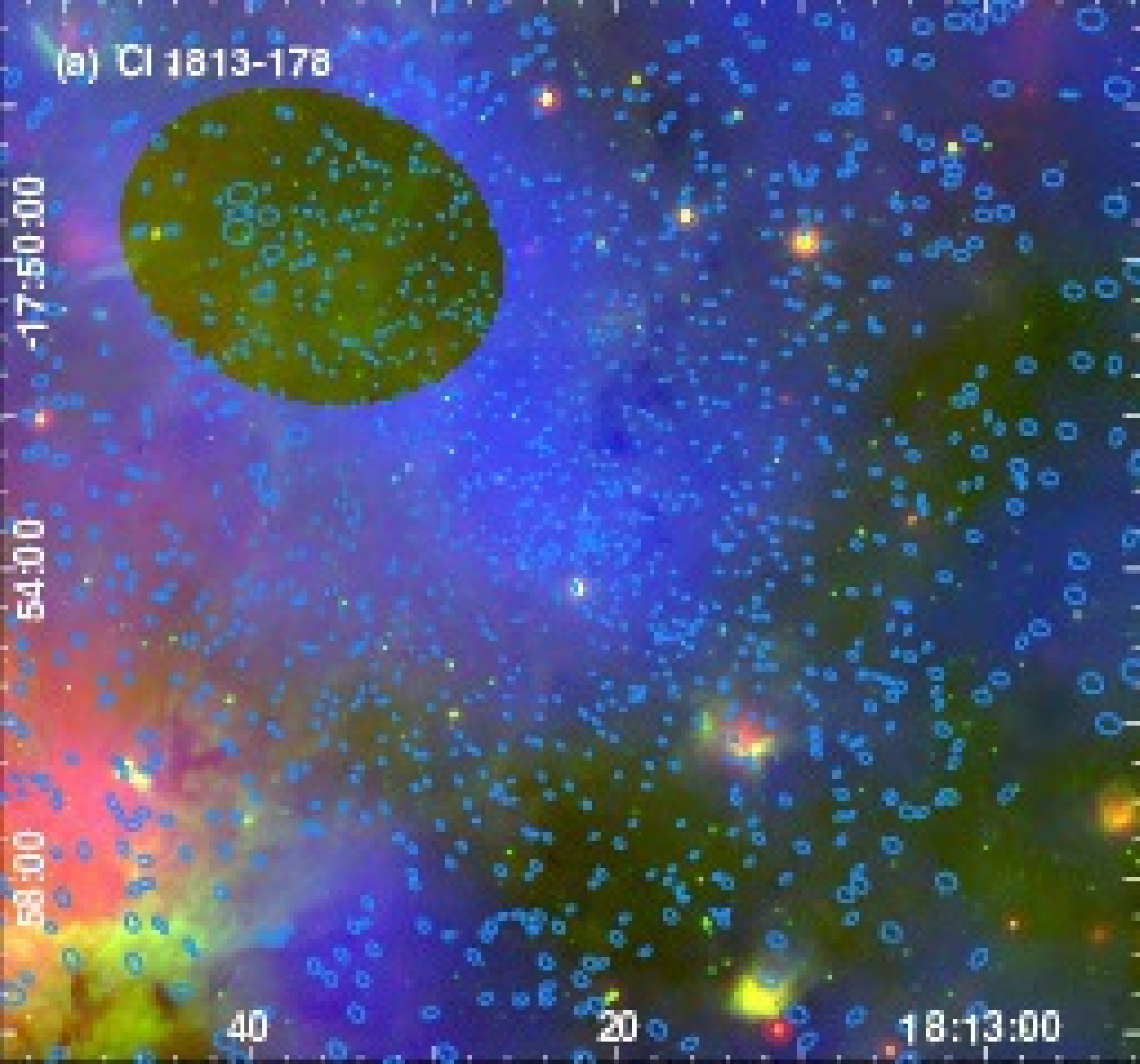}
\includegraphics[width=0.5\textwidth]{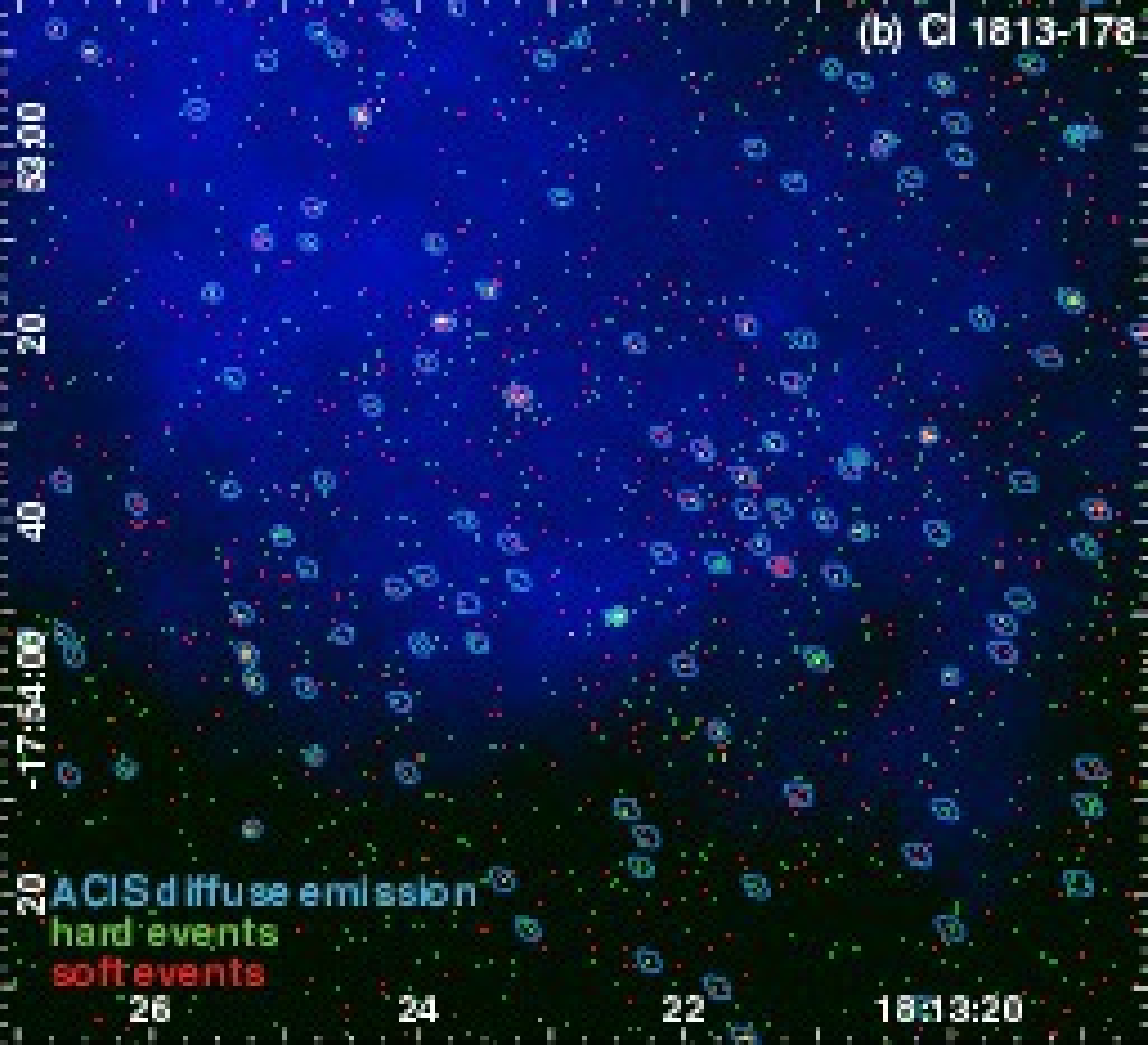}
\caption{Cl~1813-178.
(a) Zoomed version of Figure~\ref{W33.fig}(b) for Cl~1813-178, with point source extraction regions.
(b) ACIS event data and diffuse emission for the center of Cl~1813-178.
\label{Cl1813.fig}}
\end{figure}

\citet{Helfand07} analyzed the original 30-ks ACIS-I observation of this field (which targeted SNR G12.82$-$0.02) and catalogued 75 X-ray sources; we recover all of those sources in our analysis.  Our combined (ACIS GTO + archival) 59-ks \Chandra dataset provides a wealth of candidate members of Cl~1813-178 (Figure~\ref{Cl1813.fig}), with 572 X-ray sources within the reported cluster radius of 3.5$\arcmin$ \citep{Messineo11} and hundreds more beyond that.  

We have X-ray counterparts to at least 16 of the 25 massive stars listed in \citet[][Table~1]{Messineo11}.  They find two Wolf-Rayet (WR) stars in this cluster:  their \#4 and \#7, both of type WN7.  The first of these, \#4, is our source c807 (CXOU~J181314.20-175343.4), with 348 net counts and a median energy of 3.6~keV; a spectral fit gives $N_H = 8 \times 10^{22}$~cm$^{-2}$, $kT = 2.8$~keV, and $L_X = 1.2 \times 10^{33}$~erg~s$^{-1}$.  % Note that this source appears to have a strong emission line at 3.75~keV -- see moxc2_withgau.xcm.  This looks a little like Anderson11's source I20S in IRAS20126 -- is this line due to fluorescing the brick wall around the source?  When column is lower, these funny lines go away.
The second  of these, \#7, is our source c1281 (CXOU~J181322.48-175350.2), with 68 net counts and a median energy of 2.8~keV.  It is crowded with the faint source c1286 (CXOU~J181322.56-175350.1), with 6 net counts and median energy 3.7~keV.  A spectral fit to c1281 gives $N_H = 4 \times 10^{22}$~cm$^{-2}$, $kT = 2.6$~keV, and $L_X = 1.3 \times 10^{32}$~erg~s$^{-1}$.  Based on the high thermal plasma temperatures and X-ray luminosities for these WR stars, we concur with Messineo et al.\ that they are likely CWBs.

We also fit the X-ray spectrum of \citet{Messineo11} source \#5, which they find to be a late-O star but with an unusually hard X-ray spectrum.  This is our source c1346 (CXOU~J181323.71-175040.5), with 265 net counts and a median energy of 3.0~keV; fit parameters are $N_H = 2.0 \times 10^{22}$~cm$^{-2}$, $kT > 10$~keV, and $L_X = 3.5 \times 10^{32}$~erg~s$^{-1}$.  This spectrum is equally well fit by a power law ({\em TBabs*pow} in {\em XSPEC}) with $N_H = 2.0 \times 10^{22}$~cm$^{-2}$, slope $\Gamma = 1.5$, and $L_X = 3.6 \times 10^{32}$~erg~s$^{-1}$.  As Messineo et al.\ suggest, this source may be another CWB.  Its spectrum is so hard that it might be compared to that found for source I20S in IRAS~20126+4104 \citep{Anderson11}.

Our ACIS mosaic reaches approximately the same sensitivity in both W33 and Cl~1813-178; W33 is closer but more obscured.  As shown in all figures in this section, diffuse X-ray emission pervades the entire W33 mosaic.  As we saw in our W42 mosaic, the brightest diffuse emission here is associated with the PWN around a pulsar (PSR~J1813$-$1749) and its SNR (G12.82$-$0.02) \citep{Funk07,Helfand07,Gotthelf09}.  Some of our catalogued point sources that are superposed on this bright diffuse emission could be artifacts of image reconstruction; identifying counterparts to these X-ray sources at longer wavelengths is the best way to set aside such a concern.  Fainter diffuse emission extends to the west across the entire Cl~1813-178 ACIS pointing.  It appears at the W33~Main and Cl~1813-178 cluster centers and around the edges of the W33 ACIS-I field.  It is faint but present on the easternmost ACIS-S CCD.  The W33 MSFR probably shadows background diffuse emission across most of the W33 pointing.

%\clearpage
%-----------------------------------------------------------------------------
\subsection{NGC~7538 \label{sec:n7538}}
% NGC 7538 -- 487 point sources
% D = 2.65 kpc, so 4*pi*D^2 = 8.4041e44 cm^2.
% Spitzer data are a chopped-up mess.
% This is a cluster of clusters.

NGC~7538 is a clumpy Outer Galaxy MSFR, made up of several separate star-forming sites with a range of ages \citep{McCaughrean91,Balog04,Ojha04}, similar to W33.  It hosts the O3V star IRS6 and the O9V star IRS5 \citep{Puga10}, along with a variety of MYSOs that are well-studied in the radio \citep[e.g.,][]{Beuther17}.

\begin{figure}[htb]
\centering
\includegraphics[width=0.41\textwidth]{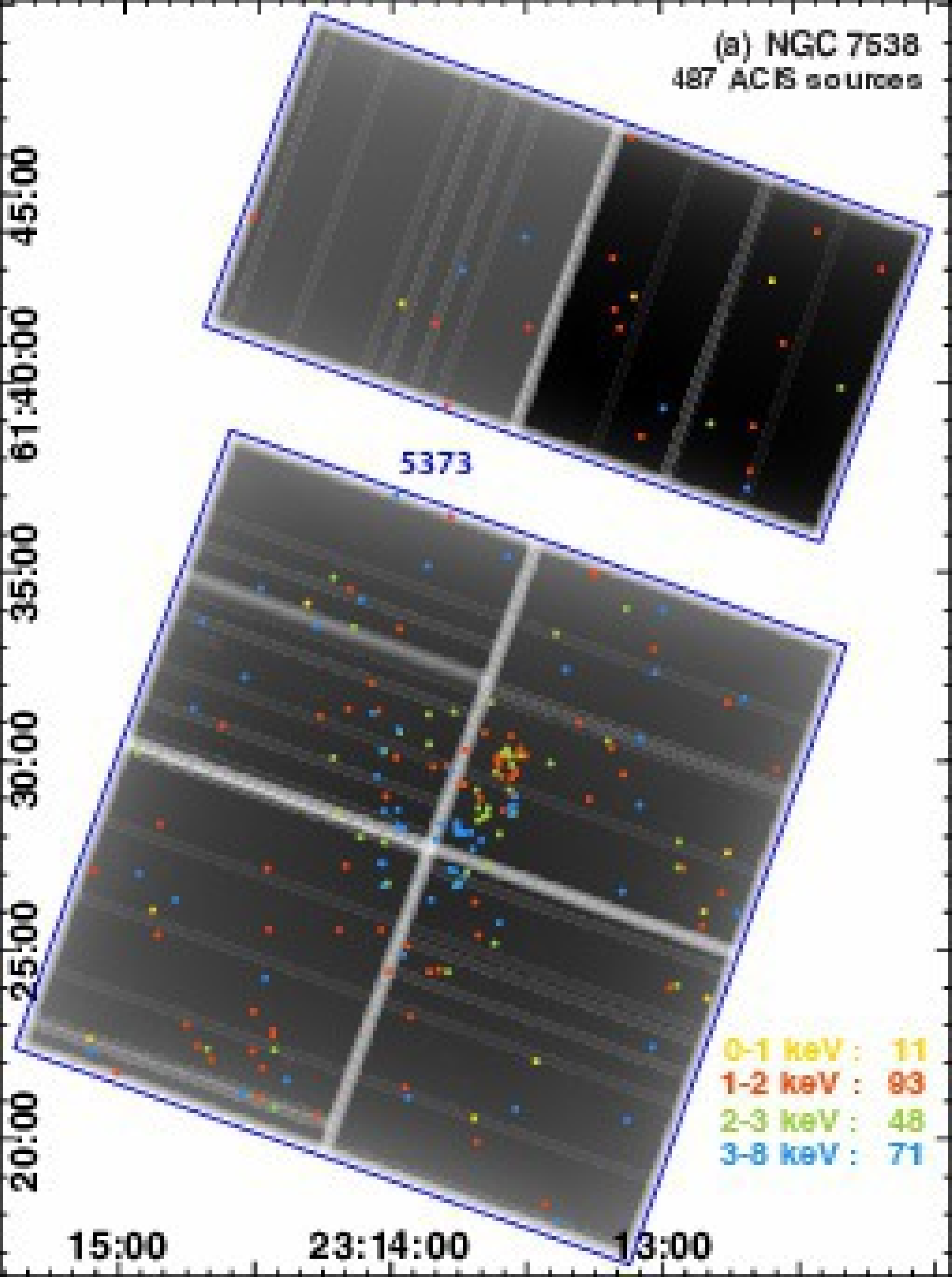}
\includegraphics[width=0.48\textwidth]{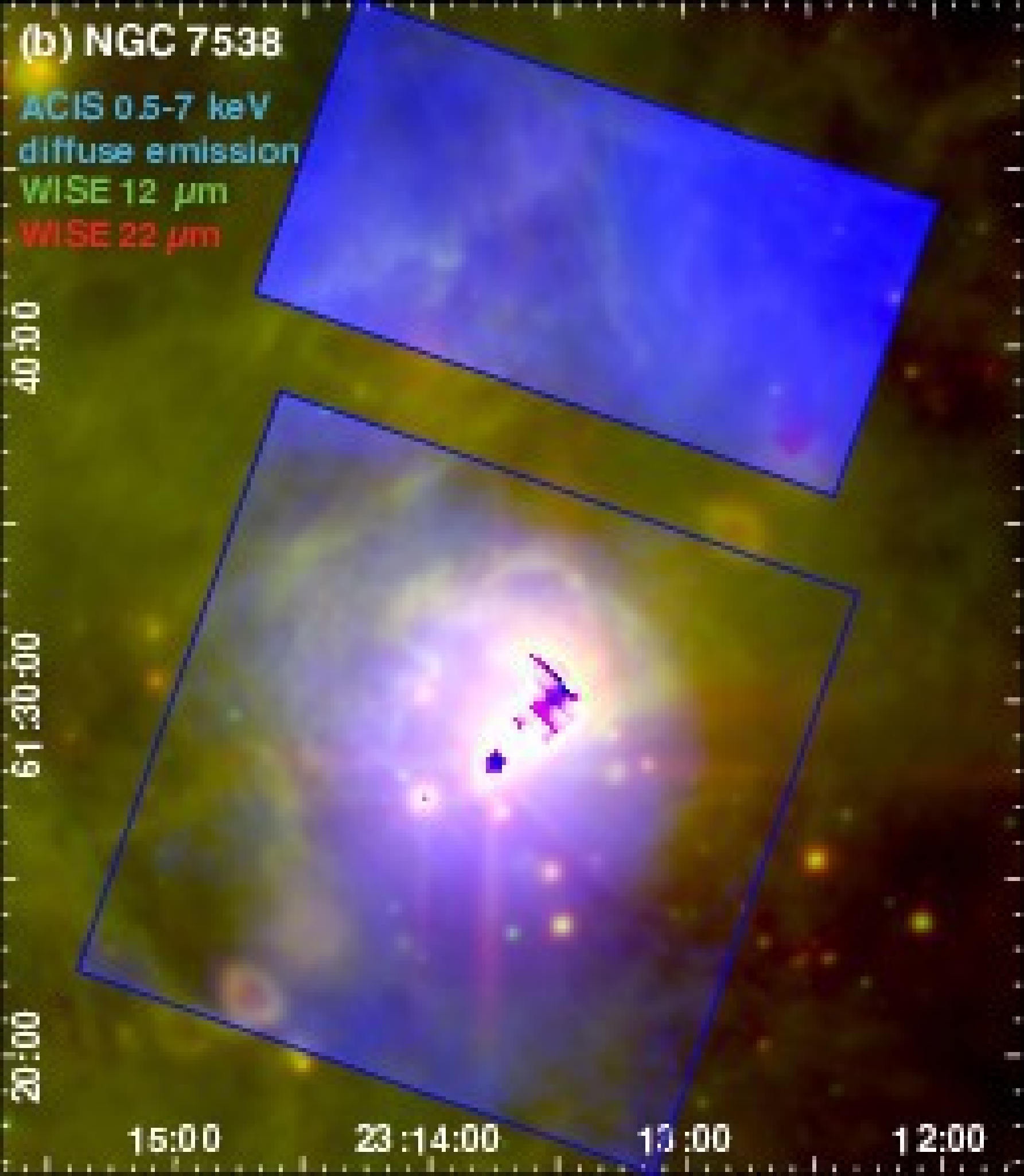}
\includegraphics[width=0.24\textwidth]{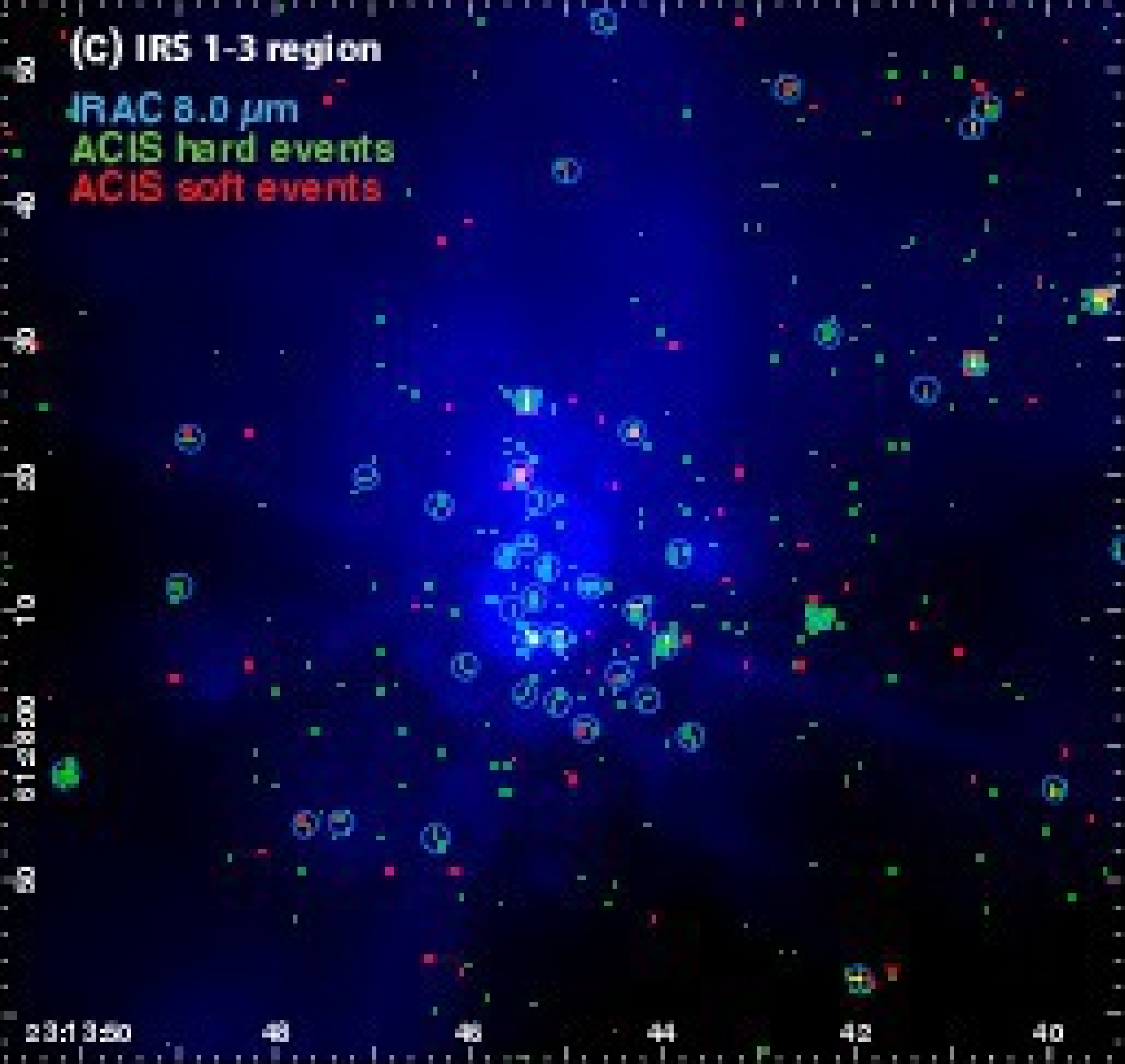}
\includegraphics[width=0.24\textwidth]{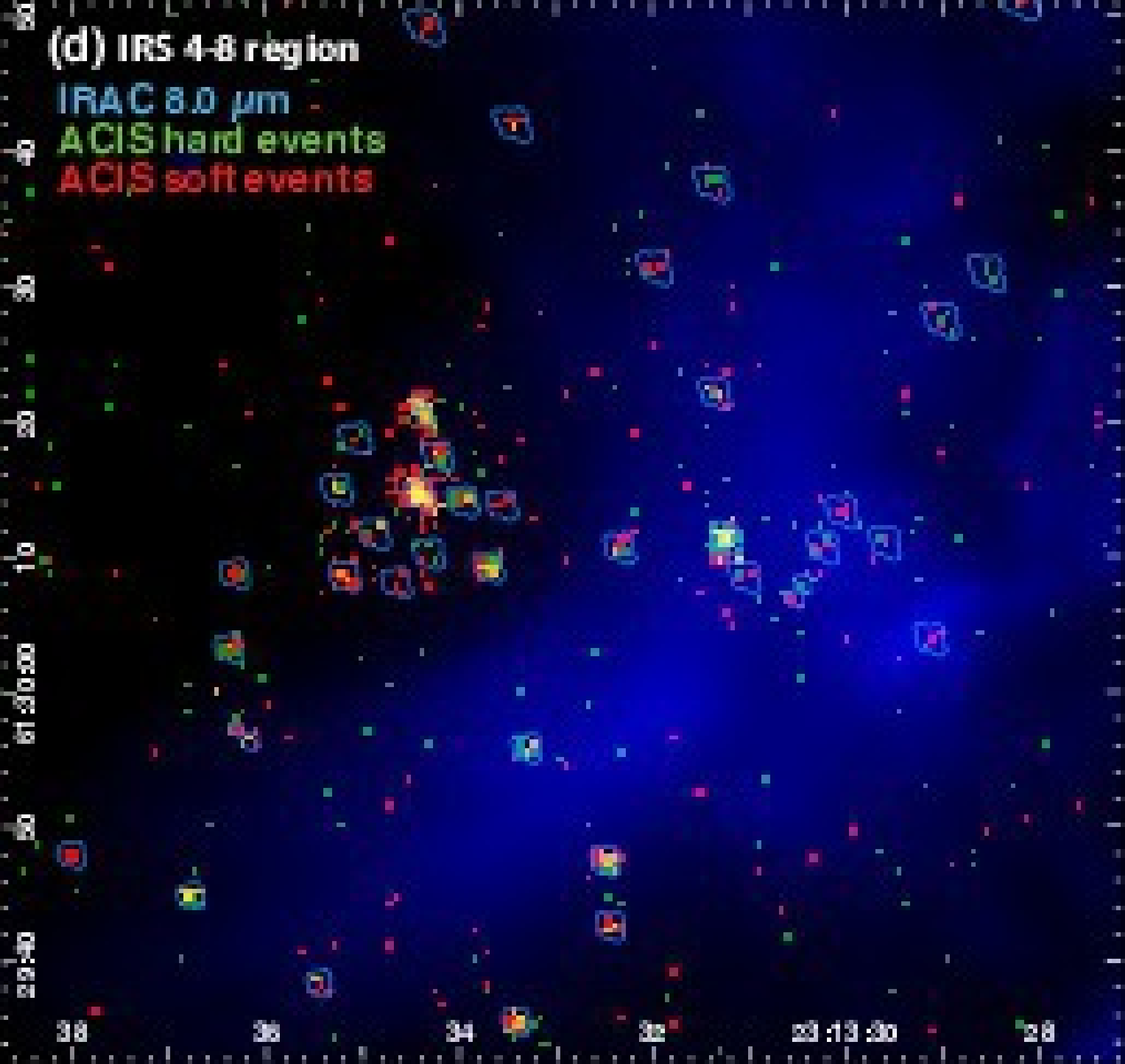}
\includegraphics[width=0.24\textwidth]{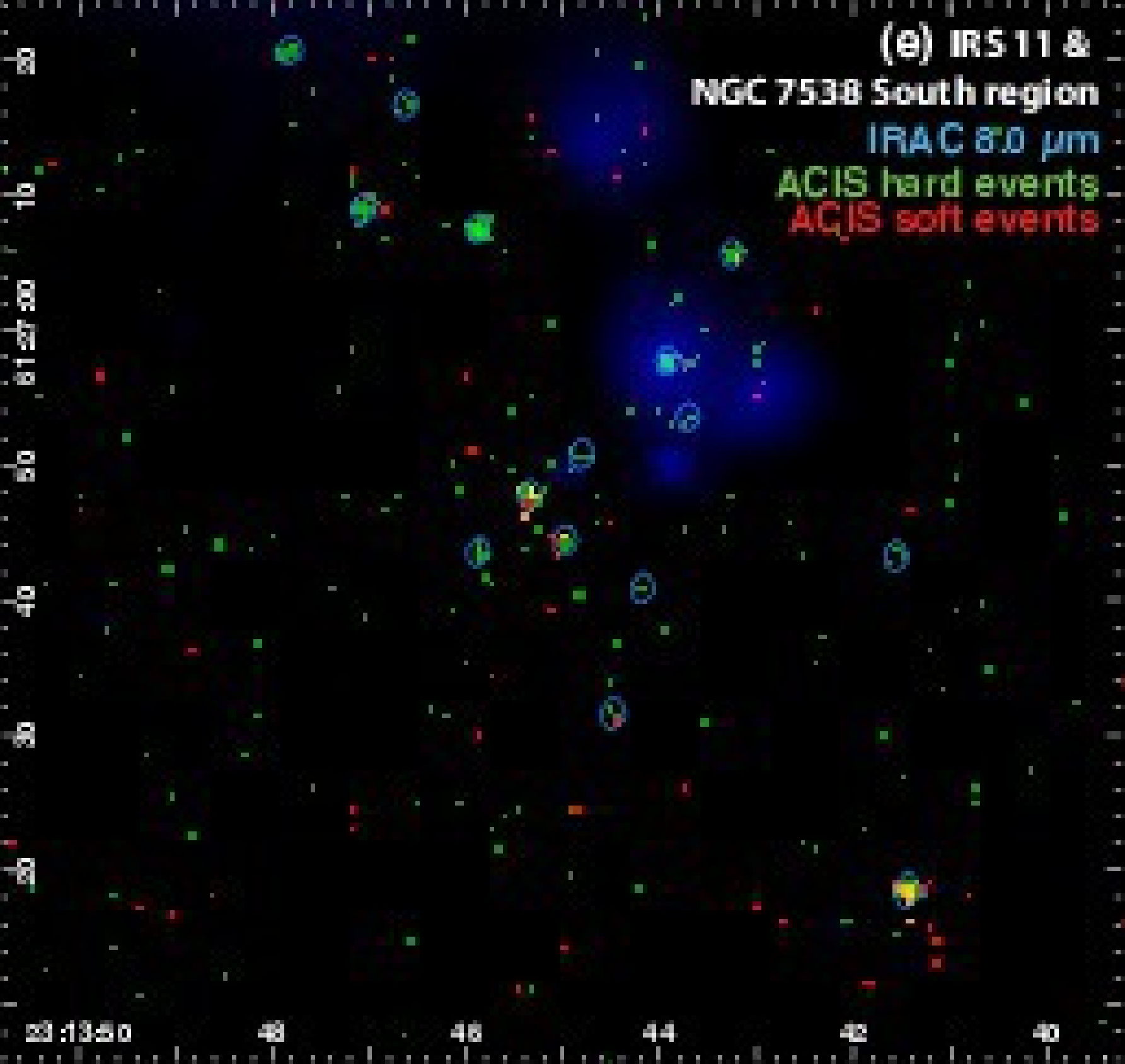}
\includegraphics[width=0.24\textwidth]{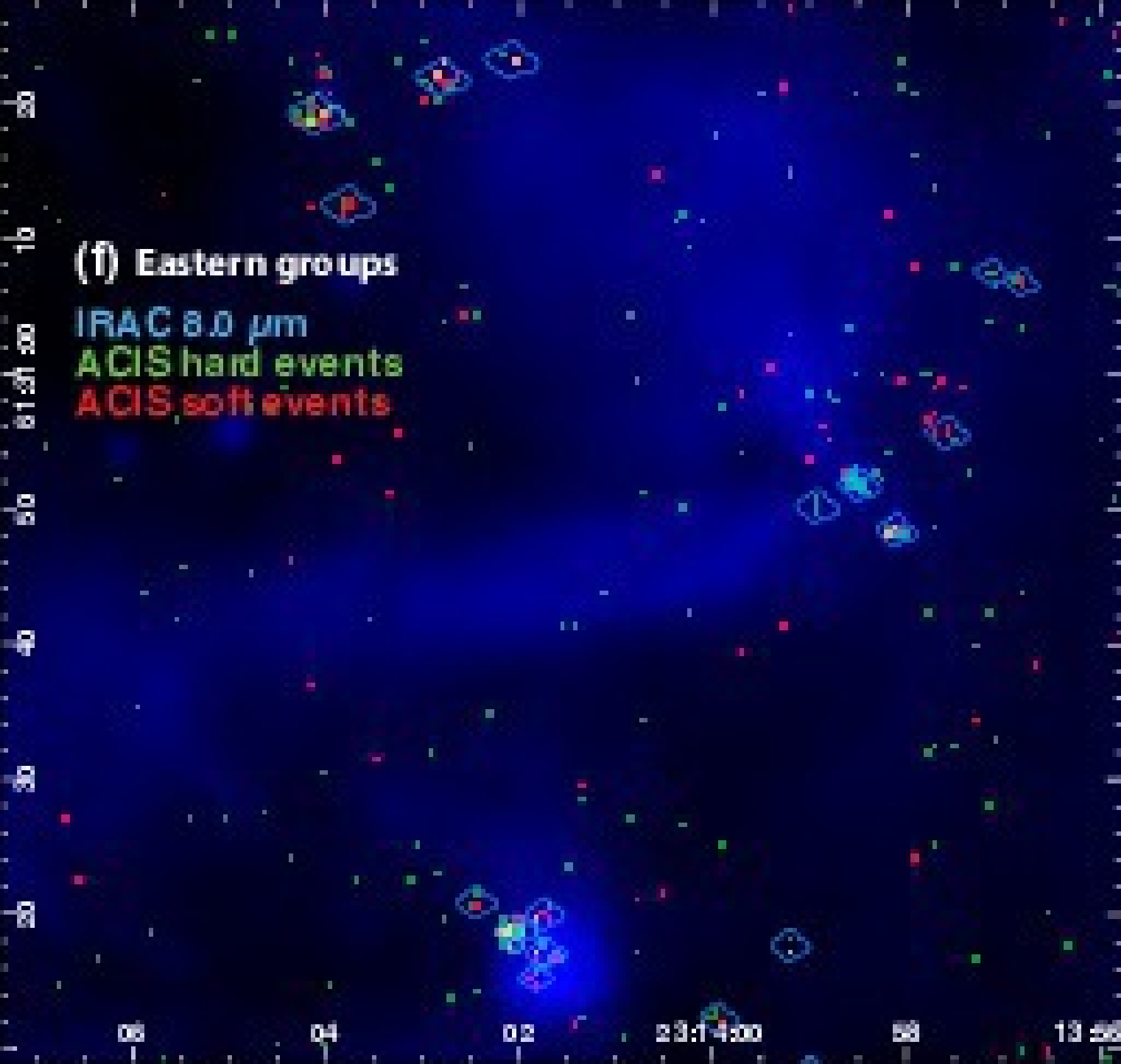}
\caption{NGC~7538.
(a) ACIS exposure map with brighter ($\geq$5 net counts) ACIS point sources overlaid; colors denote median energy for each source.  The ObsID number is shown in blue.
(b) ACIS diffuse emission in the \WISE context.  \Spitzer data have incomplete coverage of this area.
(c)--(f) ACIS event images for select star-forming clumps, with \Spitzer IRAC 8~$\mu$m images included for orientation.
(c) The IRS1--3 region.  Saturated sources dominate the center of the IRAC image.
(d) The IRS4--8 region.
(e) The IRS11 and NGC~7538 South region.
(f) The globules noted by \citet{Sharma17}.
\label{NGC7538.fig}}
\end{figure}

The \Chandra data were analyzed recently as part of two multiwavelength studies \citep{Mallick14,Sharma17}, finding 182 and 190 X-ray point sources respectively, but neither of these papers gives full catalogs for their \Chandra sources.  Mallick et al.\ tabulate 27 X-ray sources with NIR counterparts in the IRS1--3 and IRS9 regions; they provide some X-ray properties for that subset of their X-ray detections.  We recover all of these sources in our analysis and go on to find almost 500 X-ray sources and diffuse X-ray emission in this single 30-ks ACIS-I observation (Figures~\ref{NGC7538.fig}(a) and (b)).

In fact we find clumps of X-ray sources associated with all the well-known star-forming regions in NGC~7538:  IRS1--3, IRS4--8, IRS9, IRS11/NGC~7538S, and the globules described by \citet{Sharma17}.  ACIS images of most of these regions are presented in Figures~\ref{NGC7538.fig}(c)--(f); the IRS9 region is omitted because it has few X-ray sources and the IRAC 8~$\mu$m image is dominated by the single bright source IRS9 itself.  \citet{Sharma17} give short summaries of these star-forming regions; they all harbor massive stars or MYSOs, many ionizing various types of \hii regions.  Here we briefly describe X-ray counterparts to the famous IRS sources.  

IRS1--3 (Figure~\ref{NGC7538.fig}(c)).  We find many X-ray sources in this young MSFR.  \Chandra source c343 (CXOU~J231345.32+612810.9) is the counterpart to the MYSO IRS1; it has just 5 X-ray events but all have energies $>$3~keV, with a median energy of 4.4~keV.  This indicates that the X-ray source is highly obscured, as expected for this massive protostar.  \Chandra source c351 (CXOU~J231345.47+612820.1) is the counterpart to IRS2, an O9.5V star; its 8 counts have a median energy of 1.6~keV.  IRS3 has no X-ray counterpart in our data.  

IRS4--8 (Figure~\ref{NGC7538.fig}(d)).  This is the main star-forming complex in NGC~7538; it is well-populated with X-ray sources.  \Chandra source cc228 (CXOU~J231330.22+613010.7) is the counterpart to the O9V star IRS5, with 9 counts and a median energy of 1.4~keV, similar to IRS2.  \Chandra source c222 (CXOU~J231334.40+613014.6) is the counterpart to the O3V star IRS6, with 400 counts and a median energy of 1.4~keV.  A spectral fit yields $N_H = 1.7 \times 10^{22}$~cm$^{-2}$, $kT = 0.8$~keV, and $L_X = 8 \times 10^{32}$~erg~s$^{-1}$.  Sources IRS4, IRS7, and IRS8 have no X-ray counterparts.

IRS9.  The faint \Chandra source c460 (CXOU~J231401.78+612718.8) is the counterpart to the MYSO IRS9.  It has just 8 counts and an extreme median energy of 5.0~keV; event energies span the wide range of 1.5--7.5 keV.  \citet{Sandell05} find multiple outflows in this region, suggesting an embedded cluster rather than a single MYSO, but we find just a few X-ray sources here.

IRS11/NGC~7538S (Figure~\ref{NGC7538.fig}(e)).  \Chandra source c322 (CXOU~J231343.92+612657.6) may be a match to the MYSO IRS11.  It has 17 counts and a median energy of 4.0~keV, with event energies ranging over 2.6--5.0~keV.  The spectral fit is not well-constrained, but the source is certainly highly obscured, with $N_H > 20 \times 10^{22}$~cm$^{-2}$, $kT \sim 1$~keV plasma with $L_X \sim 1 \times 10^{33}$~erg~s$^{-1}$.  We do not detect an X-ray counterpart to NGC~7538S.

Globules \citep{Sharma17} (Figure~\ref{NGC7538.fig}(f)).  These host small groups of X-ray sources at the eastern edge of NGC~7538.  Sharma et al.\ noted that they resemble bright-rimmed clouds and point back to the O3 star IRS6. 

The brightest diffuse X-ray emission in the ACIS observation is far off-axis on the ACIS-S chips.  On-axis, a diffuse enhancement is coincident with the central cavity around the cluster.  Faint patches of diffuse X-rays extend north and south to the edges of the ACIS-I field, anticoincident with the \WISE mid-IR emission.  \citet{Luisi16} find ionizing radiation leaking past the primary photodissociation region (PDR) to a second PDR northeast of the main bubble, demonstrating the influence of MSFRs on the surrounding ISM.  Diffuse X-ray emission extends at least as far as the second PDR to the northeast.  It also spreads to the south and southwest, past the main NGC~7538 bubble seen in the \WISE data, once again demonstrating that the hot ISM around MSFRs is more complex and extensive than might be predicted from IR data alone.

%\clearpage
%-----------------------------------------------------------------------------
\subsection{G333 \label{sec:g333}}
% G333 -- 3935 point sources
% At 2.6 kpc, 4*pi*D^2 = 8.090e44.
% At 3.6 kpc, 4*pi*D^2 = 1.551e45.
% Consider moving this target out to 3.6 kpc, in line with recent papers.  That would increase LX's by a factor of 1.9.

The elongated G333 GMC hosts several obscured MSFRs, spread over 80~pc across its long axis \citep[e.g.,][]{Russeil05,Nguyen15}.  G333 has formed several massive young clumps and clusters, all at nearly the same time and with nearly regular spacing along the cloud.  These clusters are too young to have hosted supernovae and their wind-blown bubbles do not overlap significantly, so it is unlikely that we are witnessing a ``sequential triggering'' process \citep{Elmegreen77}.

G333 may be a more evolved version of the 80-pc-long massive molecular filament and IRDC ``Nessie'' \citep{Jackson10}.  Near-IR dust extinction maps \citep{Kainulainen11} show that IRDCs are often surrounded by molecular clouds with $>$10 times more mass; thus a massive (but not extraordinary) IRDC of $10^{4}$M$_{\odot}$ can easily represent just the densest part of a typical $10^{5}$M$_{\odot}$ GMC.  So G333 may be a rare example of a very massive IRDC with an unusually favorable sky orientation that has nearly completely collapsed into regularly-spaced sites of coeval massive star formation.  Soon supernovae in these embedded clusters will destroy the last vestiges of the original IRDC and start blowing superbubbles to feed the hot ISM, as we see in Carina \citep{Townsley11a,Townsley11b} and NGC~6357 (MOXC1).  But since the underlying elongated structure still remains in G333, we can study the transformation of a massive molecular filament into a cluster of clusters.  Similar to G352 described above but more distant (so the entire GMC can be captured in a few ACIS-I pointings), G333 provides an important observational and evolutionary link between giant molecular filaments such as Nessie and  cluster-of-clusters star-forming complexes such as Carina.

\begin{figure}[htb]
\centering
\includegraphics[width=0.495\textwidth]{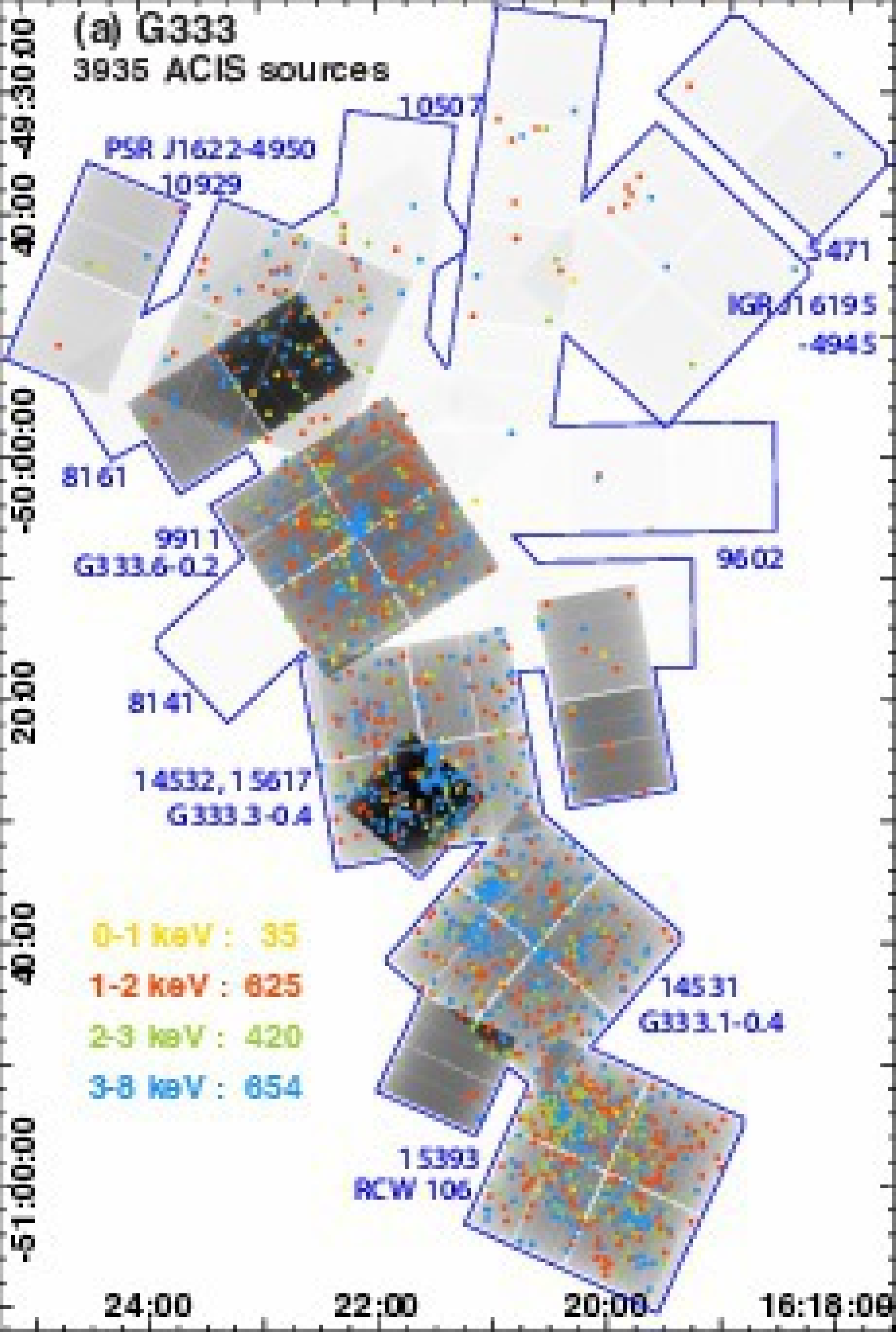}
\includegraphics[width=0.49\textwidth]{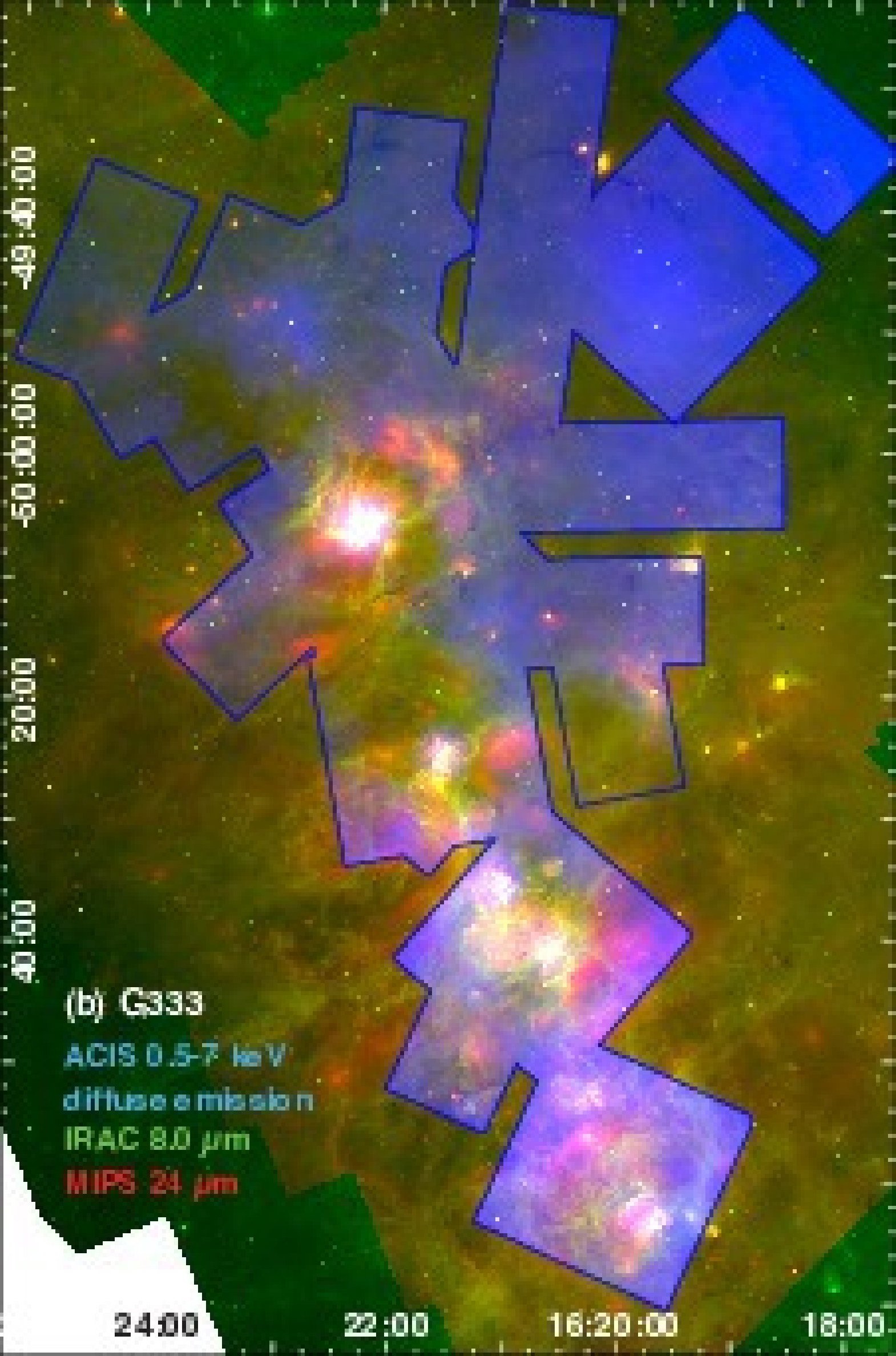}
\caption{G333.
(a) ACIS exposure map with brighter ($\geq$5 net counts) ACIS point sources overlaid; colors denote median energy for each source.  ObsID numbers and regions named in the text are shown in blue.  The image appears white for regions with $<$5~ks exposure.
(b) ACIS diffuse emission in the \Spitzer context.  
\label{G333.fig}}
\end{figure}

This study required four 60-ks ACIS-I pointings to mosaic MSFRs across the entire G333 GMC.
% A Cycle~10 GO observation of G333.6-0.2, Cycle~14 GO observations of G333.3-0.4 and G333.1-0.4, and a Cycle~14 ACIS GTO pointing on RCW~106.  
We included additional archival \Chandra data on surrounding pointings to create a wide-field, high-resolution X-ray mosaic of the entire G333 GMC (Figure~\ref{G333.fig}).  The young SNR RCW~103 lies just off the southwest edge of our G333 mosaic. We have omitted its ACIS data from our mosaic because there is a recent analysis of those data \citep{Frank15}.  It is also extremely bright and swamps the fainter diffuse X-ray emission in our mosaic.  As in MOXC1, we continue to assume a distance of 2.6~kpc to G333 \citep{Figueredo05,Roman-Lopes09}.  Recent papers \citep[e.g.,][]{Lo15,Nguyen15} use a distance of $\sim$3.6~kpc to this GMC; assuming this larger distance would increase our X-ray source luminosities by nearly a factor of two.  Below we describe each of our four main ACIS-I pointings on G333 MSFRs.

%\clearpage
\subsubsection{G333.6-0.2}  
\begin{figure}[htb]
\centering
\includegraphics[width=0.48\textwidth]{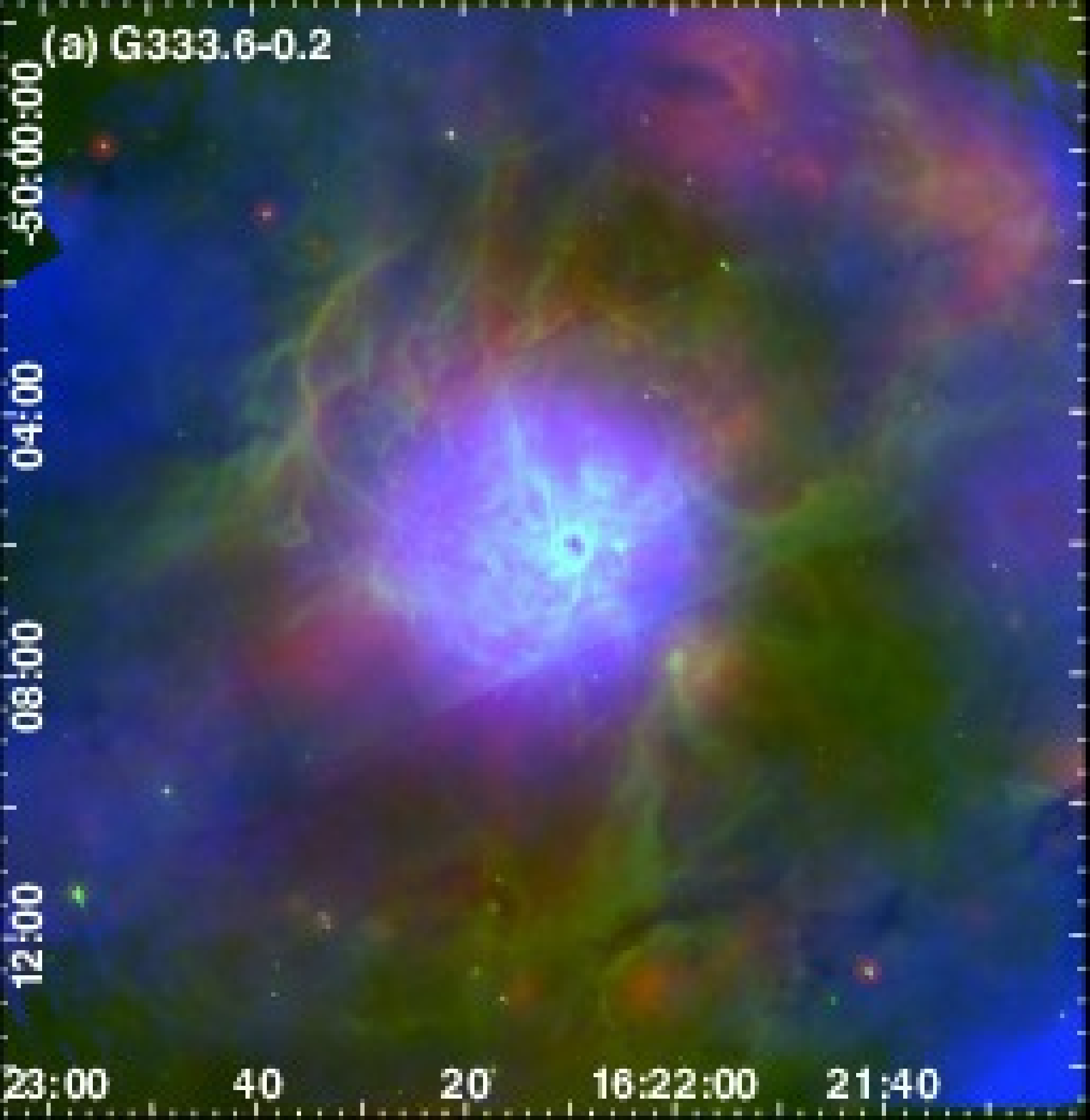}
\includegraphics[width=0.49\textwidth]{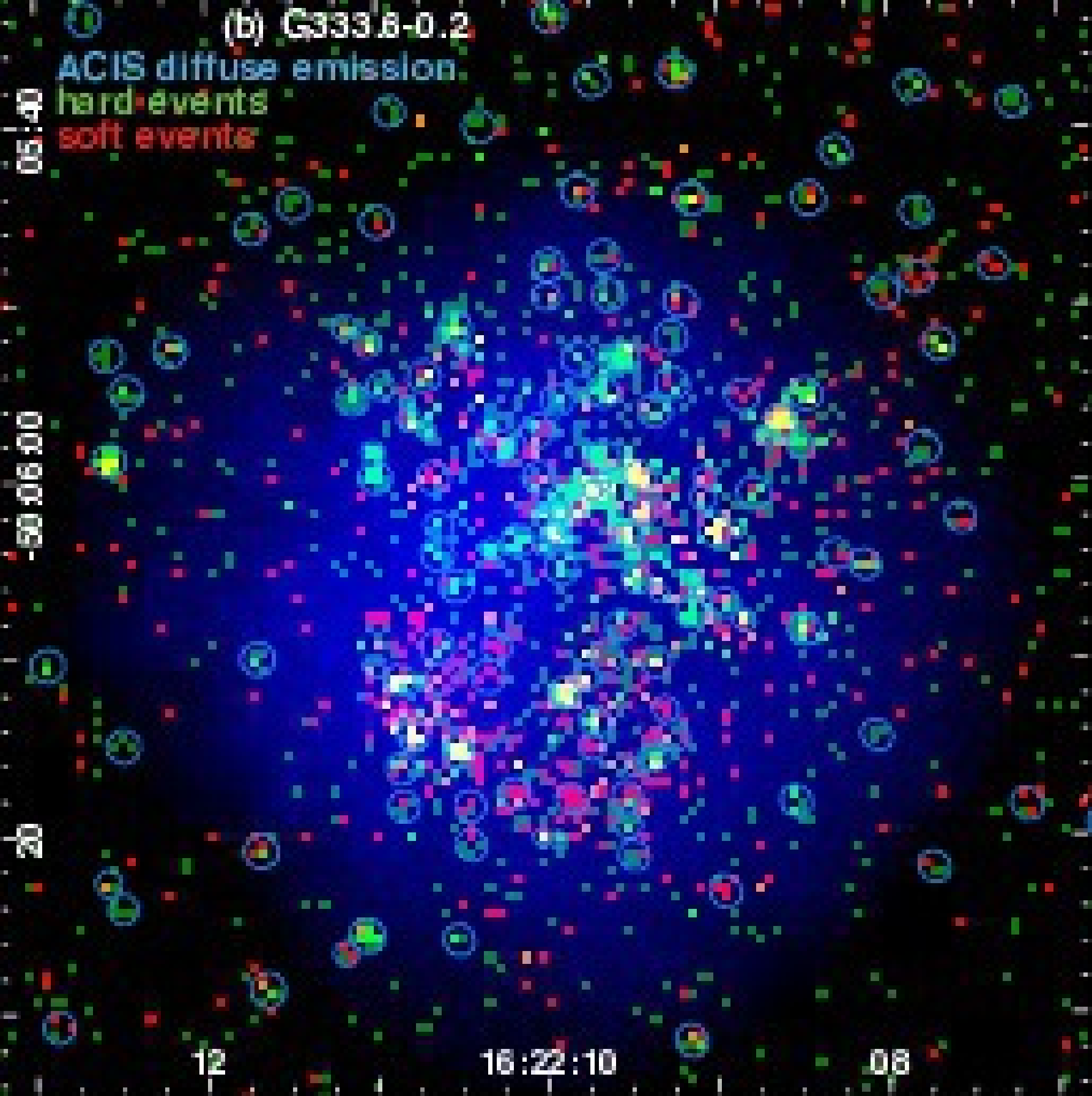}
\caption{G333.6-0.2.
(a) Zoomed version of Figure~\ref{G333.fig}(b) for G333.6-0.2.  
(b) MOXC2 ACIS event data and diffuse emission for the center of G333.6-0.2.
%(c) Comparing the MOXC1 analysis of the crowded core of G333.6-0.2 to the MOXC2 analysis.  The greyscale image shows full-band MOXC2 events; the blue polygons are MOXC2 point source extraction regions (centered on MOXC2 source positions).  Sources found by both algorithms (green) are shown with MOXC1 positions.  Magenta sources were found only by MOXC1; cyan sources were found only by MOXC2.
\label{G333.6-0.2.fig}}
\end{figure}

The most prominent MSFR in G333 is the bipolar bubble and giant \hii region G333.6-0.2 (Figure~\ref{G333.6-0.2.fig}(a)).  We presented a single-ObsID (60.1~ks) ACIS-I study of this region in MOXC1.  The MOXC2 analysis used slightly more data (61.6~ks).  In both epochs of analysis ACIS reveals over 100 X-ray sources in a central compact cluster powering the giant \hii region, a distributed population of over 500 more point sources across this ACIS-I pointing, and soft diffuse X-ray emission tracing hot plasma apparently leaking from the embedded cluster (Figure~\ref{G333.6-0.2.fig}(b)).  As we found for the main ridge in NGC~6334, MOXC1 and MOXC2 catalogs are the same for brighter sources but find different sources at the faint limit.  As before, we favor the MOXC2 results because they reflect improvements in our analysis methodologies over the years.

%Figure~\ref{G333.6-0.2.fig}(c) shows a comparison between the MOXC1 analysis of the sources found in the center of G333.6-0.2 (60.1~ks integration) and that same region with slightly more data in the MOXC2 analysis (61.6~ks integration).  In the field shown, there were 77 sources in common between the two epochs of analysis and around 20 faint sources unique to one algorithm or the other.  We believe that the MOXC2 analysis is more robust, but many factors affect our evaluation of source validity at the faint limit.  Users of our data products may want to look for longer-wavelength counterparts to both MOXC1 and MOXC2 source lists if they wish to include all possible faint X-ray sources in their multiwavelength analysis.  

%\clearpage
\subsubsection{G333.3-0.4}
\begin{figure}[htb]
\centering
\includegraphics[width=0.48\textwidth]{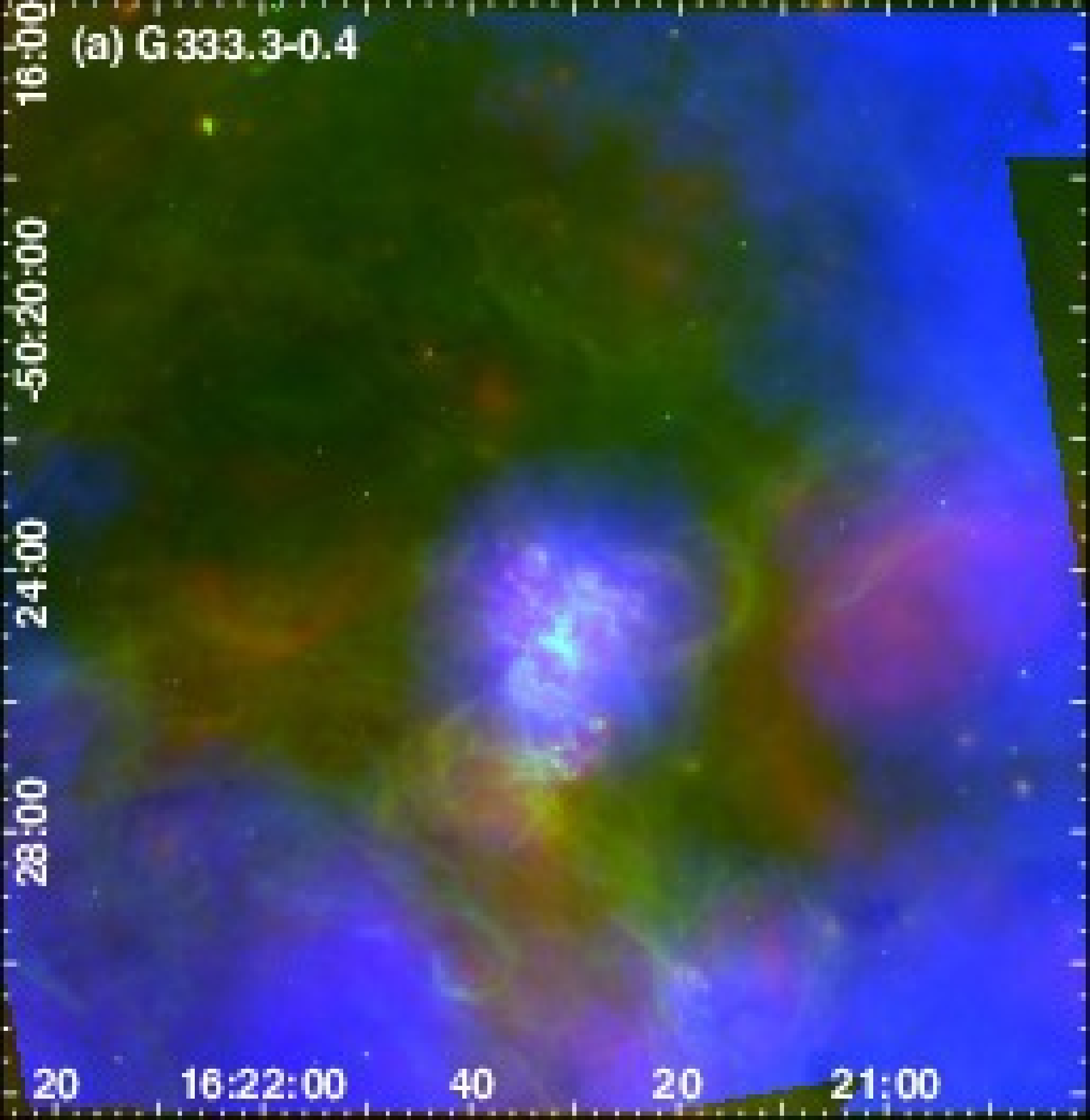}
\includegraphics[width=0.49\textwidth]{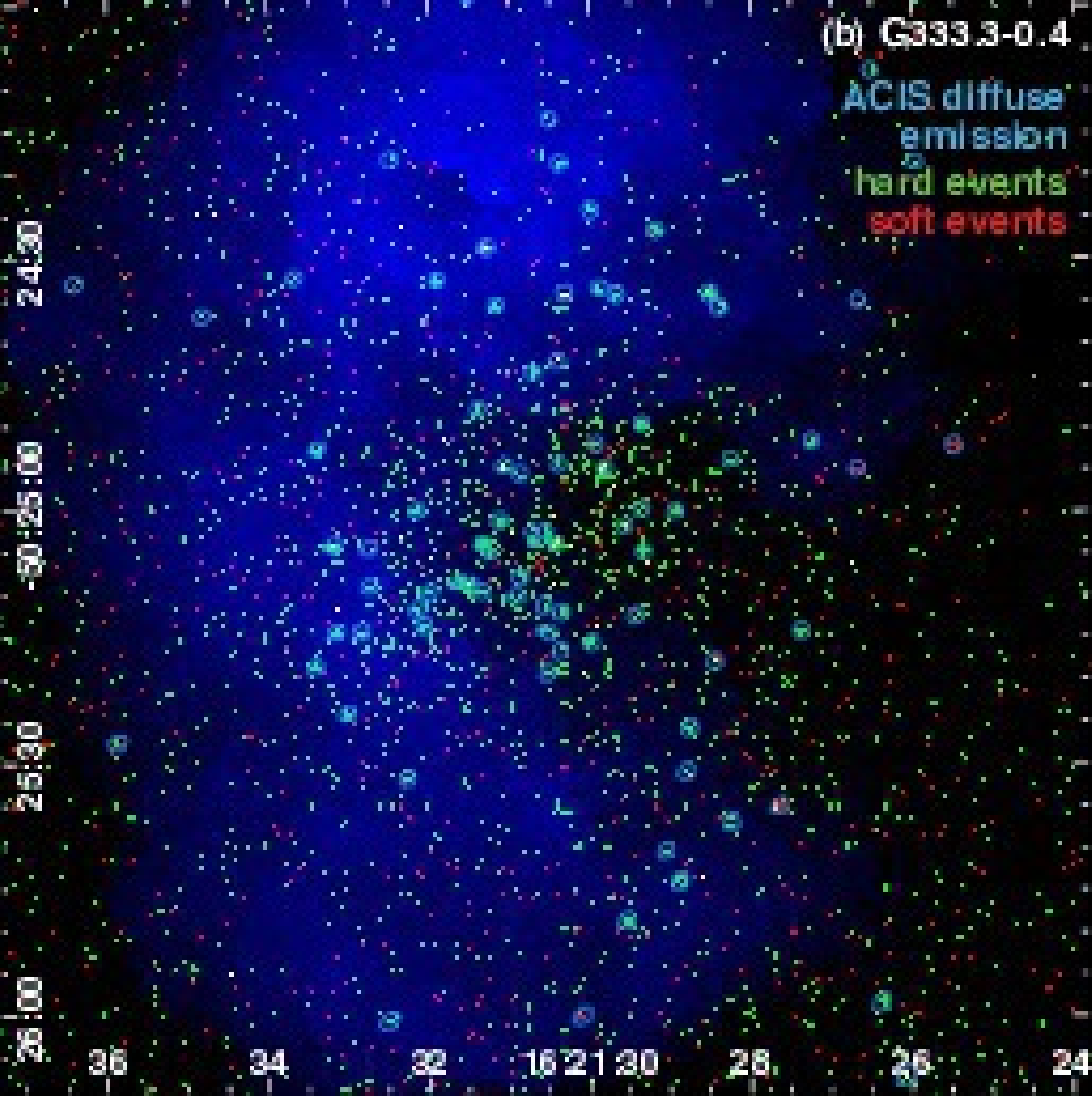}
\caption{
(a) Zoomed version of Figure~\ref{G333.fig}(b) for G333.3-0.4.
(b) ACIS event data and diffuse emission for the center of G333.3-0.4.
\label{G333.3-0.4.fig}}
\end{figure}

This pointing (Figure~\ref{G333.3-0.4.fig}) contrasts an IR-bright MSFR with the only part of the G333 GMC that is not bright at 24~$\mu$m.  The upper half of Figure~\ref{G333.3-0.4.fig}(a) shows this dark bay; we see no diffuse X-ray emission from this region.  Water masers are found even in the IR-dark part of the field, indicating ongoing massive star formation \citep{Breen07}.  
A near-IR study of G333.3-0.4 \citep{Roman-Lopes03} found 41 reddened massive stars in a 1$^{\prime}$ cluster behind $\langle A_{V}\rangle$$\sim$28~mag, with spectral types $\sim$O5.5--B5.  We find X-ray counterparts to at least 13 of those massive stars, sometimes resolving two or three X-ray sources at the location of the IR source.

The MSFR G333.3-0.4 is associated with IRAS~16177-5018, powered by an O3If$^{+}$ star at $2.6 \pm 0.7$~kpc \citep{Roman-Lopes09}.  Our X-ray counterpart to that star is c5623 (CXOU~J162131.60-502508.3), with 13 net counts and a median energy of 3.2~keV; it has two close neighbors (each $<$1.2$\arcsec$ away).  With so few counts, spectral fit parameters for c5623 are not well-constrained, but they are roughly $N_H \sim 13 \times 10^{22}$~cm$^{-2}$, $kT \sim 0.9$~keV, and $L_X \sim 4 \times 10^{32}$~erg~s$^{-1}$.
% Roman-Lopez09 gives a position for IRAS~16177-5018: 16 21 31.60 -50 25 08.3.  

In Figure~\ref{G333.3-0.4.fig}(a) we do see diffuse emission from the G333.3-0.4 cluster and around the edges of this field, sometimes filling \Spitzer bubbles.  The center of the cluster (Figure~\ref{G333.3-0.4.fig}(b)) contains a substantial number of X-ray sources; the overdensity of hard events not captured by detected sources implies that there are many more X-ray sources just below the detection limit.  Diffuse X-ray emission is seen around the cluster but is fainter in the center, perhaps shadowed by the high absorbing column in this region.  The X-ray sources are typically hard and faint, again probably due to high absorption.  No other major grouping of X-ray sources is found in this pointing.

%From proposal:  "A second bright dust core hosting an IRAS source (16174-5022) and several UCH{\scriptsize II}R candidates will be captured on the southern side of this ACIS pointing."
% SIMBAD position for IRAS 16174-5022: 16 21 13.3 -50 30 05.  No X-ray source in the vicinity.

%\clearpage
\subsubsection{G333.1-0.4}
\begin{figure}[htb]
\centering
\includegraphics[width=0.48\textwidth]{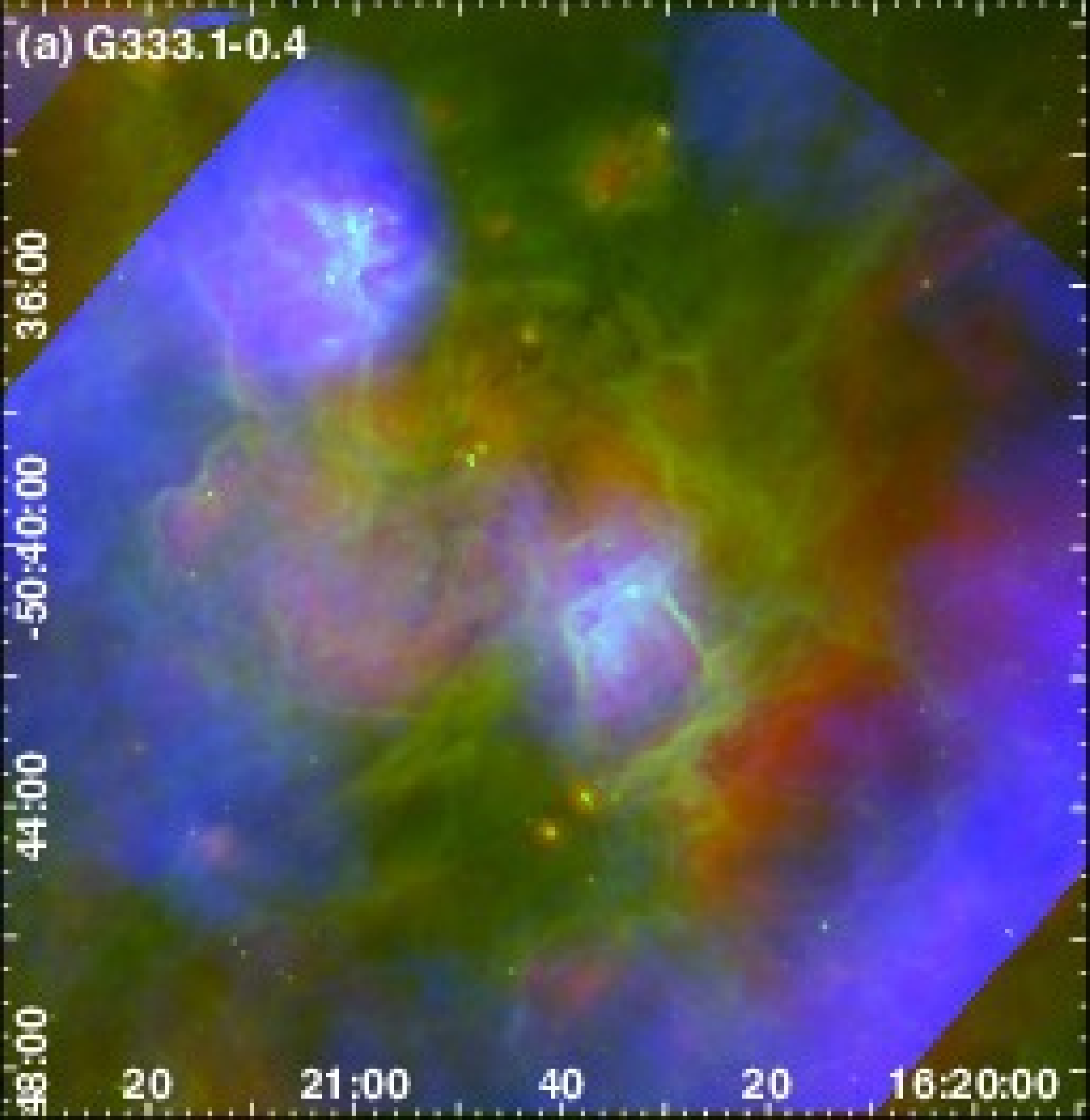}
\includegraphics[width=0.49\textwidth]{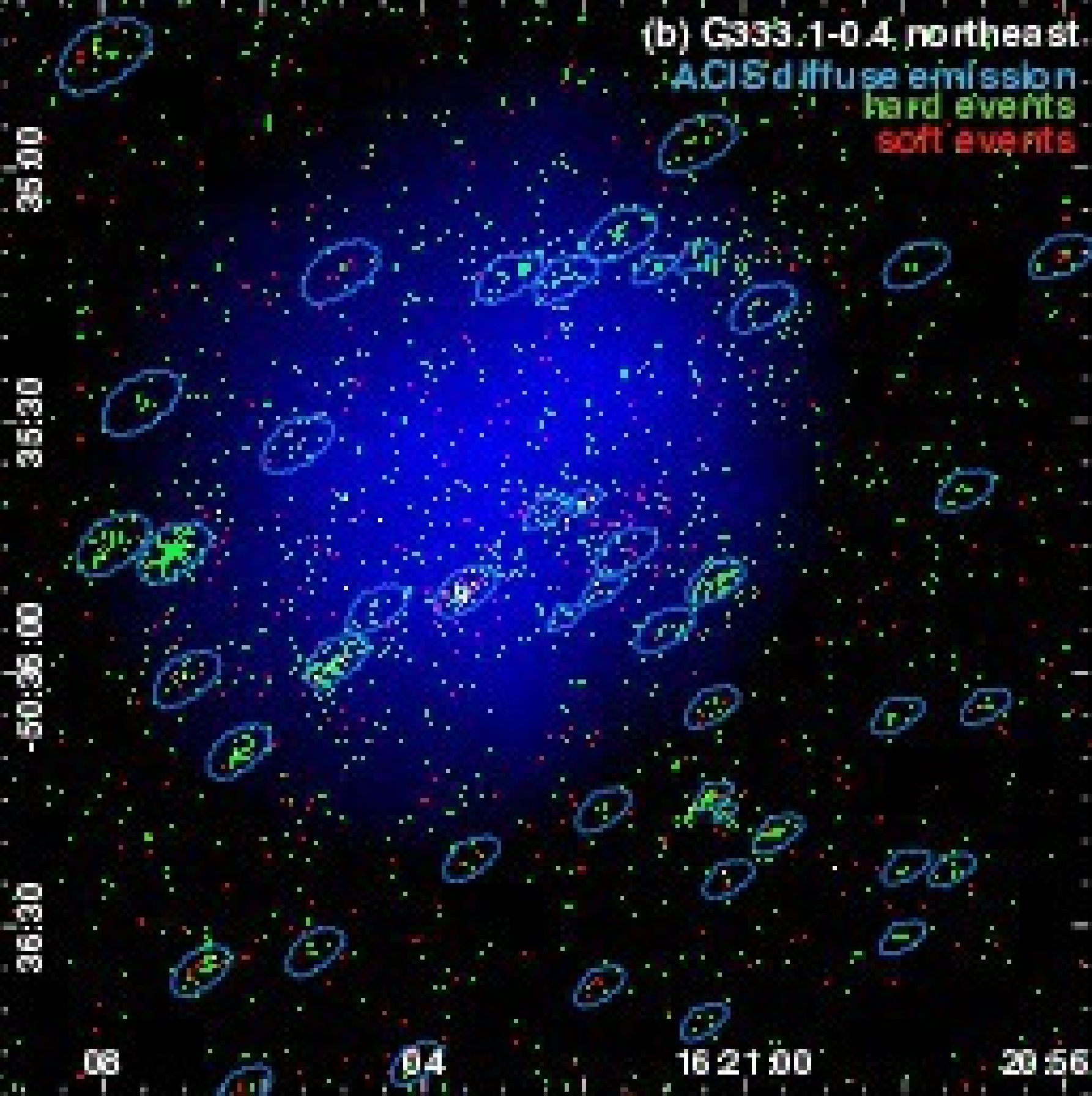}
\includegraphics[width=0.49\textwidth]{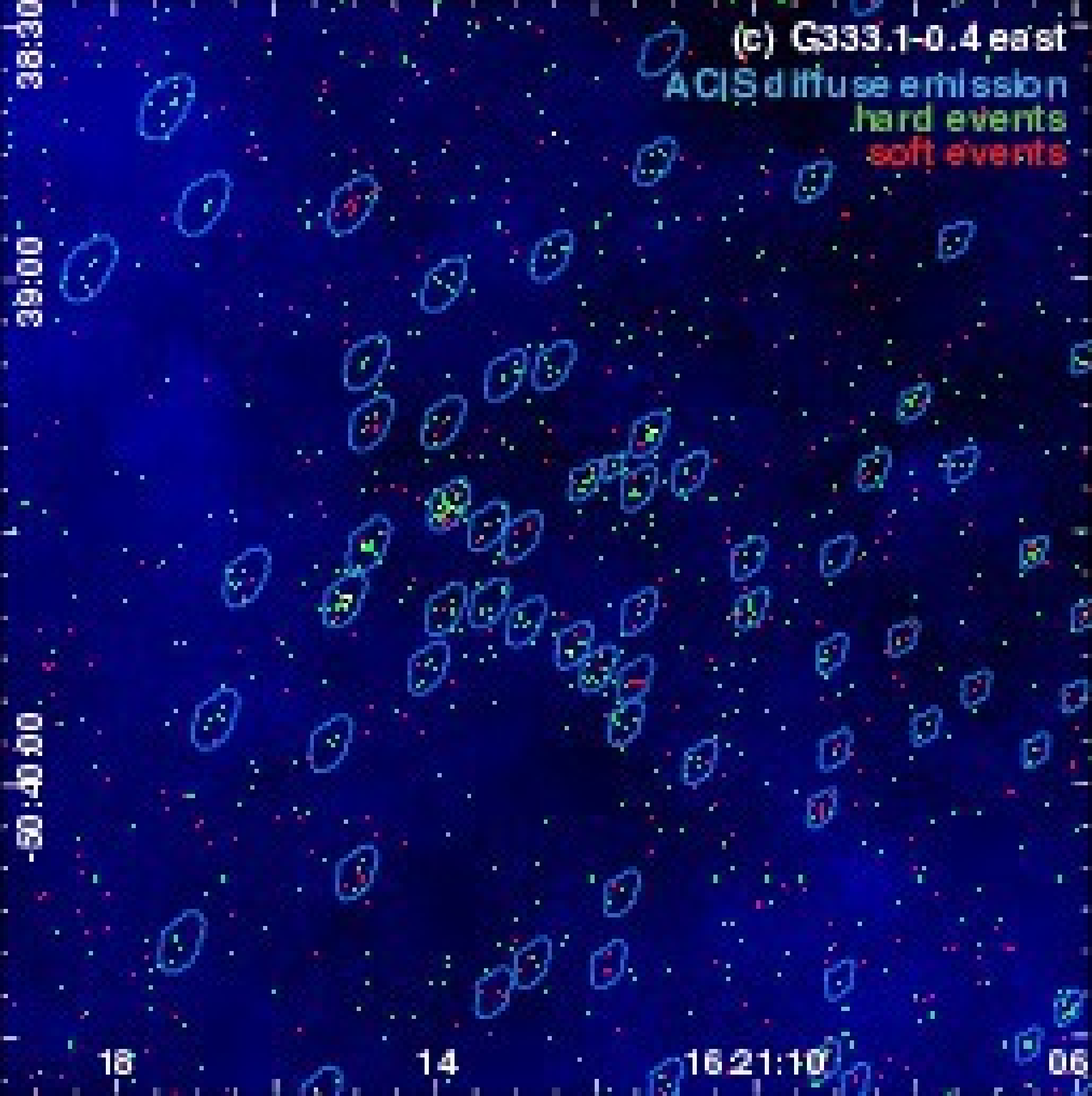}
\includegraphics[width=0.49\textwidth]{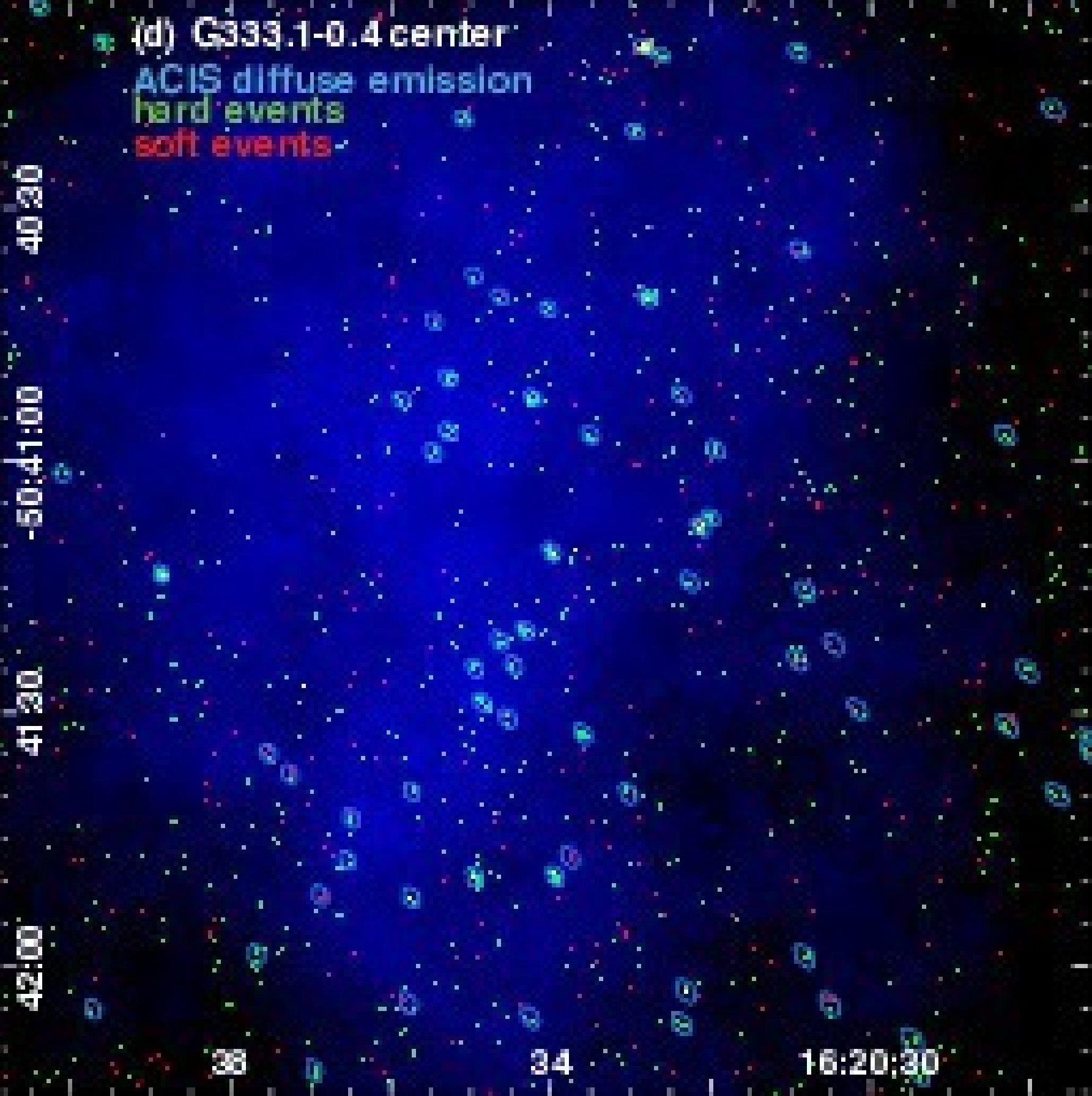}
\caption{
(a) Zoomed version of Figure~\ref{G333.fig}(b) for G333.1-0.4.
(b)--(d) ACIS event data and diffuse emission for source concentrations in G333.1-0.4.
(b) G333.1-0.4 northeast, with the brightest diffuse emission in the G333.1-0.4 pointing.
(c) G333.1-0.4 east.
(d) G333.1-0.4 center.
\label{G333.1-0.4.fig}}
\end{figure}

This pointing (Figure~\ref{G333.1-0.4.fig}) captures two powerful \hii regions, G333.12-0.45 and G333.03-0.45 (GAL~333.1-00.4 and GAL~333.0-00.4 in SIMBAD).  A near-IR study of G333.12-0.45 also finds a cluster distance of 2.6~kpc, a range of extinctions ($A_{V}$$\sim$12--32~mag), several MYSOs with spectral types O4--B4, and a total cluster mass of $>$$10^{3}$M$_{\odot}$ \citep{Figueredo05}, but no source positions are provided.  G333.12-0.45 shows direct evidence for large-scale infall \citep{Lo15}, implying that it is very young.  

The grouping of X-ray sources closest to G333.12-0.45 is shown in Figure~\ref{G333.1-0.4.fig}(b), in the northeast region of this pointing.  Bright diffuse X-ray emission is seen here.  This region is several arcminutes off-axis where ACIS PSFs are large and sensitivity is reduced, thus point sources are typically faint; the diffuse emission may be due partially to unresolved pre-MS stars from this obscured cluster.

Another group of X-ray sources is shown in Figure~\ref{G333.1-0.4.fig}(c), in the eastern part of this ACIS pointing.  Faint diffuse X-ray emission pervades the field.  There is an \UCHIIR candidate here \citep{Mookerjea04}.  These X-ray sources may be part of GLIMPSE IR cluster \#78 \citep{Mercer05}.

The \hii region G333.03-0.45 is not as well-studied as G333.12-0.45 but shows a large number of dust cores and \UCHIIR candidates \citep{Mookerjea04}.  It is as bright in Lyman continuum as the other G333 MSFRs \citep{Conti04}.  G333.03-0.45 is at the center of this ACIS pointing.  It shows diffuse X-ray emission distributed throughout a fairly loose grouping of point sources (Figure~\ref{G333.1-0.4.fig}(d)).

% Figueredo05 don't give source positions!  They say G333.1-0.4 is at 16 21 03.3 -50 36 19.
% G333.12-0.45 is listed as GAL 333.1-00.4 (Cluster of Stars) in SIMBAD, at 16 21 03.3 -50 36 19.  This is towards the bottom of the grouping of X-ray sources that I call G333.1-0.4ne.
% G333.03-0.45 is listed as GAL 333.0-00.4 (HII region) in SIMBAD, at 16 20 38.2 -50 40 21.  This is at the northeast corner of the grouping I call G333.1-0.4c.
% My G333.1-0.4e is close to the cluster Mercer~78, Mercer05 position 16 21 08 -50 39 57.

%\clearpage
\subsubsection{RCW~106}
\begin{figure}[htb]
\centering
\includegraphics[width=0.48\textwidth]{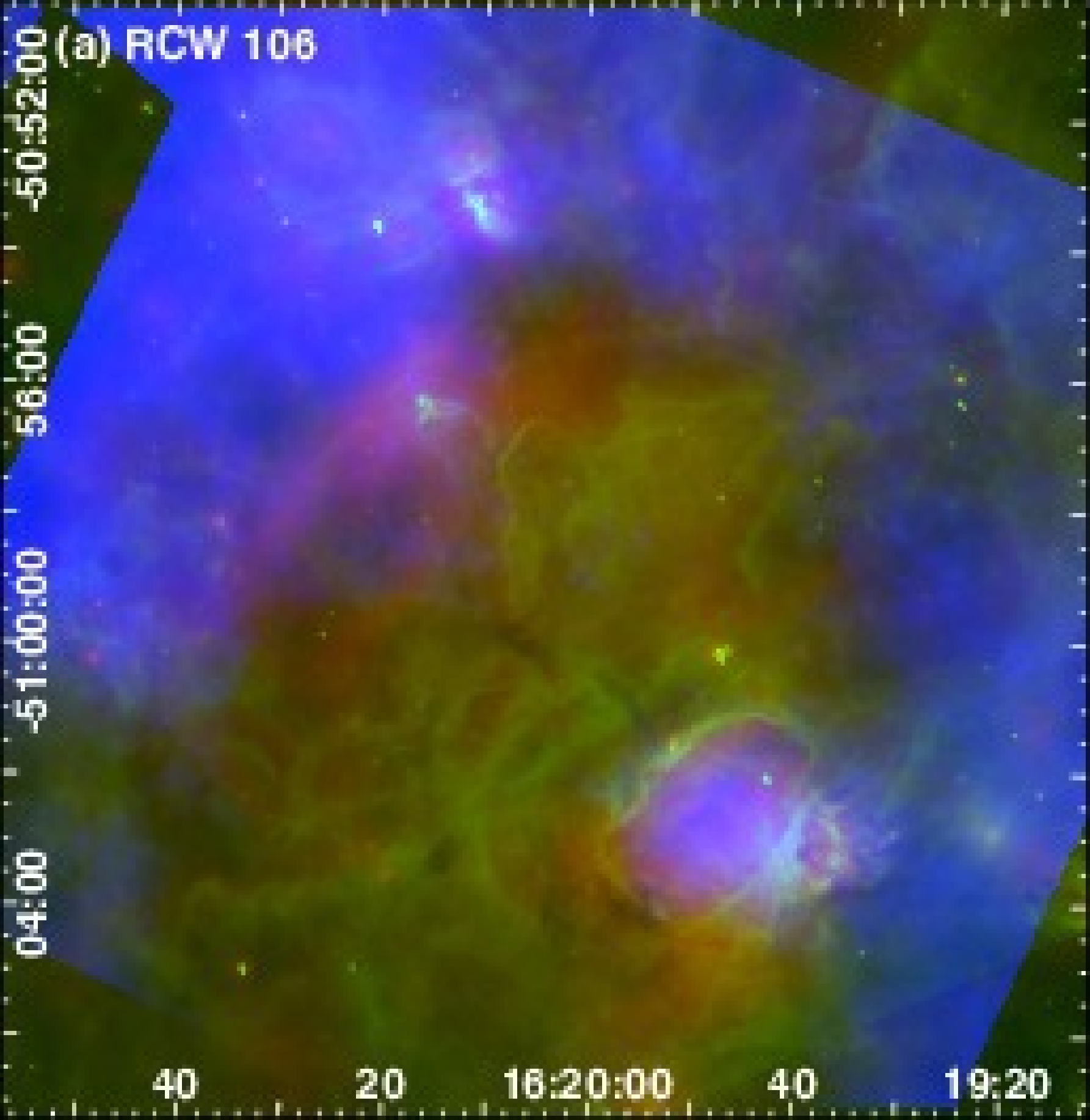}
\includegraphics[width=0.49\textwidth]{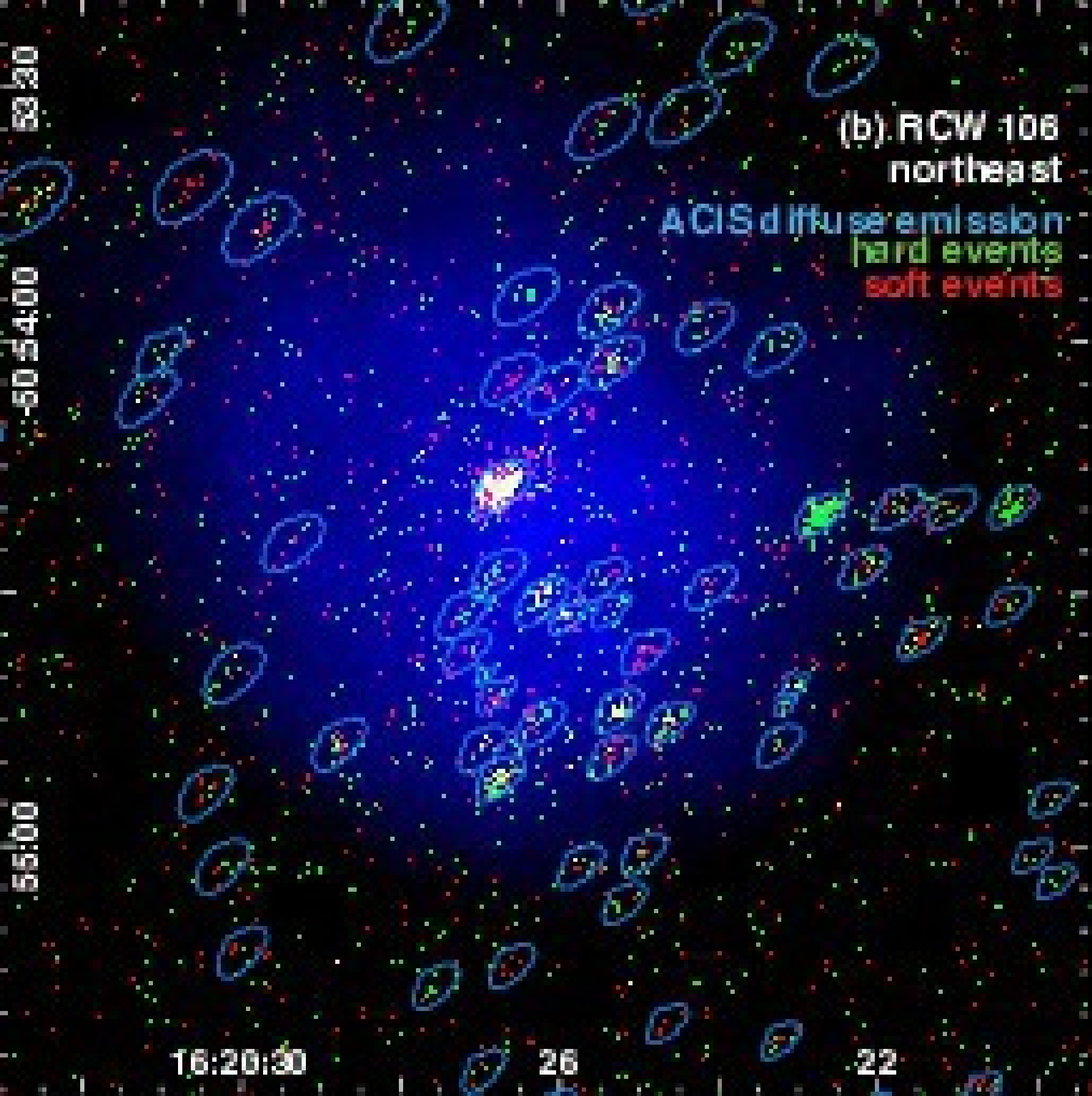}
\includegraphics[width=0.49\textwidth]{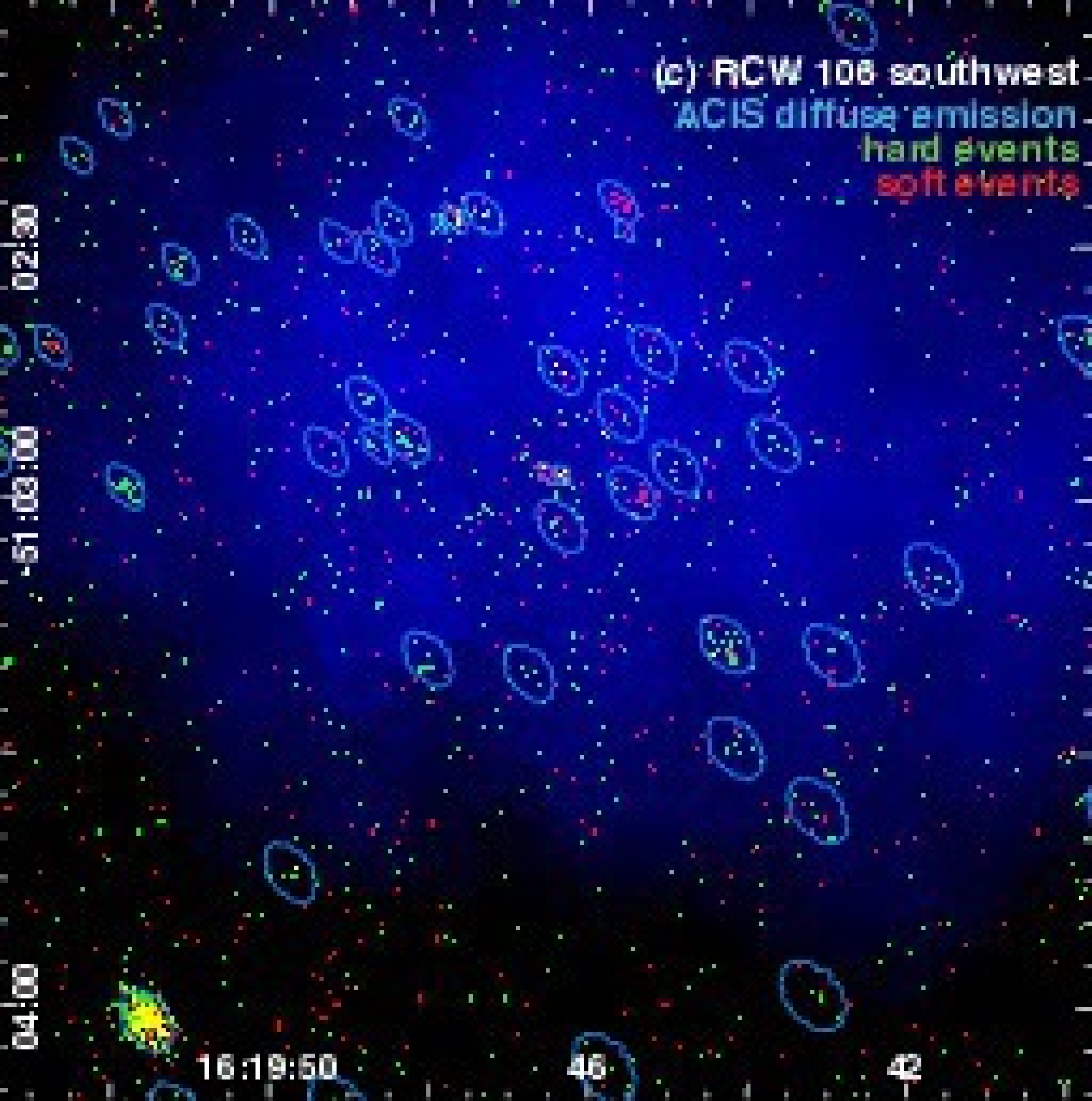}
\caption{
(a) Zoomed version of Figure~\ref{G333.fig}(b) for RCW~106.
(b)--(c) ACIS event data and diffuse emission for source concentrations in RCW~106.
(b) RCW~106 northeast.
(c) RCW~106 southwest.
\label{RCW106.fig}}
\end{figure}

The final pointing (Figure~\ref{RCW106.fig}) in our G333 mosaic of MSFRs again shows diffuse X-ray emission at the periphery and in a \Spitzer bubble, with strong shadowing at the center (Figure~\ref{RCW106.fig}(a)).  We find two main clumps of X-ray sources.  The first is on the northeast side of the pointing (Figure~\ref{RCW106.fig}(b)).  Two bright X-ray sources are found here:  c2691 (CXOU~J162026.75-505417.1; 398 net counts, median energy 1.7~keV) and c2412 (CXOU~J162022.75-505420.4; 110 net counts, median energy 4.2~keV, variable).  A spectral fit to c2691 requires two thermal plasma components, with $N_H = 2.2 \times 10^{22}$~cm$^{-2}$, $kT1 = 0.5$~keV, $kT2 > 3$~keV, and $L_X = 9.0 \times 10^{32}$~erg~s$^{-1}$.  The harder source, c2412, can be fit with a one-temperature thermal plasma, although that temperature is so high that it is not well-constrained by the ACIS data:  $N_H = 10 \times 10^{22}$~cm$^{-2}$, $kT > 10$~keV, and $L_X = 1 \times 10^{32}$~erg~s$^{-1}$.
% No famous counterpart to c2691 or c2412.
% Third brightest source is c2693 (CXOU~J162026.77-505451.8; 59 net counts, median energy 2.8~keV).

The second clump of X-ray sources is on the southwest side of the pointing, in a prominent \Spitzer bubble (Figure~\ref{RCW106.fig}(c)).  These sources may be associated with GLIMPSE IR cluster \#76 \citep{Mercer05}.
% Mercer~76, Mercer05 position 16 19 43 -51 03 37, is at the SW edge of my RCW106sw.

\clearpage
\subsubsection{Archival Pointings}
% Note -- PSR J1622-4950 is piled in the short ObsID 8161 but not in the longer ObsID 10929.

\begin{figure}[htb]
\centering
\includegraphics[width=0.49\textwidth]{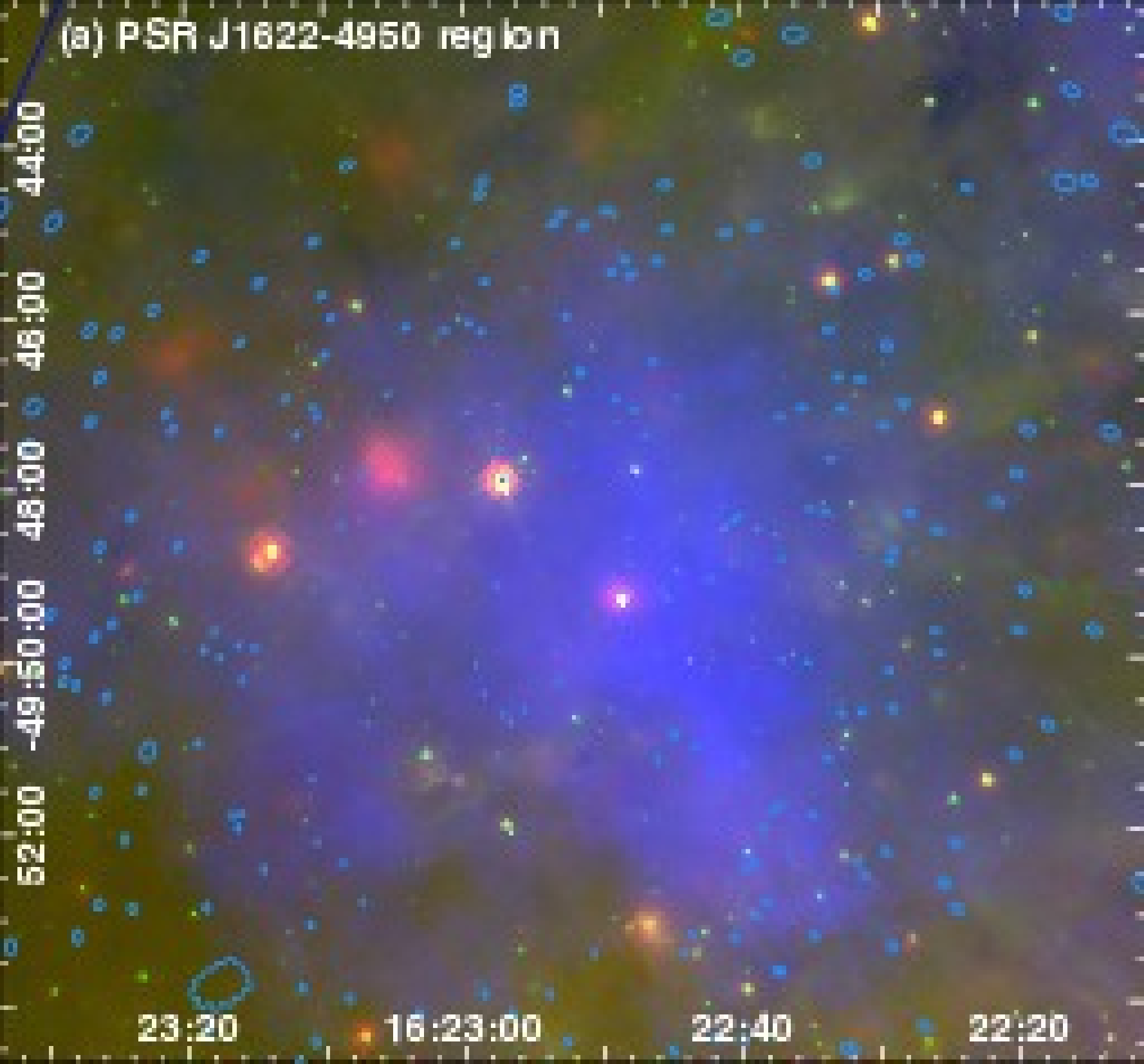}
\includegraphics[width=0.49\textwidth]{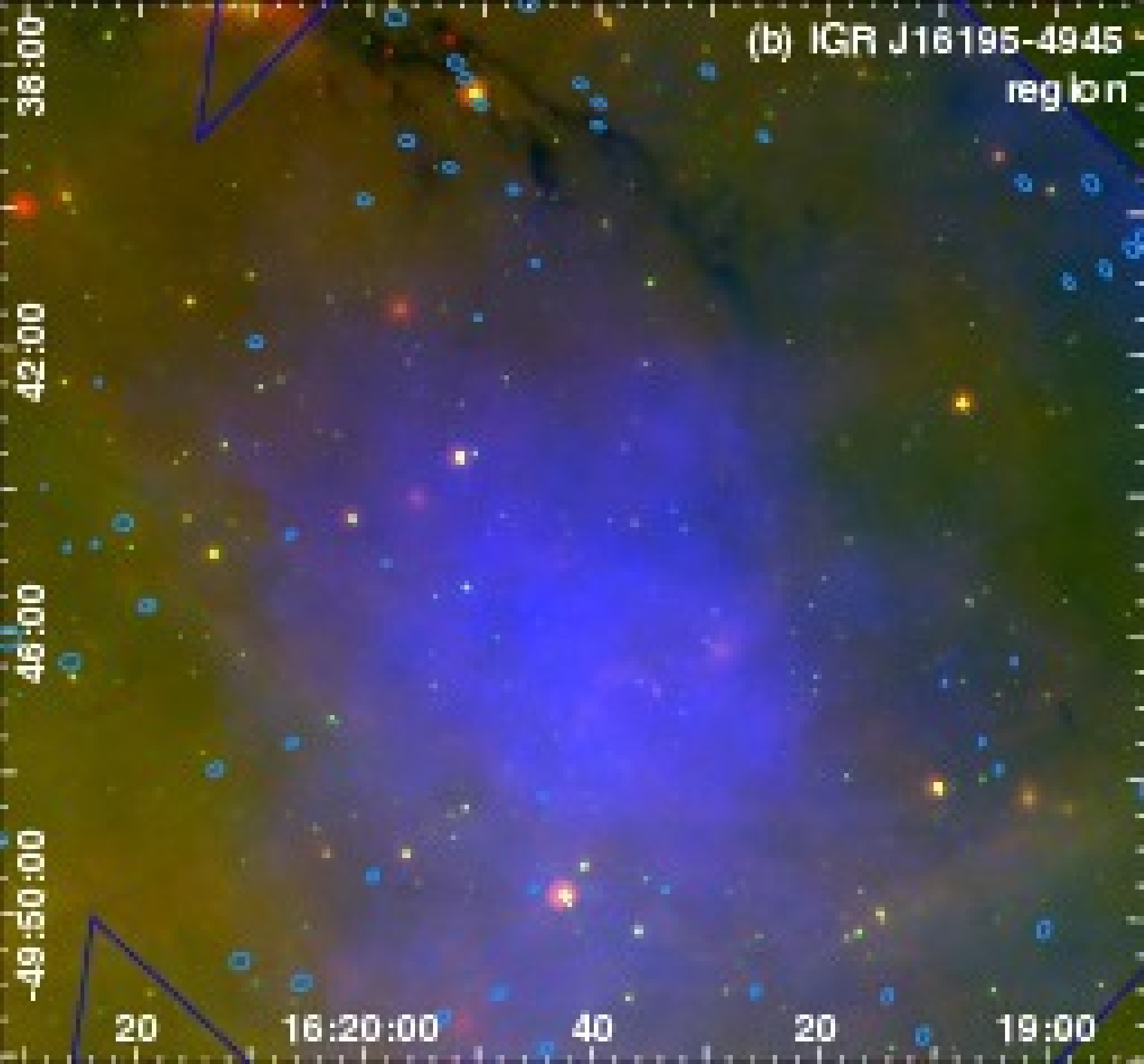}
\caption{Interesting diffuse X-ray features in archival pointings of the G333 mosaic.
(a) Zoomed version of Figure~\ref{G333.fig}(b) for the PSR~J1622-4950 pointing, with ACIS point source extraction regions.
(b) Zoomed version of Figure~\ref{G333.fig}(b) for the IGR~J16195-4945 pointing, with ACIS point source extraction regions.
\label{G333archival.fig}}
\end{figure}

We included several archival \Chandra datasets in the G333 mosaic, to broaden the field and provide a wider context for the diffuse X-ray emission.  ObsIDs 8141, 8161, 9602, and 10507 were from the ChIcAGO project \citep{Anderson14}.  ObsID~10929 re-observed a bright ChIcAGO source, which was determined to be the magnetar PSR~J1622-4950 \citep{Levin10,Anderson12}.  This is our source c8972 (CXOU~J162244.91-495052.8); it is piled up in ObsID~8161.  
% A tbabs*bbody fit to pileup-corrected spectrum is consistent with Anderson12 -- big errors on fit parameters, so it's easy to be consistent.
We find prominent diffuse X-ray emission surrounding the magnetar (Figure~\ref{G333archival.fig}(a)).  This could be the X-ray signature of its SNR; \citet{Anderson12} found a radio arc just north of this region (G333.9+0.0) and suggested that it might be a SNR associated with PSR~J1622-4950.  Our diffuse X-ray emission could be tracing hot plasma filling that radio-bright partial shell.  Additionally, Anderson et al.\ showed PSR~J1622-4950 to be coincident with a potentially nonthermal diffuse radio source (``Source A'') that may be a PWN associated with the magnetar.  Since the diffuse X-ray emission is concentrated in the vicinity of PSR~J1622-4950 and the diffuse radio source, this adds further support for the existence of a magnetar wind nebula.

ObsID~5471 targeted IGR~J16195-4945, which is a likely high-mass X-ray binary \citep{Tomsick06}.  This is our source c439 (CXOU~J161932.20-494430.6); it is piled up, but there are too few counts to perform pile-up correction.  There is bright diffuse X-ray emission in this region (Figure~\ref{G333archival.fig}(b)).  \citet{Pandey06} note extended radio emission in the vicinity of IGR~J16195-4945, so perhaps we are once again seeing X-ray emission from a SNR.

\clearpage
%-----------------------------------------------------------------------------
\subsection{AFGL~2591 \label{sec:afgl2591}}
% AFGL 2591 -- 288 point sources
% At 3.33 kpc, 4*pi*D^2 = 1.327e45.
% Region file VLA-3.reg has position from Sanna12 and a 2"-radius circle representing the region they searched for masers.
% Parkin09 do not provide a list of their 62 ACIS detections on S3.

AFGL~2591 is an embedded MYSO with prominent outflows and maser emission, surrounded by a small cluster \citep{Trinidad03}; it is seen in the direction of the Cygnus~X star-forming complex \citep{Reipurth08} but it is now believed to be substantially behind Cygnus~X (at 3.33~kpc), from parallax of its water masers \citep{Rygl12}.  The radio-bright source VLA-3 provides most of the IR emission and is thought to be a late-O-type MYSO \citep{Sanna12}.

\begin{figure}[htb]
\centering
\includegraphics[width=0.51\textwidth]{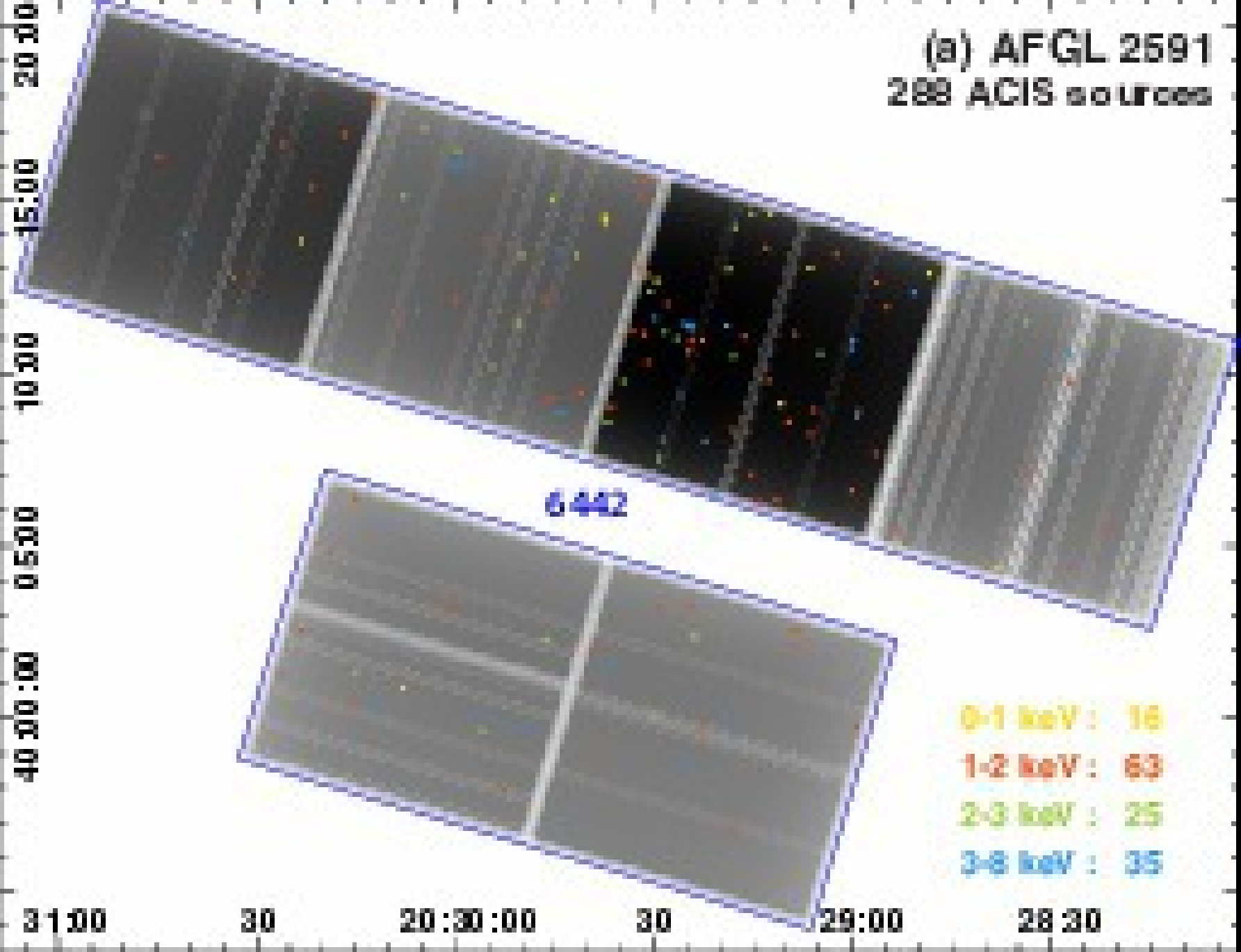}
\includegraphics[width=0.48\textwidth]{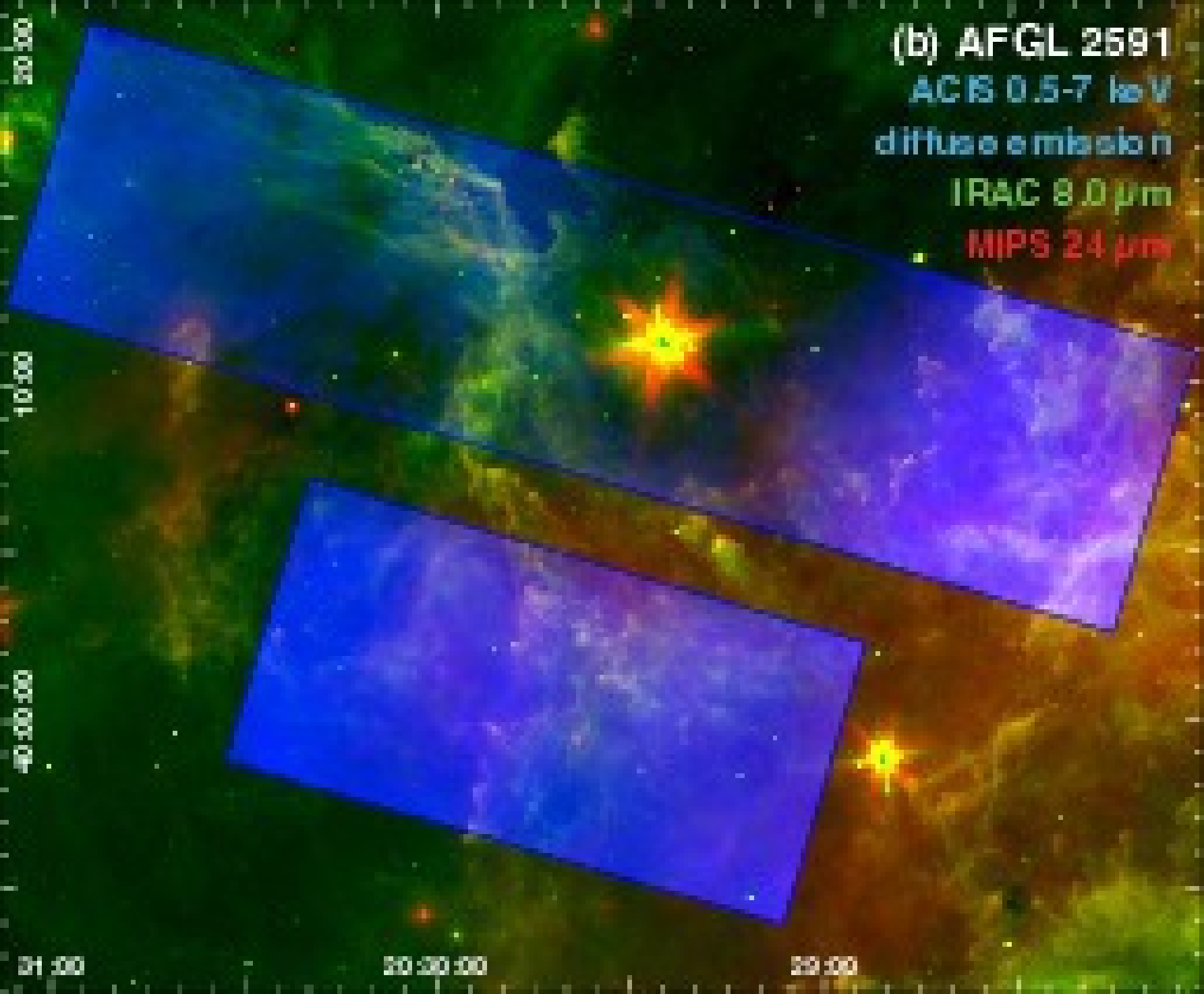}
\includegraphics[width=0.49\textwidth]{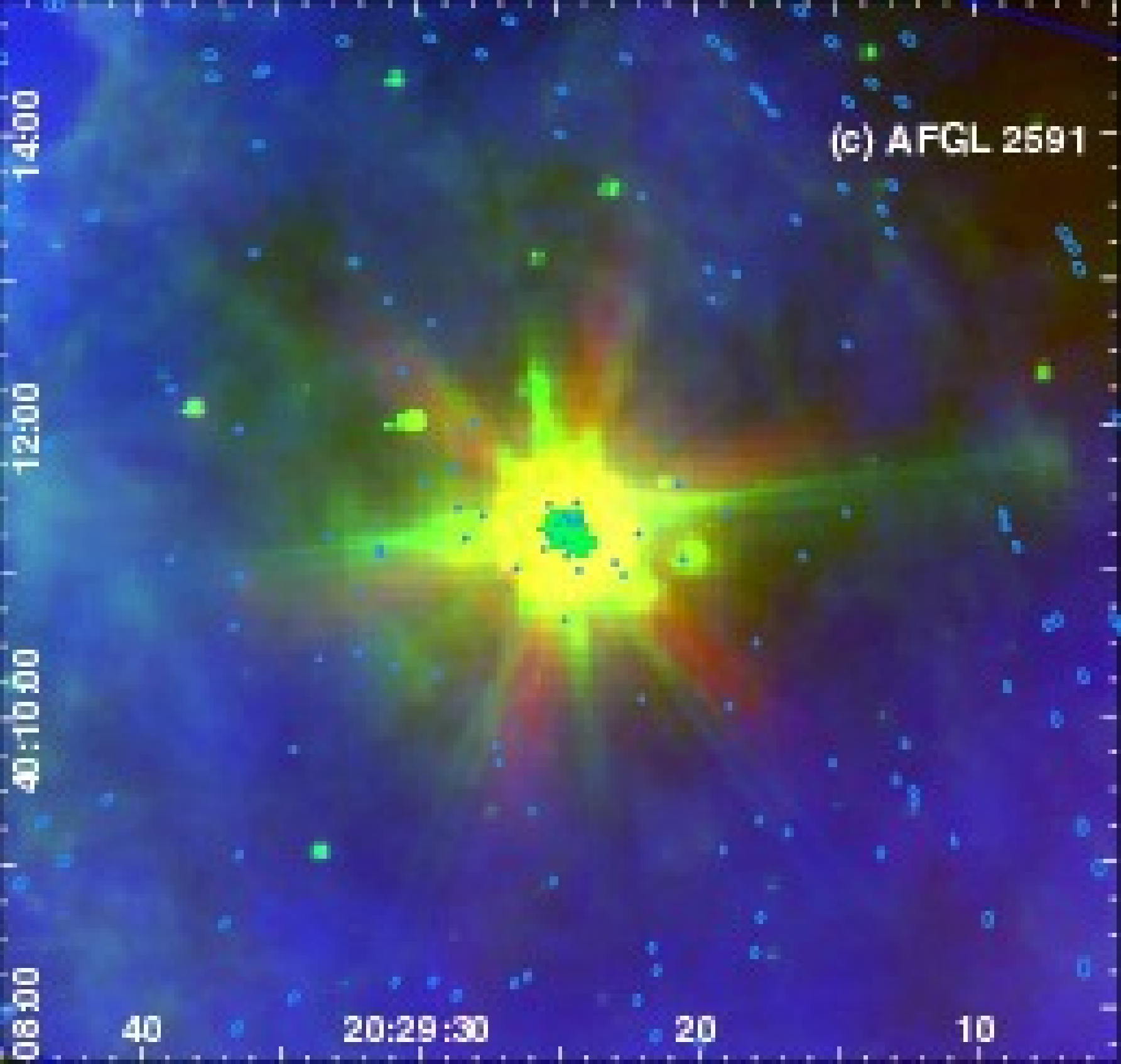}
\includegraphics[width=0.49\textwidth]{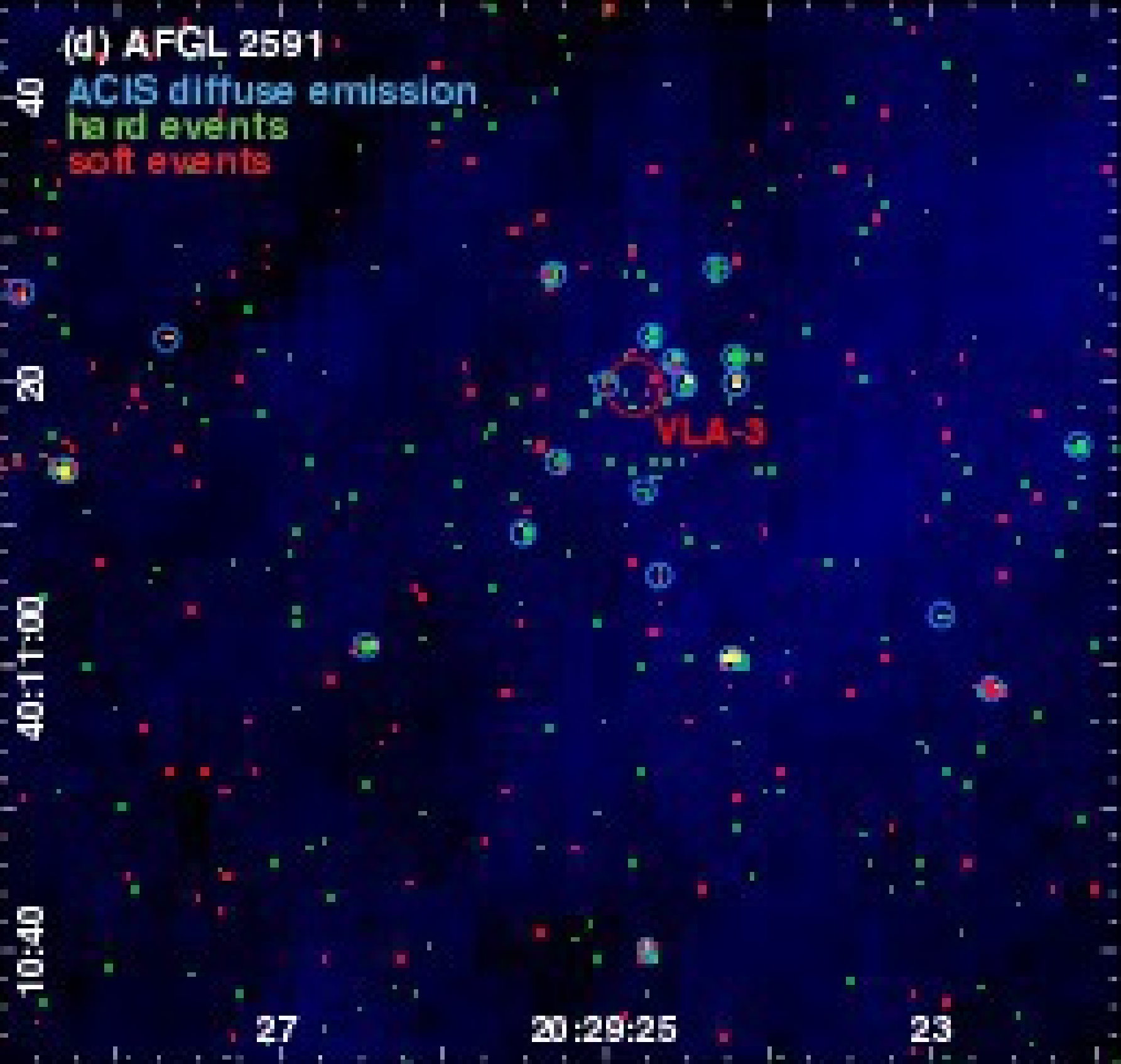}
\caption{AFGL~2591.
(a) ACIS exposure map with brighter ($\geq$5 net counts) ACIS point sources overlaid; colors denote median energy for each source.  The ObsID number is shown in blue.
(b) ACIS diffuse emission in the \Spitzer context. 
(c) Zoom of (b) for the central region around AFGL~2591, with point source extraction regions.
(d) ACIS event data and diffuse emission at the core of AFGL~2591.  The position of VLA-3 from \citet{Sanna12} is shown as a red 2\arcsec-radius circle, representing the region they searched for masers.
\label{AFGL2591.fig}}
\end{figure}

\citet{Parkin09} analyzed the \Chandra data on AFGL~2591, finding 62 sources on the S3 CCD; our analysis of the same data (Figures~\ref{AFGL2591.fig}) yields 152 sources on S3 and 288 sources in the entire observation.  In the vicinity of VLA-3, Parkin et al.\ find 4 X-ray sources but do not detect the MYSO.  We recover those 4 X-ray sources and add several more in our analysis (Figure~\ref{AFGL2591.fig}(d)), but again we do not detect VLA-3.  Parkin et al.\ interpreted a small excess of counts close to VLA-3 as possible diffuse emission from the winds of the proto-O star.  We find a few of those counts to be consistent with a point source east of VLA-3; the remaining small number of counts could come from other point sources (including VLA-3) fainter than our detection limit, from the background, or from other sources of diffuse emission.  Firmly establishing diffuse X-ray emission from VLA-3 wind interactions would require much more \Chandra data.

Extensive diffuse X-ray emission is seen in this ACIS field, including off-axis on the I2 and I3 CCDs.  Figure~\ref{AFGL2591.fig}(c) demonstrates that diffuse emission surrounds AFGL~2591; Figure~\ref{AFGL2591.fig}(d) shows that faint structures extend down to the embedded core.  Some of this diffuse emission could be associated with the foreground Cygnus~X complex; future X-ray spectral fitting to infer the absorption across the field may help us locate it along the line of sight.

\clearpage
%-----------------------------------------------------------------------------
\subsection{G34.4+0.23 \label{sec:g34}}
% G34.4+0.23 -- 565 point sources
% At 3.7 kpc, 4*pi*D^2 = 1.638e45.
% Foster14 -- there are UKIDSS data for this field.

The first compact IRDC targeted by \Chandra was G014.225-00.506 \citep{Povich10}, part of a large molecular cloud complex associated with the M17 giant \hii region known as the M17 Southwest Extension \citep{Elmegreen79}, located at a distance of 2.0~kpc \citep{Xu11}.  Using the same analysis methods as those presented here, we catalogued 840 X-ray sources in a 99-ks ACIS-I observation of G014.225-00.506 \citep{Povich16}.  The second IRDC observed by \Chandra was the more well-studied object G34.4+0.23 \citep{Rathborne05,Shepherd07}, observed with ACIS-I for just 63~ks under the assumption that it lies at a maser parallax distance of 1.56~kpc \citep{Kurayama11}, substantially closer than G014.225-00.506.  This distance was soon questioned, however \citep{Foster12}, and recent papers \citep{Foster14,Xu16} favor the kinematic distance of around 3.7~kpc \citep{Rathborne05}.

\begin{figure}[htb]
\centering
\includegraphics[width=0.49\textwidth]{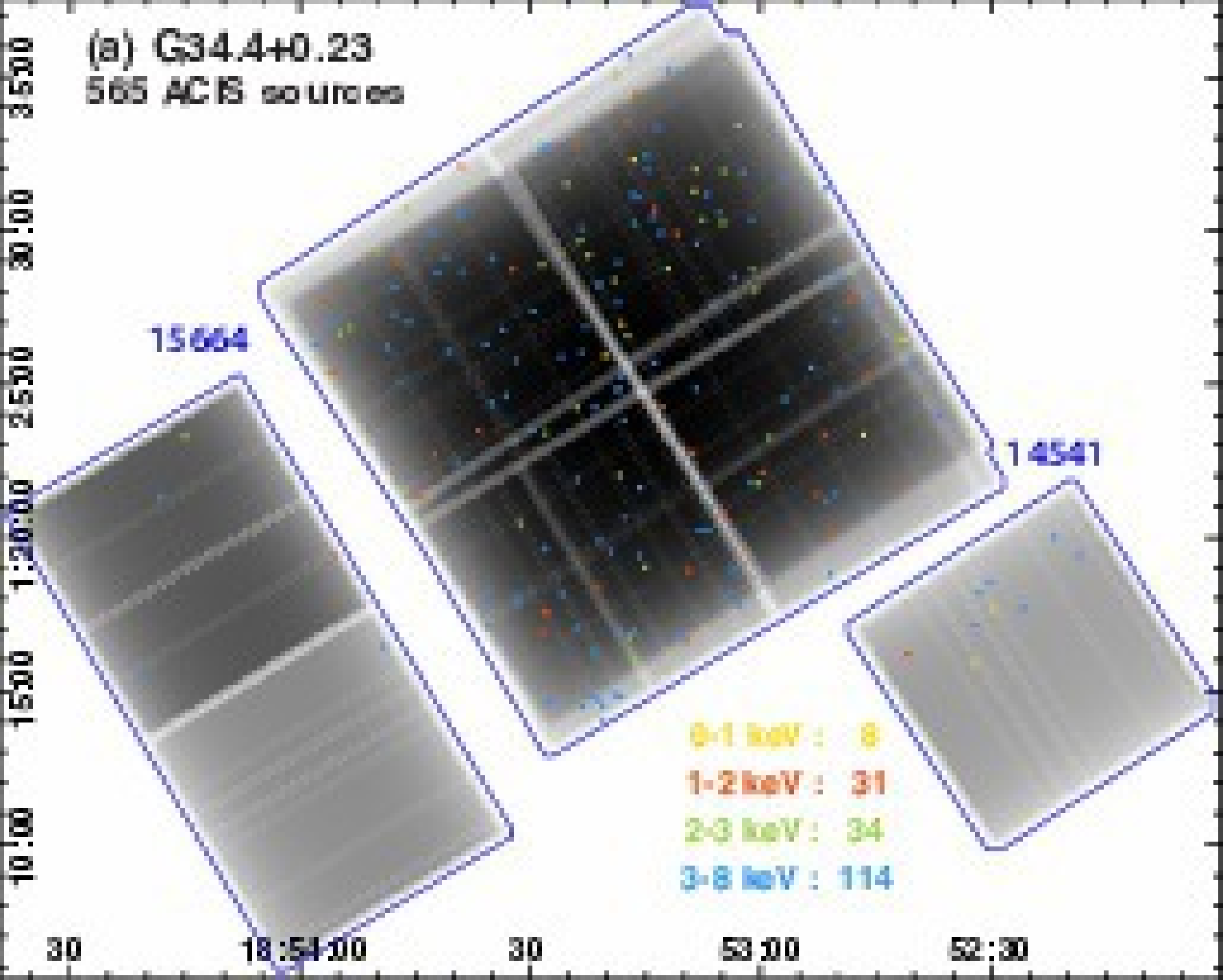}
\includegraphics[width=0.49\textwidth]{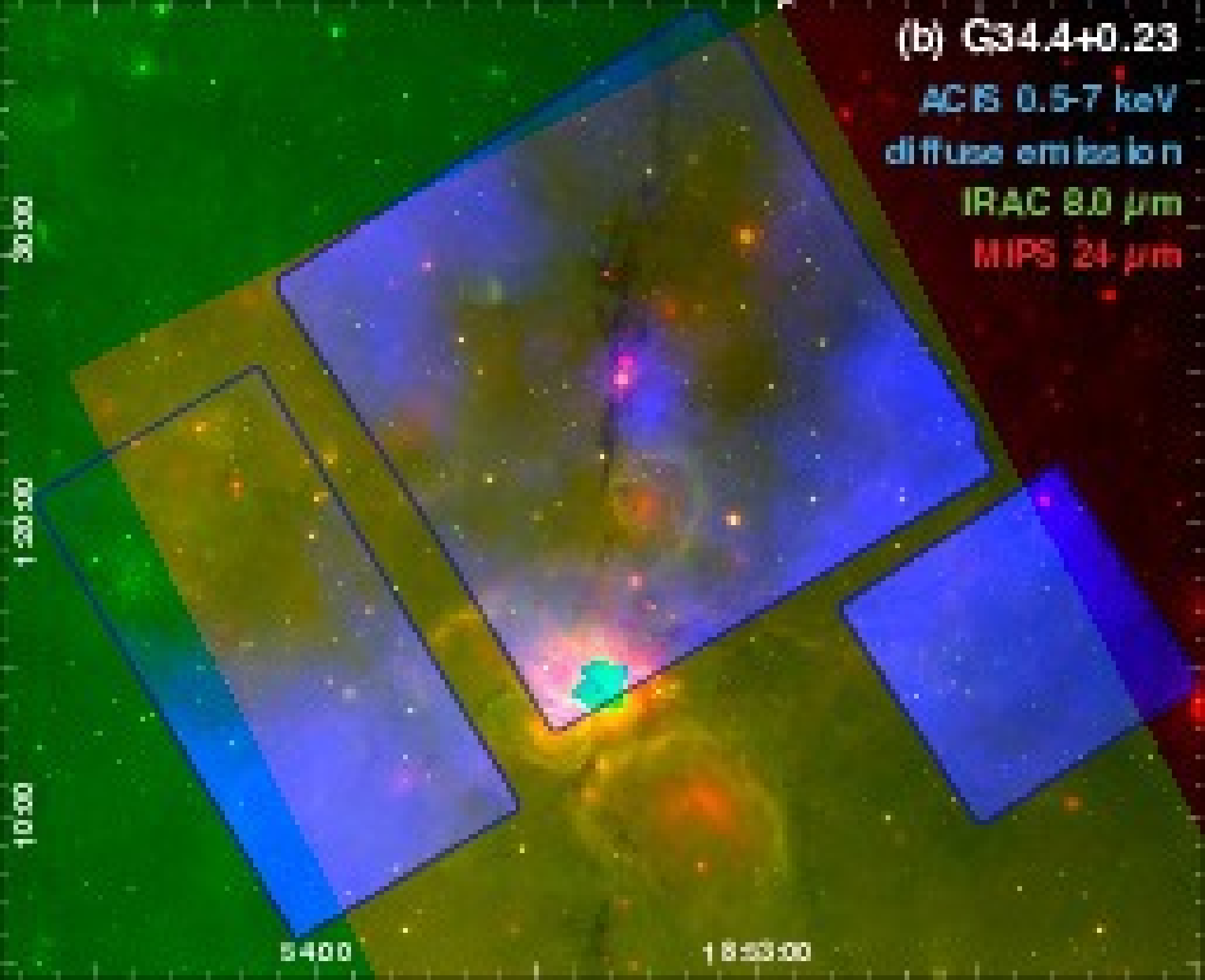}
\includegraphics[width=0.49\textwidth]{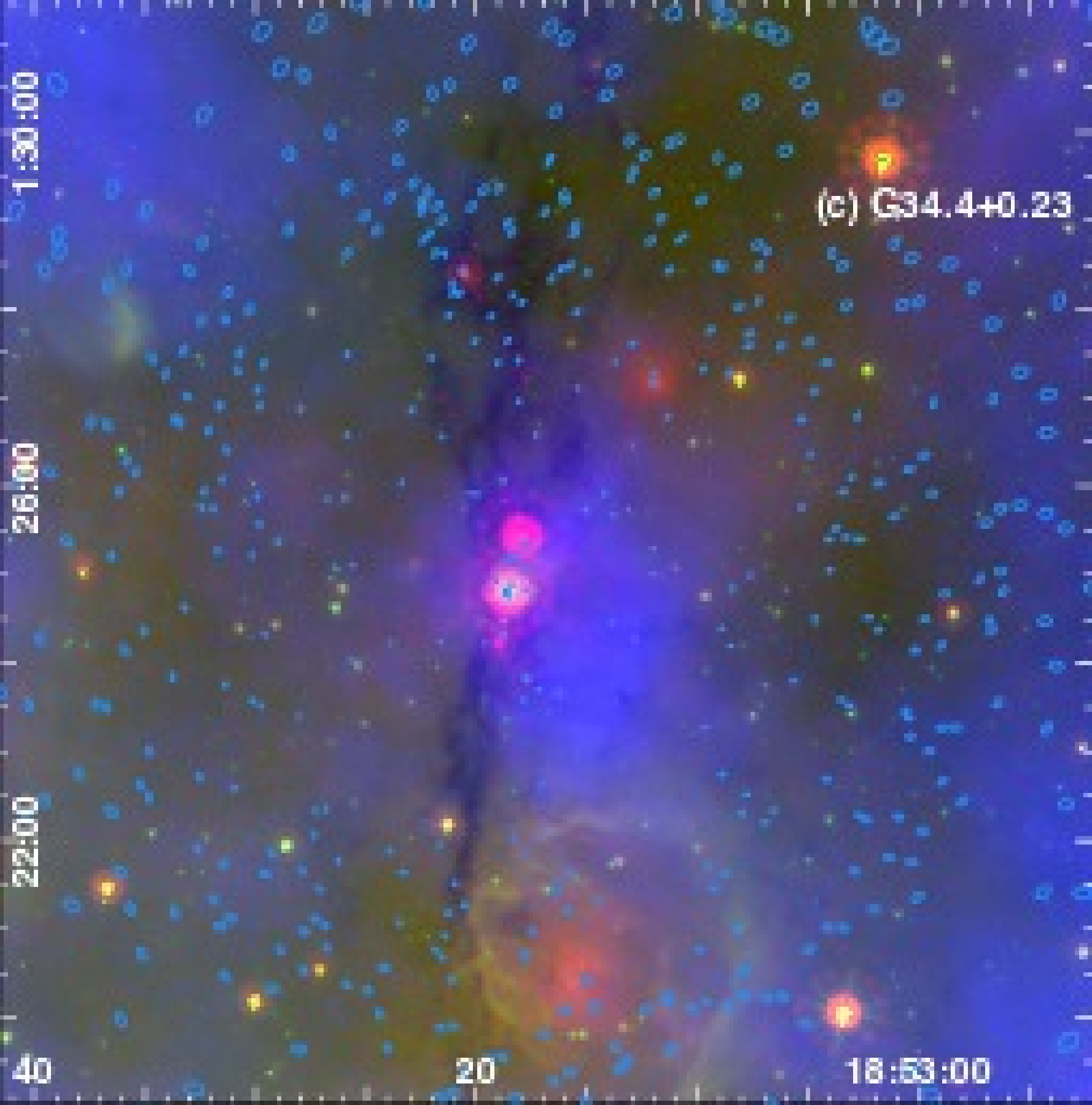}
\includegraphics[width=0.49\textwidth]{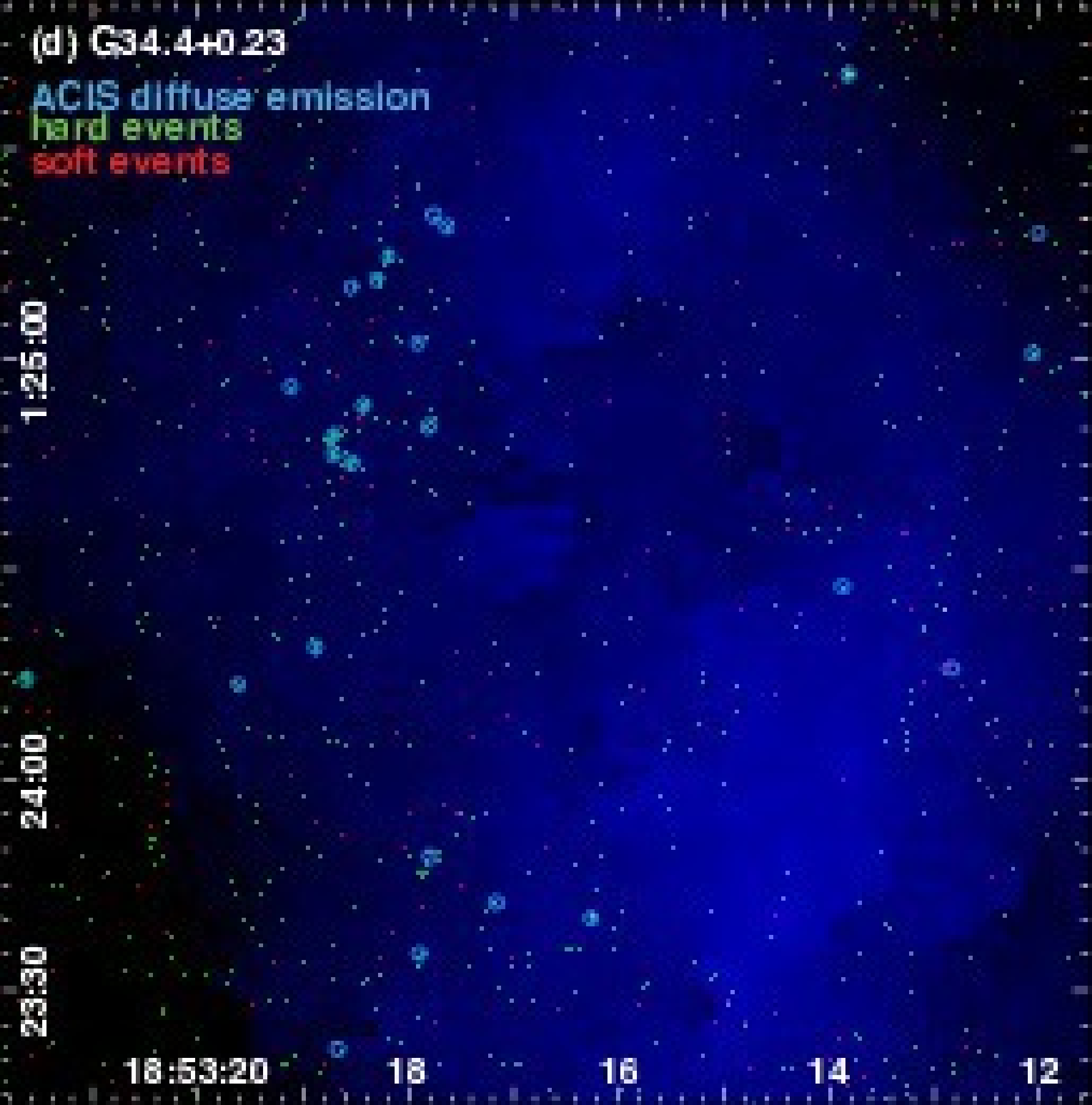}
\caption{G34.4+0.23.
(a) ACIS exposure map with brighter ($\geq$5 net counts) ACIS point sources overlaid; colors denote median energy for each source.  ObsID numbers are shown in blue.
(b) ACIS diffuse emission in the \Spitzer context. 
(c) Zoom of (b) centered on G34.4+0.23, with point source extraction regions.
(d) ACIS event data and diffuse emission at the center of G34.4+0.23.
\label{G34.fig}}
\end{figure}

This \Chandra observation (Figure~\ref{G34.fig}) reveals few X-ray sources along the G34.4+0.23 IRDC compared to G014.225-00.506 \citep{Povich16}, again suggesting that the kinematic distance is correct and that the 63-ks ACIS exposure was too short to detect much of the young stellar population embedded in this IRDC.  Nevertheless, we do find a large number of X-ray point sources in this field (Figure~\ref{G34.fig}(c)).  Some will be foreground and background contaminants, but many hundreds are likely pre-MS stars associated in some way with the star formation going on in this region.  

We have X-ray counterparts to a few of the young stellar objects listed in \citet{Shepherd07}.  In particular, we find three faint, very hard X-ray sources in close proximity to source \#11 from Shepherd et al., at the center of the \UCHIIR (Figure~\ref{G34.fig}(d)):
c284 (CXOU~J185318.71+012449.1, 8 net counts, median energy 4.1 keV);
c283 (CXOU~J185318.71+012446.3, 6 net counts, median energy 4.5 keV);
cc554 (CXOU~J185318.54+012445.2, 4 net counts, median energy 4.2 keV).

\citet{Foster14} found a distributed population of low-mass protostars associated with the IRDC but located outside the densest parts of the filament.  Perhaps our X-ray sources sample an even older and more distributed stellar population.  Further multiwavelength analysis is necessary to understand this X-ray source population and the history of star formation across this region.

The G34.4+0.23 \Chandra mosaic shows abundant diffuse X-ray emission.  A small patch at field center may contain unresolved point sources from the central proto-cluster (Figure~\ref{G34.fig}(d)).  This is largely surrounded by dark regions, perhaps due to shadowing by the IRDC and surrounding molecular material.  Diffuse emission is seen again at the field edges, especially at the east and west corners of the ACIS-I image.  As often demonstrated in other MOXC2 targets, the off-axis ACIS-S CCDs hint at more diffuse X-ray structures far off-axis.  This emission is anticoincident with moderately bright \Spitzer 8~$\mu$m emission to the southeast of G34.4+0.23.

The more developed MSFR G34.26+0.15 was partially imaged off-axis in this observation, at the southern corner of the ACIS-I array (Figure~\ref{G34.fig}(b)).  A few X-ray point sources are found there (likely limited by low sensitivity and source confusion) and the region is surrounded by diffuse X-ray emission.  A \Chandra observation centered on this MSFR would probably reveal a rich young stellar population and an interesting network of hot plasma from massive star feedback in G34.26+0.15.

\clearpage
%-----------------------------------------------------------------------------
\subsection{Westerlund~1 \label{sec:wd1}}
% Westerlund 1 -- 1721 point sources.  
% At 4.0 kpc, 4*pi*D^2 = 1.915e45.
% Clark08 find 241 srcs on the 2 Wd1 ObsIDs, using only CCDs 0,1,6,7 apparently.
% When data become available, match to Andersen17 Table 1 -- new HST study of central 4'x4'.

We finish our tour of MOXC2 MSFRs with Westerlund~1 (Wd1), the most massive YMC known in the Galaxy \citep{Andersen17}.  It is a monolithic cluster, not part of a complex of multiple MSFRs, as we have seen in many MOXC2 targets.  The center of Wd1 is known to be elongated \citep{Gennaro11}, but recent work \citep{Gennaro17} finds that its massive stars are not strongly concentrated towards the center (mass-segregated); these authors conclude that Wd1 has changed little in size or density since its formation.  Its combination of mass and age explains its uniquely large population of evolved massive stars, including red supergiants, yellow hypergiants, and WRs; Wd1 may have weathered $\sim$100 supernovae already \citep{Clark08}.  Twelve WRs \citep{Skinner06} and a magnetar \citep{Muno06a} were among the 241 point sources detected in the original 57-ks ACIS-S observation \citep{Clark08}.  

\begin{figure}[htb]
\centering
\includegraphics[width=0.48\textwidth]{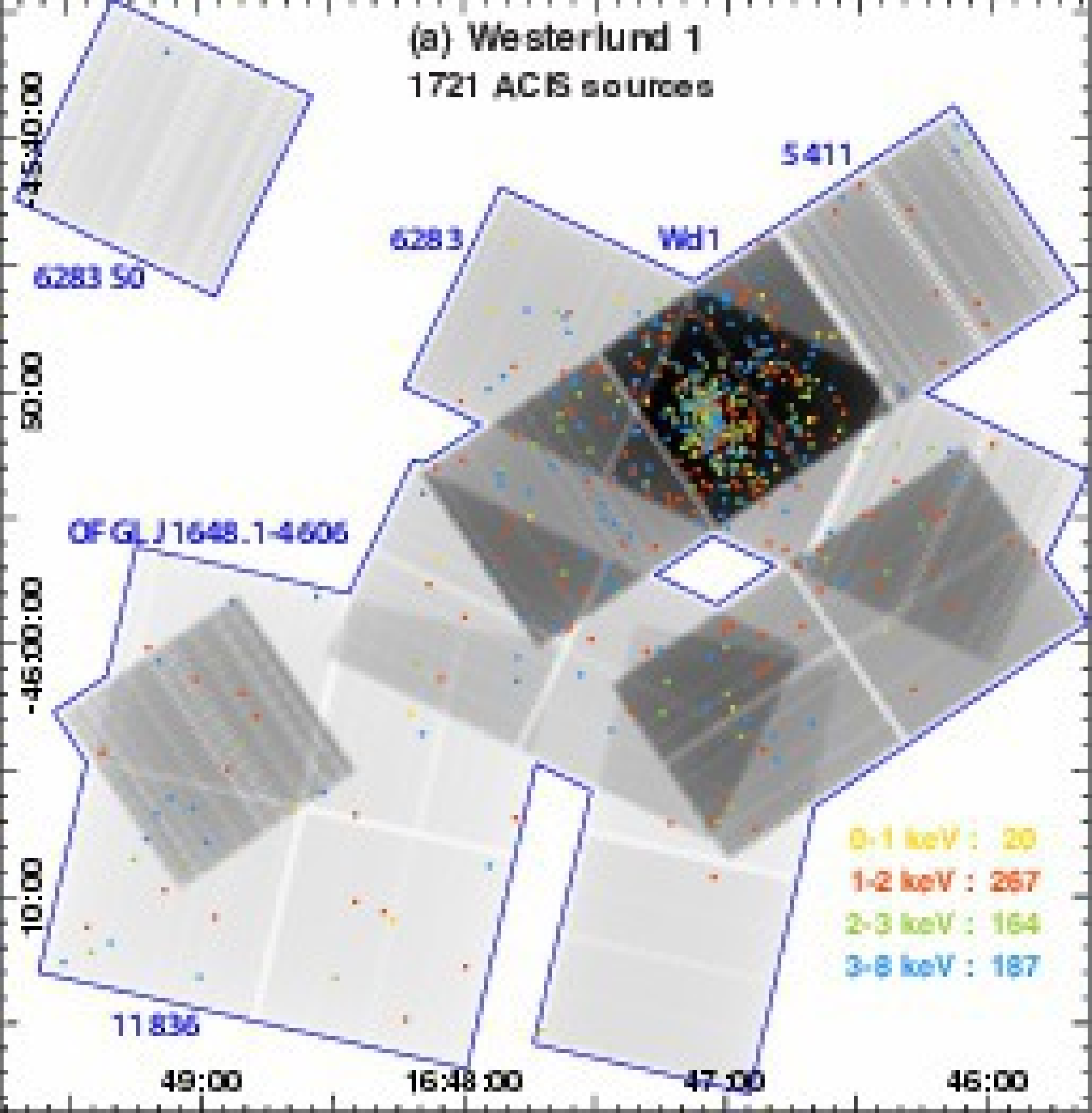}
\includegraphics[width=0.48\textwidth]{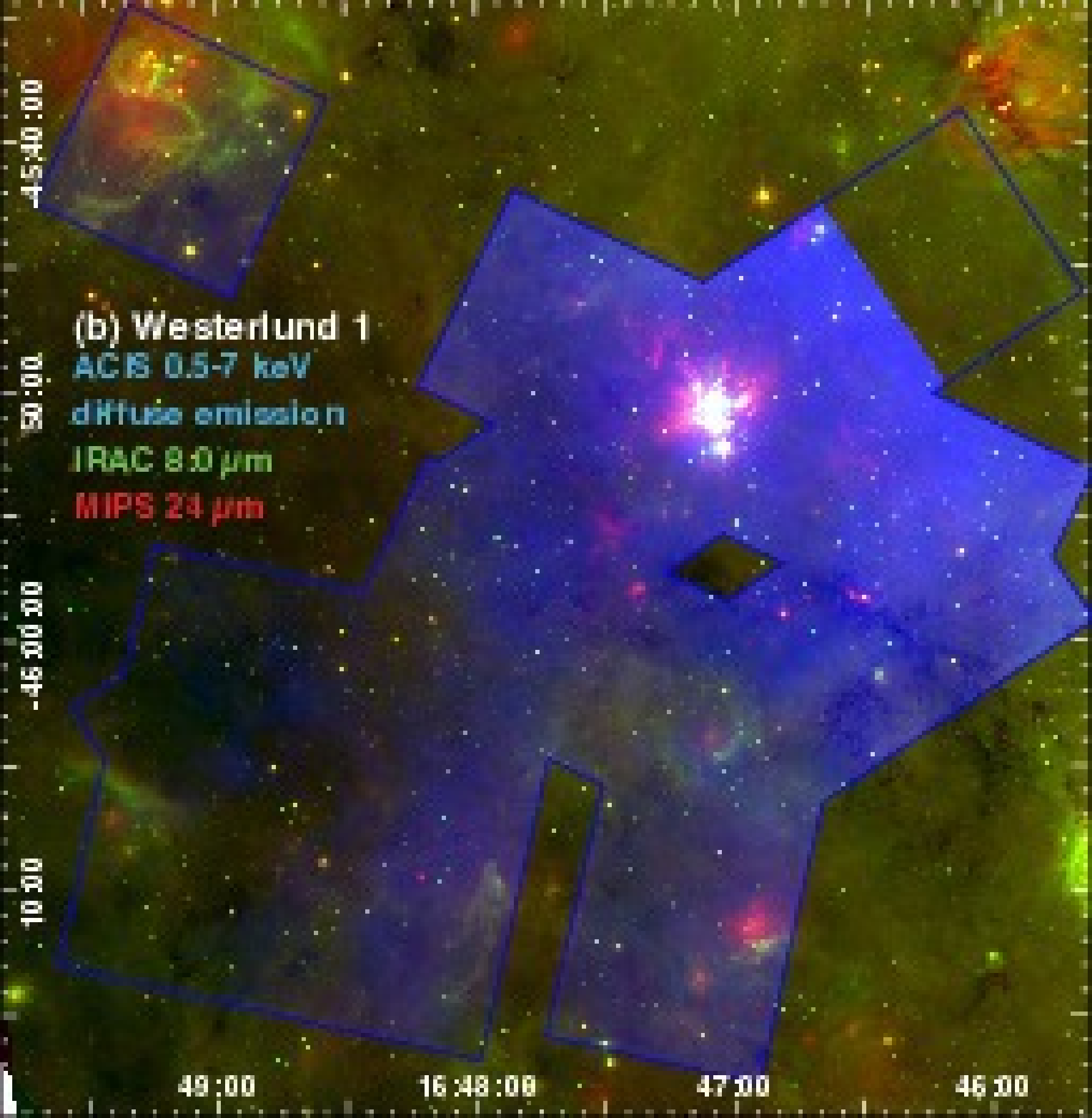}
\includegraphics[width=0.48\textwidth]{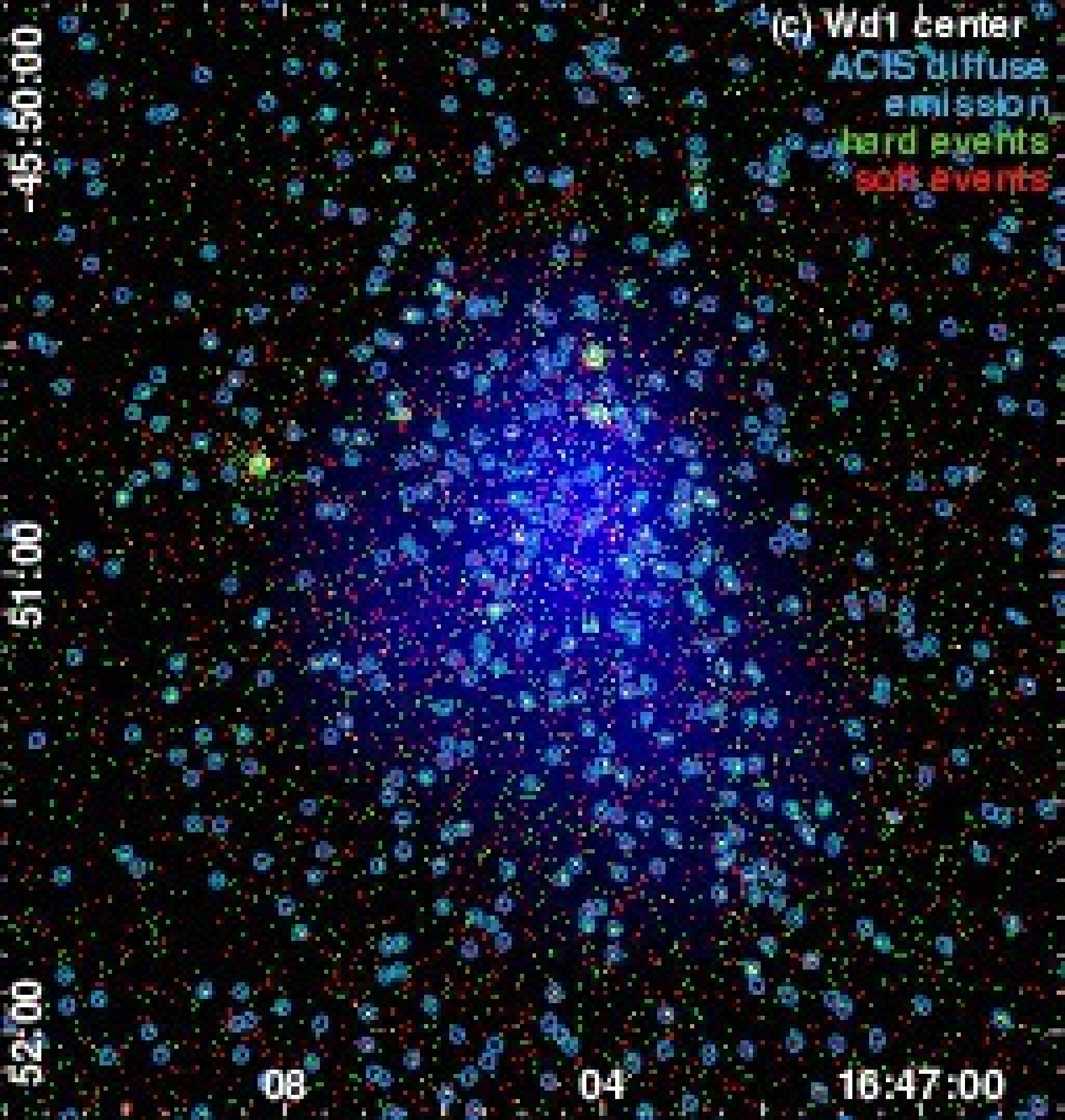}
\includegraphics[width=0.48\textwidth]{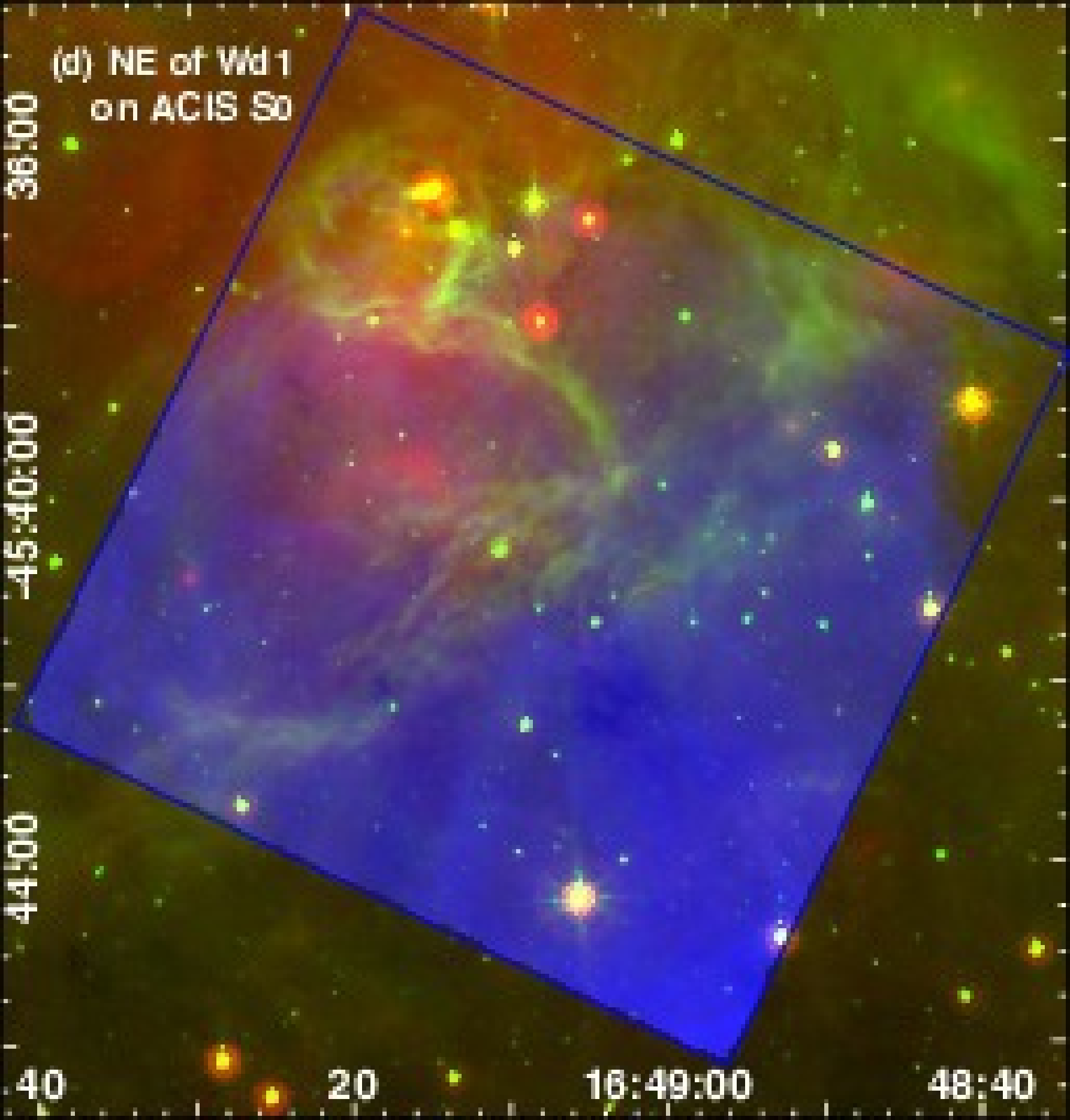}
\caption{Wd1.
(a) ACIS exposure map with brighter ($\geq$5 net counts) ACIS point sources overlaid; colors denote median energy for each source.  ObsID numbers and regions named in the text are shown in blue.
(b) ACIS diffuse emission in the \Spitzer context. 
(c) ACIS event data and diffuse emission on the Wd1 cluster core.
(d) Zoomed version of (b) for the far off-axis S0 chip in the 19-ks ObsID~6283.
\label{Wd1.fig}}
\end{figure}

We recover all but six of the X-ray sources tabulated in \citet{Clark08}.  We find many hundreds more X-ray sources and diffuse emission across our ACIS mosaic (Figure~\ref{Wd1.fig}). The magnetar (CXOU~J164710.2-455216) suffers from photon pile-up in these observations.  \citet{Clark08} presented X-ray spectral fits for the brightest ACIS sources, so we will not report further spectral fitting results here.

Wd1 is four times as massive as NGC~3603 (featured in MOXC1) but eight times lower in central density \citep{Figer08}.  This and the relatively short \Chandra exposure on Wd1 means that our X-ray point source list suffers much less confusion than we found in NGC~3603 (MOXC1).  After excising resolved point sources, the ACIS event data still show a substantial overdensity of counts at the center of Wd1 (Figure~\ref{Wd1.fig}(c)); a longer \Chandra observation would undoubtedly resolve out a large number of additional X-ray point sources in the cluster core.

%Kavanagh11 note "the bright soft source to the southwest (seen in red) is the foreground star HD~151018 (16 46 56.1174 -45 53 14.312), an O9Iab star."  Clark08 present X-ray spectral fit parameters in their Table 2.

The second pointing, to the south of Wd1 in our ACIS mosaic (ObsID~11836, 10~ks), is an ACIS-I GTO observation of 0FGL~J1648.1-4606, part of a snapshot survey of unidentified Fermi/LAT GeV sources.  PSR~J1648-4611 in this field is listed as a non-detection in Table 2 of \citet{Kargaltsev12}.  This source remains undetected in our analysis.

Diffuse X-ray emission from Wd1 was studied by \citet{Muno06b} ({\em Chandra}) and \citet{Kavanagh11} ({\em XMM}).  In our ACIS mosaic (Figure~\ref{Wd1.fig}(b)), we see diffuse emission across much of the field, including the cluster center (Figure~\ref{Wd1.fig}(c)).  The unresolved X-ray emission at the northwest corner of the mosaic is due in large part to the bright LMXB GX340+0, located off the field.  We have masked the brightest of this scattered light from our diffuse images.  Northeast of Wd1, far off-axis on the S0 chip of ObsID~6283, we have another serendipitous detection of diffuse X-ray emission apparently associated with a bubble structure in the \Spitzer data (Figure~\ref{Wd1.fig}(d)).

\clearpage
%=============================================================================
\section{SUMMARY \label{sec:summary}}

Expanding on the results presented in MOXC1 \citep{Townsley14}, with MOXC2 we have assembled and analyzed 13 {\em Chandra}/ACIS Galactic Plane mosaics featuring 16 MSFRs.  These mosaics combine \Chandra archival, GO, and GTO datasets to form the most complete high-resolution X-ray picture possible of the MSFRs and their surrounding environments.  Major software and procedural changes for MOXC2 have made it possible to include far off-axis data from ACIS-I observations and archival data using the ACIS-S imaging configuration.  This broadens the fields covered by our mosaics and the MSFRs we can access; most importantly, these changes reveal the ubiquity of hot plasmas pervading the Plane and {\em Chandra}'s amazing ability to trace their diffuse X-ray emission.

We present properties for more than 18,000 X-ray point sources found in these fields.  From past experience, most of these sources will turn out to be pre-MS MSFR members; \Chandra will increase the census of young stars in every MOXC2 MSFR.  We find likely CWBs, IMPS, and the ionizing sources of well-known radio UCH{\scriptsize II}Rs; these sources are rare and stand out because of their extremes in X-ray luminosity, variability, and hard spectra.  They demonstrate that X-ray emission turns on early in the star formation process, at least in some cases, bombarding cold molecular environments with hard radiation and shocks.

We excised the X-ray point sources from each mosaic and made adaptively-smoothed images from the remaining X-ray emission.  We find that this unresolved emission appears to be truly diffuse in many cases, tracing hot plasma in and around these MSFRs.  This hot plasma appears to pervade cluster centers, fill \Spitzer bubbles, and leak out of perforated \hii region boundaries.  Although MOXC2 MSFRs display a wide variety of diffuse X-ray morphologies and surface brightnesses, none lacks diffuse X-ray emission.  With detailed spectral studies, pressures and densities of these hot plasmas will give physical insights into the birth of the hot ISM.

We hope that our X-ray point source catalogs will be useful to the wider star formation community for understanding young stellar populations in and around MSFRs.  The next step is to match these X-ray source lists to longer-wavelength datasets in order to deduce which X-ray sources are cluster members and, in turn, to study multiwavelength properties of those cluster members.  For diffuse X-ray emission, bright fields need to be tessellated, spectra from each tessellate extracted and fitted, and maps generated of spectral fit parameters \citep{Townsley11b}.  Even for faint fields, a rough spectral characterization should be possible for the diffuse emission pervading every clump and cluster of young stars in MOXC2.

The \Chandra Cycle~18 archival program MSFRs Across the Galaxy in X-rays (MAGiX) is underway; it will provide ACIS source lists and map diffuse X-ray emission for another 16 Galactic MSFRs.  MAGiX2 is a new \Chandra Cycle~19 archival program that will perform similar ACIS data analysis on a further 13 Galactic MSFRs plus three more in the Magellanic Clouds.  With these and other ongoing \Chandra programs, by the end of the decade we will add $\sim$40 MSFRs analyzed with similar methods to the body of work amassed already by COUP, CCCP, MYStIX, MOXC1, SFiNCs, MOXC2, and other \Chandra projects.  We hope that these efforts demonstrate the benefit of \Chandra observations to MSFR studies and facilitate the use of \Chandra data by the multiwavelength star formation community.

Finally, MOXC2 shows that there are still many important observations for \Chandra to make on MSFRs.  Coverage is shallow or absent for a surprising number of famous MSFRs that are mainstays of longer-wavelength star formation science.  We need to explore younger regions (IRDCs) and older regions (red supergiant clusters), rich MSFR complexes near (G352) and far (W31), molecular filaments anchoring the spiral arms (Nessie) and cauldrons of massive star evolution (Wd1).  Much discovery awaits us.

% =============================================================================
\acknowledgements Acknowledgments:  
We appreciate the time and effort donated by our anonymous referee to comment on this paper.  
This work was supported by {\em Chandra X-ray Observatory} general observer grants 
GO3-14002X,   % AO14:  G333 (co-I's Broos, Povich, Feigelson, Naylor) (AO10 G333.6-0.2 co-I's were Gemma Anderson, Broos, Feigelson)
GO5-16015X,   % Generations, AO16:  W33, Cl 1813-178, W42, RSGC1 (co-I's Broos, Povich)
and GO6-17132X   % AO17:  GM24 (co-I's Broos, Povich)
(PI:  L.\ Townsley),
and GO9-0155X (PI: Bryan Gaensler), % ChIcAGO
and by the Penn State ACIS Instrument Team Contract SV4-74018.  All of these were issued by the \Chandra X-ray Center, which is operated by the Smithsonian Astrophysical Observatory for and on behalf of NASA under contract NAS8-03060.
The ACIS Guaranteed Time Observations included here were selected by the ACIS Instrument Principal Investigator, Gordon P.\ Garmire, of the Huntingdon Institute for X-ray Astronomy, LLC, which is under contract to the Smithsonian Astrophysical Observatory; Contract SV2-82024.
M.S.P.\ is supported by the National Science Foundation through grant CAREER-1454333.
T.N.\ was supported for the final part of this work through a Leverhulme Trust Research Project Grant.
This research used data products from the \Chandra Data Archive, software provided by the \Chandra X-ray Center in the application package {\it CIAO}, and {\it SAOImage DS9} software developed by the Smithsonian Astrophysical Observatory.  This research also used data products from the {\em Spitzer Space Telescope}, operated by the Jet Propulsion Laboratory (California Institute of Technology) (JPL/CalTech) under a contract with NASA, and data products from the {\em Wide-field Infrared Survey Explorer} ({\em WISE}), which is a joint project of the University of California, Los Angeles and JPL/CalTech, funded by NASA.  This research used NASA's Astrophysics Data System Bibliographic Services, and the SIMBAD database and VizieR catalog access tool provided by CDS, Strasbourg, France.

% See http://journals.aas.org/authors/aastex/facility.html
\vspace{5mm}
\facilities{CXO (ACIS), Spitzer (IRAC, MIPS), WISE}.

\software{
	{\em ACIS Extract} \citep{Broos10,AE12,AE16},
	\CIAO\ \citep{Fruscione06},
	\DSnine\ \citep{Joye03},
	\MARX\ \citep{Davis12}.
}

% =============================================================================
% APPENDICES
%\appendix
% =============================================================================

% =============================================================================
% Bibliography
% =============================================================================
% BIBLIOGRAPHY
% =============================================================================

\end{document}